\documentclass[showpacs,preprintnumbers,amsmath,amssymb]{revtex4}
\usepackage{amssymb}
\usepackage{amsfonts}

\usepackage{graphicx}
\usepackage{dcolumn}
\usepackage{bm}

\begin{document}

\title{Quantum gravity and mass of gauge field: a four-dimensional unified quantum theory}

\author{Chang-Yu Zhu$^1$ }

\affiliation{%
$^1$Department of Physics, Zhengzhou University, Zhengzhou, Henan
450052, China
}%

\author{Heng Fan$^{2}$ }
\affiliation{%
$^2$Institute of Physics, Chinese Academy of Sciences, Beijing
100190, China
}%
\date{\today}

\begin{abstract}
We present in detail a four-dimensional unified quantum theory. In
this theory, we identify three class of parameters,
coordinate-momentum, spin and gauge, as all and as the only
fundamental parameters to describe quantum fields. The
coordinate-momentum is formulated by the general relativity in
four-dimensional space-time. This theory satisfies the general
covariance condition and the general covariance derivative operator
is given. In a unified and combined description, the matter fields,
gravity field and gauge fields satisfy Dirac equation, Einstein
equation and Yang-Mills equation in operator form. In the framework
of our theory, we mainly realize the following aims: (1) The gravity
field is described by a quantum theory, the graviton is massless, it
is spin-2; (2) The mass problem of gauge theory is solved. Mass
arises naturally from the gauge space and thus Higgs mechanism is
not necessary; (3) Color confinement of quarks is explained; (4)
Parity violation for weak interactions is obtained; (5) Gravity will
cause CPT violation; (6) A dark energy solution of quantum theory is
presented. It corresponds to Einstein's cosmological constant. We
propose that the candidate for dark energy should be gluon which is
one of the elementary particles.
\end{abstract}

\pacs{00., 12.10.-g, 03.65.-w, 04.60.-m, 11.15.-q, 95.36.+x}
\maketitle


\tableofcontents

\section{Introduction}
Quantum mechanics and the relativity theory
\cite{einstein1,einstein2} are well-known great discoveries in the
last century. Both of them are so successful that they play the
central roles in modern science and even in our life. In quantum
mechanics, in order to explain a finer structure in the spectrum of
hydrogen atom due to the spin of the electron, Dirac \cite{dirac}
introduced his equation, the Dirac equation,  by adding spin
corrections to the Schr\"odinger equation \cite{schrodinger}. Dirac
equation is a quantum theory while satisfying invariance under
special relativity. It describes a single-particle obeying both
relativity and quantum mechanics and thus unifies the theory of
special relativity and quantum mechanics.

Dirac equation is extended as a quantum field theory which provides
an unified description for both fields and particles
\cite{QFT1,QFT2,QFT3,QFT4,QFT5}. Later, Yang and Mills \cite{YM}
introduced gauge field theories. Then the Standard Model and the
related theories
\cite{standard1,standard2,standard3,standard4,standard5,nambu,kmmatrix,thooft}
are proposed for elementary particles and the interactions between
them, see Ref.\cite{Nobel} for complete references. Up to then, the
electromagnetic force, weak and strong nuclear forces are merged
into a unified model. While only gravity which is one of the four
basic forces in nature remains outside of this unified framework.

In the past decades, much effort has been put into studying how to
combine quantum mechanics with general relativity into a quantum
theory of gravity. This is because that on the one hand, we would
like to find a unified final theory \cite{weinberg}, and on the
other hand, this theory can describe systems where both quantum
mechanics and general relativity are important, for example in black
hole or the early stage of universe.


Our start point is the general relativity, particles and fields are
set of ``events'' which is a concept coined by Einstein to emphasize
space-time. We propose a quantum theory of general relativity, an
``event'' is described by not only a complete set of canonical
coordinates and canonical momentum of 4D space in operator form
$\hat{x}^{\alpha }, \hat{p}_{\alpha }$, but also by spin operators
$\hat {s}_{\mu \nu }$ and gauge charge operators $\hat {T}_{a}$.
These operators are covariance and constitute a Lie algebra. In our
work, events are quantum states of elementary particles. We then
define a covariance derivative operator in the sense of event with
vierbein and connections of general relativity. Thus Einstein
equation of general relativity, Dirac equation and Yang-Mill
equation are all reformulated by the covariant operators. This
provides a unified framework of quantum mechanics and general
relativity.

We next clarify four questions concerning about our theory. Firstly,
what is the structure of the theory? Secondly, what is new in the
theory? Thirdly, why is the theory correct? Fourthly, any results
from this theory and is this theory necessary?

Our theory can be simply summarized as (1) representations, (2)
equations and (3) observable quantities.  Consequently the structure
of our theory can be described similarly as:  (1)Representations: In
the language of quantum field theory, ``particles'' are dealt like
``fields''. Fields are represented by quantum states with three
class of independent parameters, coordinate-momentum, spin and
gauge. The forces are represented as operators. (2) Equations: All
fields, gravity field, matter fields and gauge fields, which already
have the representations, satisfy three fundamental equations,
Einstein equation, Dirac equation and Yang-Mills equation, all in
operator form which is the standard of quantum theory. (3)
Observable quantities: With equations and representations, we may
construct quantities which are related with observable interactions
and observable physical quantities such as current densities,
particle production rate density and scattering of particles.

As is well-known, it is obvious not easy to find a unified quantum
theory. Then what is new in our theory?

In the framework of general relativity, there is no absolute space
and time. The space and time are dealt symmetrically and constitute
together to be a four-dimensional space-time. In comparison, in
conventional theory, time is generally special which is different
from the three-dimensional space. For example, the evolution of
particles is depicted by a world-line in three-dimension space as
time passes. In order to propose a unified theory of general
relativity and quantum mechanics, we should start directly from the
four-dimensional space-time. In our theory, coordinate-momentum
space is the four-dimensional space of the general relativity. In
this sense, we mean that our theory is a four-dimensional theory.

A quantum theory compatible with the general relativity should
satisfy the general covariance condition. Our theory is a general
covariant theory where three fundamental equations, quantum states,
operators, current densities and all other quantities satisfy the
general covariance condition. So instead of gauge invariance
condition for conventional quantum field theory, one new feature of
our theory is that it satisfies the general gauge covariance
condition. Due to the general covariance condition, mass can not be
defined through momentum like $\hat {p}_{\mu }\hat {p}^{\mu }=m^2$.
Mass-matrices $\hat {m}, \hat {M}$ are defined in the gauge space
which is another new feature of our theory. Since mass naturally
arises from the gauge space, the Higgs mechanism seems not necessary
in our theory. In particular, for massless particles, the gauge
condition is arbitrary such as for electromagnetic field, but for
massive particles, the gauge will not be arbitrary.

Next, we would like to show the evidences that our theory is
correct.

The most important point that our theory is correct is that all
fundamentals of our theory, more or less, are well accepted while
all results of our theory agree well with the basic physics facts
and are reasonable. The key result of our theory is that with
suitable representations, three fundamental equations, Dirac
equation, Einstein equation and Yang-Mills equation in operator form
are all self-consistently presented. This is not a coincidence. We
all agree that those three equations are the cornerstones of modern
physics. We do not change the spirit of those fundamental equations,
but present them in a unified and combined form. From our theory the
graviton is found to be massless which can be considered as the
benchmark to test a quantum theory with general relativity. The
gauge charge representations of our theory agree with the basic
results of the elementary particles scattering process obtained by
Feynman diagram method. Our results agree with the Standard Model
except that we do not have Higgs mechanism.

Finally, we would like to point out that our theory is powerful.
Some important results can be explained and obtained easily in this
theory such as color confinement, parity violation for weak
interactions, gravity can cause CPT violation, a dark energy of
universe solution. With a completely new platform, we expect that
answers may be found for some other unsolvable problems. So our
theory is not only necessary, but also we hope that it may provide a
new foundation for quantum physics.

\subsection{Brief explanation of the concepts in this work}
In quantum field theory, particles and fields are generally dealt in
an unified description. In principle, we accept this framework.
However, there are also some subtle differences for particles and
fields in this work. We consider that fields are described as the
vectors, while particles are the quantum states acting as basis of
the vectors. So fields are the expansions in terms of the particle
states in the superposition form like $|\Psi \rangle =\int
\widetilde {\Psi }^{st}(p)|e_{st}(p)d^4p$. Particles as the elements
constitute the fields as the whole. The motion equations of fields
are described by the fields equations in operator form, in this work
these equations are three fundamental equations, Dirac equation,
Yang-Mills equation and Einstein equation. The evolution of fields
can be explained as propagation, scattering, creation and
annihilation of particles while all of these processes should rely
on those three fundamental equations.

The quantization of fields generally means finding a complete
set of canonical coordinate-momentums and changing them as the
corresponding operators. In this work, we still keep the spirit of
this concept, additionally, the three fundamental equations are
in operator form as generally expected.



\section{Coordinate and momentum}
\subsection{General covariance and commutation relation}
In the framework of general relativity, the space-time is in 4D
space. We consider a 4D quantum mechanics, the coordinate and
momentum are denoted as $\hat {x}^{\mu }$ and $\hat {p}^{\nu }$,
$(\mu ,\nu =0,1,2,3)$, which as usual satisfy the relations
\begin{eqnarray}
[\hat {x}^{\mu }, \hat {p}_{\nu }]&=&-i\delta _{\nu }^{\mu },
\end{eqnarray}
\begin{eqnarray}
[\hat {x}^{\mu }, \hat {x}^{\nu }]&=& [\hat {p}_{\mu }, \hat
{p}_{\nu }]=0.
\end{eqnarray}
Suppose $F(\hat {x})$ is an analytic function of 4D space-time
coordinate $\hat {x}$, i.e., $F(\hat {x})$ can be expanded in terms
of $\hat {x}$ by Taylor expansion. We can prove that $F(\hat {x})$
satisfies the commutation relation
\begin{eqnarray}
[\hat {p}_{\mu }, F(\hat {x})]=i\frac{\partial }{\partial \hat
{x}^{\mu }}F(\hat {x}).
\end{eqnarray}

The coordinate-momentum algebra is constituted by coordinate $\hat
{x}^{\mu }$, the momentum $\hat {p}_{\mu }$ and the unit operator 1,
\begin{eqnarray}
A_{xp}=\{ \hat {Z}_{xp}: \hat {Z}_{xp}=a_{\mu }\hat {x}^{\mu
}+b^{\mu }\hat {p}_{\mu }+\alpha ; a_{\mu }, b^{\mu }, \alpha \in
R\}.
\end{eqnarray}
This algebra is a 9D Lie algebra.

The coordinate-momentum group $G_{xp}$ is the Lie group
corresponding to Lie algebra $A_{xp}$,
\begin{eqnarray}
G_{xp}=\{ \hat {U}_{xp}: \hat {U}_{xp}=\exp [i(a_{\mu }\hat {x}^{\mu
}+b^{\mu }\hat {p}_{\mu }+\alpha )];a_{\mu }, b^{\mu }, \alpha \in
R\}
\end{eqnarray}
The group $G_{xp}$ is a 9D non-Abelian Lie group manifold. Four
elements of coordinate $\hat {x}$, four elements of momentum $\hat
{p}$ and the unit identity 1 constitute the generators of this
group. The geometry structure of coordinate-momentum group $G_{xp}$
can provide the geometry structures for quantum space-time and
quantum energy-momentum.

We next define the general covariance law in quantum case. In our
quantum theory, $\hat {x}^{\mu }$ and $\hat {p}_{\mu }$ are not
orthogonal basis for coordinate and momentum, respectively. They are
general covariance coordinate and momentum. Coordinate $\hat
{x}^{\mu }$ should satisfy the quantum general coordinate
transformation and inversion transformation which are written as,
\begin{eqnarray}
\hat {x}'^{\mu }&=&\hat {x}'^{\mu }(\hat {x}), \nonumber \\
\hat {x}^{\mu }&=&\hat {x}^{\mu }(\hat {x}').
\end{eqnarray}
Corresponding to general coordinate transformation, the
transformation and its inversion for general momentum are,
\begin{eqnarray}
\hat {p}_{\mu }'&=&\frac {\partial \hat {x}^{\nu }}{\partial \hat
{x}'^{\mu }}\hat {p}_{\nu },
\\
\hat {p}_{\mu }&=&\frac {\partial \hat {x}'^{\nu }}{\partial \hat
{x}^{\mu }}\hat {p}'_{\nu }.
\end{eqnarray}
We can find that for general coordinate transformation, the
commutation relations of coordinate and momentum are invariant,
\begin{eqnarray}
[\hat {x}'^{\mu }, \hat {p}'_{\nu }]&=&-i\delta _{\nu }^{\mu },
\end{eqnarray}
\begin{eqnarray}
[\hat {x}'^{\mu }, \hat {x}'^{\nu }]&=& [\hat {p}'_{\mu }, \hat
{p}'_{\nu }]=0.
\end{eqnarray}
The proof of this result can be found by simple calculations,
\begin{eqnarray}
[\hat {x}'^{\mu }, \hat {p}'_{\nu }]&=&\left[ \hat {x}'^{\mu }(\hat
{x}), \frac {\partial \hat {x}^{\lambda }}{\partial \hat {x}'^{\nu
}}\hat {p}_{\lambda } \right] \nonumber \\
&=&\frac {\partial \hat {x}^{\lambda }}{\partial \hat {x}'^{\nu
}}[ \hat {x}'^{\mu }(\hat {x}), \hat {p}_{\lambda } ] \nonumber \\
&=&-i\frac {\partial \hat {x}^{\lambda }}{\partial \hat {x}'^{\nu }}
\frac {\partial \hat {x}'^{\mu }}{\partial \hat {x}'^{\lambda
}}\nonumber \\
&=& -i\delta _{\nu }^{\mu }.
\end{eqnarray}

\begin{eqnarray}
[\hat {p}'_{\mu }, \hat {p}'_{\nu }]&=&\left[ \frac {\partial \hat
{x}^{\kappa }}{\partial \hat {x}'^{\mu }}\hat {p}_{\kappa } , \frac
{\partial \hat {x}^{\lambda }}{\partial \hat {x}'^{\nu
}}\hat {p}_{\lambda } \right] \nonumber \\
&=&\frac {\partial \hat {x}^{\lambda }}{\partial \hat {x}'^{\nu
}}\left[ \frac {\partial \hat {x}^{\kappa }}{\partial \hat {x}'^{\mu
}}, \hat {p}_{\lambda }\right] \hat {p}_{\kappa }+ \frac {\partial
\hat {x}^{\kappa }}{\partial \hat {x}'^{\mu }}\left[ \hat
{p}_{\kappa }, \frac {\partial \hat {x}^{\lambda }}{\partial \hat
{x}'^{\nu }}\right] \hat {p}_{\lambda }
\nonumber \\
&=& -\frac {\partial \hat {x}^{\lambda }}{\partial \hat {x}'^{\nu
}}\frac {\partial }{\partial \hat {x}^{\lambda }} \left( \frac
{\partial \hat {x}^{\kappa }}{\partial \hat {x}'^{\mu }}\right) \hat
{p}_{\kappa }+ \frac {\partial \hat {x}^{\kappa }}{\partial \hat
{x}'^{\mu }}\frac {\partial }{\partial \hat {x}^{\kappa }} \left(
\frac {\partial \hat {x}^{\lambda }}{\partial \hat {x}'^{\nu
}}\right) \hat
{p}_{\lambda }\nonumber \\
&=& 0.
\end{eqnarray}
When an equation is unchanged under the quantum general coordinate
transformation, we say this equation is quantum general covariance.
In our theory, we demand that all physical equations, operators and
quantities be quantum general covariance. This can be looked as a
generalization of the classical general covariance to quantum case.
The quantum general covariance is a condition to has an unified
description of general relativity and quantum mechanics.

The three types of particles, matter particles, gauge particles and
graviton, all have their coordinate-momentum representations. Next,
we will consider their coordinate-momentum properties.

\subsection{Coordinate-momentum representation of the matter particles }
The coordinate-momentum basis for matter particles can be denoted as
\begin{eqnarray}
|x\rangle =|x^0\rangle \otimes |x^1\rangle \otimes |x^2\rangle
\otimes |x^3\rangle ,
\end{eqnarray}
where $|x^0\rangle , |x^1\rangle , |x^2\rangle , |x^3\rangle $ are
eigenvectors of $\hat {x}^0, \hat {x}^1, \hat {x}^2,\hat {x}^3$, and
the eigenvalues are $x^0, x^1, x^2, x^3\in R$, respectively,
\begin{eqnarray}
\hat {x}^{\mu }|x\rangle =x^{\mu }|x\rangle .
\end{eqnarray}
For momentum operator, we have,
\begin{eqnarray}
\hat {p}_{\mu }|x\rangle =\int _{R^4}i\frac {\partial }{\partial
x^{\mu }}\delta ^4(x-x')|x\rangle d^4x'.
\end{eqnarray}
The adjoint of basis $|x\rangle $ is denoted as $\langle
x|=\overline{|x\rangle }$, and they satisfy the relations,
\begin{eqnarray}
\langle x'|x\rangle =\delta ^4(x-x'), \\
\int _{R^4}|x\rangle \langle x|d^4x=1.
\end{eqnarray}
Similarly, for momentum basis, we have,
\begin{eqnarray}
|p\rangle =|p_0\rangle \otimes |p_1\rangle \otimes |p_2\rangle
\otimes |p_3\rangle ,
\end{eqnarray}
where $|p_{\mu }\rangle $ are eigenvectors of $\hat {p}_{\mu }$, and
the eigenvalues are $p_{\mu }\in R$, $(\mu =0,1,2,3)$, respectively,
\begin{eqnarray}
\hat {p}_{\mu }|p\rangle &=&p_{\mu }|p\rangle ,
\\
\hat {x}^{\mu }|p\rangle &=&\int _{R^4}i\frac {\partial }{\partial
p_{\mu }}\delta ^4(p-p')|p'\rangle d^4p'.
\end{eqnarray}
The adjoint basis $\langle p|=\overline{|p\rangle }$ satisfies the
relations,

\begin{eqnarray}
\langle p'|p\rangle =\delta ^4(p-p'), \\
\int _{R^4}|p\rangle \langle p|d^4p=1.
\end{eqnarray}
The transformation between coordinate and momentum takes the form
\begin{eqnarray}
\langle x|p\rangle =\langle p|x\rangle ^*=(2\pi )^{-2}\exp (-ipx).
\end{eqnarray}

\subsection{Comparison between the quantum mechanics and the unified theory}
The main difference between our unified theory and the quantum
mechanics is that the time parameter is dealt in a different way. In
our theory, space and time are symmetric and should satisfy the
general covariance condition in the framework of general relativity.
While in quantum mechanics, the time is used to describe the motion
of particles, in principle, the time is in a different position from
space parameters. Here for completeness, we briefly list some
results of quantum mechanics.

In quantum mechanics, the 3D coordinate and 3D momentum parameters
are in symmetric positions, and should take the following form
\begin{eqnarray}
[\hat {x}_i,\hat {p}_j]&=&i\delta _{ij}, \\
\hat {x}_i|\vec {x}\rangle &=&x_i|\vec {x}\rangle ,\\
\hat {p}_i|\vec {p}\rangle &=&p_i|\vec {p}\rangle ,\\
\langle \vec {x}|\vec {x'}\rangle &=&\delta ^3(\vec {x}-\vec {x'}),\\
\langle \vec {p}|\vec {p'}\rangle &=&\delta ^3(\vec {p}-\vec
{p'}),\\
\int |\vec {x}\rangle \langle \vec {x}|d^3\vec {x}&=&1, \\
\int |\vec {p}\rangle \langle \vec {p}|d^3\vec {p}&=&1, \\
\langle \vec {x}|\vec {p}\rangle &=&(2\pi )^{-\frac {3}{2}}\exp
(-i\vec {p}\cdot \vec {x}).
\end{eqnarray}
If we consider the role of time, we need to introduce the moving
pictures depending on time, Heisenberg picture, or instead the
Sch\"odinger picture. The time-translation operator is introduced as
\begin{eqnarray}
u=\exp (-i\hat {H}t),
\end{eqnarray}
where $\hat {H}$ is the Hamiltonian. The time evolution of the
operators such as $\hat {x}_i(t), \hat {p}_i(t)$ and the quantum
states $|\vec {x},t\rangle ,|\vec {p},t\rangle $ are defined as
\begin{eqnarray}
\hat {x}_i(t)&=&\exp (i\hat {H}t)\hat {x}_i\exp (-i\hat {H}t),
\nonumber \\
\hat {p}_i(t)&=&\exp (i\hat {H}t)\hat {p}_i\exp (-i\hat {H}t),
\\
|\vec {x}, t\rangle &=&\exp (-i\hat {H}t)|\vec {x}\rangle ,
\nonumber
\\
|\vec {p}, t\rangle &=&\exp (-i\hat {H}t)|\vec {p}\rangle .
\end{eqnarray}
When the time $t$ are the same, operator $\hat {x}_i(t), \hat
{p}_i(t)$ and the quantum state $|\vec {x},t\rangle ,|\vec
{p},t\rangle $ satisfy the relations,
\begin{eqnarray}
[\hat {x}_i(t),\hat {p}_j(t)]&=&i\delta _{ij}, \\
\hat {x}_i(t)|\vec {x}, t\rangle &=&x_i|\vec {x}, t\rangle ,\\
\hat {p}_i(t)|\vec {p}, t\rangle &=&p_i|\vec {p}, t\rangle ,\\
\langle \vec {x},t|\vec {x'},t\rangle &=&\delta ^3(\vec {x}-\vec {x'}),\\
\langle \vec {p},t|\vec {p'},t\rangle &=&\delta ^3(\vec {p}-\vec
{p'}),\\
\int |\vec {x},t\rangle \langle \vec {x},t|d^3\vec {x}&=&1, \\
\int |\vec {p},t\rangle \langle \vec {p},t|d^3\vec {p}&=&1, \\
\langle \vec {x},t|\vec {p},t\rangle &=&(2\pi )^{-\frac {3}{2}}\exp
(-i\vec {p}\cdot \vec {x}).
\end{eqnarray}

In quantum mechanics, the transformation between eigenvectors with
different time can be represented as the path-integral:
\begin{eqnarray}
\langle \vec {x}',t'|\vec {x},t\rangle &=& \langle  \vec {x}'|\exp
[-i\hat {H}(t-t')]|\vec {x}\rangle \nonumber \\
&=&\int D\vec {x}\exp \left[ i\int ' L(\vec {x}, \dot{\vec
{x}},t)dt\right] ,
\end{eqnarray}
where $L(\vec {x}, \dot{\vec {x}},t)$ is the Lagrangian of the
system; $D\vec {x}$ means to make integral for all possible paths.

Here let us list the differences between our theory and the standard
quantum mechanics,
\begin{enumerate}
\item Time is dealt differently, as we already mentioned.

\item Energy is dealt differently. In our theory, the energy is
dealt as an element of 4D energy-momentum tensor. Thus the energy
and momentum are in symmetric positions. In quantum mechanics,
energy is used for the Hamiltonian which is a function of
coordinate, momentum and time. The Hamiltonian determines the motion
property of the system. Hamiltonian is symmetric with the 3D
momentum.

\item In our theory, we have an unified orthogonal equation
\begin{eqnarray}
\langle x'|x\rangle =\delta ^4(x-x').
\end{eqnarray}
For comparison, in quantum mechanics, the orthogonal equations are
satisfied when time are equal. For different time, the path-integral
are used,
\begin{eqnarray}
\langle \vec {x}',t'|\vec {x},t\rangle =\left\{ \begin{array}{l}
\delta ^3(\vec {x}-\vec {x}'), ~~~~t'=t\\
\int D\vec {x}\exp \left[ i\int 'L(\vec {x}, \dot{\vec
{x}},t)dt\right], ~~~~t'\not =t \end{array}\right.
\end{eqnarray}

\item In our theory, we have the general covariance condition.
\end{enumerate}

\subsection{Coordinate-momentum state of the matter particles}
A coordinate-momentum state of matter particles can be expanded in
terms of coordinate $|x\rangle $ or momentum $|p \rangle $,
\begin{eqnarray}
|\Psi \rangle =\int _{R^4}\Psi (x)|x\rangle d^4x=\int
_{R^4}\widetilde {\Psi }(p)|p\rangle d^4p,
\end{eqnarray}
where $\Psi (x)$ and $\widetilde {\Psi }(p)$ are coordinate state
functions and momentum state functions. They are complex functions
and connected through Fourier transformation,
\begin{eqnarray}
\Psi (x)=\langle x|\Psi \rangle =(2\pi )^{-2}\int _{R^4}\widetilde
{\Psi }(p)\exp (-ipx)d^4p,
\end{eqnarray}
\begin{eqnarray}
\widetilde {\Psi }(p)=\langle p|\Psi \rangle =(2\pi )^{-2}\int
_{R^4}\Psi (x)\exp (ipx)d^4x.
\end{eqnarray}

The actions of coordinate and momentum operators on the state take
the form
\begin{eqnarray}
\hat {x}^{\mu }|\Psi \rangle =\int _{R^4}x^{\mu }\Psi (x)|x\rangle
d^4x= \int _{R^4}-i\frac {\partial }{\partial p_{\mu }}\widetilde
{\Psi }(p)|p\rangle d^4p,
\end{eqnarray}

\begin{eqnarray}
\hat {p}_{\mu }|\Psi \rangle =\int _{R^4}i\frac {\partial }{\partial
x^{\mu }}\Psi (x)|x\rangle d^4x= \int _{R^4}p_{\mu }\widetilde {\Psi
}(p) |p\rangle d^4p.
\end{eqnarray}

The calculations  concerning about adjoint state $\langle \Psi |$ of
state $|\Psi \rangle $ are standard, the results are listed below:
\begin{eqnarray}
\langle \Psi |=\overline{|\Psi \rangle }&=&\int _{R^4}\Psi
^*(x)\langle x|d^4x \nonumber \\
&=&\int _{R^4}\widetilde {\Psi }^*(p)\langle p|d^4p,
\end{eqnarray}

\begin{eqnarray}
\Psi ^*(x)=\langle \Psi |x\rangle &=&(2\pi )^{-2}\int
_{R^4}\widetilde
{\Psi }^*(p)\exp (ipx)d^4p, \\
\widetilde {\Psi }^*(p)=\langle \Psi |p\rangle &=&(2\pi )^{-2}\int
_{R^4}\Psi ^*(x)\exp (-ipx)d^4x.
\end{eqnarray}
The action of the coordinate and momentum operators on the adjoint
state can be represented as
\begin{eqnarray}
\langle \Psi |\hat {x}^{\mu } =\int _{R^4}x^{\mu }\Psi ^*(x)\langle
x|d^4x= \int _{R^4}i\frac {\partial }{\partial p_{\mu }}\widetilde
{\Psi } (p)\langle p|d^4p,
\end{eqnarray}

\begin{eqnarray}
\langle \Psi |\hat {p}_{\mu }=\int _{R^4}-i\frac {\partial
}{\partial x^{\mu }}\Psi ^*(x)\langle x|d^4x= \int _{R^4}p_{\mu
}\widetilde {\Psi }^*(p) \langle p|d^4p.
\end{eqnarray}

The inner product can be defined as
\begin{eqnarray}
\langle \Psi |\Phi \rangle =\langle \Phi |\Psi \rangle ^* =\int
_{R^4}\Psi ^*(x)\Phi (x)d^4x=\int _{R^4}\widetilde {\Psi
}^*(p)\widetilde {\Phi }^(p)d^4p.
\end{eqnarray}

The representation space of coordinate-momentum for matter particles
is a linear space in support of  coordinate basis $|x\rangle $ and
momentum basis $|p\rangle $. So the space of the representation
$V_{xp}(M)$ of coordinate and momentum for matter particles is
\begin{eqnarray}
V_{xp}(M)&=&\left\{ |\Psi \rangle : |\Psi \rangle =\int _{R^4}\Psi
(x)|x\rangle d^4x\right\} \nonumber \\
&=&\left\{ |\Psi \rangle : |\Psi \rangle =\int _{R^4}\widetilde
{\Psi } (p)|p\rangle d^4p\right\}
\end{eqnarray}
Similarly for adjoint representation, we have
\begin{eqnarray}
\overline{V}_{xp}(M)&=&\left\{ \langle \Psi | : \langle \Psi | =\int
_{R^4}\Psi ^*
(x)\langle x| d^4x\right\} \nonumber \\
&=&\left\{ \langle \Psi | : \langle \Psi | =\int _{R^4}\widetilde
{\Psi }^* (p)\langle p|d^4p\right\}
\end{eqnarray}

\subsection{Matrix representation of coordinate-momentum}
The matrix representation of coordinate-momentum operators on their
basis can be written respectively as
\begin{eqnarray}
\hat {x}^{\mu }&=&\int _{R^4}\int _{R^4}x^{\mu }\delta
^4(x-x')|x'\rangle \langle x|d^4x'd^4x \nonumber \\
&=&\int _{R^4}\int _{R^4}-i\frac {\partial }{\partial p_{\mu
}}\delta ^4(p-p')|p'\rangle \langle p|d^4p'd^4p;
\end{eqnarray}
\begin{eqnarray}
\hat {p}_{\mu }&=&\int _{R^4}\int _{R^4}i\frac {\partial }{\partial
x^{\mu }}\delta
^4(x-x')|x'\rangle \langle x|d^4x'd^4x \nonumber \\
&=&\int _{R^4}\int _{R^4}p_{\mu }\delta ^4(p-p')|p'\rangle \langle
p|d^4p'd^4p;
\end{eqnarray}
The adjoint matrices are
\begin{eqnarray}
\overline{\hat {x}^{\mu }}&=&(\hat {x}^{\mu })^{\dagger }=\hat
{x}^{\mu }, \\
\overline{\hat {p}_{\mu }}&=&(\hat {p}_{\mu })^{\dagger }=\hat
{p}_{\mu },
\end{eqnarray}

We now consider the general operator, its matrix representation and
the corresponding representation space. A linear operator on
coordinate-momentum space can be represented as,
\begin{eqnarray}
\hat {A}&=&\int _{R^4}\int _{R^4}A(x',x)|x'\rangle \langle
x|d^4x'd^4x= \int _{R^4}\int _{R^4}\widetilde {A}(p',p)|p'\rangle
\langle p|d^4p'd^4p;
\end{eqnarray}
$A(x',x)$ and $A(p',p)$ are elements of the coordinate-momentum
representation matrices of operator $\hat {A}$,
\begin{eqnarray}
A(x',x)=\langle x'|\hat {A}|x\rangle =(2\pi )^{-4}\int _{R^4}\int
_{R^4}\widetilde {A}(p',p)\exp (ipx-ip'x')d^4p'd^4p,
\end{eqnarray}
\begin{eqnarray}
\widetilde {A}(p',p)=\langle p'|\hat {A}|p\rangle =(2\pi )^{-4}\int
_{R^4}\int _{R^4}A(x',x)\exp (-ipx+ip'x')d^4x'd^4x,
\end{eqnarray}
The space of the representation is the space in support of
$|x'\rangle \langle x|$ and $|p'\rangle \langle p|$,
\begin{eqnarray}
O_{xp}&=&\left\{ \hat {A}:  \hat {A}=\int _{R^4}\int
_{R^4}A(x',x)|x'\rangle \langle x|d^4x'd^4x\right\} \nonumber \\
&=&\left\{ \hat {A}:  \hat {A}=\int _{R^4}\int _{R^4}\widetilde
{A}(p',p)|p'\rangle \langle p|d^4p'd^4p\right\} .
\end{eqnarray}
The space $O_{xp}$ can be represented as the direct product of
$V_{xp}(M)$ and $\overline{V}_{xp}(M)$,
\begin{eqnarray}
O_{xp}=V_{xp}(M)\otimes \overline{V}_{xp}(M)
\end{eqnarray}

\subsection{Representation of coordinate-momentum of the gauge
particles and the graviton}

The gauge particles and the graviton have property of
coordinate-momentum. Since they are related with forces, there are
operators related to them. For operators, the representation basis
is
\begin{eqnarray}
\hat {\varepsilon }(x)=\delta ^4(\hat {x}-x)=\int _{R^4}\delta
^4(x'-x)|x'\rangle \langle x'|d^4x'=|x\rangle \langle x|.
\end{eqnarray}
The coordinate basis $\hat {\varepsilon }(x)$ has the following
properties:
\begin{eqnarray}
&&\overline{\hat {\varepsilon }(x)}=\hat {\varepsilon }(x) \nonumber
\\
&&[\hat {x}^{\mu }, \hat {\varepsilon}(x)]=0, \nonumber
\\
&&[\hat {p}_{\nu }, \hat {\varepsilon}(x)]=\int _{R^4}i\frac
{\partial }{\partial x'^{\nu }}\delta ^4(x'-x)\hat
{\varepsilon}(x)d^4x', \nonumber
\\
&&\hat {\varepsilon }(x)\hat {\varepsilon }(x')=\hat {\varepsilon
}(x')\hat {\varepsilon }(x)=\delta ^4(x-x')\hat {\varepsilon }(x),
\nonumber \\
&&{\rm tr}\hat {\varepsilon }(x)={\rm tr}\left( |x\rangle \langle
x|\right)=1, \nonumber \\
&&\langle \hat {\varepsilon }(x),\hat {\varepsilon }(x')\rangle
={\rm tr}[\overline{\hat {\varepsilon }(x)}\hat {\varepsilon
}(x')]=\delta ^4(x-x'),\nonumber \\
&&\hat {\varepsilon }(x)|x'\rangle =\delta ^4(x-x')|x'\rangle ,
\nonumber \\
&&\langle \Psi |\hat {\varepsilon }(x)|\Phi \rangle =\Psi ^*(x)\Phi
(x).
\end{eqnarray}
The definition of the momentum basis and its properties are listed
below,
\begin{eqnarray}
&&\hat {\varepsilon }(p)=(2\pi )^{-2}\exp (-ip\hat {x}) =(2\pi
)^{-2}\int _{R^4}\exp (-ip\hat {x})|p'\rangle \langle p'|d^4p'
\nonumber \\
&&=(2\pi )^{-2}\int _{R^4}|(p+p')\rangle \langle p'|d^4p'. \nonumber
\\
&&\overline{\hat {\varepsilon }(p)}=\hat {\varepsilon }(-p)
\nonumber
\\
&&[\hat {x}^{\mu }, \hat {\varepsilon}(p)]=0, \nonumber
\\
&&[\hat {p}_{\mu }, \hat {\varepsilon}(p)]=p_{\mu }\hat {\varepsilon
}(p), \nonumber
\\
&&\hat {\varepsilon }(p)\hat {\varepsilon }(p')=\hat {\varepsilon
}(p')\hat {\varepsilon }(p)=(2\pi )^{-2}\hat {\varepsilon }(p'+p),
\nonumber \\
&&{\rm tr}\hat {\varepsilon }(p)=(2\pi )^2\delta ^4(p), \nonumber \\
&&\langle \hat {\varepsilon }(p'),\hat {\varepsilon }(p)\rangle
={\rm tr}[\overline{\hat {\varepsilon }(p')}\hat {\varepsilon
}(p)]=\delta ^4(p-p'),\nonumber \\
&&\hat {\varepsilon }(p)|p'\rangle =(2\pi )^{-2}|p'+p\rangle ,
\nonumber \\
&&\langle \Psi |\hat {\varepsilon }(p)|\Phi \rangle =\widetilde
{\Psi }^*(p)*\widetilde {\Phi }(p),
\end{eqnarray}
where $*$ means convolution,
\begin{eqnarray}
\widetilde {\Psi }^*(p)*\widetilde {\Phi }(p)=(2\pi )^{-2}\int
_{R^4}\widetilde {\Psi }^*(p')\widetilde {\Phi }(p-p')d^4p'.
\end{eqnarray}

The transformations between basis of coordinate and momentum are
\begin{eqnarray}
\langle \hat {\varepsilon }(x),  \hat {\varepsilon }(p)\rangle &=&
\langle \hat {\varepsilon }(p),  \hat {\varepsilon }(x)\rangle
^*=(2\pi )^{-2}\exp (-ip\hat {x}), \nonumber \\
\hat {\varepsilon }(x)&=&(2\pi )^{-2}\int _{R^4}\exp (ipx)\hat
{\varepsilon }(p)d^4p; \nonumber \\
\hat {\varepsilon }(p)&=&(2\pi )^{-2}\int _{R^4}\exp (-ipx)\hat
{\varepsilon }(x)d^4x.
\end{eqnarray}

For gauge particles and the graviton, their coordinate-momentum
operator $\hat {F}$ can be expanded by the basis of $\hat
{\varepsilon}(x)$ and $\hat {\varepsilon}(p)$,
\begin{eqnarray}
\hat {F}=\int _{R^4}F(x)\hat {\varepsilon}(x)d^4x=\int
_{R^4}\widetilde {F}(p)\hat {\varepsilon}(p)d^4p ,
\end{eqnarray}
where $F(x)$ and $\widetilde {F}(x)$ are coordinate and momentum
functions, they are related by the Fourier transformation
\begin{eqnarray}
F(x)&=&(2\pi )^{-2}\int _{R^4}\widetilde {F}(p)\exp (-ixp)d^4p,
\nonumber
\\
\widetilde {F}(p)&=&(2\pi )^{-2}\int _{R^4}F(x)\exp (ixp)d^4x.
\end{eqnarray}
The adjoint operator of $\hat {F}$ is,
\begin{eqnarray}
\overline{\hat {F}}=\int _{R^4}F^*(x)\hat {\varepsilon}(x)d^4x= \int
_{R^4}\widetilde {F}^*(-p)\hat {\varepsilon}(p)d^4p.
\end{eqnarray}

Next let us list some properties of the coordinate-momentum
operators for gauge particles and graviton:

\begin{enumerate}
\item Commutation relations,
\begin{eqnarray}
[\hat {x}^{\mu }, F(\hat {x})]=0,
\end{eqnarray}
\begin{eqnarray}
[\hat {p}_{\mu }, F(\hat {x})]&=&i\frac {\partial }{\partial \hat
{x}^{\mu }}F(\hat {x})=\int _{R^4}i\frac {\partial }{\partial x^{\mu
}}F(x)\hat
{\varepsilon }(x)d^4x\nonumber \\
&=&\int _{R^4}p_{\mu }\widetilde {F}(p)\hat {\varepsilon }(p)d^4p
\end{eqnarray}

\item Product between operators of gauge particles and graviton
\begin{eqnarray}
\hat {F}\hat {G}=\hat {G}\hat {F}=\int _{R^4}F(x)G(x)\hat
{\varepsilon }(x)d^4x=\int _{R^4}\widetilde {F}(p)*\widetilde
{G}(p)\hat {\varepsilon }(p)d^4p,
\end{eqnarray}
where the convolution takes the form
\begin{eqnarray}
\widetilde {F}^*(p)*\widetilde {G}(p)=(2\pi )^{-2}\int
_{R^4}\widetilde {F }^*(p')\widetilde {G}(p-p')d^4p'.
\end{eqnarray}

\item Trace.
\begin{eqnarray}
{\rm tr}\hat {F}=\int _{R^4}F(x)d^4x=(2\pi )^2\widetilde {F}(0)
\end{eqnarray}

\item Inner product.
\begin{eqnarray}
\langle F(\hat {x}),G(\hat {x})\rangle =\int
_{R^4}F^*(x)G(x)d^4x=\int _{R^4}\widetilde {F}^*(p)\widetilde
{G}(p)d^4p,
\end{eqnarray}

\item Action of operators on the state.
\begin{eqnarray}
\hat {F}|\Psi \rangle =\int _{R^4}F(x)\Psi (x)|x\rangle d^4x =\int
_{R^4}\widetilde {F}(p)*\widetilde {\Psi } (p)|p\rangle d^4p,
\end{eqnarray}
where convolution is used. Similarly, we have equation for adjoint
case,
\begin{eqnarray}
\langle \Psi |\hat {F}=\int _{R^4}\Psi ^*(x)F(x)\langle x|d^4x =\int
_{R^4}\widetilde {\Psi }^*(p)*\widetilde {F} (p)\langle p|d^4p,
\end{eqnarray}

\item Representation space for the coordinate-momentum of gauge
particles and graviton,
\begin{eqnarray}
V_{xp}(A)=V_{xp}(G)&=&\left\{ \hat {F}:\hat {F}=\int _{R^4}F(x)\hat
{\varepsilon}(x)d^4x \right\} \nonumber \\
&=&\left\{ \hat {F}:\hat {F}=\int _{R^4}\widetilde {F}(p)\hat
{\varepsilon}(p)d^4p \right\}
\end{eqnarray}

\item Action multiplication of two matter fields is defined as
\begin{eqnarray}
|\Psi \rangle \circ \langle \Phi |=\int _{R^4}\Phi ^*(x)\Psi (x)\hat
{\varepsilon }(x)d^4x=\int _{R^4}\widetilde {\Phi }^*(p)*\widetilde
{\Psi }(p)\hat {\varepsilon }(p)d^4p.
\end{eqnarray}
It defines the action fields. There are several properties for this
action multiplication,
\begin{eqnarray}
\overline{|\Psi \rangle \circ \langle \Phi |}=|\Phi \rangle \circ
\langle \Psi |,
\end{eqnarray}
\begin{eqnarray}
{\rm tr}(|\Psi \rangle \circ \langle \Phi |)=\langle \Phi |\Psi
\rangle ,
\end{eqnarray}
\begin{eqnarray}
F(\hat {x})(|\Psi \rangle \circ \langle \Phi |)&=&\left( F(\hat
{x})|\Psi \rangle \right) \circ \langle \Phi |
\nonumber \\
&=&|\Psi \rangle \circ \left( \langle \Phi |F(\hat {x})\right)
\nonumber \\
&=&(|\Psi \rangle \circ \langle \Phi |)F(\hat {x}),
\end{eqnarray}
\begin{eqnarray}
[\hat {p}_{\mu }, |\Psi \rangle \circ \langle \Phi |]=(\hat {p}_{\mu
}|\Psi \rangle )\circ \langle \Phi |-|\Psi \rangle \circ (\langle
\Phi |\hat {p}_{\mu }).
\end{eqnarray}

\end{enumerate}

The partial derivation operator $\hat {\partial }_{\mu } $ is
defined as,
\begin{eqnarray}
\hat {\partial }_{\mu } =-i\hat {p}_{\mu }.
\end{eqnarray}
It's action on a state and its adjoint can be calculated as,
\begin{eqnarray}
\hat {\partial }_{\mu }|\Psi \rangle &=&\int _{R^4}\frac {\partial
}{\partial x^{\mu }}\Psi (x)|x\rangle d^4x=\int _{R^4}-ip_{\mu
}\widetilde {\Psi }(p)|p\rangle d^4p, \nonumber \\
\langle \Psi |\hat {\partial }_{\mu }&=&-\int _{R^4}\frac {\partial
}{\partial x^{\mu }}\Psi ^*(x)\langle x|d^4x=\int _{R^4}-ip_{\mu
}\widetilde {\Psi }^*(p)\langle p|d^4p.
\end{eqnarray}
The commutation relations for gauge particles and the graviton and
the Leibniz rule are presented as,
\begin{eqnarray}
[\hat {\partial }_{\mu }, F(\hat {x})]&=&\frac {\partial }{\partial
\hat {x}^{\mu }}F(\hat {x})=\int _{R^4}\frac {\partial }{\partial
\hat {x}^{\mu }}F(x)\hat {\varepsilon }(x)d^4x \nonumber \\
&=&\int _{R^4}-ip_{\mu }\widetilde {F}(p)\hat {\varepsilon }(p)d^4p
\end{eqnarray}
\begin{eqnarray}
[\hat {\partial }_{\mu }, \hat {F}\hat {G}]&=&[\hat {\partial }_{\mu
}, \hat {F}]\hat {G}+\hat {F}[\hat {\partial }_{\mu }, \hat {G}]
\\
\hat {\partial }_{\mu }\left( \hat {F}|\Psi \rangle \right) &=&[\hat
{\partial }_{\mu }, \hat {F}]|\Psi \rangle +\hat {F}\left( \hat
{\partial }_{\mu }|\Psi \rangle \right) ,
\\
\left( \langle \Psi |\hat {F}\right)\hat {\partial }_{\mu } &=&
\left( \langle \Psi |\hat {\partial }_{\mu }\right) \hat{F}-\langle
\Psi |[\hat {\partial }_{\mu }, \hat {F}],
\end{eqnarray}
\begin{eqnarray}
[\hat {\partial }_{\mu }, \langle \Phi |\Psi \rangle ]&=&\langle
\Phi |\left( \hat {\partial }_{\mu }|\Psi \rangle \right)-\left(
\langle \Phi |\hat {\partial }_{\mu }\right) |\Psi \rangle =0
\end{eqnarray}

\section{Spin}
\subsection{Spin algebra}
In our theory, the spin is one of the fundamental parameters of
fields and operators. In 4D space-time, we use notations $\hat
{s}_{\alpha \beta }$, $(\alpha ,\beta =0,1,2,3)$ denote the spin
operators which are antisymmetric $\hat {s}_{\alpha \beta }=-\hat
{s}_{\beta \alpha }$ and satisfy the commutation relation
\begin{eqnarray}
[\hat {s}_{\alpha \beta }, \hat {s}_{\rho \sigma }]=-i(\eta _{\alpha
\rho}\hat {s}_{\beta \sigma }-\eta _{\beta \rho}\hat {s}_{\alpha
\sigma }+\eta _{\alpha \sigma}\hat {s}_{\rho \beta }-\eta _{\beta
\sigma}\hat {s}_{\rho \alpha }),
\end{eqnarray}
where $\eta _{\alpha \beta }$ is the Minkowski metric defined as
\begin{eqnarray}
\eta _{00}&=&1, \nonumber \\
\eta _{ij}&=&-\delta _{ij}, \nonumber \\
\eta _{0i}&=&\eta _{i0}=0, \nonumber \\
\eta ^{\alpha \beta }&=&\eta _{\alpha \beta}
\end{eqnarray}

The spin algebra is represented as the direct summation of two Lie
 $su(2)$ algebras. The new-defined 6 operators constitute a basis for
spin algebra $A_s$
\begin{eqnarray}
\hat {M}_i&=&\frac {1}{4}\varepsilon _{ijk}\hat {s}_{jk}+\frac
{i}{2}\hat {s}_{0i}, \nonumber \\
\hat {N}_i&=&\frac {1}{4}\varepsilon _{ijk}\hat {s}_{jk}-\frac
{i}{2}\hat {s}_{0i},
\end{eqnarray}
where $i,j,k=1,2,3$,$\varepsilon _{ijk}$ is the Levi-Civita symbol.
$\hat {M}_i$ and $\hat {N}_i$ satisfy the commutation relations
\begin{eqnarray}
[\hat {M} _i, \hat {M} _j]&=&i\varepsilon _{ijk}\hat {M} _k,
\end{eqnarray}
\begin{eqnarray}
[\hat {N} _i, \hat {N} _j]&=&i\varepsilon _{ijk}\hat {N} _k,
\end{eqnarray}
\begin{eqnarray}
[\hat {M} _i, \hat {N} _j]&=&0.
\end{eqnarray}
One can find that three $\hat {M}_i$ and three $\hat {M}_i$
constitute independently a $su(2)$ algebra, named usually as
left-spin algebra and right-spin algebra, respectively. Spin algebra
can be represented as the direct summation of two algebras $su(2)_L$
and $su(2)_R$.
\begin{eqnarray}
A_S=su(2)_L\oplus su(2)_R
\end{eqnarray}

The spin angular-momentum operator $\hat {s}_{\alpha \beta }$ can be
represented in terms of $\hat {M}_i$ and  $\hat {N}_i$ as
\begin{eqnarray}
\hat {s}_{ij}&=&\varepsilon _{ijk}(\hat {M}_k+\hat {N}_k), \nonumber
\\
\hat {s}_{0i}&=&-i(\hat {M}_i-\hat {N}_i)
\end{eqnarray}

\subsection{Irreducible representations of left and right-spin
algebras}

Left-spin algebra is a $su(2)$ algebra, the irreducible
representation space is $V_L(j_1)$, where $j_1=0, \frac {1}{2}, 1,
\frac {3}{2},...$, the dimension is

\begin{eqnarray}
{\rm dim}V_L(j_1)=2j_1+1.
\end{eqnarray}
The basis of the representation is denoted as $|j_1,m_1\rangle $,
$(m_1=-j_1, -j_1+1, ...,j_1 )$, it is the eigenvector of the
operator $\hat {M}_3$,
\begin{eqnarray}
\hat {M}_3|j_1,m_1\rangle =m_1|j_1,m_1\rangle .
\end{eqnarray}

Similarly for right-spin algebra, the irreducible representation
space is denoted as $V_R(j_2), (j_2=0,\frac {1}{2}, 1, \frac
{3}{2},... )$, and the dimension is ${\rm dim}V_L(j_1)=2j_1+1$. For
representation basis $|j_2,m_2\rangle , (m_2=-j_2, -j_2+1, ...,j_2
)$, we have
\begin{eqnarray}
\hat {N}_3|j_2,m_2\rangle =m_2|j_2,m_2\rangle .
\end{eqnarray}

The irreducible representation of spin algebra $A_S$ can be
represented as the direct product of the irreducible representations
of left-spin algebra and right-spin algebra,
\begin{eqnarray}
V_S(j_1,j_2)=V_L(j_1)\otimes V_R(j_2).
\end{eqnarray}
Apparently, the dimension of the irreducible representation is
\begin{eqnarray}
{\rm dim}V_S(j_1,j_2)=(2j_1+1)(2j_2+1).
\end{eqnarray}
The basis of the representation space of spin algebra can be simply
taken as
\begin{eqnarray}
|j_1,m_1;j_2,m_2\rangle =|j_1,m_1\rangle \otimes |j_2,m_2\rangle .
\end{eqnarray}
It is known that those basis are the eigenvectors for $\hat
{s}_{12}, \hat {s}_{03}$
\begin{eqnarray}
\hat {s}_{12}|j_1,m_1;j_2,m_2\rangle
&=&(m_1+m_2)|j_1,m_1;j_2,m_2\rangle , \nonumber \\
\hat {s}_{03}|j_1,m_1;j_2,m_2\rangle
&=&-i(m_1-m_2)|j_1,m_1;j_2,m_2\rangle ,
\end{eqnarray}

\subsection{Spin of matter particles}
Representation space for spin of matter particles $V_S(M)$ is the
Dirac spinor space, or simply spinor-space. The spinor space can be
represented as the direct summation of two irreducible
representation space:
\begin{eqnarray}
V_S(M)=V_S(\frac {1}{2}, 0)\otimes V_S(0, \frac {1}{2}),
\end{eqnarray}
where $V_S(\frac {1}{2}, 0)$ is the left-spin space, and $V_R(\frac
{1}{2}, 0)$ is the right-spin space. The total dimension of spinor
space is
\begin{eqnarray}
{\rm dim}V_S(M)={\rm dim}V_S(\frac {1}{2},0)+{\rm dim}V_R(0,\frac
{1}{2})=4
\end{eqnarray}
The basis for spin of matter particles is denoted as $|e_s\rangle $
(s=1,2,3,4), the definition for this basis by standard basis
$|j_1,m_1; j_2,m_2\rangle $ is
\begin{eqnarray}
|e_1\rangle &=&|\frac {1}{2},\frac {1}{2};0,0\rangle , \nonumber \\
|e_2\rangle &=&|\frac {1}{2},-\frac {1}{2};0,0\rangle ,\nonumber
\\
|e_3\rangle &=&|0,0;\frac {1}{2},\frac {1}{2}\rangle ,\nonumber
\\
|e_4\rangle &=&|0,0;\frac {1}{2},-\frac {1}{2}\rangle ,
\end{eqnarray}
where $|e_1\rangle ,|e_2\rangle $ are the basis of left-spin space
$V_S(\frac {1}{2}, 0)$, and similarly $|e_3\rangle ,|e_3\rangle $
are the basis of right-spin space $V_R(0,\frac {1}{2})$. The spin
state of the matter particle is represented as
\begin{eqnarray}
V_S(M)=\{ |\Psi \rangle :|\Psi \rangle =\Psi ^s|e_s\rangle \} ,
\end{eqnarray}
where $\Psi ^s=\langle e_s|\Psi \rangle  $, and $\langle e_s|$ is
the adjoint of $|e_s\rangle $.

Here the adjoint of $|e_s\rangle $ is defined as
\begin{eqnarray}
\langle e_s|=\overline{|e_s\rangle }.
\end{eqnarray}
The inner-product for $|e_s\rangle $ and its adjoint is defined as
\begin{eqnarray}
\langle e_s|e_{s'}\rangle =(\gamma ^0)_{ss'}.
\end{eqnarray}
Thus the spinor metric defined by the inner product of basis, which
is the parity matrix, takes the form
\begin{eqnarray}
\hat {g}&=&\hat {\gamma }^0 \nonumber \\
&=&\hat {\sigma }_1\otimes \hat {\sigma }_0=\left(
\begin{array}{cc}0&\sigma _0\\
\sigma _0&0\end{array}\right) ,
\end{eqnarray}
where $\sigma _0$ is the identity in 2D. Also please note the
lowering or rising of the indices can be realized by the metric,
\begin{eqnarray}
\langle e^s|=g^{ss'}\langle e_{s'}|=g^{ss'}\overline{|e_{s'}\rangle
},
\end{eqnarray}
and
\begin{eqnarray}
\langle e^s|e_{s'}\rangle =\delta ^s_{s'},
\\
|e_s\rangle \otimes \langle e^s|=\hat {I}.
\end{eqnarray}
The spin of matter particles is the spinor in spinor space $V_S(M)$
which can be expanded in basis of spinor,
\begin{eqnarray}
|\Psi \rangle =\Psi ^s|e_s\rangle ,
\end{eqnarray}
where $\Psi ^s=\langle e^s|\Psi \rangle $. The adjoint of $|\Psi
\rangle $ is $\rangle \Psi |$ which has the form,
\begin{eqnarray}
\langle \Psi |=\overline{|\Psi \rangle }=\Psi _s^*\langle e^s|,
\end{eqnarray}
where $\Psi _s^*=\langle \Psi |e_s\rangle =\langle e_s|\Psi \rangle
^*=g_{ss'}\Psi ^{*s}$. The inner product of the two spinors $|\Psi
\rangle $ and $|\Phi \rangle $ is defined as,
\begin{eqnarray}
\langle \Psi |\Phi \rangle &=&\langle \Phi |\Psi \rangle ^* =\Psi
_s^*\Phi ^s \nonumber \\
&=&g_{ss'}\Psi ^{*s}\Phi ^{s'}.
\end{eqnarray}

\subsection{Matrix representation of the spin}
Spin operator $\hat {s}_{\alpha \beta}$ with basis $|e_s\rangle $
can be represented by matrix as
\begin{eqnarray}
\hat {s}_{\alpha \beta}&=&(\hat {s}_{\alpha
\beta})_{s'}^s|e_s\rangle \otimes \langle e^{s'}|, \nonumber \\
(\hat {s}_{\alpha \beta})_{s'}^s&=&\langle e^s|\hat {s}_{\alpha
\beta }|e_{s'}\rangle .
\end{eqnarray}

Explicitly, the spin angular-momentum operators are $4\times 4$
matrices with the form:
\begin{eqnarray}
\hat {s}_{ij}=\frac {1}{2}\varepsilon _{ijk}\left( \begin{array}{cc}
\sigma ^k&0\\
0&\sigma ^k\end{array}\right) , \nonumber \\
\hat {s}_{0k}=-\frac {i}{2}\varepsilon _{ijk}\left(
\begin{array}{cc}
\sigma ^k&0\\
0&-\sigma ^k\end{array}\right) ,
\end{eqnarray}
where $\sigma ^k$ are the $2\times 2$ Pauli matrices.

The Hermitian conjugate of the spin operators are
\begin{eqnarray}
\hat {s}_{ij}^{\dagger }&=&\hat {s}_{ij}, \nonumber \\
\hat {s}_{0k}^{\dagger }&=&-\hat {s}_{0k}.
\end{eqnarray}
Also we have
\begin{eqnarray}
[\hat {g}, \hat {s}_{ij}]&=&0,\nonumber \\
\{ \hat {g}, \hat {s}_{0k}\} &=&0.
\end{eqnarray}
The adjoint of these operators are
\begin{eqnarray}
\overline{\hat {s}_{ij}}&=&\hat {g}^{-1}\hat {s}_{ij}^{\dagger
}\hat {g}=\hat {s}_{ij}, \nonumber \\
\overline{\hat {s}_{0k}}&=&\hat {g}^{-1}\hat {s}_{0k}^{\dagger }\hat
{g}=\hat {s}_{0k},
\end{eqnarray}
those equation can be written as a concise form
\begin{eqnarray}
\overline{\hat {s}_{\alpha \beta }}&=&\hat {g}^{-1}\hat {s}_{\alpha
\beta }^{\dagger }\hat {g}=\hat {s}_{\alpha \beta }.
\end{eqnarray}

The chiral charge $\hat {\gamma }^5$ in basis $|e_s\rangle $ is
defined as
\begin{eqnarray}
\hat {\gamma }^5=\left( \begin{array}{cc} \hat {\sigma }_0&0\\
0&\hat {\sigma }_0\end{array}\right) .
\end{eqnarray}
We also have the properties
\begin{eqnarray}
(\hat {\gamma }^5)^{\dagger }&=&\hat {\gamma }^5, \nonumber \\
\overline{\hat {\gamma }^5}&=&\hat {g}^{-1}(\hat {\gamma
}^5)^{\dagger }\hat {g}=-\hat {\gamma }^5,\nonumber \\
(\hat {\gamma }^0)^{\dagger }&=&\hat {\gamma }^0, \nonumber \\
\overline{\hat {\gamma }^0}&=&\hat {g}^{-1}(\hat {\gamma
}^0)^{\dagger }\hat {g}=\hat {\gamma }^0
\end{eqnarray}

The left-chiral projector and the right-chiral projector are defined
as
\begin{eqnarray}
\hat {P}_L&=&\frac {1}{2}(1+\hat {\gamma }^5), \nonumber \\
\hat {P}_R&=&\frac {1}{2}(1-\hat {\gamma }^5),
\end{eqnarray}
These two operators project a 4D space to 2D chiral left or right
spaces, respectively, and they have the properties,
\begin{eqnarray}
\hat {P}_L+\hat {P}_R&=&1,\nonumber \\
\hat {P}_L\hat {P}_L&=&\hat {P}_L,\nonumber \\
\hat {P}_R\hat {P}_R&=&\hat {P}_R,\nonumber \\
\hat {P}_L\hat {P}_R&=&\hat {P}_R\hat {P}_L=0.
\end{eqnarray}

\subsection{Dirac matrices}
There are 16 Dirac matrices with $4\times 4$ acting on 4D space.
Those Dirac matrices are divided into 5 groups. 1 rank-0
antisymmetric metric $\hat {\gamma }$, 4 rank-1 antisymmetric
metrics $\hat {\gamma }^{\alpha _1}$, 6 rank-2 antisymmetric metrics
$\hat {\gamma }^{\alpha _1\alpha _2}$, 4 rank-3 antisymmetric
metrics $\hat {\gamma }^{\alpha _1\alpha _2\alpha _3}$ and 1 rank-4
antisymmetric metric $\hat {\gamma }^{\alpha _1\alpha _2\alpha
_3\alpha _4}$. Those 16 matrices can be represented as the following
by basis $|e_s\rangle $,
\begin{eqnarray}
\hat {\gamma }&=&\sigma _0\otimes \sigma _0=\left(
\begin{array}{cc}\hat {\sigma }_0&0\\
0&\hat {\sigma }_0\end{array}\right) ,\nonumber \\
\hat {\gamma }^0&=&\sigma _1\otimes \sigma _0=\left(
\begin{array}{cc}0&\hat {\sigma }_0\\
\hat {\sigma }_0&0\end{array}\right) ,\nonumber \\
\hat {\gamma }^i&=&i\sigma _2\otimes \sigma _i=\left(
\begin{array}{cc}0&\hat {\sigma }_i\\
-\hat {\sigma }_i&0\end{array}\right) ,\nonumber \\
\hat {\gamma }^{0i}&=&i\sigma _3\otimes \sigma _i=i\left(
\begin{array}{cc}\hat {\sigma }_i&0\\
0&-\hat {\sigma }_i\end{array}\right) ,\nonumber \\
\hat {\gamma }^{ij}&=&\varepsilon _{ijk}\sigma _0\otimes \sigma
_k=\varepsilon _{ijk}\left(
\begin{array}{cc}\hat {\sigma }_k&0\\
0&\hat {\sigma }_k\end{array}\right) ,\nonumber \\
\hat {\gamma }^{0ij}&=&\varepsilon _{ijk}\sigma _1\otimes \sigma
_k=\varepsilon _{ijk}\left(
\begin{array}{cc}0&\hat {\sigma }_k\\
\hat {\sigma }_k&0\end{array}\right) ,\nonumber \\
\hat {\gamma }^{123}&=&-i\sigma _2\otimes \sigma _0=\left(
\begin{array}{cc}0&-\hat {\sigma }_0\\
\hat {\sigma }_0&0\end{array}\right) ,\nonumber \\
\hat {\gamma }^{0123}&=&-i\sigma _3\otimes \sigma _0=-i\left(
\begin{array}{cc}\hat {\sigma }_0&0\\
0&-\hat {\sigma }_0\end{array}\right) ,\nonumber \\
\end{eqnarray}
where $\hat {\sigma }_0$ is a $2\times 2$ identity, $\hat {\sigma
}_i$, (i=1,2,3) are Pauli matrices.

We next introduce an unified notations to denote Dirac matrices $\hat
{\gamma }^{\alpha }$,
\begin{eqnarray}
\hat {\gamma }^{\alpha _1...\alpha _p}=i^{\frac {1}{2}p(p-1)}\frac
{1}{p!}\delta ^{\alpha _1...\alpha _p}_{\rho _1...\rho _p}\hat
{\gamma }^{\rho _1}...\hat {\gamma }^{\rho _p},
\end{eqnarray}
where $p=0,1,2,3,4$, $\delta ^{\alpha _1...\alpha _p}_{\rho
_1...\rho _p}$ are the generalized Kronecker matrices defined as
\begin{eqnarray}
\delta ^{\alpha _1...\alpha _p}_{\rho _1...\rho _p}=\left|
\begin{array}{ccc}\delta ^{\alpha _1}_{\rho _1}&\cdots &
\delta ^{\alpha _1}_{\rho _p}\\
\vdots &\vdots &\vdots\\
\delta ^{\alpha _p}_{\rho _1}&\cdots &
\delta ^{\alpha _p}_{\rho _p}\\
\end{array}\right| .
\end{eqnarray}
It will be 1 when the upper indices and the lower indices are even
permutations, -1 for odd permutations, and 0 for other cases.

The Dirac matrices with upper indices are called inversion Dirac
matrices. By spin metric $\eta _{\alpha \beta }$, the covariance
Dirac matrices can be defined as
\begin{eqnarray}
\hat {\gamma }_{\alpha _1...\alpha _p}=\eta _{\alpha _1\beta
_1}\cdots \eta _{\alpha _p\beta _p}\hat {\gamma }^{\beta _1...\beta
_p}.
\end{eqnarray}
Compared with the representation matrices of $\hat {s}_{\alpha \beta
}$, we can find
\begin{eqnarray}
\hat {s}_{\alpha \beta }=\frac {1}{2}\eta _{\alpha \rho }\eta
_{\beta \sigma }\hat {\gamma }^{\rho \sigma }=\frac {1}{2}\hat
{\gamma }_{\alpha \beta }.
\end{eqnarray}
We next present some properties of the matrices $\hat {\alpha }$:

\begin{enumerate}
\item Anti-commuting relation
\begin{eqnarray}
\hat {\gamma }^{\alpha }\hat {\gamma }^{\beta }+\hat {\gamma
}^{\beta }\hat {\gamma }^{\alpha }=2\eta ^{\alpha \beta }\hat {I}
\end{eqnarray}

\item Commuting relation
\begin{eqnarray}
\hat {\gamma }^{\alpha }\hat {\gamma }^{\beta }-\hat {\gamma
}^{\beta }\hat {\gamma }^{\alpha }=-4i\hat {s} ^{\alpha \beta }.
\end{eqnarray}

\item Product relation
\begin{eqnarray}
\hat {\gamma }^{\alpha }\hat {\gamma }^{\beta }=g^{\alpha
\beta}-2i\hat {s}^{\alpha \beta }
\end{eqnarray}

\item Unitary relation
\begin{eqnarray}
(\hat {\gamma }^{\alpha })^{\dagger }=(\hat {\gamma }^{\alpha
})^{-1}=\hat {\gamma }_{\alpha }=\eta _{\alpha \beta }\hat {\gamma
}^{\beta }
\end{eqnarray}

\item Adjoint matrices
\begin{eqnarray}
\overline{\hat {\gamma }^{\alpha }}=\hat {\gamma }^0(\hat {\gamma
}^{\alpha })^{\dagger }(\hat {\gamma }^0)^{-1}=-\hat {\gamma
}^{\alpha }
\end{eqnarray}

\item Commuting relation with spin operator $\hat {s}_{\alpha \beta
}$,
\begin{eqnarray}
[\hat {s}_{\alpha \beta }, \hat {\gamma }^{\gamma }]=i(\delta
_{\beta }^{\gamma }\hat {\gamma }_{\alpha }-\delta _{\alpha
}^{\gamma }\hat {\gamma }_{\beta })=i\eta _{\rho \tau }\delta
_{\alpha \beta }^{\rho \sigma }\delta ^{\gamma }_{\sigma }\hat
{\gamma }^{\tau }.
\end{eqnarray}

\item Trace is zero
\begin{eqnarray}
{\rm tr}\hat {\gamma }^{\alpha }=0.
\end{eqnarray}

\item Orthogonal condition
\begin{eqnarray}
\langle \hat {\gamma }_{\alpha }, \hat {\gamma }^{\beta }\rangle
={\rm tr}[\overline{\hat {\gamma }_{\alpha }}\hat {\gamma }^{\beta
}]=-4\delta _{\alpha }^{\beta }.
\end{eqnarray}

\end{enumerate}

The properties of the Dirac matrices:
\begin{enumerate}
\item Unitary
\begin{eqnarray}
(\hat {\gamma }^{\alpha _1\cdots \alpha _p})^{\dagger }=(\hat
{\gamma }^{\alpha _1\cdots \alpha _p})^{-1}=\hat {\gamma }_{\alpha
_1\cdots \alpha _p}=\eta _{\alpha _1\beta _1}\cdots \eta _{\alpha
_p\beta _p}\hat {\gamma }^{\beta _1\cdots \beta _p}
\end{eqnarray}

\item Adjoint matrices
\begin{eqnarray}
\overline{(\hat {\gamma }^{\alpha _1\cdots \alpha _p})}= \hat
{\gamma }^0(\hat {\gamma }^{\alpha _1\cdots \alpha _p}) (\hat
{\gamma }^0)^{-1}=(-1)^p\hat {\gamma }^{\alpha _1\cdots \alpha _p}
\end{eqnarray}

\begin{eqnarray}
\overline{(\hat {\gamma }_{\alpha _1\cdots \alpha _p})}= \hat
{\gamma }^0(\hat {\gamma }_{\alpha _1\cdots \alpha _p}) (\hat
{\gamma }^0)^{-1}=(-1)^p\hat {\gamma }_{\alpha _1\cdots \alpha _p}
\end{eqnarray}

\item Commutation relation with spin $\hat {s}_{\alpha \beta }$,

\begin{eqnarray}
[\hat {s}_{\alpha \beta }, \hat {\gamma }^{\alpha _1\cdots \alpha
_p}]&=&i\eta _{\rho \tau }\delta _{\alpha \beta }^{\rho \sigma
}(\delta _{\sigma }^{\alpha _1}\hat {\gamma }^{\tau \cdots \alpha
_p}+\cdots +\delta _{\sigma }^{\alpha _p}\hat {\gamma }^{\alpha _1
\cdots \tau }),
\end{eqnarray}
\begin{eqnarray}
[\hat {s}_{\alpha \beta }, \hat {\gamma }_{\alpha _1\cdots \alpha
_p}]&=&-i\eta _{\rho \tau }\delta _{\alpha \beta }^{\rho \sigma
}(\delta ^{\tau }_{\alpha _1}\hat {\gamma }_{\sigma \cdots \alpha
_p}+\cdots +\delta ^{\tau }_{\alpha _p}\hat {\gamma }_{\alpha _1
\cdots \sigma }),
\end{eqnarray}

\item Trace
\begin{eqnarray}
{\rm tr}\hat {\gamma }&=&4 \nonumber \\
{\rm tr}\hat {\gamma }^{\alpha _1\cdots \alpha _p}&=&0, ~~(p\not= 0)
\end{eqnarray}

\item Orthogonality
\begin{eqnarray}
\langle \hat {\gamma }_{\alpha _1\cdots \alpha _p},\hat {\gamma
}^{\beta _1\cdots \beta _q}={\rm tr}[\overline{\hat {\gamma
}_{\alpha _1\cdots \alpha _p}}\hat {\gamma }^{\beta _1\cdots \beta
_p}]=(-1)^p4\delta _{pq}\delta _{\alpha _1\cdots \alpha _p}^{\beta
_1\cdots \beta _p}.
\end{eqnarray}
\end{enumerate}

\subsection{Tensor representation of the spin algebra}
The tensor representation of the spin algebra can be listed from
rank-0 to rank-4, we then use a general form to summarize these
representations.

The rank-0 anti-symmetric tensor is actually a scalar
representation, the representation space is denoted as
$AT_s(0)=V_S(0,0)$, the dimension of the space is ${\rm
dim}[AT_S(0)]={\rm dim}[V_S(0,0)]=1$. The representation basis is
$\hat {\gamma }$, so the 0-order antisymmetric tensor is represented
as
\begin{eqnarray}
\hat {K}=K\hat {\gamma }
\end{eqnarray}.

Similarly for rank-1, we have
\begin{eqnarray}
AT_S(1)&=&V_S(\frac {1}{2}, \frac {1}{2})
\\
{\rm dim}[AT_S(1)]&=&{\rm dim}[V_S(\frac {1}{2}, \frac {1}{2})]=4
\\
\hat {K}&=&K_{\alpha }\hat {\gamma }^{\alpha }.
\end{eqnarray}
And further for rank-2, the results are,
\begin{eqnarray}
AT_S(2)&=&A_S=V_S(0,1)\oplus V_S(1,0)
\\
{\rm dim}[AT_S(2)]&=&{\rm dim}A_S={\rm dim}[V_S(0,1)\oplus
V_S(1,0)]=6
\\
\hat {K}&=&\frac {1}{2!}K_{\alpha \beta }\hat {\gamma }^{\alpha
\beta }.
\end{eqnarray}
The rank-3 results are,
\begin{eqnarray}
AT_S(3)&=&V_S(\frac {1}{2},\frac {1}{2})
\\
{\rm dim}[AT_S(3)]&=&{\rm dim}[V_S(\frac {1}{2},\frac {1}{2})]=4
\\
\hat {K}&=&\frac {1}{3!}K_{\alpha \beta \gamma }\hat {\gamma
}^{\alpha \beta \gamma }.
\end{eqnarray}
And for rank-4, we have
\begin{eqnarray}
AT_S(4)&=&V_S(0,0)
\\
{\rm dim}[AT_S(4)]&=&{\rm dim}[V_S(0,0)]=1
\\
\hat {K}&=&\frac {1}{4!}K_{\alpha \beta \gamma \delta }\hat {\gamma
}^{\alpha \beta \gamma \delta }.
\end{eqnarray}

The summarized form for rank-$p$ tensor can be written in the
following, the space is $AT_S(p)$,
\begin{eqnarray}
{\rm dim}[AT_S(p)]&=&\frac {4!}{p!(4-p)!}
\\
\hat {K}&=&\frac {1}{p!}K_{\alpha _1\cdots \alpha _p}\hat {\gamma
}^{\alpha _1\cdots \alpha _p}.
\end{eqnarray}

The anti-symmetric tensor representation of the spin algebra can be
represented as a direct summation of the rank-0 to rank-4
anti-symmetric tensor.
\begin{eqnarray}
AT_S=\sum _{p=0}^4 \oplus AT_S(p).
\end{eqnarray}
The dimension of the space can be calculated as
\begin{eqnarray}
{\rm dim}[AT_S]=\sum _{p=0}^4{\rm dim}[AT_S(p)]=\sum _{p=0}^4\frac
{4!}{p!(4-p)!}=16
\end{eqnarray}
The representation space is the direct summation of spinor space and
the adjoint spinor space:
\begin{eqnarray}
AT_S=V_S\otimes \overline{V_S}
\end{eqnarray}

We should note that the space for spin of the gauge particles is the
16D space for spin antisymmetric tensor,
\begin{eqnarray}
V_S(A)=AT_S.
\end{eqnarray}

\subsection{Mixed tensor representation of the spin algebra}
The (p,q) mixed tensor representation of the spin algebra is the
direct product of the rank-$p$ and rank-$q$ antisymmetric tensor
representations,
\begin{eqnarray}
MT_S(p,q)=AT_S(p)\otimes AT_S(q).
\end{eqnarray}
The dimension of the representation space is
\begin{eqnarray}
{\rm dim}[MT_S(p,q)]={\rm dim}[AT_S(p)]\times {\rm dim}[AT_S(q)]
=\frac {4!4!}{p!(4-p)!q!(4-q)!}.
\end{eqnarray}
The basis of the representation is
\begin{eqnarray}
\hat {\gamma }^{\alpha _1\cdots \alpha _p}_{\beta _1\cdots \beta _q}
=\hat {\gamma }^{\alpha _1\cdots \alpha _p}\otimes \hat {\gamma
}_{\beta _1\cdots \beta _q},
\end{eqnarray}
which is a $16\times 16$ matrix. The commutation relation takes the
form
\begin{eqnarray}
[\hat {s}_{\alpha \beta }, \hat {\gamma }^{\alpha _1\cdots \alpha
_p}_{\beta _1\cdots \beta _q}]=i\eta _{\rho \tau }\delta ^{\rho
\sigma }_{\alpha \beta }(\delta _{\sigma }^{\alpha _1}\hat {\gamma
}^{\tau \cdots \alpha _p}_{\beta _1\cdots \beta _q}+\cdots +\delta
_{\sigma }^{\alpha _p}\hat {\gamma }^{\alpha _1 \cdots \tau
}_{\sigma \cdots \beta _q}-\delta _{\beta _1 }^{\tau }\hat {\gamma
}^{\alpha _1\cdots \alpha _p}_{\sigma \cdots \beta _q}-\cdots
-\delta ^{\tau }_{\beta _q}\hat {\gamma }^{\alpha _1 \cdots \alpha
_p}_{\beta _1\cdots \sigma })
\end{eqnarray}
This $(p,q)$ mixed tensor representation includes all tensor
representation of spin algebra, while rank-$p$ antisymmetric tensor
representation can be dealt as a $(p,0)$ or $(0,p)$ form mixed
tensor representation. The $(p,q)$ mixed tensor takes the form
\begin{eqnarray}
\hat {K}=\frac {1}{p!q!}K_{\alpha _1\cdots \alpha _p}^{\beta
_1\cdots \beta _q}\hat {\gamma }^{\alpha _1\cdots \alpha _p}_{\beta
_1\cdots \beta _q}
\end{eqnarray}

The mixed tensor representation of the spin algebra is represented
as the direct summation of all $(p,q)$ mixed tensor representations,
\begin{eqnarray}
MT_S=\sum _{p,q=0}^4\oplus MT_S(p,q).
\end{eqnarray}
The dimension is
\begin{eqnarray}
{\rm dim}[MT_S]=\sum _{p,q=0}^4{\rm dim}[MT_S(p,q)]=\sum
_{p,q=0}^4\frac {4!4!}{p!(4-p)!q!(4-q)!}=256.
\end{eqnarray}
The basis is $\hat {\gamma }^{\alpha _1\cdots \alpha _p}_{\beta
_1\cdots \beta _q}$, $(p,q=0,1,2,3,4)$. The mixed tensor space is a
direct summation of two antisymmetric tensor space,
\begin{eqnarray}
V_S(G)=MT_S
\end{eqnarray}
We should note that the space for spin of the gravity particle is
the 256D mixed spin tensor space.
\begin{eqnarray}
V_S(G)=MT_S
\end{eqnarray}

\subsection{Some fundamental calculations of the spin tensor}

\begin{enumerate}
\item Exterior product of the antisymmetric spin tensor.
Similar as in differential geometry, the exterior product of two
basis  can be defined as
\begin{eqnarray}
\hat {\gamma }^{\alpha _1\cdots \alpha _p} \wedge \hat {\gamma
}^{\beta _1\cdots \beta _q}=\hat {\gamma }^{\alpha _1\cdots \alpha
_p\beta _1\cdots \beta _q}
\end{eqnarray}
So the exterior product of two antisymmetric tensors will take the
form
\begin{eqnarray}
\hat {K}\wedge \hat {H}=(\frac {1}{p!}K_{\alpha _1\cdots \alpha _p}
\hat {\gamma }^{\alpha _1\cdots \alpha _p})\wedge (\frac
{1}{q!}H_{\beta _1\cdots \beta _q} \hat {\gamma }^{\beta _1\cdots
\beta _q})=(\frac {1}{p!q!}K_{\alpha _1\cdots \alpha _p}H_{\beta
_1\cdots \beta _q} \hat {\gamma }^{\alpha _1\cdots \alpha _p\beta
_1\cdots \beta _q})
\end{eqnarray}
Thus the exterior product of a $p$-form antisymmetric tensor and a
$q$-form antisymmetric tensor is a $p+q$-order antisymmetric tensor.
The 16D antisymmetric spin tensor space is closed for exterior
product.

\item Tensor product for the antisymmetric spin tensor.

The tensor product obeys the following role,
\begin{eqnarray}
\hat {\gamma }^{\alpha _1\cdots \alpha _p} \otimes \hat {\gamma
}_{\beta _1\cdots \beta _q}=\hat {\gamma }^{\alpha _1\cdots \alpha
_p}_{\beta _1\cdots \beta _q}
\end{eqnarray}

\begin{eqnarray}
\hat {K}\otimes \hat {H}&=&(\frac {1}{p!}K_{\alpha _1\cdots \alpha
_p} \hat {\gamma }^{\alpha _1\cdots \alpha _p})\otimes (\frac
{1}{q!}H^{\beta _1\cdots \beta _q} \hat {\gamma }_{\beta _1\cdots
\beta _q}) \nonumber \\
&=&(\frac {1}{p!q!}K_{\alpha _1\cdots \alpha _p}H^{\beta _1\cdots
\beta _q} \hat {\gamma }^{\alpha _1\cdots \alpha _p}_{\beta _1\cdots
\beta _q}).
\end{eqnarray}

\item Scalar product of rank-1 spin tensor.
The scalar product of two rank-1 Dirac matrices is defined as,
\begin{eqnarray}
\hat {\gamma }^{\alpha }\cdot \hat {\gamma }^{\beta }=\eta ^{\alpha
\beta }.
\end{eqnarray}
It is known that $\eta ^{\alpha \beta }$ is the Minkowski vierbein,
also named as free vierbein, so $\hat {\gamma }^{\alpha }$ are
orthogonal. The scalar product of two 1-form spin tensor is written
as:
\begin{eqnarray}
K\cdot L=\eta ^{\alpha \beta }K_{\alpha }L_{\beta }.
\end{eqnarray}

\item Contraction calculation for mixed tensor.
With the form of free vierbein $\eta ^{\alpha \beta }$, we can
define the contraction calculation for mixed tensor. First, the
contraction of the $(p,q)$ mixed tensor basis $(p,q\ge 1)$ is
defined as
\begin{eqnarray}
{\rm con}\hat {\gamma }^{\alpha _1\cdots \alpha _p }_{\beta _1\cdots
\beta _q}=\delta ^{\alpha _p}_{\beta _q}{\gamma }^{\alpha _1\cdots
\alpha _{p-1} }_{\beta _1\cdots \beta _{q-1}}.
\end{eqnarray}
So we can define the contraction of $(p,q)$ mixed tensor, $(p,q\ge
1)$,

\begin{eqnarray}
{\rm con}\hat {K}={\rm con}(\frac {1}{p!q!}K_{\alpha _1\cdots \alpha
_p}^{\beta _1\cdots \beta _q} \hat {\gamma }^{\alpha _1\cdots \alpha
_p}_{\beta _1\cdots \beta _q})=(\frac {1}{p!q!}K_{\alpha _1\cdots
\alpha _p}^{\beta _1\cdots \beta _q} \delta ^{\alpha _p}_{\beta
_q}\hat {\gamma }^{\alpha _1\cdots \alpha _{p-1}}_{\beta _1\cdots
\beta _{q-1}}).
\end{eqnarray}
After the contraction calculation, the $(p,q)$ mixed tensor is
reduced as a $(p-1,q-1)$-tensor.

\item Lowering and rising of the indices of the spin tensor.
By using the product of the inversion Minkowski vierbein $\eta
^{\alpha \beta }$ and covariance Minkowski vierbein $\eta _{\alpha
\beta }$, we can realize the lowering and rising of the indices for
spin tensor, we use the antisymmetric tensor as the examples,
\begin{eqnarray}
\hat {\gamma }_{\alpha _1\cdots \alpha _p}&=&\eta _{\alpha _1\beta
_1}\cdots \eta _{\alpha _p\beta _p}\hat {\gamma }^{\beta _1\cdots
\beta _p}
\\
K^{\alpha _1\cdots \alpha _p}&=&\eta ^{\alpha _1\beta _1}\cdots \eta
^{\alpha _p\beta _p}K_{\beta _1\cdots \beta _p}
\\
\hat {K}&=&\frac {1}{p!}K_{\alpha _1\cdots \alpha _p}\hat {\gamma
}^{\alpha _1\cdots \alpha _p} =\frac {1}{p!}K^{\alpha _1\cdots
\alpha _p}\hat {\gamma }_{\alpha _1\cdots \alpha _p}.
\end{eqnarray}
For lowering or rising of the indices, the rank of the indices
remains unchanged. One spin tensor may have different indices
representation, they may be used for different purposes.

\end{enumerate}

\section{Gauge charges}
As we pointed out that gauge is one fundamental parameter in our
theory, we next consider the properties of gauge charges. Also we
would like to emphasize that mass matrices are defined in gauge
space in our theory. The conventional gauge theory generally
satisfies the gauge invariant condition. Our unified theory
satisfies the general covariance condition which releases the
previous restriction. Thus mass arises naturally.

What we study is the all elementary particles and the interactions
between them. The elementary particles are divided into three
families as matter particles ($\Psi $), gauge particles ($A$) and
graviton ($g$). They satisfy Dirac equation, Yang-Mills equation and
Einstein equation, respectively. In this work, we will present an
unified quantum theory for those equations

Matter particles ($\Psi $) include quarks and leptons each class
have 6 types depending on mass and electric-charge. Each type of
lepton also has isospin singlet state and isospin doublet state, so
leptons have 12 states, altogether. Each flavor of quark has three
color states, each color has isospin singlet state and isospin
doublet state. So quarks have 36 states. Thus matter particles have
48 states. Note that the spin of matter particles is $1/2$ described
by a 4D Dirac spinor

Gauge particles ($A$) are divided into photons, three types of weak
gauge bosons and eight types of gluons. Photon has no
mass, and gluons are generally assumed massless. However, the massive
gluons are allowed in this theory. We will find that gluon might be related
with dark energy of universe, its mass is very small.
The masses of weak gauge bosons are $m_Z, m_{W_+}, m_{W_-}$
corresponding to three types of bosons $Z_0, W_+, W_-$,
respectively. We also have $m_{W_+}=m_{W_-}=m_Z\cos \theta _w$,
where $\theta _w$ is Weinberg angle. The spin of gauge particles is
1 corresponding to a 4D vector.

There is only one type of graviton ($g$), its mass and gauge charge
are zeroes, the spin is 2 in tensor representation.

\subsection{Gauge algebra and gauge charges}

There are 12 gauge charges, including hypercharge $\hat{Y}$, isospin
charges $\hat{I}_i, (i=1,2,3)$ and color charges $\hat{\lambda} _p,
(p=1,2,...,8)$. Hypercharge $\hat{Y}$ is the generator of gauge
group $U(1)$. Three isospin charges $\hat{I}_i,(i=1,2,3)$ are
generators of gauge group $SU(2)$, they constitute a basis for
algebra $su(2)$, and eight color charges $\hat{\lambda } _p,
(p=1,2,...,8)$ are generators of gauge group $SU(3)$ and constitute
a set of basis.

Those generators satisfy the commutation relations
\begin{eqnarray}
[\hat{I}_i,\hat{I}_j]=i\epsilon _{ijk}\hat{I}_k,
\end{eqnarray}
\begin{eqnarray}
[\hat{\lambda }_p,\hat{\lambda }_q]=if_{pq}^r\hat{\lambda }_r,
\end{eqnarray}
\begin{eqnarray}
[\hat{Y}, \hat{I}_i]=[\hat{Y}, \hat{\lambda }_p]=[\hat
{I}_i,\hat{\lambda }_p]=0,
\end{eqnarray}
where $\epsilon _{ijk}$ is Levi-Civita symbol and is the structure
constant for algebra $su(2)$, $f_{pq}^r$ are structure constants of
algebra $su(3)$ group. They are completely symmetric for three
indices, the non-zero elements are
\begin{eqnarray}
&&f_{12}^3=1,\nonumber \\
&&f_{14}^7=-f_{15}^6=f_{24}^6=f_{25}^7=f_{34}^5=-f_{36}^7=1,\nonumber \\
&&f_{45}^8=f_{67}^8=\frac {\sqrt{3}}{2}.
\end{eqnarray}
We can define square of the isospin charges $\hat{I}^2$ and the
color charges $\hat{\lambda }^2$:
\begin{eqnarray}
\hat{I}^2=\sum _{i=1}^3\hat{I}^2_i,\\
\hat{\lambda }^2=\sum _{p=1}^8\hat{\lambda }^2_p.
\end{eqnarray}
It can be checked that $\hat{I}^2$ and $\hat{\lambda }^2$ commute
with all elements of gauge charge
\begin{eqnarray}
&&[\hat{I}^2, \hat{Y}]=[\hat{I}^2, \hat{I}_i]=[\hat {I}^2,
\hat{\lambda }_p]=0, \\
&&[\hat{\lambda }^2, \hat{Y} ]=[\hat{\lambda }^2,
\hat{I}_i]=[\hat{\lambda }^2, \hat{\lambda }_p]=0,
\end{eqnarray}

We use notation $\hat{t}_a, (a=1,2,...,12)$ to represent those 12
gauge charges as

\begin{eqnarray}
\hat{t}_1&=&g_1\frac {\hat {Y}}{2}, \\
\hat{t}_{1+i}&=&g_2\hat {I}_i, ~~~(i=1,2,3), \\
 \hat {t}_{4+p}&=&g_3\hat
{\lambda }_p, ~~~(p=1,2,...,8),
\end{eqnarray}
 where $g_1,g_2$ and $g_3$ are
coefficients of hypercharge, isospin charges and color charges,
respectively. So the commutation relations take a concise form as
\begin{eqnarray}
[\hat{t}_a,\hat{t}_b]=id_{ab}^c\hat {t_c},
\end{eqnarray}
where coefficients $d_{ab}^c$ are defined as
\begin{eqnarray}
d_{1+i,1+j}^{1+k}=g_2\epsilon _{ijk},~~(i,j,k=1,2,3)
\\
d_{4+p,4+q}^{4+r}=g_3f_{pq}^r,~~(p,q,r=1,2,...,8)
\end{eqnarray}
and $d_{ab}^c=0$ elsewhere. Since $\epsilon _{ijk}$ and $f_{rs}^t$
are completely antisymmetric, $d_{ab}^c$ are also completely
antisymmetric
\begin{eqnarray}
d_{ab}^c=d_{bc}^a=d_{ca}^b=-d_{cb}^a=-d_{ba}^c=-d_{ac}^b
\end{eqnarray}

Those gauge charges $\hat{t}_a$ can be denoted as gauge bosons in
Cartan-Weyl basis $\hat {T}_a$, (a=1,2,...,12),
\begin{eqnarray}
\hat{T}_1&=&\hat{t}_1\cos \theta _w+\hat {t}_4\sin \theta
_w,\nonumber \\
\hat{T}_2&=&-\hat{t}_1\sin \theta _w+\hat {t}_4\cos \theta _w,\nonumber \\
\hat{T}_3&=&\frac {1}{\sqrt {2}}(\hat {t}_2+i\hat {t}_3),\nonumber \\
\hat{T}_4&=&\frac {1}{\sqrt {2}}(\hat {t}_2-i\hat {t}_3),\nonumber \\
\hat{T}_5&=&\hat {t}_7,\nonumber \\
\hat{T}_6&=&\hat {t}_{12},\nonumber \\
\hat{T}_7&=&\frac {1}{\sqrt {2}}(\hat {t}_5+i\hat {t}_6),\nonumber \\
\hat{T}_8&=&\frac {1}{\sqrt {2}}(\hat {t}_5-i\hat {t}_6),\nonumber \\
\hat{T}_9&=&\frac {1}{\sqrt {2}}(\hat {t}_8+i\hat {t}_9),\nonumber \\
\hat{T}_{10}&=&\frac {1}{\sqrt {2}}(\hat {t}_8-i\hat {t}_9),\nonumber \\
\hat{T}_{11}&=&\frac {1}{\sqrt {2}}(\hat {t}_{10}+i\hat {t}_{11}),\nonumber \\
\hat{T}_{12}&=&\frac {1}{\sqrt {2}}(\hat {t}_{10}-i\hat {t}_{11}),
\end{eqnarray}
where $\theta _w$ is the Weinberg angle.  The transformation from
orthogonal gauge charges $\hat {t}_a$ to eigen-gauge charges $\hat
{T}_a$ is unitary,
\begin{eqnarray}
\hat {T}_a=\hat {L}_a^b\hat {t}_b,
\end{eqnarray}
as we can find that the transformation matrix satisfy
\begin{eqnarray}
\hat {L}^{-1}=\hat {L}^{\dagger },
\end{eqnarray}
where super-indices $\dagger $ means the hermitian conjugation
(complex conjugation plus matrix transposition). So the inverse
transformation takes the form
\begin{eqnarray}
\hat {t}_a=\hat {L}_a^{\dagger b}\hat {T}_b,
\end{eqnarray}
The eigen-gauge charges satisfy the relation,
\begin{eqnarray}
[\hat {T}_a, \hat {T}_b]=C_{ab}^c\hat {T}_c.
\end{eqnarray}
The coefficients can be found to be
\begin{eqnarray}
&&C_{1,3}^3=-C_{1,4}^4=C_{3,4}^1=g_2\sin \theta _w, ~~~
C_{2,3}^3=-C_{2,4}^4=C_{3,4}^2=g_2\cos \theta _w,\nonumber \\
&&C_{5,7}^7=-C_{5,8}^8=C_{7,8}^5=g_3,~~~~~~
C_{6,7}^7=-C_{6,8}^8=C_{7,8}^6=0,\nonumber \\
&&C_{5,9}^9=-C_{5,10}^{10}=C_{9,10}^5=-\frac {1}{2}g_3,~~~
C_{6,9}^9=-C_{6,10}^{10}=C_{9,10}^6=\frac {\sqrt {3}}{2}g_3,\nonumber \\
&&C_{5,11}^{11}=-C_{5,12}^{12}=C_{11,12}^5=-\frac {1}{2}g_3,~~~
C_{6,11}^{11}=-C_{6,12}^{12}=C_{11,12}^6=-\frac {\sqrt {3}}{2}g_3,\nonumber \\
&&C_{7,9}^{12}=-C_{9,11}^{8}=C_{11,7}^{10}=\frac {\sqrt
{2}}{2}g_3,~~~ C_{8,10}^{11}=C_{10,12}^{7}=C_{12,8}^9=-\frac {\sqrt
{2}}{2}g_3. \label{coup-constant}
\end{eqnarray}

Those 12 gauge charges $\hat {t}_a, a=1,\cdots ,12,$ (or $\hat
{T}_a$), constitute the gauge algebra,
\begin{eqnarray}
A_g=\{ \hat {Z}_g:\hat {Z}_g=\theta ^a\hat {t}_a\}=u(1)\oplus
su(2)\oplus su(3).
\end{eqnarray}
The gauge algebra is a 12-dimension Lie algebra, it is constituted
by direct summation of an Abel algebra $u(1)$ and two simple Lie
algebras $su(2)$ and $su(3)$.

Consequently, corresponding to gauge algebra $A_g$, we have the
gauge group $G_g$ which can be obtained by the exponential of the
gauge algebra,
\begin{eqnarray}
G_g=\{ \hat {U}_g:\hat {U}_g=\exp (i\theta _a\hat
{t}_a)\}=U(1)\otimes SU(2)\otimes SU(3).
\end{eqnarray}

For gauge algebra $A_g$, we introduce the gauge metric tensor:
\begin{eqnarray}
\hat{G}=g^{ab}\hat {t}_a\otimes \hat {t}_b=G^{ab}\hat {T}_a\otimes
\hat{T}_b.
\end{eqnarray}
Gauge metric tensor is a generalization of Lie algebra Cartan metric
tensor. For simplicity, we introduce the ortho-normal metric for
orthogonal gauge charges $\hat {t}_a$,
\begin{eqnarray}
g_{ab}=g^{ab}=\delta _{ab}.
\end{eqnarray}
Due to the transformation for $\hat {t}_a$ to $\hat {T}_a$ and the
explicit form of $g^{ab}$, we can find that the gauge metric tensors
$G^{ab}$ and $G_{ab}$ take the form
\begin{eqnarray}
&&G^{ab}=G_{ab},\nonumber \\
&&G_{1,1}=G_{2,2}=G_{3,4}=G_{4,3}=1, \nonumber \\
&&G_{5,5}=G_{6,6}=G_{7,8}=G_{8,7}=G_{9,10}=G_{10,9}=G_{11,12}=G_{12,11}=1,
\nonumber \\
&&G_{ab}=0,~~~{\rm elsewhere}.
\end{eqnarray}
According to the gauge metric tensor, the scalar product of the
gauge charges are defined as:
\begin{eqnarray}
&&\hat{t}_a\cdot \hat {t}_b=\hat{t}_b\cdot \hat {t}_a=g_{ab}=\delta
_{ab}; \nonumber \\
&&\hat{T}_a\cdot \hat {T}_b=\hat{T}_b\cdot \hat {T}_a=\hat G_{ab} \nonumber \\
&&\hat {Y}\cdot \hat {Y}=4g_1^{-2},\nonumber \\
&&\hat {I}_i\cdot \hat {I}_j=g_2^{-2}\delta _{ij},\nonumber \\
&&\hat {\lambda }_p\cdot \hat {\lambda }_q=g_3^{-2}\delta _{pq},\nonumber \\
&&\hat {Y}\cdot \hat {I}_i=\hat {Y}\cdot \hat {\lambda }_p=\hat
{I}_i \cdot \hat {\lambda }_p=0.
\end{eqnarray}
The rising and lowering the indices by the gauge metric tensor can
be written as, for example,
\begin{eqnarray}
&&\hat{t}^a=g^{ab}\hat {t}_b=\hat {t}_a, \nonumber \\
&&\hat{T}^a=G^{ab}\hat {T}_b.
\end{eqnarray}

\subsection{Irreducible representation of the gauge algebra}
The irreducible representation space of algebra $u(1)$ is denoted as
$V_1(Y)$, where $Y$ is the eigenvalue of the hypercharge $\hat {Y}$.
Here $Y$ can be arbitrary real number, and the dimension of space
$V_1(Y)$ is,
\begin{eqnarray}
{\rm dim}V_1(Y)=1.
\end{eqnarray}

We denote the irreducible representation space of algebra $su(2)$ as
$V_2(I)$, $I$ is the maximal eigenvalue of the third isospin
operator $\hat {I}_3$. $I$ can be non-negative integer and
half-integer. The dimension of space $V_2(I)$ is
\begin{eqnarray}
{\rm dim}V_2(I)=2I+1.
\end{eqnarray}

Similarly the space of the irreducible representation of algebra
$su(3)$ is denoted as $V_3(m,n)$, where $m,n$ are non-negative. The
dimension of space $V_3(m,n)$ is
\begin{eqnarray}
{\rm dim}V_3(m,n)=\frac {1}{2}(m+1)(n+1)(m+n+2).
\end{eqnarray}

Gauge algebra $A_g$ is the direction summation of algebras $u(1)$,
$su(2)$ and $su(3)$, and its irreducible representation is the
tensor product of the irreducible representations of $u(1), su(2)$
and $su(3)$. Denote the irreducible representation space of algebra
$A_g$ as $V_g(Y,I,m,n)$ and it takes the form
\begin{eqnarray}
V_g(Y,I,m,n)=V_1(Y)\otimes V_2(I)\otimes V_3(m,n).
\end{eqnarray}
Thus the dimension of space $V_g(Y,I,m,n)$ is
\begin{eqnarray}
{\rm dim}V_g(Y,I,m,n)&=&{\rm dim}V_1(Y)\times {\rm dim}V_2(I)\times
{\rm dim}V_3(m,n)\nonumber \\
&=&\frac {1}{2}(2I+1)(m+1)(n+1)(m+n+2)
\end{eqnarray}

\subsection{Gauge representations of the matter particles,
representation spaces and gauge states}

Matter particles include leptons and quarks. We next consider their
gauge representations respectively.

(i), Space of the gauge representations for leptons. Leptons can be
divided into three generations, the gauge representation space for
each generation is the same. The representation space is
\begin{eqnarray}
V_g(-1,\frac {1}{2}, 0, 0)\oplus V_g(0,0, 0, 0)\oplus V_g(-2,0, 0,
0).
\end{eqnarray}
The dimension for each generation of leptons in the gauge
representation is $2+1+1=4$. It is corresponding to the fact that
each generation of leptons includes four states as: isospin doublet
state for leptons with electric-charge, isospin singlet state of
leptons with electric-charge, isospin doublet state of neutrino and
isospin singlet state of neutrino.

Gauge representation space for all leptons can be denoted as the
direct summation of the gauge representation spaces for three
generations of lepton:

\begin{eqnarray}
V_g(l)&=&[V_g(-1,\frac {1}{2}, 0, 0)\oplus V_g(0,0, 0, 0)\oplus
V_g(-2,0, 0, 0)] \nonumber \\
&&\oplus [V_g(-1,\frac {1}{2}, 0, 0)\oplus V_g(0,0, 0, 0)\oplus
V_g(-2,0, 0, 0)]
\nonumber \\
&&\oplus [V_g(-1,\frac {1}{2}, 0, 0)\oplus V_g(0,0, 0, 0)\oplus
V_g(-2,0, 0, 0)]
\end{eqnarray}
The total dimension is
\begin{eqnarray}
{\rm dim}V_g(l)=(2+1+1)\times 3=12.
\end{eqnarray}
The total dimension 12 corresponds to 12 different leptons.

(ii), Space of the gauge representations for quarks. Quarks also
have three generations, the representation space for each generation
of quarks is the same. Each generation can be represented as
\begin{eqnarray}
V_g(\frac {1}{3},\frac {1}{2}, 1, 0)\oplus V_g(\frac {4}{3},0, 1,
0)\oplus V_g(-\frac {2}{3},0, 1, 0).
\end{eqnarray}
One can check that the dimension is $6+3+3=12$, it corresponds to
that there are 12 gauge states for each generation of quark.

The gauge representation space for three generations of quarks is
denoted as the direction summation as
\begin{eqnarray}
V_g(q)&=&[V_g(\frac {1}{3},\frac {1}{2}, 1, 0)\oplus V_g(\frac
{4}{3},0, 1,
0)\oplus V_g(-\frac {2}{3},0, 1, 0)] \nonumber \\
&&\oplus [V_g(\frac {1}{3},\frac {1}{2}, 1, 0)\oplus V_g(\frac
{4}{3},0, 1, 0)\oplus V_g(-\frac {2}{3},0, 1, 0)
\nonumber \\
&&\oplus V_g(\frac {1}{3},\frac {1}{2}, 1, 0)\oplus V_g(\frac
{4}{3},0, 1, 0)\oplus V_g(-\frac {2}{3},0, 1, 0).
\end{eqnarray}
The dimension of quark gauge representation space is
\begin{eqnarray}
{\rm dim}V_g(q)=(6+3+3)\times 3=36.
\end{eqnarray}
The dimension $36$ corresponds to 36 kind of quarks.

So in together, the gauge representation space of matter particles
is denoted as
\begin{eqnarray}
V_g(M)=V_g(l)\oplus V_g(q),
\end{eqnarray}
The total dimension is
\begin{eqnarray}
{\rm dim}V_g(M)=12+36=48.
\end{eqnarray}
The total dimension corresponds to 48 gauge states.

The gauge state basis of matter particles is represented as:
\begin{eqnarray}
|e_t\rangle ,
\end{eqnarray}
where $t=1,2,...,48$ is the indices of matter particles. For each
kind of matter particles, there is a corresponding gauge basis.

Matter particles in gauge representation space $V_g(M)$ can be
represented by gauge state basis as
\begin{eqnarray}
V_g(M)=\left\{ |\Psi \rangle : |\Psi \rangle =\Psi ^t|e_t\rangle
\right\} .
\end{eqnarray}

The adjoint representation of matter particles takes the form
\begin{eqnarray}
\langle e_t|=\overline{|e_t\rangle }.
\end{eqnarray}
The inner product of basis and its adjoint is,
\begin{eqnarray}
\langle e_t|e_{t'}\rangle =\delta _{tt'}.
\end{eqnarray}
And also,
\begin{eqnarray}
|e_t\rangle \langle e^t|=\hat {I}.
\end{eqnarray}
This is the property of the metric for the gauge state space.

The gauge state $|\Psi \rangle $ of matter particles can be
represented as the the superposition of the gauge basis
\begin{eqnarray}
|\Psi \rangle =\Psi ^t|e_t\rangle ,
\end{eqnarray}
where
\begin{eqnarray}
\Psi ^t=\langle e^t|\Psi \rangle .
\end{eqnarray}
The adjoint of $|\Psi \rangle $ is $\langle \Psi |$, it has the form
\begin{eqnarray}
\langle \Psi |=\overline{|\Psi \rangle } =\Psi _t^*\langle e^t|.
\end{eqnarray}
The inner product has the form,
\begin{eqnarray}
\langle \Psi |\Phi \rangle &=&\langle \Phi |\Psi \rangle ^*
\nonumber \\
&=&\Psi _t^*\Phi ^t=\delta _{tt'}\Psi ^{*t}\Phi ^{t'}.
\end{eqnarray}

\subsection{Elementary particles and their classification}
Here we list the properties of electric-charge and the mass of
matter particles:

\vskip 1truecm

\begin{tabular}{|c|c|c|c|c|c|c|}
  \hline
  generations& lepton & electric-charge & mass & quark & electric-charge & mass \\
  \cline{1-7}
  first generation& e& -1 & $m_e$ & d & $-\frac {1}{3}$ & $m_d$ \\
  \cline{2-7}
   & $\nu _e$& 0& $m_{\nu e}$  & $u$ & $+\frac {2}{3}$ & $m_u$ \\
   \cline{1-7}
second generation& $\mu $ & -1 & $m_{\mu }$ & $s$ & $-\frac {1}{3}$ & $m_s$ \\
\cline{2-7}
 & $\nu _{\mu }$& 0& $m_{\nu \mu }$  & $c$ & $+\frac {2}{3}$ & $m_c$ \\
\cline{1-7}
third generation& $\tau $ & -1 & $m_{\tau }$ & $b$ & $-\frac {1}{3}$ & $m_b$ \\
\cline{2-7}

  & $\nu _{\tau }$& 0& $m_{\nu \tau }$  & $t$ & $+\frac {2}{3}$ & $m_t$ \\
 \hline
\end{tabular}
\vskip 1truecm

As we mentioned, in our theory, in order to consider properties of
the involvements of the matter particles into the action force, the
matter particles are divided into 48 classes. There are 6 classes of
leptons according to the above table, besides mass and
electric-charge, there are isospin singlet and isospin doublet, so
there are 12 classes of leptons. There are 6 classes of quarks in
the above table, we can also consider the isospin singlet and
isospin doublet and three colors for quarks. So quarks are divided
into 36 types. So matter particles have 48 classes. The explicit
classification of those particles are presented explicitly in the
following tables.

Color charge represents the quantum number of the involvement of
particles into color interactions. Color charges of particles are
represented by two parameters $(\lambda _3, \lambda _8)$. According
to color charge, matter particles are divided as leptons ($l$) and
quarks ($q$). The color charge of leptons is $(0,0)$, that means the
color charge of them are zeroes, and they are not involved into the
color force. The color charges of quarks are three types, ($\frac
{1}{2}, \frac {\sqrt {3}}{6}$), ($-\frac {1}{2}, \frac {\sqrt
{3}}{6}$), ($0, \frac {\sqrt {3}}{3}$). Usually they are called red,
green and blue colors, respectively. So quarks can be red quark
($q_r$), green quark ($q_g$) and blue quark ($q_b$).

Electric charge $Q$ represents the quantum number of particles in
interaction of electromagnetism. The leptons are divided into
neutral-leptons ($l_0$) and electric-charged leptons ($l_{-1}$). The
neutral-lepton is the neutrino. The electric-charge of the $l_{-1}$
is $-1$. Quarks can be divided into positive-electric-charge quarks
($q_{+\frac {2}{3}}$) and negative-electric-charge quarks
($q_{-\frac {1}{3}}$) according to their electric charge. The
positive electric-charge is $+\frac {2}{3}$, the negative
electric-charge is $-\frac {1}{3}$.

The isospin charge $I_3$ represents the quantum number of the
particles in weak interactions. The isospin charge can take three
values $+\frac {1}{2}, -\frac {1}{2}$ and $0$, where $I_3=\pm \frac
{1}{2}$ represents the isospin doublet $(\Psi _D)$, and $I_3=0$
represent the isospin singlet. The matter particles with same color
charge and electric-charge can have isospin singlet and isospin
doublet. The isospin doublet are in the weak interactions, while
isospin singlet is not involved in the weak interactions.

The hypercharge and the weak-charge can be represented by
electric-charge and the isospin charge as,
\begin{eqnarray}
Y&=&2(Q-I_3), \nonumber \\
Z&=&I_3-Q\sin ^2\theta _{w},
\end{eqnarray}
where $\sin \theta _w=\frac {g_1}{\sqrt {g_1^2+g_2^2}}$, and
$g_1,g_2$ are the coupling constants between hypercharge and the
isospin charges, $\theta _w$ is the Weinberg angle. Weak-charge $Z$
represents the quantum number of particles involved in the
interactions of $Z_0$ particles.

Note there are only two independent parameters in four quantum
numbers, electric-charge $Q$, isospin charge $I_3$, hypercharge $Y$
and the weak-charge $Z$. Usually the hypercharge $Y$ and isospin
charge $I_3$ are chosen as the independent parameters, and
electric-charge $Q$ and weak-charge $Z$ are represented as,
\begin{eqnarray}
Q&=&I_3+\frac {1}{2}Y, \nonumber \\
Z&=&I_3\cos ^2\theta _w-\frac {1}{2}Y\sin ^2\theta _w.
\end{eqnarray}

Masses of the particles are the quantum numbers representing the
property of the symmetry breaking of the isospin in weak
interactions. Matter particles with the same color-charge,
electric-charge and isospin charge can be divided into three
generations according to their masses.

Three generations of neutrinos are called respectively the
electric-neutrino, $\mu $ neutrino and $\tau $ neutrino represented
as $\nu _e, \nu _{\mu }, \nu _{\tau }$. The masses are $m_{\nu e}$,
$m_{\nu \mu }$ and $m_{\nu \tau }$. It is believed previously that
the masses of the three generations of neutrinos are zeroes. Later
experiments showed that the masses are small but not zeroes. Three
generations of electric-leptons are named respectively electron,
$\mu $ particle and $\tau $ particle represented as $e,\mu ,\tau$.
The masses are $m_{u}, m_c, m_t$.

Three generations of positive-electric quarks are $u,c,t$ with
masses $m_u,m_c,m_t$. The negative-electric quarks are $d,s,b$ with
masses $m_d,m_s,m_b$. The masses of matter particles differ only
depending on electrical-charge and generation, but not depending on
color charge and isospin charge. That means matter particles with
the same electrical-charge and generation but different color
charges and isospin charges, their masses will be the same.

Depending on color-charge,electrical-charge, isospin-charge and
mass, the matter particles are divided into 48 classes. The
representation of those particles constitute a 48-dimensional gauge
space. The gauge numbers of those particles can be found in the
table next.

\vskip 1truecm

\begin{tabular}{|c|c|c|c|c|}
  \hline
  elementary particle& gauge charge ($Y,I_3,\lambda _3,\lambda _8$)
  & first generation& second generation& third generation \\
  \cline{1-5}
  neutrino doublet& ($-1, \frac {1}{2},0,0$)& $\nu _{eD}(1)$ & $\nu _{\mu D}(2)$ &
  $\nu _{\tau D}(3)$\\
  \cline{1-5}
neutrino singlet& ($0,0,0,0$)& $\nu _{eS}(4) $& $\nu _{\mu S}(5)$ &
  $\nu _{\tau S}(6)$\\
\cline{1-5} electrical-lepton doublet& ($-1,-\frac {1}{2},0,0$)&
$e_D(7) $& $\mu _D(8)$ &
  $\tau _D(9)$\\
\cline{1-5} electrical-lepton singlet& ($-2,0,0,0$)& $e_S(10) $&
$\mu _S(11)$ &
  $\tau _S(12)$\\
  \cline{1-5} red-(+quark) doublet& ($\frac {1}{3},\frac {1}{2},\frac {1}{2},
  \frac {\sqrt {3}}{6}$)& $u_{rD}(13) $&
$c_{rD}(14)$ &
  $t_{rD}(15)$\\
  \cline{1-5} red-(+quark) singlet& ($\frac {4}{3},0,\frac {1}{2},
  \frac {\sqrt {3}}{6}$)& $u_{rS}(16) $&
$c_{rS}(17)$ &
  $t_{rS}(18)$\\
  \cline{1-5} red-(-quark) doublet& ($\frac {1}{3},-\frac {1}{2},\frac {1}{2},
  \frac {\sqrt {3}}{6}$)& $d_{rD}(19) $&
$s_{rD}(20)$ &
  $b_{rD}(21)$\\
  \cline{1-5} red-(-quark) singlet& ($-\frac {2}{3},0,\frac {1}{2},
  \frac {\sqrt {3}}{6}$)& $d_{rS}(22) $&
$s_{rS}(23)$ &
  $b_{rS}(24)$\\
  \cline{1-5} green-(+quark) doublet& ($\frac {1}{3},\frac {1}{2},-\frac {1}{2},
  \frac {\sqrt {3}}{6}$)& $u_{gD}(25) $&
$c_{gD}(26)$ &
  $t_{gD}(27)$\\
  \cline{1-5} green-(+quark) singlet& ($\frac {4}{3},0,-\frac {1}{2},
  \frac {\sqrt {3}}{6}$)& $u_{gS}(28) $&
$c_{gS}(29)$ &
  $t_{gS}(30)$\\
   \cline{1-5} green-(-quark) doublet& ($\frac {1}{3},-\frac {1}{2},-\frac {1}{2},
  \frac {\sqrt {3}}{6}$)& $d_{gD}(31) $&
$s_{gD}(32)$ &
  $b_{gD}(33)$\\
  \cline{1-5} green-(-quark) singlet& ($-\frac {2}{3},0,-\frac {1}{2},
  \frac {\sqrt {3}}{6}$)& $d_{gS}(34) $&
$s_{gS}(35)$ &
  $b_{gS}(36)$\\
  \cline{1-5} blue-(+quark) doublet& ($\frac {1}{3},\frac {1}{2},0,
  \frac {\sqrt {3}}{3}$)& $u_{bD}(37) $&
$c_{bD}(38)$ &
  $t_{bD}(39)$\\
  \cline{1-5} blue-(-quark) singlet& ($\frac {4}{3},0,0,
  \frac {\sqrt {3}}{3}$)& $u_{bS}(40) $&
$c_{bS}(41)$ &
  $t_{bS}(42)$\\
  \cline{1-5} blue-(+quark) doublet& ($\frac {1}{3},-\frac {1}{2},0,
  \frac {\sqrt {3}}{3}$)& $d_{bD}(43) $&
$s_{bD}(44)$ &
  $b_{bD}(45)$\\
  \cline{1-5} blue-(-quark) singlet& ($-\frac {2}{3},0,0,
  \frac {\sqrt {3}}{3}$)& $d_{bS}(46) $&
$s_{bS}(47)$ &
  $b_{bS}(48)$\\
 \hline
\end{tabular}
\vskip 1truecm

The spin of the matter particles is $\frac {1}{2}$ corresponding to
fermions. The spin is represented by $s_3$ which take values $\pm
1/2$.

Gauge particles are spin-1 corresponding to bosons. The gauge forces
are divided as electromagnetic force, weak-interaction force and
color force, correspondingly the gauge particles are photons $\gamma
$, weak interaction bosons $W_{\pm },Z_0$ and gluons
$g_{(i)},i=1,2,\cdots ,8$. One photon and three weak bosons
corresponds to 4 generators of electric-weak gauge group
$U(1)\otimes SU(2)$. Gluons are responsible for color force, eight
gluons corresponds to 8 generators of group $SU(3)$. The following
table shows the quantum numbers of gauge particles, where masses of
gluons are equal and take value $m$.

\vskip 1truecm

\begin{tabular}{|c|c|c|c|c|c|c|c|c|c|c|c|c|}
  \hline
  &1&2&3&4&5&6&7&8&9&10&11&12\\
  \cline{1-13}
&$\gamma $&$Z_0$&$W_+$&$W_-$&$g_{(1)}$
&$g_{(2)}$&$g_{(3)}$&$g_{(4)}$&$g_{(5)}$&$g_{(6)}$&$g_{(7)}$&$g_{(8)}$\\
  \cline{1-13}
mass&0&$m_Z$&$m_{W_+}$&$m_{W_-}$&m&m&m&m&m&m&m&m\\
\cline{1-13}
$Y$&0&0&0&0&0&0&0&0&0&0&0&0\\
\cline{1-13}
$I_3$&0&0&+1&-1&0&0&0&0&0&0&0&0\\
\cline{1-13} $\lambda _3$&0&0&0&0&0&0&+1&-1&$-\frac {1}{2}$&$+\frac
{1}{2}$
&$-\frac {1}{2}$&$+\frac {1}{2}$\\
\cline{1-13}
$\lambda _8$&0&0&0&0&0&0&0&0&$+\frac {\sqrt {3}}{2}$
&$-\frac {\sqrt {3}}{2}$&$-\frac {\sqrt {3}}{2}$
&$+\frac {\sqrt {3}}{2}$\\
\hline
\end{tabular}
\vskip 1truecm

There is one kind of graviton, it has complicated spin
representation, its mass, electric-charge and color charge are all
zeroes. Also we do not assume there exist Higgs particles.

\subsection{Matrix representations of the gauge charges}
In gauge basis $|e_t\rangle $ of matter particles, gauge charges are
represented as $48\times 48$ matrices. As we know, those gauge
charges are hypercharge $\hat{Y}$, isospin charges $\hat{I}_i,
(i=1,2,3)$ and color charges $\hat{\lambda} _p, (p=1,2,...,8)$. We
will present explicitly the matrix representations of those gauge
charges.

\subsubsection{Matrix representation of hypercharge, isospin charges and color charges}
We will next use the following notations: $\hat {0}$ is a $3\times
3$ matrix with all elements zeros, $\hat {I}$ is an $3\times 3$
identity matrix, $\hat {U}_l$ and  $\hat {U}_q$ are the
Kobayashi-Maskawa mixed matrices of leptons and quarks which will be
presented later.

\begin{eqnarray}
\hat {Y}=\left( \begin{array}{cccc} -\hat {I} & \hat {0} & \hat {0}
& \hat {0} \\
\hat {0} & \hat {0} & \hat {0} & \hat {0}\\
\hat {0} & \hat {0} & -\hat {I} & \hat {0}\\
\hat {0} & \hat {0} & \hat {0} & -2\hat {I}
\end{array}\right) \oplus \hat {I}\otimes
\left( \begin{array}{cccc} \frac {1}{3}\hat {I} & \hat {0} & \hat
{0}
& \hat {0} \\
\hat {0} & \frac {4}{3}\hat {I} & \hat {0} & \hat {0}\\
\hat {0} & \hat {0} & \frac {1}{3}\hat {I} & \hat {0}\\
\hat {0} & \hat {0} & \hat {0} & -\frac {2}{3}\hat {I}
\end{array}\right)
\end{eqnarray}

\begin{eqnarray}
\hat {I}_1=\frac {1}{2}\left\{ \left( \begin{array}{cccc} \hat {0} &
\hat {0} & \hat {U}_l
& \hat {0} \\
\hat {0} & \hat {0} & \hat {0} & \hat {0}\\
\hat {U}_l^{\dagger } & \hat {0} & \hat {0} & \hat {0}\\
\hat {0} & \hat {0} & \hat {0} & \hat {0}
\end{array}\right) \oplus \hat {I}\otimes
\left( \begin{array}{cccc} \hat {0} & \hat {0} & \hat {U}_q^{\dagger
}
& \hat {0} \\
\hat {0} & \hat {0} & \hat {0} & \hat {0}\\
\hat {U}_q & \hat {0} & \hat {0} & \hat {0}\\
\hat {0} & \hat {0} & \hat {0} & \hat {0}
\end{array}\right) \right\} ,
\end{eqnarray}

\begin{eqnarray}
\hat {I}_2=\frac {1}{2}\left\{ \left( \begin{array}{cccc} \hat {0} &
\hat {0} & -i\hat {U}_l
& \hat {0} \\
\hat {0} & \hat {0} & \hat {0} & \hat {0}\\
i\hat {U}_l^{\dagger } & \hat {0} & \hat {0} & \hat {0}\\
\hat {0} & \hat {0} & \hat {0} & \hat {0}
\end{array}\right) \oplus \hat {I}\otimes
\left( \begin{array}{cccc} \hat {0} & \hat {0} & -i\hat
{U}_q^{\dagger }
& \hat {0} \\
\hat {0} & \hat {0} & \hat {0} & \hat {0}\\
i\hat {U}_q & \hat {0} & \hat {0} & \hat {0}\\
\hat {0} & \hat {0} & \hat {0} & \hat {0}
\end{array}\right) \right\} ,
\end{eqnarray}

\begin{eqnarray}
\hat {I}_3=\frac {1}{2}\left\{ \left( \begin{array}{cccc} \hat {I} &
\hat {0} & \hat {0}
& \hat {0} \\
\hat {0} & \hat {0} & \hat {0} & \hat {0}\\
\hat {0} & \hat {0} & -\hat {I} & \hat {0}\\
\hat {0} & \hat {0} & \hat {0} & \hat {0}
\end{array}\right) \oplus \hat {I}\otimes
\left( \begin{array}{cccc} \hat {I} & \hat {0} & \hat {0}
& \hat {0} \\
\hat {0} & \hat {0} & \hat {0} & \hat {0}\\
\hat {0} & \hat {0} & -\hat {I} & \hat {0}\\
\hat {0} & \hat {0} & \hat {0} & \hat {0}
\end{array}\right) \right\} ,
\end{eqnarray}

\begin{eqnarray}
\hat {\lambda }_p= \left( \begin{array}{cccc} \hat {0} & \hat {0} &
\hat {0}
& \hat {0} \\
\hat {0} & \hat {0} & \hat {0} & \hat {0}\\
\hat {0} & \hat {0} & -\hat {I} & \hat {0}\\
\hat {0} & \hat {0} & \hat {0} & \hat {0}
\end{array}\right) \oplus (\hat {\lambda }_p)_{3\times 3}\otimes
\left( \begin{array}{cccc} \hat {I} & \hat {0} & \hat {0}
& \hat {0} \\
\hat {0} & \hat {I} & \hat {0} & \hat {0}\\
\hat {0} & \hat {0} & \hat {I} & \hat {0}\\
\hat {0} & \hat {0} & \hat {0} & \hat {I}
\end{array}\right) ,
\end{eqnarray}
where $(\hat {\lambda }_p)_{3\times 3}$ are $3\times 3$ Gell-Mann
matrices which take the forms,

\begin{eqnarray}
\hat {\lambda }_1=\frac {1}{2}\left[ \begin{array}{ccc} 0&1&0\\
1&0&0 \\
0&0&0\end{array}\right], ~~~
\hat {\lambda }_2=\frac {1}{2}\left[ \begin{array}{ccc} 0&-i&0\\
i&0&0 \\
0&0&0\end{array}\right], \nonumber
\end{eqnarray}

\begin{eqnarray}
\hat {\lambda }_3=\frac {1}{2}\left[ \begin{array}{ccc} 1&0&0\\
0&-1&0 \\
0&0&0\end{array}\right], ~~~
\hat {\lambda }_4=\frac {1}{2}\left[ \begin{array}{ccc} 0&0&1\\
0&0&0 \\
1&0&0\end{array}\right], \nonumber
\end{eqnarray}

\begin{eqnarray}
\hat {\lambda }_5=\frac {1}{2}\left[ \begin{array}{ccc} 0&0&-i\\
0&0&0 \\
i&0&0\end{array}\right], ~~~
\hat {\lambda }_6=\frac {1}{2}\left[ \begin{array}{ccc} 0&0&0\\
0&0&1 \\
0&1&0\end{array}\right], \nonumber
\end{eqnarray}

\begin{eqnarray}
\hat {\lambda }_7=\frac {1}{2}\left[ \begin{array}{ccc} 0&0&0\\
0&0&-i \\
0&i&0\end{array}\right], ~~~
\hat {\lambda }_8=\frac {1}{2\sqrt {3}}\left[ \begin{array}{ccc} 1&0&0\\
0&1&0 \\
0&0&-2\end{array}\right]. \nonumber
\end{eqnarray}

We can find the matrices of $\hat {I}^2$ and $\hat {\lambda }^2$
take the form
\begin{eqnarray}
\hat {I}^2=\frac {3}{4}\left\{ \left( \begin{array}{cccc} \hat {I} &
\hat {0} & \hat {0}
& \hat {0} \\
\hat {0} & \hat {0} & \hat {0} & \hat {0}\\
\hat {0} & \hat {0} & \hat {I} & \hat {0}\\
\hat {0} & \hat {0} & \hat {0} & \hat {0}
\end{array}\right) \oplus \hat {I}\otimes
\left( \begin{array}{cccc} \hat {I} & \hat {0} & \hat {0}
& \hat {0} \\
\hat {0} & \hat {0} & \hat {0} & \hat {0}\\
\hat {0} & \hat {0} & \hat {I} & \hat {0}\\
\hat {0} & \hat {0} & \hat {0} & \hat {0}
\end{array}\right) \right\} ,
\end{eqnarray}

\begin{eqnarray}
\hat {\lambda }^2=\frac {4}{3}\left\{ \left( \begin{array}{cccc}
\hat {0} & \hat {0} & \hat {0}
& \hat {0} \\
\hat {0} & \hat {0} & \hat {0} & \hat {0}\\
\hat {0} & \hat {0} & \hat {0} & \hat {0}\\
\hat {0} & \hat {0} & \hat {0} & \hat {0}
\end{array}\right) \oplus \hat {I}\otimes
\left( \begin{array}{cccc} \hat {I} & \hat {0} & \hat {0}
& \hat {0} \\
\hat {0} & \hat {I} & \hat {0} & \hat {0}\\
\hat {0} & \hat {0} & \hat {I} & \hat {0}\\
\hat {0} & \hat {0} & \hat {0} & \hat {I}
\end{array}\right) \right\} .
\end{eqnarray}

The operators $\hat {Y}, \hat {I}, \hat {\lambda }_p$ are unitary,
and the following properties are satisfied,
\begin{eqnarray}
\hat {Y}^{\dagger }&=&\hat {Y}, \nonumber \\
\hat {I}^{\dagger }&=&\hat {I}, \nonumber \\
\hat {\lambda }_p^{\dagger }&=&\hat {\lambda }_p, \nonumber \\
\langle \hat {Y}, \hat {Y}\rangle &=&{\rm tr}(\overline{\hat
{Y}}\hat
{Y})=40, \nonumber \\
\langle \hat {I}_i, \hat {I}_j\rangle &=&{\rm tr}(\overline{\hat
{I}_i}\hat
{I}_j)=6\delta _{ij}, \nonumber \\
\langle \hat {\lambda }_p, \hat {\lambda }_q\rangle &=&{\rm
tr}(\overline{\hat {\lambda }_p}\hat
{\lambda }_q)=6\delta _{pq}, \nonumber \\
\langle \hat {Y}, \hat {I}_i\rangle &=& \langle \hat {Y}, \hat
{\lambda }_p\rangle = \langle \hat {I}_i, \hat {\lambda }_p\rangle
=0.
\end{eqnarray}

\subsubsection{Matrix representation of the eigen-gauge charges}
Due to the transformation between orthogonal gauge charges $\hat
{t}_a$ and the eigen-gauge charges $\hat {T}_a$ and the matrices
representations of hypercharge $\hat {Y}$, isospin charges $\hat
{I}_i$ and the color charges $\hat {\lambda }_p$, the matrix
representations of the eigen-gauge charges take the form

\begin{eqnarray}
\hat {T}_1=e\left\{ \left( \begin{array}{cccc} \hat {0} & \hat {0} &
\hat {0}
& \hat {0} \\
\hat {0} & \hat {0} & \hat {0} & \hat {0}\\
\hat {0} & \hat {0} & -\hat {I} & \hat {0}\\
\hat {0} & \hat {0} & \hat {0} & -\hat {I}
\end{array}\right) \oplus \hat {I}\otimes
\left( \begin{array}{cccc} \frac {2}{3}\hat {I} & \hat {0} & \hat
{0}
& \hat {0} \\
\hat {0} & \frac {2}{3}\hat {I} & \hat {0} & \hat {0}\\
\hat {0} & \hat {0} & -\frac {1}{3}\hat {I} & \hat {0}\\
\hat {0} & \hat {0} & \hat {0} & -\frac {1}{3}\hat {I}
\end{array}\right) \right\} ,
\end{eqnarray}

\begin{eqnarray}
\hat {T}_2&=&\frac {\sqrt {g_1^2+g_2^2}}{2}\left\{ \left(
\begin{array}{cccc} \hat {I} & \hat {0} & \hat {0}
& \hat {0} \\
\hat {0} & \hat {0} & \hat {0} & \hat {0}\\
\hat {0} & \hat {0} & (-1+2\sin ^2\theta _w)\hat {I} & \hat {0}\\
\hat {0} & \hat {0} & \hat {0} & 2\sin ^2\theta _w\hat {I}
\end{array}\right)\right.
\nonumber \\
&&\left. \oplus \hat {I}\otimes \left( \begin{array}{cccc} (1-\frac
{4}{3}\sin ^2\theta _w)\hat {I} & \hat {0} & \hat {0}
& \hat {0} \\
\hat {0} & -\frac {4}{3}\sin ^2\theta _w\hat {I} & \hat {0} & \hat {0}\\
\hat {0} & \hat {0} & (-1+\frac {2}{3}\sin ^2\theta _w)\hat {I} & \hat {0}\\
\hat {0} & \hat {0} & \hat {0} & \frac {2}{3}\sin ^2\theta _w\hat
{I}
\end{array}\right) \right\} ,
\end{eqnarray}

\begin{eqnarray}
\hat {T}_3=\frac {g_2}{\sqrt {2}}\left\{ \left( \begin{array}{cccc}
\hat {0} & \hat {0} & \hat {U}_l
& \hat {0} \\
\hat {0} & \hat {0} & \hat {0} & \hat {0}\\
\hat {0} & \hat {0} & \hat {0} & \hat {0}\\
\hat {0} & \hat {0} & \hat {0} & \hat {0}
\end{array}\right) \oplus \hat {I}\otimes
\left( \begin{array}{cccc} \hat {0} & \hat {0} & \hat {U}_q^{\dagger
}
& \hat {0} \\
\hat {0} & \hat {0} & \hat {0} & \hat {0}\\
\hat {0} & \hat {0} & \hat {0} & \hat {0}\\
\hat {0} & \hat {0} & \hat {0} & \hat {0}
\end{array}\right) \right\} ,
\end{eqnarray}

\begin{eqnarray}
\hat {T}_4=\frac {g_2}{\sqrt {2}}\left\{ \left( \begin{array}{cccc}
\hat {0} & \hat {0} & \hat {0}
& \hat {0} \\
\hat {0} & \hat {0} & \hat {0} & \hat {0}\\
\hat {U}_l^{\dagger } & \hat {0} & \hat {0} & \hat {0}\\
\hat {0} & \hat {0} & \hat {0} & \hat {0}
\end{array}\right) \oplus \hat {I}\otimes
\left( \begin{array}{cccc} \hat {0} & \hat {0} & \hat {0}
& \hat {0} \\
\hat {0} & \hat {0} & \hat {0} & \hat {0}\\
\hat {U}_q & \hat {0} & \hat {0} & \hat {0}\\
\hat {0} & \hat {0} & \hat {0} & \hat {0}
\end{array}\right) \right\} ,
\end{eqnarray}

\begin{eqnarray}
\hat {T}_{4+p}=g_3\left\{ \left(
\begin{array}{cccc} \hat {0} & \hat {0} & \hat {0}
& \hat {0} \\
\hat {0} & \hat {0} & \hat {0} & \hat {0}\\
\hat {0} & \hat {0} & \hat {0} & \hat {0}\\
\hat {0} & \hat {0} & \hat {0} & \hat {0}
\end{array}\right) \oplus (\hat {\Lambda }_p)_{3\times 3}\otimes
\left( \begin{array}{cccc} \hat {I} & \hat {0} & \hat {0}
& \hat {0} \\
\hat {0} & \hat {I} & \hat {0} & \hat {0}\\
\hat {0} & \hat {0} & \hat {I} & \hat {0}\\
\hat {0} & \hat {0} & \hat {0} & \hat {I}
\end{array}\right) \right\}
\end{eqnarray}

Here matrix $(\hat {\Lambda }_p)_{3\times 3}$ are the representation
of algebra $su(3)$ in Cartan-Weyl basis:

\begin{eqnarray}
\hat {\Lambda }_1=\frac {1}{2}\left[ \begin{array}{ccc} 1&0&0\\
0&-1&0 \\
0&0&0\end{array}\right], ~~~
\hat {\Lambda }_2=\frac {1}{2\sqrt {3}}\left[ \begin{array}{ccc} 1&0&0\\
0&1&0 \\
0&0&-2\end{array}\right], \nonumber
\end{eqnarray}

\begin{eqnarray}
\hat {\Lambda }_3=\frac {1}{\sqrt {2}}\left[ \begin{array}{ccc} 0&1&0\\
0&0&0 \\
0&0&0\end{array}\right], ~~~
\hat {\Lambda }_4=\frac {1}{\sqrt {2}}\left[ \begin{array}{ccc} 0&0&0\\
1&0&0 \\
0&0&0\end{array}\right], \nonumber
\end{eqnarray}

\begin{eqnarray}
\hat {\Lambda }_5=\frac {1}{\sqrt {2}}\left[ \begin{array}{ccc} 0&0&0\\
0&0&1 \\
0&0&0\end{array}\right], ~~~
\hat {\Lambda }_6=\frac {1}{\sqrt {2}}\left[ \begin{array}{ccc} 0&0&0\\
0&0&0 \\
0&1&0\end{array}\right], \nonumber
\end{eqnarray}

\begin{eqnarray}
\hat {\Lambda }_7=\frac {1}{\sqrt {2}}\left[ \begin{array}{ccc} 0&0&0\\
0&0&0 \\
1&0&0\end{array}\right], ~~~
\hat {\lambda }_8=\frac {1}{\sqrt {2}}\left[ \begin{array}{ccc} 0&0&1\\
0&0&0 \\
0&0&0\end{array}\right].
\end{eqnarray}

One can check that we have the following property:
\begin{eqnarray}
\hat {T}_a^{\dagger }&=&G^{ab}\hat {T}_b=\hat {T}^a, \\
\overline{\hat {T}}_a&=&\hat {T}^a.
\end{eqnarray}

\subsection{Mass matrices of the matter particles}
Mass is of fundamental for us. One feature of our theory is that the
mass matrices are defined in gauge space. Conventionally, the gauge
invariant is necessary for gauge theory, thus mass can only be
created by gauge symmetry breaking caused by Higgs mechanism. In
comparison for our theory, weak interaction gauge space does not
satisfy gauge invariant, but satisfy the general covariance
condition. Thus mass can be represented as covariance matrix in
gauge space. We denote the mass matrix in gauge basis $|e_t\rangle $
as:

\begin{eqnarray}
\hat {m}=\left( \begin{array}{cccc} \hat {0} & \hat {m}_{l0} & \hat
{0}
& \hat {0} \\
\hat {m}_{l0} & \hat {0} & \hat {0} & \hat {0}\\
\hat {0} & \hat {0} & \hat {0} & \hat {m}_{l-}\\
\hat {0} & \hat {0} & \hat {m}_{l-} & \hat {0}
\end{array}\right) \oplus \hat {I}\otimes
\left( \begin{array}{cccc} \hat {0} & \hat {m}_{q+} & \hat {0}
& \hat {0} \\
\hat {m}_{q+} & \hat {0} & \hat {0} & \hat {0}\\
\hat {0} & \hat {0} & \hat {0} & \hat {m}_{q-}\\
\hat {0} & \hat {0} & \hat {m}_{q-} & \hat {0}
\end{array}\right) ,
\label{mmass}
\end{eqnarray}
where $\hat {m}_{l0},\hat {m}_{l-},\hat {m}_{q+}$ and $\hat
{m}_{q-}$ are mass matrices of neutral-lepton, electric-lepton,
positive-electric-quark and negative-electric-quark, respectively.

Suppose the masses of three neutrinos are $m_{\nu e}, m_{\nu \mu},
m_{\nu \tau }$, the mass matrices take the form
\begin{eqnarray}
\hat {m}_{l0}=\left( \begin{array}{ccc} m_{\nu e}&0&0\\
0&m_{\nu \mu} &0\\
0&0&m_{\nu \tau }\end{array} \right)
\end{eqnarray}
The three electric-leptons have masses $m_{e}, m_{\mu }, m_{\tau }$,
we denote
\begin{eqnarray}
\hat {m}_{l-}=\left( \begin{array}{ccc} m_{e}&0&0\\
0&m_{\mu} &0\\
0&0&m_{\tau }\end{array} \right).
\end{eqnarray}
Also we denote three positive-electric-quarks have masses $m_{u},
m_{c}, m_{t}$, three negative-electric-quarks have masses $m_{d},
m_{s}, m_{b}$, and the matrices as the following forms
\begin{eqnarray}
\hat {m}_{q+}=\left( \begin{array}{ccc} m_{u}&0&0\\
0&m_{c} &0\\
0&0&m_{t}\end{array} \right).\\
\hat {m}_{q-}=\left( \begin{array}{ccc} m_{d}&0&0\\
0&m_{s} &0\\
0&0&m_{b}\end{array} \right).
\end{eqnarray}
The coupling constant of the electric-charge is
\begin{eqnarray}
e=\frac {g_1g_2}{\sqrt {g_1^2+g_2^2}}
\end{eqnarray}
The coupling constant of the weak charges is
\begin{eqnarray}
g_Z=\sqrt {g_1^2+g_2^2}
\end{eqnarray}
The Weinberg angle takes the form
\begin{eqnarray}
\cos \theta _w=\frac {g_2}{\sqrt {g_1^2+g_2^2}},\\
\sin \theta _w=\frac {g_1}{\sqrt {g_1^2+g_2^2}}
\end{eqnarray}
The angles between weak interactions of three generations of quarks
are denoted by $\theta _1, \theta _2, \theta _3$, the PC broken
symmetry factor of weak interaction is $\delta _1$. And the
Kobayashi-Maskawa matrices \cite{kmmatrix} for weak interaction are defined as
\begin{eqnarray}
\hat {U}_q=\left( \begin{array}{ccc} c_1&-s_1c_2&-s_1c_2\\
s_1c_3&c_1c_2c_3+s_2s_3e_1&c_1s_2c_3-c_2s_3e_1\\
s_1c_3&c_1c_2s_3-s_2c_3e_1&c_1s_2s_3+c_2c_3e_1\end{array} \right)
\end{eqnarray}
where $s_i=\sin \theta _i, c_i=\cos \theta _i, (i=1,2,3)$, $e_1=\exp
(-i\delta _1)$.

Similarly for three generations of leptons, three angles of weak
interaction are denoted as $\theta _4, \theta _5, \theta _6$, the
broken symmetry factor is $\delta _2$, and the Kobayashi-Kaskawa
matrices take the form
\begin{eqnarray}
\hat {U}_l=\left( \begin{array}{ccc} c_4&-s_4c_5&-s_4c_5\\
s_4c_6&c_4c_5c_6+s_5s_6e_2&c_4s_5c_5-c_5s_6e_2\\
s_4c_6&c_4c_5s_6-s_5c_6e_2&c_4s_5s_6+c_5c_6e_2\end{array} \right)
\end{eqnarray}
where similarly, $s_j=\sin \theta _j, c_j=\cos \theta _j,
(j=4,5,6)$, $e_2=\exp (-i\delta _2)$.

The commutation relation between gauge charges $\hat {T}_a$ and the
mass matrices of matter particles is

\begin{eqnarray}
[\hat {T}_a, \hat {m}]=\hat {m}_a
\end{eqnarray}

where
\begin{eqnarray}
\hat {m}_1&=&0,
\\
\hat {m}_2&=&\frac {g_Z}{2}\left\{ \left( \begin{array}{cccc} \hat
{0} & \hat {m}_{l0} & \hat {0}
& \hat {0} \\
-\hat {m}_{l0} & \hat {0} & \hat {0} & \hat {0}\\
\hat {0} & \hat {0} & \hat {0} & -\hat {m}_{l-}\\
\hat {0} & \hat {0} & \hat {m}_{l-} & \hat {0}
\end{array}\right) \oplus \hat {I}\otimes
\left( \begin{array}{cccc} \hat {0} & \hat {m}_{q+} & \hat {0}
& \hat {0} \\
-\hat {m}_{q+} & \hat {0} & \hat {0} & \hat {0}\\
\hat {0} & \hat {0} & \hat {0} & -\hat {m}_{q-}\\
\hat {0} & \hat {0} & \hat {m}_{q-} & \hat {0}
\end{array}\right) \right\} , \\
\hat {m}_3&=&\frac {g_2}{\sqrt {2}}\left\{ \left(
\begin{array}{cccc} \hat {0} & \hat {0} & \hat {0}
& \hat {U}_l\hat {m}_{l0} \\
\hat {0} & \hat {0} & -\hat {m}_{l-}\hat {U}_q & \hat {0}\\
\hat {0} & \hat {0} & \hat {0} & \hat {0}\\
\hat {0} & \hat {0} & \hat {0} & \hat {0}
\end{array}\right) \oplus \hat {I}\otimes
\left( \begin{array}{cccc} \hat {0} & \hat {0} & \hat {0}
& \hat {U}_q^{\dagger }\hat {m}_{q+} \\
\hat {0} & \hat {0} & -\hat {m}_{q-}\hat {U}_q^{\dagger } & \hat {0}\\
\hat {0} & \hat {0} & \hat {0} & \hat {0}\\
\hat {0} & \hat {0} & \hat {0} & \hat {0}
\end{array}\right) \right\} ,
\\
\hat {m}_4&=&\frac {g_2}{\sqrt {2}}\left\{ \left(
\begin{array}{cccc} \hat {0} & \hat {0} & \hat {0}
&\hat {0} \\
\hat {0} & \hat {0} & \hat {0}& \hat {0}\\
\hat {0} & \hat {U}_l^{\dagger }\hat {m}_{l-} & \hat {0} & \hat {0}\\
-\hat {m}_{l0}\hat {U}_l^{\dagger } & \hat {0} & \hat {0} & \hat {0}
\end{array}\right) \oplus \hat {I}\otimes
\left( \begin{array}{cccc} \hat {0} & \hat {0} & \hat {0}
& \hat {0} \\
\hat {0} & \hat {0} & \hat {0} & \hat {0}\\
\hat {0} & \hat {U}_q\hat {m}_{q-} & \hat {0} & \hat {0}\\
-\hat {m}_{q+}\hat {U}_q & \hat {0} & \hat {0} & \hat {0}
\end{array}\right) \right\} ,\\
\hat{m}_{4+p}&=&0
\end{eqnarray}
In the above commutation relations, mass matrices commute with
electric-charge and color charge, while does not commute with weak
charges.

Also we have $\hat {m}^{\dagger }=\hat {m}$ and $\overline{\hat
{m}}=-\hat {m}$.

Here are some comments about the mass matrix.  On gauge basis of
matter particles $|e_t\rangle $, the matrix representation of the
following operators are diagonal: hypercharge $\hat {Y}$, the third
element of the isospin $\hat {I}_3$, the third element of the color
charge $\hat {\lambda }_3$, the eighth element of the color charge
$\hat {\lambda }_8$. So basis $|e_t\rangle $ is the common
eigenvector of those operators. The mass matrix $\hat {m}$ is
quasi-diagonal. The eigenvalues are $Y$, $I_3$, $\lambda _3$,
$\lambda _8$, $m$ corresponding to $\hat {Y}$, $\hat {I}_3$, $\hat
{\lambda }_3$, $\hat {\lambda }_8$, $\hat {m}$, respectively. Those
quantities are hypercharge, third element of the isospin, third and
eighth elements of the color charge and mass of the matter
particles.

\subsection{Projection operators}
Here let us define some projectors of gauge space related with
isospin singlet and isospin doublet.

(1)Projection operators $\hat {P}_{D}, \hat {P}_{S}$ for isospin
singlet and doublet are written as
\begin{eqnarray}
\hat {P}_{D}&=&\frac {4}{3}\hat {I}^2
\\
\hat {P}_{S}&=&1-\frac {4}{3}\hat {I}^2.
\end{eqnarray}
The singlet projector and the doublet projector satisfy the
equations,
\begin{eqnarray}
\hat {P}_{D}+\hat {P}_{S}&=&1 \nonumber \\
\hat {P}_{D}\hat {P}_{D}&=&\hat {P}_{D} \nonumber \\
\hat {P}_{S}\hat {P}_{S}&=&\hat {P}_{S} \nonumber \\
\hat {P}_{D}\hat {P}_{S}&=&0
\end{eqnarray}

(2)Similarly, lepton projector and quark projector are written as
the following and have the properties,
\begin{eqnarray}
\hat {P}_{l}&=&1-\frac {3}{4}\hat {\lambda }^2
\\
\hat {P}_{q}&=&\frac {3}{4}\hat {\lambda }^2,
\\
\hat {P}_{l}+\hat {P}_{q}&=&1 \nonumber \\
\hat {P}_{l}\hat {P}_{l}&=&\hat {P}_{l} \nonumber \\
\hat {P}_{q}\hat {P}_{q}&=&\hat {P}_{q} \nonumber \\
\hat {P}_{l}\hat {P}_{q}&=&0
\end{eqnarray}
Those projectors will be useful when we consider later Dirac
equation.

\subsection{Eigenvectors of the weak interaction}
Gauge charges of the weak interaction are $\hat {T}_3$ and $\hat
{T}_4$. On the gauge charge basis of matter particles, $\hat {T}_3$
and $\hat {T}_4$ include the Kobayashi-Maskawa matrices $\hat {U}_l$
and $\hat {U}_q$, so $|e_t\rangle $ is not the eigenvector for weak
interaction. The basis can be changed by unitary transformation to
the eigenvector of the weak interaction,
\begin{eqnarray}
|e_{t'}'\rangle =\hat {U}_{t'}^t|e_t\rangle ,
\end{eqnarray}
where this $\hat {U}$ is a $48\times 48$ unitary matrix and can be
represented as
\begin{eqnarray}
\hat {U}&=&\left( \begin{array}{cccc} \hat {U}_l^{\dagger } &\hat
{0}&\hat {0}&\hat {0} \\
\hat {0}&\hat {U}_l^{\dagger }&\hat {0}&\hat {0} \\
\hat {0}&\hat {0}&\hat {I}&\hat {0} \\
\hat {0}&\hat {0}&\hat {0}&\hat {I} \end{array}\right) \oplus \hat
{I}\otimes \left( \begin{array}{cccc} \hat {I}&\hat
{0}&\hat {0}&\hat {0} \\
\hat {0}&\hat {I}&\hat {0}&\hat {0} \\
\hat {0}&\hat {0}&\hat {U}_q^{\dagger }&\hat {0} \\
\hat {0}&\hat {0}&\hat {0}&\hat {U}_q^{\dagger } \end{array}\right)
\nonumber \\
\hat {U}\hat {U}^{\dagger }&=&1.
\end{eqnarray}

On the basis of the eigenvector of weak interaction $|e_{t'}'\rangle
, (t'=1,2,...,48)$, operators $\hat {I}_1, \hat {I}_2, \hat {T}_3,
\hat {T}_4$ and $\hat {m}$ can be represented as
\begin{eqnarray}
\hat {I_1'}=\hat {U}\hat {I_1}\hat {U}^{\dagger }=\frac
{1}{2}\left\{ \left(
\begin{array}{cccc} \hat {0} &\hat
{0}&\hat {I}&\hat {0} \\
\hat {0}&\hat {0}&\hat {0}&\hat {0} \\
\hat {I}&\hat {0}&\hat {0}&\hat {0} \\
\hat {0}&\hat {0}&\hat {0}&\hat {0} \end{array}\right) \oplus \hat
{I}\otimes \left( \begin{array}{cccc} \hat {0}&\hat
{0}&\hat {I}&\hat {0} \\
\hat {0}&\hat {0}&\hat {0}&\hat {0} \\
\hat {I}&\hat {0}&\hat {0}&\hat {0} \\
\hat {0}&\hat {0}&\hat {0}&\hat {0} \end{array}\right) \right\} ,
\end{eqnarray}

\begin{eqnarray}
\hat {I_2'}=\hat {U}\hat {I_2}\hat {U}^{\dagger }=\frac
{1}{2}\left\{ \left(
\begin{array}{cccc} \hat {0} &\hat
{0}&-i\hat {I}&\hat {0} \\
\hat {0}&\hat {0}&\hat {0}&\hat {0} \\
i\hat {I}&\hat {0}&\hat {0}&\hat {0} \\
\hat {0}&\hat {0}&\hat {0}&\hat {0} \end{array}\right) \oplus \hat
{I}\otimes \left( \begin{array}{cccc} \hat {0}&\hat
{0}&-i\hat {I}&\hat {0} \\
\hat {0}&\hat {0}&\hat {0}&\hat {0} \\
i\hat {I}&\hat {0}&\hat {0}&\hat {0} \\
\hat {0}&\hat {0}&\hat {0}&\hat {0} \end{array}\right) \right\} ,
\end{eqnarray}

\begin{eqnarray}
\hat {T_3'}=\hat {U}\hat {T_3}\hat {U}^{\dagger }=\frac {g_2}{\sqrt
{2}}\left\{ \left(
\begin{array}{cccc} \hat {0} &\hat
{0}&\hat {I}&\hat {0} \\
\hat {0}&\hat {0}&\hat {0}&\hat {0} \\
\hat {0}&\hat {0}&\hat {0}&\hat {0} \\
\hat {0}&\hat {0}&\hat {0}&\hat {0} \end{array}\right) \oplus \hat
{I}\otimes \left( \begin{array}{cccc} \hat {0}&\hat
{0}&\hat {I}&\hat {0} \\
\hat {0}&\hat {0}&\hat {0}&\hat {0} \\
\hat {0}&\hat {0}&\hat {0}&\hat {0} \\
\hat {0}&\hat {0}&\hat {0}&\hat {0} \end{array}\right) \right\} ,
\end{eqnarray}

\begin{eqnarray}
\hat {T_4'}=\hat {U}\hat {T_4}\hat {U}^{\dagger }=\frac {g_2}{\sqrt
{2}}\left\{ \left(
\begin{array}{cccc} \hat {0} &\hat
{0}&\hat {0}&\hat {0} \\
\hat {0}&\hat {0}&\hat {0}&\hat {0} \\
\hat {I}&\hat {0}&\hat {0}&\hat {0} \\
\hat {0}&\hat {0}&\hat {0}&\hat {0} \end{array}\right) \oplus \hat
{I}\otimes \left( \begin{array}{cccc} \hat {0}&\hat
{0}&\hat {0}&\hat {0} \\
\hat {0}&\hat {0}&\hat {0}&\hat {0} \\
\hat {I}&\hat {0}&\hat {0}&\hat {0} \\
\hat {0}&\hat {0}&\hat {0}&\hat {0} \end{array}\right) \right\} ,
\end{eqnarray}

\begin{eqnarray}
\hat {m}'=\hat {U}\hat {m}\hat {U}^{\dagger }=\left(
\begin{array}{cccc} \hat {0} & \hat {U}_l^{\dagger }\hat {m}_{l0}\hat {U}_l
& \hat {0}
& \hat {0}\\
\hat {U}_l^{\dagger }\hat {m}_{l0}\hat {U}_l& \hat {0} & \hat {0} & \hat {0}\\
\hat {0} & \hat {0} & \hat {0} & \hat {m}_{l-}\\
\hat {0} & \hat {0} & \hat {m}_{l-} & \hat {0}
\end{array}\right) \oplus \hat {I}\otimes
\left( \begin{array}{cccc} \hat {0} & \hat {m}_{q+} & \hat {0}
& \hat {0}\\
\hat {m}_{q+} & \hat {0} & \hat {0} & \hat {0}\\
\hat {0} & \hat {0} & \hat {0} & \hat {U}_q^{\dagger }\hat {m}_{q-}\hat {U}_q\\
\hat {0} & \hat {0} & \hat {U}_q^{\dagger }\hat {m}_{q-}\hat {U}_q &
\hat {0}
\end{array}\right),
\end{eqnarray}

Operators $\hat {Y}, \hat {I}_3, \hat {\lambda }_p, \hat {T}_1, \hat
{T}_2, \hat {T}_{4+p}$ are invariant under the unitary
transformation $\hat {U}$.

\subsection{The correspondence between gauge particles and gauge charges}
The 12 gauge particles corresponds to 12 elements of the gauge
charge $\hat {T}_a$. Photon $\gamma $ corresponds to gauge charge
$\hat {T}_1$, weak interaction bosons $Z_0, W_+, W_-$ correspond to
gauge charges $\hat {T}_2, \hat {T}_3, \hat {T}_4$, gluons $g_{(p)},
(p=1,2,...,8)$ correspond to gauge charges $\hat {T}_{4+p}$.

The representation of the gauge basis can take the form of 12
eigen-gauge charges, they satisfy the equation,
\begin{eqnarray}
[\hat {T}_a, \hat {T}_b]=C_{ab}^c\hat {T}_c,
\end{eqnarray}
The elements of the gauge charges can be the structure constants
\begin{eqnarray}
(\hat {T}_a)^c_b=C_{ab}^c.
\end{eqnarray}

The following properties can be checked by direct calculations,

\begin{eqnarray}
[\hat {Y}, \hat {T}_a]&=&Y\hat {T}_a,
\end{eqnarray}
\begin{eqnarray}
[\hat {I}_3, \hat {T}]&=&I_3\hat {T}_a,
\end{eqnarray}
\begin{eqnarray}
[\hat {\lambda }_3, \hat {T}_a]&=&\lambda _3\hat {T}_a,
\end{eqnarray}
\begin{eqnarray}
[\hat {\lambda }_8, \hat {T}_a]&=&\lambda _8\hat {T}_a,
\end{eqnarray}

\subsection{Mass matrix of the gauge particles}
The mass matrix of the gauge particles is defined as gauge tensor in
eigen-gauge charges $\hat {T}_a$,
\begin{eqnarray}
\hat {M}&=&M^a_b\hat {T}_a\otimes \hat {T}^b. \label{Mmass}
\end{eqnarray}
We consider respectively the weak interaction bosons part $\hat
{M}_{weak}$ and the gluon part $\hat {M}_{gluon}$, so
\begin{eqnarray}
\hat {M}=\hat {M}_{weak}+\hat {M}_{gluon}.
\end{eqnarray}

The mass matrix elements concerning about weak interaction bosons
can be written as,
\begin{eqnarray}
\hat {M}_{weak} &=&m_Z^2\hat {T}_2\otimes \hat {T}^2+m_W^2\hat
{T}_3\otimes \hat {T}^3+m_W^2\hat {T}_4\otimes \hat {T}^4,
\end{eqnarray}
where $m_Z=91188MeV$, is the mass of particle $Z_0$, $m_W=80398MeV$,
is the mass of particles $W_+, W_-$, they are connected through the
Weinberg angle as
\begin{eqnarray}
m_W=m_Z\cos \theta _w.
\end{eqnarray}
Due to the transformation between $\hat {t}_a$ and $\hat {T}_a$, we
have
\begin{eqnarray}
\hat {M}_{weak}&=&m^{ab}\hat {t}_a\otimes \hat {t}_b \nonumber \\
&=&m_Z^2(\sin ^2\theta _w\hat {t}_1\otimes \hat {t}_1-2\sin \theta
_w\cos \theta _{w}\hat {t}_1\otimes \hat {t}_4+\cos ^2\theta _w\hat
{t}_4\otimes \hat {t}_4) \nonumber \\
&&+m_Z^2\cos ^2\theta _w(\hat {t}_2\otimes \hat {t}_2+\hat
{t}_3\otimes \hat {t}_3) \nonumber \\
&=&m_Z^2\cos ^2\theta _w({\rm tg}^2\theta _w\hat {t}_1\otimes \hat
{t}_1-2{\rm tg}\theta _w\hat {t}_1\otimes \hat {t}_4+\hat
{t}_2\otimes \hat {t}_2+\hat {t}_3\otimes \hat {t}_3 +\hat
{t}_4\otimes \hat {t}_4) .
\end{eqnarray}
Note if $\hat {t}_1=0$, operators $\hat {t}_2, \hat {t}_3, \hat
{t}_4$ are symmetric for $\hat {M}$, that means $\hat {M}$ are
symmetric for isospin group $SU(2)$, and also $m_W=m_Z\cos \theta
_W$ is necessary for this symmetry.

In this work, we assume the masses of gluons are all $m$, thus we
have
\begin{eqnarray}
\hat {M}_{gluon}&=&M_{(4+p)}^{(4+p)}\hat {T}_{4+p}\otimes \hat
{T}^{4+p} \nonumber \\
&=&m^2\hat {T}_{4+p}\otimes \hat {T}^{4+p}
\end{eqnarray}
where $p=1,2,\cdots ,8$, $M_{(4+p)}^{(4+p)}=m^2$, and $\hat
{T}_{4+p}$ corresponds to color charges. This result will be useful
in studying the dark energy. Note that the $SU(3)$ symmetry does not
be broken by this mass matrix.

\subsection{Gauge representation for gravity field}

Gauge representation of gravity is 1D identity denoted as
\begin{eqnarray}
V_g(G)=T_g(0)=V_g(0,0,0,0)
\end{eqnarray}
We can choose the gauge basis for graviton as 1, gauge charges $\hat
{T}_a$ commute with 1, thus we can say the gauge charge of graviton
is zero.

\section{Representation theory}
In this work, particles or fields are defined by three class of
parameters: coordinate-momentum, spin and gauge charges. The
coordinates-momentum are independent with spin and gauge bosons.
Fields or particles and the interactions are represented as a vector
for matter fields or operator for force fields in a direct product
space by three class of spaces: coordinate-momentum space, spin
space and gauge space, $V(M)=V_{xp}\otimes V_S(M)\otimes V_g(M)$.
The quantum state $|e_{st}\rangle $ of a matter particle is defined
as the direct (tensor) product in coordinate-basis, spin-basis and
gauge-basis, $|e_{st}(x)\rangle =|x\rangle \otimes |e_s\rangle
\otimes |e_t\rangle $, where $x\in R^4, s=1,2,3,4$ and
$t=1,2,...,48$. It is the common-eigenstate of 10 operators $\hat
{x}^0,\hat {x}^1, \hat {x}^2, \hat {x}^3, \hat {\gamma }_5,\hat
{s}_{12}, \hat {Y}, \hat{I}_3, \hat {\lambda }_3$ and $\hat {\lambda
}_8$. In this representation, coordinate state can be changed to
momentum state $|x\rangle \rightarrow |p\rangle $ and we have
$|e_{st}(p)\rangle =|p\rangle \otimes |e_s\rangle \otimes
|e_t\rangle $. The quantum state of matter-particle $|\psi \rangle $
can then be expanded in either coordinate state $|e_{st}(x)\rangle $
or momentum state $|e_{st}(p)\rangle $, Please note in our
representation, spin, gauge and general coordinate-momentum are
dealt in the same positions.

Before proceed, we would like to briefly summarize our
representation results. The framework of this theory is that all
fields and particles are described by their representations with
three properties: coordinate-momentum, spin and gauge. Their
properties will be governed by three fundamental equations.

The quantum parameters are coordinate $\hat {x}^{\mu }$, momentum
$\hat {p}_{\mu }$, spin $\hat {s}_{\alpha \beta }$ and gauge charges
$\hat {T}_a$ (or $\hat {t}_a$), altogether there are 26 parameters.
The commutation relations for those parameters are:

\begin{eqnarray}
[\hat {x}^{\mu }, \hat {p}_{\nu }]&=&-i\delta ^{\mu }_{\nu },
\end{eqnarray}
\begin{eqnarray}
[\hat {x}^{\mu }, \hat {x}^{\nu }]&=&[\hat {p}_{\mu }, \hat {p}_{\nu
}]=0,
\end{eqnarray}
\begin{eqnarray}
[\hat {s}_{\alpha \beta }, \hat {s}_{\rho \sigma }]=-i(\eta _{\alpha
\rho}\hat {s}_{\beta \sigma }-\eta _{\beta \rho}\hat {s}_{\alpha
\sigma }+\eta _{\alpha \sigma}\hat {s}_{\rho \beta }-\eta _{\beta
\sigma}\hat {s}_{\rho \alpha }),
\end{eqnarray}
\begin{eqnarray}
[\hat {T}_a, \hat {T}_b]=C^c_{ab}\hat {T}_c,
\end{eqnarray}
\begin{eqnarray}
[\hat {x}^{\mu }, \hat {s}_{\alpha \beta }]=[\hat {x}^{\mu }, \hat
{T}_a]=[\hat {p}_{\mu }, \hat {s}_{\alpha \beta }]= [\hat {p}_{\mu
}, \hat {T}_a]=[\hat {s}_{\alpha \beta }, \hat {T}_a]=0.
\end{eqnarray}

The 26 quantum parameters and the unit constitute a Lie algebra $A$,
\begin{eqnarray}
A=\left\{ \hat {Z}: \hat {Z}=a_{\mu }\hat {x}^{\mu }+b^{\mu }\hat
{p}_{\mu }+\alpha +\frac {1}{2}\Gamma ^{\alpha \beta }\hat
{s}_{\alpha \beta }+\theta ^a\hat {T}_a\right\} .
\end{eqnarray}
This algebra is a direct summation of three algebras,
coordinate-momentum algebra $A_{xp}$, spin algebra $A_S$ and gauge
algebra $A_g$,
\begin{eqnarray}
A=A_{xp}\oplus A_S\oplus A_{g}.
\end{eqnarray}
The group corresponding to this Lie algebra is written as
\begin{eqnarray}
G=\left\{ \hat {U}:\hat {U}=\exp [i(a_{\mu }\hat {x}^{\mu }+b^{\mu
}\hat {p}_{\mu }+\alpha +\frac {1}{2}\omega ^{\alpha \beta }\hat
{s}_{\alpha \beta }+\xi ^a\hat {T}_a)]\right\} .
\end{eqnarray}
This group is a direct product of groups of coordinate-momentum,
spin and gauge
\begin{eqnarray}
G=G_{xp}\otimes G_S\otimes G_g.
\end{eqnarray}

The representation space is simply the direct product of three
representation spaces corresponding to coordinate-momentum, spin and
gauge respectively.

We next list respectively the representation of (i) matter
particles, (ii) the gauge particles and (iii) the graviton.

\subsection{Representation of matter particles}
The representation of matter particles is in the total direct
product space
\begin{eqnarray}
V(M)=V_{xp}(M)\otimes V_S(M)\otimes V_g(M).
\end{eqnarray}
The adjoint has similar representation,
\begin{eqnarray}
\overline{V}(M)=\overline{V}_{xp}(M)\otimes \overline{V}_S(M)\otimes
\overline{V}_g(M).
\end{eqnarray}
The operators are acting on those spaces,
\begin{eqnarray}
O(M)=V(M)\otimes \overline{V}(M).
\end{eqnarray}

The basis of matter particles is a direct product of three class of
basis, coordinate, spin and gauge
\begin{eqnarray}
|e_{st}(x)\rangle =|x\rangle \otimes |e_s\rangle \otimes |e_t\rangle
,
\end{eqnarray}
as we already know that $x\in R^4; s=1,2,3,4; t=1,2,...,48$. Recall
the metrics of spin space and gauge space, the adjoint of the basis
takes the form
\begin{eqnarray}
\langle e^{st}(x)|=(\hat {\gamma
}^0)^{ss'}\overline{|e_{s't}(x)\rangle }=\langle x|\otimes \langle
e^s|\otimes e^t|.
\end{eqnarray}
The normalization and the complete conditions are
\begin{eqnarray}
\langle e^{st}(x)|e_{s't'}(x')\rangle &=&\delta ^s_{s'}\delta
^t_{t'}\delta ^4(x-x'), \nonumber \\
\int _{R^4}|e_{st}(x)\rangle \langle e^{st}(x)|d^4x&=&1.
\end{eqnarray}

Similarly for momentum representation, we also have
\begin{eqnarray}
|e_{st}(p)\rangle =|p\rangle \otimes |e_s\rangle \otimes |e_t\rangle
.
\end{eqnarray}
\begin{eqnarray}
\langle e^{st}(p)|=\langle p|\otimes \langle e^s|\otimes e^t|.
\end{eqnarray}
The normalization and the complete conditions are
\begin{eqnarray}
\langle e^{st}(p)|e_{s't'}(p')\rangle &=&\delta ^s_{s'}\delta
^t_{t'}\delta ^4(p-p'), \nonumber \\
\int _{R^4}|e_{st}(p)\rangle \langle e^{st}(p)|d^4p&=&1.
\end{eqnarray}

The transformation elements between coordinate and momentum take the
form
\begin{eqnarray}
\langle e^{st}(x)|e_{s't'}(p)\rangle =(2\pi )^{-2}\delta
^s_{s'}\delta ^t_{t'}\exp (-ipx).
\end{eqnarray}

The general quantum state of the matter particles can be expanded in
terms of those basis, either in coordinate or in momentum basis,

\begin{eqnarray}
|\Psi \rangle =\int _{R^4}\Psi ^{st}(x)|e_{st}(x)\rangle d^4x =\int
_{R^4}\widetilde {\Psi }^{st}(p)|e_{st}(p)\rangle d^4p,
\label{state}
\end{eqnarray}
where coefficients $\Psi ^{st}(x)$ in the expansion are defined as
\begin{eqnarray}
\Psi ^{st}(x)&=&\langle e^{st}(x)|\Psi \rangle =(2\pi )^{-2}\int
_{R^4}\widetilde {\Psi }^{st}(p)\exp (-ipx)d^4p, \nonumber \\
\widetilde {\Psi }^{st}(p)&=&\langle e^{st}(p)|\Psi \rangle =(2\pi
)^{-2}\int _{R^4}\Psi ^{st}(x)\exp (ipx)d^4x.
\end{eqnarray}

Thus all quantum states of matter particles constitute the
representation space $V(M)$ which is simply written as,
\begin{eqnarray}
V(M)&=&\{ |\Psi \rangle :|\Psi \rangle =\int _{R^4}\Psi
^{st}(x)|e_{st}(x)\rangle d^4x \} \nonumber \\
&=&\{ |\Psi \rangle :|\Psi \rangle =\int _{R^4}\widetilde {\Psi
}^{st}(p)|e_{st}(p)\rangle d^4p\} .
\end{eqnarray}
The adjoint states $\langle \Psi |$ can be similarly written as,
\begin{eqnarray}
\langle \Psi |=\overline{|\Psi \rangle }=\int _{R^4}\Psi
_{st}^*(x)\langle e^{st}(x)|d^4x =\int _{R^4}\widetilde {\Psi
}_{st}^*(p)\langle e^{st}(p)|d^4p,
\end{eqnarray}
note that the coefficients have the metric of spin space,
\begin{eqnarray}
\Psi ^*_{st}(x)&=&\langle \Psi |e_{st}(x)\rangle =(\hat {\gamma
}^0)_{ss'}\Psi ^{*s't}(x) , \\
\widetilde {\Psi }^*_{st}(p)&=&\langle \Psi |e_{st}(p)\rangle =(\hat
{\gamma }^0)_{ss'}\Psi ^{*s't}(p).
\end{eqnarray}
The adjoint representation space is constituted by the adjoint
states $\overline{V}(M)=\{ \langle \Psi |\}$.

The operator representation space can be considered to be
constructed by two spaces $V(M)\otimes \overline{V}(M)$ and is
denoted as,

\begin{eqnarray}
O(M)&=&\left\{ \hat {A}: \hat {A}=\int _{R^4}\int
_{R^4}A^{st}_{s't'}(x',x)|e^{s't'}(x')\rangle \langle
e_{st}(x)|d^4x'd^4x\right\} \nonumber \\
&=&\left\{ \hat {A}: \hat {A}=\int _{R^4}\int _{R^4}\widetilde
{A}^{st}_{s't'}(p',p)|e^{s't'}(p')\rangle \langle
e_{st}(p)|d^4p'd^4p\right\}
\end{eqnarray}

The inner product of two vectors is represented as

\begin{eqnarray}
\langle \Phi |\Psi \rangle ={\rm tr}\left( |\Psi \rangle \langle
\Phi |\right) &=&\int _{R^4}\Phi _{st}^*(x)\Psi ^{st}(x)d^4x
\nonumber \\ &=&\int _{R^4}\widetilde {\Phi }_{st}^*(p)\widetilde
{\Psi }^{st}(p)d^4p
\end{eqnarray}

\subsection{Representation of gauge particles}
Representation theory for gauge particles: The representation space
of gauge particles is also a direct product of coordinate-momentum
representation space, spin representation space and the gauge
representation space,
\begin{eqnarray}
V(A)=V_{xp}(A)\otimes V_S(A)\otimes V_g(A).
\end{eqnarray}
The basis of $V_{xp}(A)$ for coordinate and momentum are $\hat
{\varepsilon}(x)$ and $\hat {\varepsilon}(p)$, where $x\in R^4, p\in
R^4$. The basis $V_S(A)$ of spin are $\hat {\gamma }^{\alpha
_1\cdots \alpha _p}$, where $\alpha _i=0,1,2,3$ are orthogonal
space-time indices, $p=0,1,2,3$ is the rank of the tensor. $\hat
{T}_a$ is the basis of $V_g(A)$, where $a=1,2,\cdots ,12 $. So the
total basis can be written as
\begin{eqnarray}
\hat {\varepsilon}_a^{\alpha _1\cdots \alpha _p}(x)=\hat
{\varepsilon }(x)\otimes \hat {\gamma }^{\alpha _1\cdots \alpha
_p}\otimes \hat {T}_a,\\
\hat {\varepsilon}_a^{\alpha _1\cdots \alpha _p}(p)=\hat
{\varepsilon }(p)\otimes \hat {\gamma }^{\alpha _1\cdots \alpha
_p}\otimes \hat {T}_a.
\end{eqnarray}
The operator of gauge particles is represented as,
\begin{eqnarray}
\hat {X}&=&\frac {1}{p!}\hat {X}^a_{\alpha _1\cdots \alpha _p}
\otimes \hat {\gamma }^{\alpha _1\cdots \alpha _p} \otimes \hat
{T}_a \nonumber \\
&=&\frac {1}{p!}\int _{R^4}X^a_{\alpha _1\cdots \alpha _p} (x)\hat
{\varepsilon }_a^{\alpha _1\cdots \alpha _p}(x)d^4x
\nonumber \\
&=&\frac {1}{p!}\int _{R^4}\widetilde {X}^a_{\alpha _1\cdots \alpha
_p} (p)\hat {\varepsilon }_a^{\alpha _1\cdots \alpha _p}(p)d^4p.
\end{eqnarray}
The coordinate-momentum functions $X^a_{\alpha _1\cdots \alpha _p}
(x)$ and $\widetilde {X}^a_{\alpha _1\cdots \alpha _p} (p)$ are
related through Fourier transformation,
\begin{eqnarray}
X^a_{\alpha _1\cdots \alpha _p} (x)&=&(2\pi )^{-2}\int
_{R^4}\widetilde {X}^a_{\alpha _1\cdots \alpha _p} (p)\exp
(-ixp)d^4p,
\\
\widetilde {X}^a_{\alpha _1\cdots \alpha _p} (p)&=&(2\pi )^{-2}\int
_{R^4}X^a_{\alpha _1\cdots \alpha _p} (p)\exp (ixp)d^4x,
\end{eqnarray}

\subsection{Representation of graviton}
The representation space of graviton also includes
coordinate-momentum, spin and gauge,
\begin{eqnarray}
V(G)=V_{xp}(G)\otimes V_S(G)\otimes V_g(G).
\end{eqnarray}
Here we would like to point out that $V_{xp}$ is an infinite
dimensional space, $V_S(G)$ is a 256 dimensional mixed tensor space,
$V_G(G)$ is a 1-dimensional gauge representation space.

$V_{xp}(G)$ has the basis $\hat {\varepsilon}(x)$ and $\hat
{\varepsilon}(p)$. The basis of $V_S(G)$ is $\hat {\gamma }^{\alpha
_1\cdots \alpha _p}_{\beta _1\cdots \beta _q}$, where $\alpha
_i,\beta _i=0,1,2,3$ are indices of the orthogonal space-time,
$p,q=0,1,2,3,4$ are ranks of the spin mixed tensor.

So the basis of graviton for coordinate and momentum are,
\begin{eqnarray}
\hat {\varepsilon }^{\alpha _1\cdots \alpha _p}_{\beta _1\cdots
\beta _q}(x)=\hat {\varepsilon}(x)\otimes \hat {\gamma }^{\alpha
_1\cdots \alpha _p}_{\beta _1\cdots \beta _q}, \\
\hat {\varepsilon }^{\alpha _1\cdots \alpha _p}_{\beta _1\cdots
\beta _q}(x)=\hat {\varepsilon}(p)\otimes \hat {\gamma }^{\alpha
_1\cdots \alpha _p}_{\beta _1\cdots \beta _q}.
\end{eqnarray}

The operator representation for graviton is
\begin{eqnarray}
\hat {Y}&=&\frac {1}{p!q!}\hat {Y}_{\alpha _1\cdots \alpha
_p}^{\beta _1\cdots \beta _q}\otimes \hat {\gamma }^{\alpha _1\cdots
\alpha _p}_{\beta _1\cdots \beta _q}\nonumber \\
&=&\frac {1}{p!q!}\int _{R^4}Y_{\alpha _1\cdots \alpha _p}^{\beta
_1\cdots \beta _q}(x)\hat {\varepsilon }^{\alpha _1\cdots \alpha
_p}_{\beta _1\cdots \beta _q}(x)d^4x \nonumber \\
&=&\frac {1}{p!q!}\int _{R^4}\widetilde {Y}_{\alpha _1\cdots \alpha
_p}^{\beta _1\cdots \beta _q}(p)\hat {\varepsilon}^{\alpha _1\cdots
\alpha _p}_{\beta _1\cdots \beta _q}(p)d^4p .
\end{eqnarray}
The Fourier transformation connects the functions for coordinate and
momentum together,
\begin{eqnarray}
Y^{\beta _1\cdots \beta _q}_{\alpha _1\cdots \alpha _p} (x)&=&(2\pi
)^{-2}\int _{R^4}\widetilde {Y}^{\beta _1\cdots \beta _q}_{\alpha
_1\cdots \alpha _p} (p)\exp (-ixp)d^4p,
\\
\widetilde {Y}^{\beta _1\cdots \beta _q}_{\alpha _1\cdots \alpha _p}
(p)&=&(2\pi )^{-2}\int _{R^4}Y^{\beta _1\cdots \beta _q}_{\alpha
_1\cdots \alpha _p} (x)\exp (ixp)d^4x.
\end{eqnarray}

In quantum mechanics, the time evolution of a quantum state or an
operator are described by Sch\"odinger representation or Heisenberg
representation, respectively. We may notice that time and space are
not symmetric. In comparison for our work, there is no absolute time
and space in general relativity, thus coordinates are dealt
symmetrically. The state $|e_{st}\rangle $ is an event.

\section{Differential geometry, representation of gravity field and gauge fields}
For a manifold, the geometry is characterized by vierbeins and the
corresponding connections. We consider next the differential
geometry properties related with gravity field and gauge fields. Our
work, however, is actually an algebraic realization of the
differential geometry by a proper representations.

\subsection{The spin vierbein}
The spin vierbein formalism takes the form
\begin{eqnarray}
\hat{\theta }&=&\hat{\theta }_{\alpha }^{\mu }\otimes \hat {\gamma
}^{\alpha }\otimes \hat {p}_{\mu }, \nonumber \\
\hat{\theta }_{\alpha }^{\mu }&=&\int _{R^4}\widetilde {\theta
}_{\alpha }^{\mu }(x)\hat {\varepsilon }(x)d^4x=\int
_{R^4}\widetilde {\theta }_{\alpha }^{\mu }(p)\hat {\varepsilon
}(p)d^4p,
\end{eqnarray}
where $\hat {\theta }_{\alpha }^{\mu }$ are the spin vierbein
coefficients. $\theta _{\alpha }^{\mu }(x)$ and $\widetilde {\theta
}_{\alpha }^{\mu }(x)$ are coordinate functions and momentum
functions in spin vierbein formalism, they satisfy relations
\begin{eqnarray}
\widetilde {\theta }_{\alpha }^{\mu }(p)&=&(2\pi )^{-2}\int
_{R^4}\theta
_{\alpha }^{\mu }(x)\exp (ipx)d^4x, \nonumber \\
\theta _{\alpha }^{\mu }(x)&=&(2\pi )^{-2}\int _{R^4}\widetilde
{\theta }_{\alpha }^{\mu }(p)\exp (-ipx)d^4p.
\end{eqnarray}
The adjoint of the spin vierbein coefficient is defined as
\begin{eqnarray}
\overline{\hat {\theta }_{\alpha }^{\mu }}=\hat {e}_{\alpha }^{\mu }
\end{eqnarray}

Due to the spin vierbein coefficients $\hat {\theta }_{\alpha }^{\mu
}$, we can define the momentum metric as
\begin{eqnarray}
\hat {g}^{\mu \nu }&=&\eta ^{\alpha \beta }\hat {\theta }_{\alpha
}^{\mu }\hat {\theta }_{\beta }^{\nu }, \nonumber \\
\hat {g}&=&\hat {g}^{\mu \nu }\otimes \hat {p}_{\mu }\otimes \hat
{p}_{\nu },
\end{eqnarray}
where $\eta ^{\alpha \beta }$ is the free metric, i.e., Minkowski
metric. $\hat {g}^{\mu \nu }$ satisfies,
\begin{eqnarray}
\overline{\hat {g}^{\mu \nu }}&=&\hat {g}^{\mu \nu } \nonumber \\
\hat {g}^{\mu \nu }&=&\hat {g}^{\nu \mu }.
\end{eqnarray}
$\hat {g}^{\mu \nu }$ is the contra-variance metric tensor, the
covariance metric tensor $\hat {g}_{\mu \nu }$ can be defined as the
inverse of the contra-variance metric tensor $\hat {g}^{\mu \nu}$,
\begin{eqnarray}
\hat {g}_{\mu \lambda }\hat {g}^{\lambda \nu}&=&\delta ^{\nu }_{\mu
},\nonumber \\
\hat {g}_{\mu \nu }&=&\frac {\hat {A}_{\mu \nu }}{{\rm det}[\hat
{g}]},
\end{eqnarray}
where $\hat {A}_{\mu \nu}$ is the algebraic complementary minor of
elements $\hat {g}^{\mu \nu}$. The covariance metric takes the form
\begin{eqnarray}
\eta _{\alpha \beta }=\hat {g}_{\mu \nu }\hat {\theta }_{\alpha
}^{\mu }\theta _{\beta }^{\nu }.
\end{eqnarray}

The lowering or rising of the indices in momentum tensor can be
realized by multiplying momentum metric $\hat {g}^{\mu \nu }$ and
$\hat {g}_{\mu \nu }$, while for indices in spin tensor, the spin
metrics $\eta ^{\alpha \beta }$ and $\eta _{\alpha \beta }$ should
be used. For example, the case of orthogonal vierbein we have,
\begin{eqnarray}
\hat {e}^{\alpha }_{\mu }&=&\eta ^{\alpha \beta }\hat {g}_{\mu
\nu }\hat {\theta }^{\nu }_{\beta },\\
\hat {\gamma }_{\alpha }&=&\eta _{\alpha \beta }\hat {\gamma
}^{\beta }, \\
\hat {e}&=&
\hat {e}^{\alpha }_{\mu }\otimes \hat {\gamma }_{\alpha }\otimes
dx^{\mu },
\end{eqnarray}
where the orthogonal vierbein operator $\hat {\theta }_{\alpha
}^{\mu }$ and the dual orthogonal vierbein operator $\hat {\theta
}^{\alpha }_{\mu }$ satisfy the relation,
\begin{eqnarray}
\hat {\theta }_{\alpha }^{\mu }\hat {e}^{\beta }_{\mu
}&=&\delta _{\alpha }^{\beta }, \nonumber \\
\hat {\theta }_{\alpha }^{\mu }\hat {e}^{\alpha }_{\nu }&=&\delta
_{\nu }^{\mu }.
\end{eqnarray}

The spin of the vierbein formalism $\hat {\theta }_{\alpha }=\hat
{\theta }_{\alpha }^{\mu }\otimes \hat {p}_{\mu }$ satisfies the
commutation relations
\begin{eqnarray}
[\hat {\theta }_{\alpha }, \hat {\theta }_{\beta }]=i\hat
{f}_{\alpha \beta }^{\gamma }\hat {\theta }_{\gamma },
\end{eqnarray}
where $\hat {f}_{\alpha \beta }^{\gamma }$ are the structure
coefficients in spin vierbein formalism and are represented as
\begin{eqnarray}
\hat {f}_{\alpha \beta }^{\gamma }=(\hat {\theta }_{\alpha }^{\mu
}\partial _{\mu }\hat {\theta }_{\beta }^{\nu }-\hat {\theta
}_{\beta }^{\mu }\partial _{\mu }\hat {\theta }_{\alpha }^{\nu
})\hat {e}_{\nu }^{\gamma }.
\end{eqnarray}
Note that the structure coefficients here may not be confused as the
structure constants of $su(3)$ appeared previously. The structure
coefficients are antisymmetric, have adjoint, and satisfy Jacobi
equation:
\begin{eqnarray}
&&\hat {f}_{\alpha \beta }^{\gamma }=-\hat {f}_{\beta \alpha
}^{\gamma }
\\
&&\overline{\hat {f}_{\alpha \beta }^{\gamma }}=\hat {f}_{\alpha
\beta }^{\gamma }\\
&&\hat {\theta }_{\alpha }^{\mu }\partial _{\mu }\hat {f}^{\rho
}_{\beta \gamma }+\hat {\theta }_{\beta }^{\mu }\partial _{\mu }\hat
{f}^{\rho }_{\gamma \alpha }+\hat {\theta }_{\gamma }^{\mu }\partial
_{\mu }\hat {f}^{\rho }_{\alpha \beta }+\hat {f}^{\sigma }_{\beta
\gamma }\hat {f}^{\rho }_{\alpha \sigma }+\hat {f}^{\sigma }_{\gamma
\alpha }\hat {f}^{\rho }_{\beta \sigma }+\hat {f}^{\sigma }_{\alpha
\beta }\hat {f}^{\rho }_{\gamma \sigma }=0,
\end{eqnarray}
where we have used the equation,
\begin{eqnarray}
[\hat {\theta }_{\alpha },[\hat {\theta }_{\beta }, \hat {\theta
}_{\gamma }]]+[\hat {\theta }_{\beta },[\hat {\theta }_{\alpha },
\hat {\theta }_{\gamma }]]+[\hat {\theta }_{\gamma },[\hat {\theta
}_{\alpha }, \hat {\theta }_{\beta }]]=0.
\end{eqnarray}

\subsection{Connection of gravity}
The gravity connection is defined as
\begin{eqnarray}
\hat {\Gamma }=\frac {1}{2}\hat {\Gamma }_{\alpha }^{\rho \sigma
}\otimes \hat {\gamma }^{\alpha }\otimes \hat {s}_{\rho \sigma },
\end{eqnarray}
\begin{eqnarray}
\hat {\Gamma }_{\alpha }^{\rho \sigma }=\frac {1}{2}(\hat
{f}_{\alpha }^{\rho \sigma }+\hat {f}_{\alpha }^{\rho ,\sigma }-\hat
{f}_{\alpha }^{\sigma ,\rho }),
\end{eqnarray}
where we have used the notations
\begin{eqnarray}
\hat {f}_{\tau }^{\rho \sigma }&=&\eta ^{\rho \alpha }\eta ^{\sigma
\beta }\eta _{\tau \gamma }\hat {f}_{\alpha \beta }^{\gamma },
\nonumber \\
\hat {f}_{\tau }^{\rho ,\sigma }&=&\eta ^{\sigma \beta }\hat
{f}_{\tau \beta }^{\rho }.
\end{eqnarray}

The gravity connections are antisymmetric, have adjoint and satisfy
no-torsion condition,
\begin{eqnarray}
\hat {\Gamma }^{\alpha \beta }_{\gamma }&=&-\hat {\Gamma }^{\beta
\alpha }_{\gamma }\\
\overline{\hat {\Gamma }^{\alpha \beta }_{\gamma }}&=&\hat {\Gamma
}^{\alpha \beta }_{\gamma },\\
\hat {\Gamma }^{\alpha }_{\gamma ,\beta }-\hat {\Gamma }^{\alpha
}_{\beta ,\gamma }&=&\hat {f}^{\alpha }_{\beta \gamma },
\end{eqnarray}
the last equation can lead to the result that the torsion is zero,
\begin{eqnarray}
\hat {T}^{\gamma }_{\alpha \beta }=\hat {\Gamma }^{\gamma }_{\alpha
,\beta }-\hat {\Gamma }^{\gamma }_{\beta ,\alpha }-\hat {f}^{\gamma
}_{\alpha \beta }=0.
\end{eqnarray}

Here let us discuss the spin of graviton defined by $\hat {\gamma
}^{\alpha }\otimes \hat {s}_{\rho \sigma }$. As we know the
representation space of $\hat {\gamma }^{\alpha }$ is 4-dimensional
$V_S(\frac {1}{2},\frac {1}{2})$, the representation space of $\hat
{s}_{\rho \sigma }$ is 6-dimensional $V_S(0,1)\oplus V_S(1,0)$. So
the spin representation space of graviton is expressed as,
\begin{eqnarray}
&&V_S(\frac {1}{2},\frac {1}{2})\otimes [V_S(0,1)\oplus V_S(1,0)]
\nonumber \\
&=& V_S(\frac {1}{2},\frac {1}{2})\oplus V_S(\frac {1}{2},\frac
{3}{2})\oplus V_S(\frac {1}{2},\frac {1}{2})\oplus V_S(\frac
{3}{2},\frac {1}{2}).
\end{eqnarray}
The spin is defined as the maximal eigenvalues of $\hat {s}_3$. Here
we can find ${\rm max}(s_3)=\frac {1}{2}+\frac {3}{2}=2$, thus the
graviton is spin-2.

\subsection{Gauge connection}

The gauge connection is defined as
\begin{eqnarray}
\hat {A}=\hat {A}_{\alpha }^a\otimes \hat {\gamma }^{\alpha }\otimes
\hat {T}_a,
\end{eqnarray}
similarly we have
\begin{eqnarray}
\hat {A}_{\alpha }^a=\int _{R^4}A_{\alpha }^a(x)\hat {\varepsilon
}(x)d^4x=\int _{R^4}\widetilde {A}_{\alpha }^a(p)\hat {\varepsilon
}(p)d^4p,
\end{eqnarray}
where $A_{\alpha }^a(x)$ and $A_{\alpha }^a(x)$ are coordinate and
momentum functions, respectively. The relation is
\begin{eqnarray}
\widetilde {A}_{\alpha }^a(p)&=&(2\pi )^{-2}\int _{R^4}A^a_{\alpha
}(x)\exp
(ipx)d^4x, \nonumber \\
A_{\alpha }^a(x)&=&(2\pi )^{-2}\int _{R^4}\widetilde {A}^a_{\alpha
}(p)\exp (-ipx)d^4p, \nonumber \\
\end{eqnarray}
The adjoint of the gauge connection is,
\begin{eqnarray}
\overline{\hat {A}^a_{\alpha }}=G_{ab}\hat {A}^b_{\alpha }.
\end{eqnarray}

\subsection{Definition of the general covariance derivative operator}

Now we define the covariance derivative operator as
\begin{eqnarray}
\hat {D}_{\alpha }=-i\hat {\theta }_{\alpha }^{\mu }\otimes \hat
{p}_{\mu }+\frac {i}{2}\hat {\Gamma }_{\alpha }^{\rho \sigma
}\otimes \hat {s}_{\rho \sigma }-i\hat {A}_{\alpha }^a\otimes \hat
{T}_a. \label{deri-operator}
\end{eqnarray}
This operator has connections of gravity, connections of gauge and
spin vierbein. Thus it can describe all force fields. As in quantum
mechanics, when acting on operators,  it is represented in the form
of commutating calculation, when acting on matter fields, it is
represented as an operator acting on quantum states.

Here we list the properties of this newly defined general covariance
derivative operator,
\begin{eqnarray}
&&\overline{\hat {D}_{\alpha }}=-\hat {D}_{\alpha },
\\
&&\hat {D}_{\alpha }|e_{st}(x)\rangle =\int _{R^4}\left[ \theta
_{\alpha }^{\mu }(x')\frac {\partial }{\partial x^{\mu }}+\frac
{i}{2}\Gamma _{\alpha }^{\rho \sigma }(x')\hat {s}_{\rho \sigma
}-iA^a_{\alpha }(x')\hat {T}_a\right] \delta
^4(x-x')|e_{st}(x')d^4x',\\
&&\langle e^{st}(x)|\hat {D}_{\alpha }=\int _{R^4}d^4x'\langle
e^{st}(x)|\left[ -\theta _{\alpha }^{\mu }(x')\frac {\partial
}{\partial x^{\mu }}+\frac {i}{2}\Gamma _{\alpha }^{\rho \sigma
}(x')\hat {s}_{\rho \sigma }-iA^a_{\alpha }(x')\hat {T}_a\right]
\delta ^4(x-x'),\\
&&\hat {D}_{\alpha }|e_{st}(p)\rangle =(2\pi )^{-2}\int _{R^4}\left[
-i\widetilde {\theta }_{\alpha }^{\mu }(p')p_{\mu }+\frac
{i}{2}\widetilde {\Gamma }_{\alpha }^{\rho \sigma }(p')\hat
{s}_{\rho \sigma
}-i\widetilde {A}^a_{\alpha }(p')\hat {T}_a\right] |e_{st}(p+p')d^4p',\\
&&\langle e^{st}(p)|\hat {D}_{\alpha }=(2\pi )^{-2}\int
_{R^4}d^4p'\langle e^{st}(p+p')|\left[ -i\widetilde {\theta
}_{\alpha }^{\mu }(p')p_{\mu }+\frac {i}{2}\widetilde {\Gamma
}_{\alpha }^{\rho \sigma }(p')\hat {s}_{\rho \sigma }-i\widetilde
{A}^a_{\alpha }(p')\hat
{T}_a\right] \\
&&[\hat {D}_{\alpha }, \hat {\varepsilon}(x)]=\int _{R^4}\theta
_{\alpha }^{\mu }(x')\frac {\partial }{\partial x^{\mu }}\delta
^4(x-x')\hat {\varepsilon }(x')d^4x',\\
&&[\hat {D}_{\alpha }, \hat {\varepsilon}(p)]=(2\pi )^{-2}\int
_{R^4}-i\widetilde {\theta }_{\alpha }^{\mu }(p')p_{\mu }\hat
{\varepsilon }(p+p')d^4p',\\
&&[\hat {D}_{\alpha }, \hat {\gamma }^{\alpha _1\cdots \alpha
_p}]=-\hat {\Gamma }_{\alpha ,\rho }^{\alpha _1}\hat {\gamma }^{\rho
\cdots \alpha _p}-\cdots -\hat {\Gamma }_{\alpha ,\rho }^{\alpha
_p}\hat {\gamma }^{\alpha _1\cdots \rho },
\\
&&[\hat {D}_{\alpha }, \hat {\gamma }_{\alpha _1\cdots \alpha
_p}]=\hat {\Gamma }_{\alpha ,\alpha _1}^{\rho }\hat {\gamma }_{\rho
\cdots \alpha _p}+\cdots +\hat {\Gamma }_{\alpha ,\alpha _p }^{\rho
}\hat {\gamma }_{\alpha _1\cdots \rho },
\\
&&[\hat {D}_{\alpha },\hat {T}_b]=-iC^c_{ab}A^a_{\alpha }\hat {T}_c,
\\
&&\hat {D}_{\alpha }(a|\Psi \rangle +b|\Phi \rangle )=a\hat
{D}_{\alpha }|\Psi \rangle +b\hat {D}_{\alpha }|\Phi \rangle , \\
&&[\hat {D}_{\alpha }, a\hat {A}+b\hat {B}]=a[\hat {D}_{\alpha },
\hat {A}]+b[\hat {D}_{\alpha },\hat {B}],\\
&&\hat {D}_{\alpha }(\hat {A}|\Psi \rangle )=[\hat {D}_{\alpha },
\hat {A}]|\Psi \rangle +\hat {A}(\hat {D}_{\alpha }|\Psi \rangle ),
\\
&&[\hat {D}_{\alpha }, \hat {A}\hat {B}]=[\hat {D}_{\alpha }, \hat
{A}]\hat {B}+\hat {A}[\hat {D}_{\alpha }, \hat {B}],\\
&&[\hat {D}_{\alpha }, \hat {A}\otimes \hat {B}]=[\hat {D}_{\alpha
}, \hat
{A}]\otimes \hat {B}+\hat {A}\otimes [\hat {D}_{\alpha }, \hat {B}],\\
&&[\hat {D}_{\alpha }, \hat {A}\cdot \hat {B}]=[\hat {D}_{\alpha },
\hat
{A}]\cdot \hat {B}+\hat {A}\cdot [\hat {D}_{\alpha }, \hat {B}],\\
&&[\hat {D}_{\alpha }, \hat {A}\wedge \hat {B}]=[\hat {D}_{\alpha },
\hat {A}]\wedge \hat {B}+\hat {A}\wedge [\hat {D}_{\alpha }, \hat
{B}],\\
&& [\hat {D}_{\alpha }, {\rm con}(\hat {A})]={\rm con}([\hat
{D}_{\alpha }, \hat {A}]),\\
&&[\hat {D}_{\alpha }, [\hat {A},\hat {B}]]=[[\hat {D}_{\alpha },
\hat
{A}], \hat {B}]+[\hat {A}, [\hat {D}_{\alpha }, \hat {B}],\\
&&\langle \Phi |\left( \hat {D}_{\alpha }|\Psi \rangle
\right)-\left( \langle \Phi |\hat {D}_{\alpha }\right) |\Psi \rangle
=-\langle \Phi
|\partial _{\mu }\hat {\theta }_{\alpha }^{\mu }|\Psi \rangle ,\\
&&\langle \Phi |\hat {\varepsilon}(x)\left( \hat {D}_{\alpha }|\Psi
\rangle \right)-\left( \langle \Phi |\hat {D}_{\alpha }\right)\hat
{\varepsilon }(x)|\Psi \rangle =\theta _{\alpha }^{\mu }(x)\partial
_{\mu }\langle \Phi
|\hat {\varepsilon }(x)|\Psi \rangle ,\\
&&\langle \Phi |\hat {\varepsilon}(p)\left( \hat {D}_{\alpha }|\Psi
\rangle \right)-\left( \langle \Phi |\hat {D}_{\alpha }\right)\hat
{\varepsilon }(p)|\Psi \rangle =\widetilde {\theta }_{\alpha }^{\mu
}(p)*\left( p_{\mu }\langle \Phi
|\hat {\varepsilon }(x)|\Psi \rangle \right) ,\\
&&\langle [\hat {D}_{\alpha }, \hat {A}],\hat {B}\rangle +\langle
[\hat {A}, [\hat {D}_{\alpha }, \hat {B}]\rangle =0.
\end{eqnarray}

The general covariance derivative operator $\hat {D}_{\alpha }$ can
act on the matter fields. Let's next see how the calculation can be
made explicitly.

As we have already seen, the matter fields can be described in
several forms below, respectively,
\begin{eqnarray}
|\Psi \rangle =|e^{st}\rangle \otimes |e_s\rangle \otimes
|e_t\rangle =\int _{R^4}\Psi ^{st}(x)|e_{st}(x)\rangle d^4x=\int
_{R^4}\widetilde {\Psi }^{st}(p)|e_{st}(x)\rangle d^4p.
\end{eqnarray}
By applying the operator  $\hat {D}_{\alpha }$, we have,
\begin{eqnarray}
\hat {D}_{\alpha }|\Psi \rangle &=&|D_{\alpha }\Psi ^{st}\rangle
\otimes |e_s\rangle \otimes |e_t\rangle \nonumber \\
&=&\int _{R^4}D_{\alpha }\Psi ^{st}(x)|e_{st}(x)\rangle d^4x
\nonumber \\
&=&\int _{R^4}D_{\alpha }\widetilde {\Psi }^{st}(p)|e_{st}(x)\rangle
d^4p,
\end{eqnarray}
where the calculation should be in form
\begin{eqnarray}
&&\hat {D}_{\alpha }|\Psi \rangle = (-i\hat {\theta }_{\alpha }^{\mu
}\hat {p}_{\mu }+\frac {i}{2}\hat {\Gamma }_{\alpha }^{\rho \sigma
}\hat {s}_{\rho \sigma }-i\hat {A}_{\alpha }^a\hat {T}_a)|\Psi
\rangle ,\\
&&|D_{\alpha }\Psi ^{st}\rangle =\hat {\theta }_{\alpha }^{\mu
}|\partial _{\mu }\Psi ^{st}\rangle +\frac {i}{2}\hat {\Gamma
}_{\alpha }^{\rho \sigma }(\hat {s}_{\rho \sigma })^s_{s'}|\Psi
^{s't}\rangle -i\hat {A}_{\alpha }^a(\hat {T}_a)^t_{t'}|\Psi
^{st'}\rangle ,\\
&&D_{\alpha }\Psi ^{st}= \theta _{\alpha }^{\mu }
\partial _{\mu }\Psi ^{st}+\frac {i}{2}\hat {\Gamma }_{\alpha }^{\rho \sigma
}(\hat {s}_{\rho \sigma })^s_{s'}\Psi ^{s't}-iA_{\alpha }^a(\hat
{T}_a)^t_{t'}\Psi ^{st'},\\
&&D_{\alpha }\widetilde {\Psi }^{st}= -i\widetilde {\theta }_{\alpha
}^{\mu }*[p_{\mu }\widetilde {\Psi }^{st}]+\frac {i}{2}\widetilde
{\Gamma }_{\alpha }^{\rho \sigma }*[(\hat {s}_{\rho \sigma
})^s_{s'}\widetilde {\Psi }^{s't}]-i\widetilde {A}_{\alpha }^a*[(\hat
{T}_a)^t_{t'}\widetilde {\Psi }^{st'}].\\
\end{eqnarray}

Similar results are for the adjoint state of matter fields when we
apply the operator $\hat {D}_{\alpha }$,

\begin{eqnarray}
\langle \Psi |\hat {D}_{\alpha }&=&\langle e^t| \otimes \langle
e^s|\otimes \rangle \langle D_{\alpha }\Psi _{st}|
\nonumber \\
&=&\int _{R^4}d^4x\langle e^{st}(x)|D_{\alpha }\Psi _{st}^*(x)
\nonumber \\
&=&\int _{R^4}d^4p\langle e^{st}(x)|D_{\alpha }\widetilde {\Psi
}^*_{st}(p),
\end{eqnarray}
where
\begin{eqnarray}
&&\langle \Psi |\hat {D}_{\alpha }= \langle \Psi |(-i\hat {p}_{\mu
}\hat {\theta }_{\alpha }^{\mu }+\frac {i}{2}\hat {s}_{\rho \sigma
}\hat {\Gamma }_{\alpha }^{\rho \sigma
}-i\hat {T}_a\hat {A}_{\alpha }^a),\\
&&\langle D_{\alpha }\Psi _{st}| =\langle \partial _{\mu }\Psi
_{st}|\hat {\theta }_{\alpha }^{\mu } -\frac {i}{2} \langle \Psi
_{s't}|\hat {\Gamma }_{\alpha }^{\rho \sigma }(\hat {s}_{\rho \sigma
})_s^{s'}-i\langle \Psi _{st'}|\hat {A}_{\alpha }^a(\hat {T}_a)_t^{t'},\\
&&D_{\alpha }\Psi ^*_{st}= -\theta _{\alpha }^{\mu }
\partial _{\mu }\Psi ^*_{st}+\frac {i}{2}\hat {\Gamma }_{\alpha }^{\rho \sigma
}(\hat {s}_{\rho \sigma })_s^{s'}\Psi ^*_{s't}-iA_{\alpha }^a(\hat
{T}_a)_t^{t'}\Psi ^*_{st'},\\
&&D_{\alpha }\widetilde {\Psi }^*_{st}= i\widetilde {\theta
}_{\alpha }^{\mu }*(p_{\mu }\widetilde {\Psi }^*_{st})-\frac
{i}{2}\widetilde {\Gamma }_{\alpha }^{\rho \sigma }*[(\hat {s}_{\rho
\sigma })_s^{s'}\widetilde {\Psi }^*_{s't}]-i\widetilde {A}_{\alpha
}^a*[(\hat {T}_a)_t^{t'}\widetilde {\Psi }^*_{st'}].
\end{eqnarray}

For gauge operator $\hat {X}$, the covariance derivation is
\begin{eqnarray}
&&\hat {X}=\frac {1}{p!}\hat {X}^a_{\alpha _1\cdots \alpha
_p}\otimes \hat {\gamma }^{\alpha _1\cdots \alpha _p}\otimes \hat
{T}_a,
\\
&&[\hat {D}_{\alpha }, \hat {X}]=\frac {1}{p!}(D_{\alpha }\hat
{X}^a_{\alpha _1\cdots \alpha _p})\otimes \hat {\gamma }^{\alpha
_1\cdots \alpha _p}\otimes \hat {T}_a
\\
&&D_{\alpha }\hat {X}^a_{\alpha _1\cdots \alpha _p}=\hat {\theta
}_{\alpha }^{\mu }\partial _{\mu }\hat {X}^a_{\alpha _1\cdots \alpha
_p}-iC^a_{bc}\hat {A}^b_{\alpha }\hat {X}^c_{\alpha _1\cdots \alpha
_p}-\hat {\Gamma }^{\rho }_{\alpha ,\alpha _1}\hat {X}^a_{\rho
\cdots \alpha _p}-\cdots -\hat {\Gamma }^{\rho }_{\alpha ,\alpha
_p}\hat {X}^a_{\alpha _1\cdots \rho },
\end{eqnarray}
where we have used the notation $\hat {\Gamma }^{\rho }_{\alpha
,\beta }=\eta _{\beta \sigma }\hat {\Gamma }^{\rho \sigma }_{\alpha
}$.

The calculation of the derivation of the gravity field takes the
form
\begin{eqnarray}
&&\hat {Y}=\frac {1}{p!q!}\hat {Y}_{a_{\alpha _1\cdots \alpha
_p}}^{\beta _1\cdots \beta _q}\otimes \hat {\gamma }^{\alpha
_1\cdots \alpha _p}_{\beta _1\cdots \beta _q},
\\
&&[\hat {D}_{\alpha }, \hat {Y}]=\frac {1}{p!q!}\left( D_{\alpha
}\hat {Y}_{a_{\alpha _1\cdots \alpha _p}}^{\beta _1\cdots \beta
_q}\right) \otimes \hat {\gamma }^{\alpha _1\cdots \alpha _p}_{\beta
_1\cdots \beta _q},
\\
&&D_{\alpha }\hat {Y}_{a_{\alpha _1\cdots \alpha _p}}^{\beta
_1\cdots \beta _q}=\hat {\theta }_{\alpha }^{\mu }\partial _{\mu
}\hat {Y}_{a_{\alpha _1\cdots \alpha _p}}^{\beta _1\cdots \beta
_q}-\hat {\Gamma }^{\rho }_{\alpha ,\alpha _1}\hat {Y}_{\rho \cdots
\alpha _p}^{\beta _1\cdots \beta _q}-\cdots -\hat {\Gamma }^{\rho
}_{\alpha ,\alpha _p}\hat {Y}^{\beta _1\cdots \beta _q}_{\alpha
_1\cdots \rho }
\\
&&~~~~+\hat {\Gamma }^{\beta _1}_{\alpha ,\rho }\hat {Y}^{\rho
\cdots \beta _q}_{\alpha _1\cdots \alpha _p}+\cdots +\hat {\Gamma }^{\beta
_q}_{\alpha ,\rho }\hat {Y}^{\beta _1\cdots \rho }_{\alpha _1\cdots
\alpha _p}.
\end{eqnarray}

Based on the general covariance derivation, the generalized
divergence and curl can be defined. Suppose $\hat {K}$ is a $p$-form
antisymmetric spin tensor, we define
\begin{eqnarray}
\hat {D}\wedge \hat {K}&=&\hat {\gamma }\wedge [\hat {D}_{\alpha },
\hat {K}], \\
\hat {D}\cdot \hat {K}&=&\hat {\gamma }\cdot [\hat {D}_{\alpha },
\hat {K}], \\
\end{eqnarray}
where the covariance differential takes the form $\hat {D}=\hat
{\gamma }^{\alpha }\otimes \hat {D}_{\alpha }$. We can find that
$\hat {D}\wedge \hat {K}$ is a $(p+1)$-form antisymmetric spin
tensor, and $\hat {D}\cdot \hat {K}$ is a $(p-1)$-form tensor. So
these calculations can be considered as a generalization of
divergence and curl.

\subsection{Curvatures}

We then define the interaction curvature as
\begin{eqnarray}
\hat {\Omega }_{\alpha \beta }&=&i[\hat {D}_{\alpha }, \hat
{D}_{\beta }]-i\hat {f}_{\alpha \beta }^{\gamma }\hat {D}_{\gamma },
\nonumber \\
\hat {\Omega }&=&\hat {\Omega }_{\alpha \beta }\otimes \hat
{s}^{\alpha \beta }
\end{eqnarray}

The interaction curvature has adjoint, is antisymmetric and
satisfies the Bianchi identity,
\begin{eqnarray}
&&\hat {\Omega }_{\alpha \beta }=-\hat {\Omega }_{\beta \alpha } \\
&&\overline{\hat {\Omega }_{\alpha \beta }}=\hat {\Omega }_{\alpha
\beta }\\
&&\hat {D}_{\alpha }\hat {\Omega }_{\beta \gamma }+\hat {D}_{\beta
}\hat {\Omega }_{\gamma \alpha }+\hat {D}_{\gamma }\hat {\Omega
}_{\alpha \beta }=0.
\end{eqnarray}
The Bianchi identity can be represented as the covariance curl as
\begin{eqnarray}
\hat {D}\wedge \hat {\Omega }=0.
\end{eqnarray}

The proof that the interaction curvature satisfies the Bianchi
identity is presented as follows.

We start from the Jacobi relation,
\begin{eqnarray}
[\hat {D}_{\alpha },[\hat {D}_{\beta },\hat {D}_{\gamma }]]+[\hat
{D}_{\beta },[\hat {D}_{\gamma },\hat {D}_{\alpha }]]+[\hat
{D}_{\gamma },[\hat {D}_{\alpha },\hat {D}_{\beta }]]=0.
\end{eqnarray}
With the help of the definition of the interaction curvature, we
have
\begin{eqnarray}
[\hat {D}_{\alpha },\hat {\Omega }_{\beta \gamma }]&=&i[\hat
{D}_{\alpha },[\hat {D}_{\beta },\hat {D}_{\gamma }]]-i[\hat
{D}_{\alpha }, \hat {f}^{\rho }_{\beta \gamma }]\hat {D}_{\rho }
-i\hat {f}^{\rho }_{\beta \gamma }[\hat {D}_{\alpha },\hat {D}_{\rho
}]\nonumber \\
&=&i[\hat {D}_{\alpha },[\hat {D}_{\beta },\hat {D}_{\gamma }]]-i(
\hat {\theta }_{\alpha }^{\mu }\partial _{\mu }\hat {f}^{\rho
}_{\beta \gamma }+\hat {f}^{\sigma }_{\beta \gamma }\hat {f}^{\rho
}_{\alpha \sigma })\hat {D}_{\rho } -\hat {f}^{\rho }_{\beta \gamma
}\hat {\Omega }_{\alpha \rho },
\end{eqnarray}
also
\begin{eqnarray}
&&\hat {D}_{\alpha }\hat {\Omega }_{\beta \gamma }=[\hat {D}_{\alpha
},\hat {\Omega }_{\beta \gamma }]-\hat {\Gamma }^{\rho }_{\alpha
,\beta }\hat {\Omega }_{\rho \gamma }-\hat {\Gamma }^{\rho }_{\alpha
,\gamma }\hat {\Omega }_{\beta \rho } \nonumber \\
&& = i[\hat {D}_{\alpha },[\hat {D}_{\beta}, \hat {D}_{\gamma
}]-i[\hat {D}_{\alpha }, \hat {f}^{\rho }_{\beta \gamma }]\hat
{D}_{\rho } -i\hat {f}^{\rho }_{\beta \gamma }[\hat {D}_{\alpha
},\hat {D}_{\rho }]-\hat {\Gamma }^{\rho }_{\alpha ,\beta }\hat
{\Omega }_{\rho \gamma }-\hat {\Gamma }^{\rho }_{\alpha ,\gamma
}\hat {\Omega }_{\beta \rho }
\nonumber \\
&&= i[\hat {D}_{\alpha },[\hat {D}_{\beta}, \hat {D}_{\gamma }]-i(
\hat {\theta }_{\alpha }^{\mu }\partial _{\mu }\hat {f}^{\rho
}_{\beta \gamma }+\hat {f}^{\sigma }_{\beta \gamma }\hat {f}^{\rho
}_{\alpha \sigma })\hat {D}_{\rho } -\hat {f}^{\rho }_{\beta \gamma
}\hat {\Omega }_{\alpha \rho }+\hat {\Gamma }^{\rho }_{\alpha ,\beta
}\hat {\Omega }_{\gamma \rho }-\hat {\Gamma }^{\rho }_{\alpha
,\gamma }\hat {\Omega }_{\beta \rho }
\end{eqnarray}

Substituting the definition of the spin connection and structure
functions into the Jacobi identity, we have
\begin{eqnarray}
&&\hat {D}_{\alpha }\hat {\Omega }_{\beta \gamma }+
\hat {D}_{\beta }\hat {\Omega }_{\gamma \alpha }+\hat {D}_{\gamma }
\hat {\Omega }_{\alpha \beta } \nonumber \\
&=&i\left( [\hat {D}_{\alpha },[\hat {D}_{\beta },\hat {D}_{\gamma
}]]+[\hat {D}_{\beta },[\hat {D}_{\gamma },\hat {D}_{\alpha
}]]+[\hat
{D}_{\gamma },[\hat {D}_{\alpha },\hat {D}_{\beta }]]\right)  \nonumber \\
&&-\left( \hat {\theta }_{\alpha }^{\mu }\partial _{\mu }\hat
{f}^{\rho }_{\beta \gamma } +\hat {\theta }_{\beta }^{\mu }\partial
_{\mu }\hat {f}^{\rho }_{\gamma \alpha} +\hat {\theta }_{\gamma
}^{\mu }\partial _{\mu }\hat {f}^{\rho }_{\alpha \beta } +\hat
{f}^{\sigma }_{\beta \gamma }\hat {f}^{\rho }_{\alpha \sigma }+\hat
{f}^{\sigma }_{\gamma \alpha }\hat {f}^{\rho }_{\beta \sigma }+\hat
{f}^{\sigma }_{\alpha \beta
}\hat {f}^{\rho }_{\gamma \sigma }\right) \hat {D}_{\rho } \nonumber \\
&&-(\hat {f}^{\rho }_{\beta \gamma }-\hat {\Gamma }^{\rho }_{\beta ,
\gamma }+\hat {\Gamma }^{\rho }_{\gamma ,\beta })\hat {\Omega
}_{\alpha \rho }-(\hat {f}^{\rho }_{\gamma \alpha }-\hat {\Gamma
}^{\rho }_{\gamma ,\alpha }+\hat {\Gamma }^{\rho }_{\alpha ,\gamma
})\hat {\Omega }_{\beta \rho }-(\hat {f}^{\rho }_{\alpha \beta
}-\hat {\Gamma }^{\rho }_{\alpha ,\beta }+\hat {\Gamma }^{\rho
}_{\beta ,\alpha })\hat {\Omega }_{\gamma \rho } \nonumber \\
&=&0
\end{eqnarray}

Note that the interaction curvature can be represented as the
summation of gravity curvature and the gauge curvature,
\begin{eqnarray}
\hat {\Omega }_{\alpha \beta }=\frac {1}{2}\hat {R}^{\rho \sigma
}_{\alpha \beta }\otimes \hat {s}_{\rho \sigma }+\hat {F}^a_{\alpha
\beta }\otimes \hat {T}_a.
\end{eqnarray}

We next consider respectively the gravity curvature and the gauge
curvature. The gravity curvature takes the form
\begin{eqnarray}
\hat {R}_{\alpha \beta }^{\rho \sigma }=\hat {\theta }_{\alpha
}^{\mu }\partial _{\mu }\hat {\Gamma }_{\beta }^{\rho \sigma }-\hat
{\theta }_{\beta }^{\mu }\partial _{\mu }\hat {\Gamma }_{\alpha
}^{\rho \sigma }+\hat {\Gamma }_{\gamma ,\alpha }^{\rho }\hat
{\Gamma }_{\beta }^{\gamma \sigma }-\hat {\Gamma }_{\gamma ,\beta
}^{\rho }\hat {\Gamma }_{\alpha }^{\gamma \sigma }-\hat {f}_{\alpha
\beta }^{\gamma }\hat {\Gamma }_{\gamma }^{\rho \sigma },
\end{eqnarray}
\begin{eqnarray}
\hat {R}=\frac {1}{2}\hat {R}_{\alpha \beta }^{\rho \sigma }\otimes
\hat {s}^{\alpha \beta }\otimes \hat {s}_{\rho \sigma },
\end{eqnarray}
\begin{eqnarray}
\hat {\nabla }_{\alpha }=-i\hat {\theta }_{\alpha }^{\mu }\otimes
\hat {p}_{\mu }+\frac {1}{2}\hat {\Gamma }_{\alpha }^{\rho \sigma
}\otimes \hat {s}_{\rho \sigma },
\end{eqnarray}
\begin{eqnarray}
\hat {R}_{\mu \nu }=i[\hat {\nabla }_{\mu }, \hat {\nabla }_{\nu
}]-i\hat {f}_{\mu \nu }^{\lambda }\hat {\nabla }_{\lambda }.
\end{eqnarray}

The properties of the gravity curvature are listed in the following,
\begin{eqnarray}
\overline{\hat {R}_{\alpha \beta }^{\rho \sigma }}=\hat {R}_{\alpha
\beta }^{\rho \sigma },
\end{eqnarray}
\begin{eqnarray}
\hat {R}_{\alpha \beta }^{\rho \sigma }=-\hat {R}_{\alpha \beta
}^{\sigma \rho },
\end{eqnarray}
\begin{eqnarray}
\hat {R}_{\alpha \beta }^{\rho \sigma }=-\hat {R}_{\beta \alpha }^{\rho \sigma },
\end{eqnarray}
\begin{eqnarray}
\hat {R}_{\sigma ,\alpha \beta }^{\rho }+\hat {R}_{\alpha ,\beta
\sigma }^{\rho }+\hat {R}_{\beta ,\sigma \alpha }^{\rho }=0,
\end{eqnarray}
\begin{eqnarray}
\hat {R}_{\rho \sigma ,\alpha \beta }=\hat {R}_{\alpha \beta ,\rho
\sigma }
\end{eqnarray}

The Bianchi identity for gravity curvature and the contraction take
the form,
\begin{eqnarray}
\hat {D}_{\alpha }\hat {R}^{\rho \sigma }_{\beta \gamma }+\hat
{D}_{\beta }\hat {R}^{\rho \sigma }_{\gamma \alpha }+\hat
{D}_{\gamma }\hat {R}^{\rho \sigma }_{\alpha \beta }=0,
\end{eqnarray}
\begin{eqnarray}
D_{\alpha }(\hat {R}^{\alpha }_{\beta }-\frac {1}{2}\hat {R})=0,
\end{eqnarray}
where we used the notations $\hat {R}^{\alpha }_{\beta }=\hat
{R}^{\alpha \gamma }_{\beta \gamma }$, $\hat {R}=\hat {R}^{\beta
}_{\beta }=\hat {R}^{\beta \gamma }_{\beta \gamma }$. This means
\begin{eqnarray}
\hat {D}\wedge \hat {R}=0.
\end{eqnarray}
The proof that the summation of circular of indices is zero can be
like the follows,
\begin{eqnarray}
&&\hat {R}_{\sigma ,\alpha \beta }^{\rho }+\hat {R}_{\alpha ,\beta
\sigma }^{\rho }+\hat {R}_{\beta ,\sigma \alpha }^{\rho }\nonumber
\\
&=&\hat {\theta }_{\alpha }^{\mu }\partial _{\mu }\hat {\Gamma
}^{\rho }_{\beta ,\sigma }-\hat {\theta }_{\beta }^{\mu }\partial
_{\mu }\hat {\Gamma }^{\rho }_{\alpha ,\sigma }+\hat {\Gamma }^{\rho
}_{\gamma ,\alpha }\hat {\Gamma }^{\gamma }_{\beta ,\sigma }-\hat
{\Gamma }^{\rho }_{\gamma ,\beta }\hat {\Gamma }^{\gamma }_{\alpha
,\sigma } -\hat {f}^{\gamma }_{\alpha \beta }\hat {\Gamma }^{\rho
}_{\gamma
,\sigma }\nonumber \\
&&+\hat {\theta }_{\beta }^{\mu }\partial _{\mu }\hat {\Gamma
}^{\rho }_{\sigma ,\alpha }-\hat {\theta }_{\sigma }^{\mu }\partial
_{\mu }\hat {\Gamma }^{\rho }_{\beta ,\alpha }+\hat {\Gamma }^{\rho
}_{\gamma ,\beta }\hat {\Gamma }^{\gamma }_{\sigma ,\alpha }-\hat
{\Gamma }^{\rho }_{\gamma ,\sigma }\hat {\Gamma }^{\gamma }_{\beta
,\alpha } -\hat {f}^{\gamma }_{\beta \sigma }\hat {\Gamma }^{\rho
}_{\gamma
,\alpha }\nonumber \\
&&+\hat {\theta }_{\sigma }^{\mu }\partial _{\mu }\hat {\Gamma
}^{\rho }_{\alpha ,\beta }-\hat {\theta }_{\alpha }^{\mu }\partial
_{\mu }\hat {\Gamma }^{\rho }_{\sigma ,\beta }+\hat {\Gamma }^{\rho
}_{\gamma ,\sigma }\hat {\Gamma }^{\gamma }_{\alpha ,\beta }-\hat
{\Gamma }^{\rho }_{\gamma ,\alpha }\hat {\Gamma }^{\gamma }_{\sigma
,\beta } -\hat {f}^{\gamma }_{\sigma \alpha }\hat {\Gamma }^{\rho
}_{\gamma
,\beta }\nonumber \\
&=& \hat {\theta }_{\sigma }^{\mu }\partial _{\mu }\hat {f}^{\rho
}_{\alpha \beta }+\hat {\theta }_{\alpha }^{\mu }\partial _{\mu
}\hat {f}^{\rho }_{\beta ,\sigma }+\hat {\theta }_{\beta }^{\mu
}\partial _{\mu }\hat {f}^{\rho }_{\sigma \alpha }+\hat {f}^{\gamma
}_{\alpha \beta }\hat {f}^{\rho }_{\sigma \gamma }+\hat {f}^{\gamma
}_{\beta \sigma }\hat {f}^{\rho }_{\alpha  \gamma }+\hat {f}^{\gamma
}_{\sigma
\alpha }\hat {f}^{\rho }_{\beta \gamma } \nonumber \\
&=&0
\end{eqnarray}

The gauge curvature takes the form
\begin{eqnarray}
\hat {F}_{\alpha \beta }^{a}&=&\hat {\theta }_{\alpha }^{\mu
}\partial _{\mu }\hat {A}_{\beta }^a-\hat {\theta }_{\beta }^{\mu
}\partial _{\mu }\hat {A}_{\alpha }^a-iC_{bc}^a\hat {A}_{\alpha
}^b\hat {A}_{\beta }^c -\hat {f}_{\alpha \beta }^{\gamma }\hat
{A}_{\gamma }^a. \label{gauge-curvature}
\\
\hat {F}&=&\hat {F}^a_{\alpha \beta }\otimes \hat {s}^{\alpha \beta
}\otimes \hat {T}_a
\end{eqnarray}

The properties of the gauge curvature are listed in the following,
\begin{eqnarray}
&&\overline{\hat {F}^a_{\alpha \beta }}=G_{ab}\hat {F}^b_{\alpha
\beta
},\\
&&\hat {F}^a_{\alpha \beta }=-\hat {F}^a_{\beta \alpha },
\\
&&\hat {D}_{\alpha }\hat {F}_{\beta \gamma }^{a}+\hat {D}_{\beta
}\hat {F}_{\gamma \alpha }^{a}+\hat {D}_{\gamma }\hat {F}_{\alpha
\beta }^a=0.
\end{eqnarray}
In a concise form, we have
\begin{eqnarray}
\hat {D}\wedge \hat {F}=0.
\end{eqnarray}

We may also have the square of the general covariance derivation.
Some results are,
\begin{eqnarray}
[\hat {D}_{\alpha }, \hat {D}_{\beta }]|\Psi \rangle =(-i\hat
{\Omega }_{\alpha \beta }+\hat {f}^{\gamma }_{\alpha \beta }\hat
{D}_{\gamma })|\Psi \rangle .
\end{eqnarray}
For force field represented as operator $\hat {H}$, the square of
the general covariance derivation has properties,
\begin{eqnarray}
\hat {D}_{\alpha }(\hat {D}_{\beta }\hat {H})-\hat {D}_{\beta }(\hat
{D}_{\alpha }\hat {H})=-i[\hat {\Omega }_{\alpha \beta }, \hat {H}].
\end{eqnarray}
The proof of this equation is presented in the following. The Jacobi equation
takes the form
\begin{eqnarray}
[\hat {D}_{\alpha },[\hat {D}_{\beta },\hat {H}]]-[\hat {D}_{\beta
},[\hat {H}, \hat {D}_{\alpha }]]-[H,[\hat {D}_{\alpha },\hat
{D}_{\beta }]]=0
\end{eqnarray}
With the help of the equations,
\begin{eqnarray}
[\hat {D}_{\alpha },[\hat {D}_{\beta },\hat {H}]]-[\hat {D}_{\beta
},[\hat {D}_{\alpha },\hat {H}]]=[[\hat {D}_{\alpha },\hat
{D}_{\beta }], H]=-i[\hat {\Omega }_{\alpha \beta },\hat {H}]+\hat
{f}_{\alpha \beta }^{\gamma }[\hat {D}_{\gamma }, \hat {H}]
\end{eqnarray}
we can find
\begin{eqnarray}
&&\hat {D}_{\alpha }(\hat {D}_{\beta }\hat {H})-\hat {D}_{\beta
}(\hat
{D}_{\alpha }\hat {H})\nonumber \\
&=&[\hat {D}_{\alpha },[\hat {D}_{\beta },\hat {H}]]+\hat {\Gamma
}^{\gamma }_{\beta ,\alpha }[\hat {D}_{\gamma },\hat {H}]-[\hat
{D}_{\beta },[\hat {D}_{\alpha },\hat {H}]]-\hat {\Gamma }^{\gamma
}_{\alpha ,\beta }[\hat {D}_{\gamma },\hat {H}] \nonumber \\
&=&[\hat {D}_{\alpha },[\hat {D}_{\beta },\hat {H}]]-[\hat
{D}_{\beta },[\hat {D}_{\alpha },\hat {H}]]-\hat {f}_{\alpha \beta
}^{\gamma }[\hat {D}_{\gamma },\hat {H}]\nonumber \\
&=&-i[\hat {\Omega }_{\alpha \beta },\hat {H}].
\end{eqnarray}

\section{Dirac equation, Yang-Mills equation and Einstein equation}

\subsection{Dirac equation}

The Dirac equation for matter fields can be written as:
\begin{eqnarray}
(i\hat {\gamma }^{\alpha }\hat {D}_{\alpha }-\hat {m})|\Psi \rangle
=0, \label{dirac-eq}
\end{eqnarray}
where $\hat {m}$ is the mass matrix in gauge space, its eigenvalues
are masses of the corresponding elementary particles. Similarly the
adjoint matter fields also satisfy Dirac equation,
\begin{eqnarray}
\langle \Psi |(i\hat {D}_{\alpha }\hat {\gamma }^{\alpha }-\hat {m})
=0,
\end{eqnarray}

The square differential of the matter fields can be represented as
\begin{eqnarray}
(\hat {D}_{\alpha }\hat {D}^{\alpha }-\hat {s}^{\alpha \beta }\hat
{\Omega }_{\alpha \beta }+\hat {\Gamma }^{\alpha }_{\alpha ,\beta
}\hat {D}^{\beta }+A_{\alpha }^a\hat {\gamma }^{\alpha }\hat
{m}_a+\hat {m}^2)|\Psi \rangle =0,
\end{eqnarray}
or
\begin{eqnarray}
(\hat {D}_{\alpha }\hat {D}^{\alpha }-\hat {R}-\hat {F}+\hat {\Gamma
}^{\alpha }_{\alpha ,\beta }\hat {D}^{\beta }+A_{\alpha }^a\hat
{\gamma }^{\alpha }\hat {m}_a+\hat {m}^2)|\Psi \rangle &=&0.
\label{square-dirac}
\end{eqnarray}
The proof of this equation starts from the Dirac equation by
applying operator $(-i\hat {\gamma }^{\alpha }\hat {D}_{\alpha
}-\hat {m})$ on both sides,
\begin{eqnarray}
(-i\hat {\gamma }^{\alpha }\hat {D}_{\alpha }-\hat {m})(i\hat
{\gamma }^{\alpha }\hat {D}_{\alpha }-\hat {m})|\Psi \rangle =0.
\end{eqnarray}
We have

\begin{eqnarray}
(-i\hat {\gamma }^{\alpha }\hat {D}_{\alpha }-\hat {m})(i\hat
{\gamma }^{\beta }\hat {D}_{\beta }-\hat {m})=\hat {\gamma }^{\alpha
}\hat {D}_{\alpha }\hat {\gamma }^{\beta }\hat {D}_{\beta }+i\hat
{\gamma }^{\alpha }[\hat {D}_{\alpha },\hat {m}]+\hat {m}^2,
\end{eqnarray}
where
\begin{eqnarray}
\hat {\gamma }^{\alpha }\hat {D}_{\alpha }\hat {\gamma }^{\beta
}\hat {D}_{\beta }&=&\hat {\gamma }^{\alpha }\hat {\gamma }^{\beta
}\hat {D}_{\alpha }\hat {D}_{\beta }+\hat {\gamma }^{\alpha }[\hat
{D}_{\alpha },\hat {\gamma }^{\beta }]\hat {D}_{\beta } \nonumber \\
&=&\hat {D}_{\alpha }\hat {D}^{\alpha }-i\hat {s}^{\alpha \beta
}[\hat {D}_{\alpha },\hat {D}_{\beta }]-\hat {\Gamma }^{\gamma }
_{\alpha ,\beta }\hat {\gamma }^{\alpha }\hat {\gamma }^{\beta }\hat
{D}_{\gamma }
\nonumber \\
&=&\hat {D}_{\alpha }\hat {D}^{\alpha }-\hat {s}^{\alpha \beta
}(\hat {\Omega }_{\alpha \beta }+if^{\gamma }_{\alpha \beta }\hat
{D}_{\gamma })+\hat {\Gamma }_{\alpha }^{\gamma \alpha }\hat
{D}_{\gamma }+i\hat {f}^{\gamma }_{\alpha \beta } \hat {s}^{\alpha
\beta }\hat {D}_{\gamma } \nonumber \\
&=&\hat {D}_{\alpha }\hat {D}^{\alpha }-\hat {s}^{\alpha \beta }\hat
{\Omega }_{\alpha \beta }+\hat {\Gamma }^{\alpha }_{\alpha ,\gamma
}\hat {D}^{\gamma },
\end{eqnarray}
\begin{eqnarray}
i\hat {\gamma }^{\alpha }[\hat {D}_{\alpha },\hat {m}]=\hat
{A}_{\alpha }^a\hat {\gamma }^{\alpha }\hat {m}_a.
\end{eqnarray}
Then
\begin{eqnarray}
&&(-i\hat {\gamma }^{\alpha }\hat {D}_{\alpha }-\hat {m})(i\hat
{\gamma }^{\beta }\hat {D}_{\beta }-\hat {m}) \nonumber \\
&=& \hat {\gamma }^{\alpha }\hat {\gamma }^{\beta }\hat {D}_{\alpha
}\hat {D}_{\beta }+\hat {\gamma }^{\alpha }[\hat {D}_{\alpha },\hat
{\gamma }^{\beta }]\hat {D}_{\beta }+i\hat {\gamma }^{\alpha }[\hat
{D}_{\alpha },\hat {m}]+\hat {m}^2 \nonumber \\
&=&\hat {D}_{\alpha }\hat {D}^{\alpha }-i\hat {s}^{\alpha \beta
}[\hat {D}_{\alpha },\hat {D}_{\beta }]+\Gamma ^{\gamma }_{\beta
,\alpha }\hat {\gamma }^{\alpha }\hat {\gamma }^{\beta }\hat
{D}_{\gamma }-A^a_{\alpha }\hat {\gamma }^{\alpha }[\hat {T}_a,\hat
{m}]+\hat {m}^2\nonumber \\
&=&\hat {D}_{\alpha }\hat {D}^{\alpha }-\hat {s}^{\alpha \beta
}(\hat {\Omega }_{\alpha \beta }+i\hat {f}^{\gamma }_{\alpha \beta
}\hat {D}_{\gamma })-\Gamma ^{\gamma \alpha }_{\alpha }\hat
{D}_{\gamma }+i\hat {f}^{\gamma }_{\alpha \beta }\hat {s}^{\alpha
\beta }\hat {D}_{\gamma }-\hat {\gamma }^{\alpha }[\hat {T}_a,\hat
{m}]A^a_{\alpha }+\hat {m}^2 \nonumber \\
&=&\hat {D}_{\alpha }\hat {D}^{\alpha }-\hat {s}^{\alpha \beta }\hat
{\Omega }_{\alpha \beta }-\Gamma ^{\gamma \alpha }_{\alpha }\hat
{D}_{\gamma }-\hat {\gamma }^{\alpha }[\hat {T}_a,\hat
{m}]A^a_{\alpha }+\hat {m}^2,
\end{eqnarray}
where we have used the equation,
\begin{eqnarray}
\hat {\gamma }^{\alpha }\hat {\gamma }^{\beta }=g^{\alpha \beta
}-2i\hat {s}^{\alpha \beta }.
\end{eqnarray}
Thus the squared differential equation of the matter fields is
obtained. Similarly, we have
\begin{eqnarray}
\langle \Psi |(\hat {D}_{\alpha }\hat {D}^{\alpha }-\hat {\Omega
}_{\alpha \beta }\hat {s}^{\alpha \beta }-\hat {D}^{\beta }\hat
{\Gamma }^{\alpha }_{\alpha ,\beta }-A_{\alpha }^a\hat {\gamma
}^{\alpha }\hat {m}_a+\hat {m}^2)=0.
\end{eqnarray}
We define energy-momentum tensor of matter fields as
\begin{eqnarray}
t^{\alpha }_{\beta }(x)&=&\frac {i}{4}\{ \langle \Psi |\hat
{\varepsilon }(x)[(\hat {\gamma }^{\alpha }\hat {D}_{\beta }+\hat
{\gamma }_{\beta }\hat {D}^{\alpha })|\Psi \rangle ]+[\langle \Psi
|(\hat {D}_{\beta }\hat {\gamma }^{\alpha }+\hat {D}^{\alpha }\hat
{\gamma }_{\beta })]\hat {\varepsilon }(x)|\Psi \rangle \} \nonumber
\end{eqnarray}

\subsection{Yang-Mills equation and gauge condition}
The gauge fields Yang-Mills equation takes the form
\begin{eqnarray}
\hat {D}\cdot \hat {F}=\hat {J}, \label{ym-eq}
\end{eqnarray}
or explicitly it can be written as,
\begin{eqnarray}
\hat {D}_{\alpha }\hat {F}_a^{\alpha \beta }=\hat {J}_a^{\beta }.
\end{eqnarray}
Here the total current is written as,
\begin{eqnarray}
\hat {J}_a^{\beta }=\hat {j}_a^{\beta }+M_a^b\hat {A}_b^{\beta },
\label{ym-total-current}
\end{eqnarray}
$M_a^b$ are mass tensor of gauge bosons with $M_2^2=m_z^2$,
$M_3^3=M_4^4=m^2_{W}$, $M_{4+p}^{4+p}=m^2, p=1,2,\cdots ,8$ for
gluons and $M_a^b=0$ elsewhere, $j^{\beta }_a$ is the gauge current
density defined as,
\begin{eqnarray}
j^{\beta }_a(x)=\langle \Psi |\hat {\varepsilon }(x) \hat {\gamma
}^{\beta }\hat {T}_a|\Psi \rangle . \label{ym-gauge-current}
\end{eqnarray}

We have made the calculations,
\begin{eqnarray}
\hat {D}_{\alpha }\hat {F}_a^{\alpha \beta }&=& \hat {\theta
}_{\alpha }^{\mu }\partial _{\mu }\hat {F}_a^{\alpha \beta }
+C_{ab}^c\hat {A}^b_{\alpha }\hat {F}_c^{\alpha \beta }+\hat {\Gamma
}^{\alpha }_{\alpha ,\rho } \hat {F}_a^{\rho \beta }+\hat {\Gamma
}^{\beta }_{\alpha ,\rho } \hat
{F}_a^{\alpha \rho } \nonumber \\
&=&\hat {J}_a^{\beta }.
\end{eqnarray}

The conservation law for the gauge charges is
\begin{eqnarray}
\hat {D}_{\beta }\hat {J}_a^{\beta }&=&0, \nonumber \\
{\rm or}~~~~~\hat {D}\cdot \hat {J}&=&0.
\end{eqnarray}

The proof of the conservation law of gauge charges is like the
following. Due to the Yang-Mills equation,
\begin{eqnarray}
\hat {D}_{\beta }\hat {J}_a^{\beta }&=&\hat {D}_{\beta }\hat
{D}_{\alpha }\hat {F}_a^{\alpha \beta } \nonumber \\
&=&\frac {1}{2}(\hat {D}_{\beta }\hat {D}_{\alpha }-\hat {D}_{\alpha
}\hat {D}_{\beta })\hat {F}_a^{\alpha \beta }.
\end{eqnarray}
With the help of the squared covariance derivation, we have
\begin{eqnarray}
&&(\hat {D}_{\beta }\hat {D}_{\beta }-\hat {D}_{\beta }\hat
{D}_{\beta
})\hat {F}_a^{\alpha \beta }\nonumber \\
&=&C^c_{ab}\hat {F}^b_{\beta \alpha }\hat {F}^{\alpha \beta }_c
+\hat {R}^{\alpha }_{\rho ,\beta \alpha }\hat {F}_a^{\rho \beta }+
\hat {R}^{\beta }_{\rho ,\beta \alpha }\hat {F}_a^{\alpha \rho }
\nonumber \\
&=&C_{abc}\hat {F}^b_{\beta \alpha }\hat {F}^{c,\alpha \beta }+2
\hat {R}_{\alpha ,\beta }\hat {F}_a^{\alpha \beta } \nonumber \\
&=&0,
\end{eqnarray}
where we have used the properties that $C_{abc}$ are antisymmetric
while $\hat {R}_{\alpha ,\beta }$ are symmetric.

We next consider the gauge condition and the mass commutation
relation. Since the conservation law for gauge charges and the
divergence of the gauge current, we can find that the gauge
connections satisfy the condition,
\begin{eqnarray}
\hat {D}_{\alpha }(M^b_a\hat {A}^{\alpha }_b)&=&-\hat {u}_a
\nonumber \\
&=&i\left( [\hat {T}_a, \hat {m}]|\Psi \rangle \right)\langle
\overline{\Psi }|. \label{ym-mass-comm}
\end{eqnarray}
This is the gauge condition. We can see that in the present work
this condition is not artificial but a necessary condition for
Yang-Mills equation. On the other hand, this equation is a
restriction which indicates the relation between gauge mass and the
mass of matter fields.

Since mass of photon is zero, $M^1_1=0$, we also have
\begin{eqnarray}
[\hat {Q}, \hat {m}]=0.
\end{eqnarray}
The gauge condition (\ref{ym-mass-comm}) is always satisfied, so
gauge of electromagnetic is arbitrary.

The masses of weak charges $\hat {W}_+, \hat {W}_-, \hat {Z}_0$ are
not zeroes, also they do not commute with the mass matrices of
matter fields,
\begin{eqnarray}
[\hat {W}_{\pm }, \hat {m}]\not= 0,
\end{eqnarray}
\begin{eqnarray}
[\hat {Z}_0, \hat {m}]\not= 0.
\end{eqnarray}
From Eq.(\ref{ym-mass-comm}), we know that $M_a^b$ can not be zero
for weak charges $\hat {W}_{\pm },\hat {Z}_0$, that means weak
bosons have masses. Additionally, the gauge condition
(\ref{ym-mass-comm}) is a constraint.

The masses of gluons are all $m$, and we know that,
\begin{eqnarray}
[\hat {\lambda }_p, \hat {m}]=0,
\end{eqnarray}
where $p=1,2,\cdots ,8$. The gauge condition becomes as,
\begin{eqnarray}
\hat {D}_{\alpha }(M^a_a\hat {A}^{\alpha }_b)=\hat {D}_{\alpha
}(m^2\hat {A}^{\alpha }_b)=0,
\end{eqnarray}
that means,
\begin{eqnarray}
\hat {D}_{\alpha }\hat {A}^{\alpha }_b=0.
\end{eqnarray}
This is the gauge condition for gluon field.

The energy-momentum tensor of the gauge fields is
\begin{eqnarray}
\hat {\tau }^{\alpha }_{\beta }=\hat {F}_a^{\alpha \rho }\hat
{F}^a_{\beta \rho }-\frac {1}{4}\delta ^{\alpha }_{\beta }\hat
{F}_a^{\rho \sigma }\hat {F}^a_{\rho \sigma }+M^a_b\hat {A}^{\alpha
}_a\hat {A}^b_{\beta }-\frac {1}{2}\delta ^{\alpha }_{\beta
}M^a_b\hat {A}^{\rho }_a\hat {A}^b_{\rho }. \label{en-mo-gauge}
\end{eqnarray}

\subsection{Einstein equation and energy-momentum conservation law}

The Einstein equation takes the form,
\begin{eqnarray}
\hat {R}^{\alpha }_{\beta }-\frac {1}{2}\delta _{\beta }^{\alpha
}\hat{R}=-8\pi G\hat {T}^{\alpha }_{\beta }, \label{einstein-eq}
\end{eqnarray}
where $G$ is the gravity constant, $\hat {T}^{\alpha }_{\beta }$ is
the total energy-momentum tensor, $\hat {R}^{\alpha }_{\beta }=\hat
{R}^{\alpha \gamma }_{\beta \gamma }$ and $\hat {R}=\hat {R}^{\alpha
}_{\alpha }$. There is no mass of graviton appeared in Einstein
equation, we thus mean that graviton is massless. From Einstein
equation and the contracted tensor of Bianchi identity of gravity,
we can find,
\begin{eqnarray}
\hat {D}_{\alpha }(\hat {R}^{\alpha }_{\beta }-\frac {1}{2}\delta
^{\alpha }_{\beta }\hat {R})=0.
\end{eqnarray}
This turns out to be the energy-momentum conservation law,
\begin{eqnarray}
\hat {D}_{\alpha }\hat {T}^{\alpha }_{\beta }=0.
\end{eqnarray}

The energy-momentum tensor is defined as the total energy-momentum
tensors of gauge fields $\hat {\tau }^{\alpha }_{\beta }$ and matter
fields $\hat {t}^{\alpha }_{\beta }$,
\begin{eqnarray}
\hat {T}^{\alpha }_{\beta }=\hat {\tau }^{\alpha }_{\beta }+\hat
{t}^{\alpha }_{\beta }.
\end{eqnarray}
We will later show that the energy momentum conservation law can
also be proved by a direct calculations from Dirac equation and
Yang-Mills equation.

\subsection{The unification properties of Dirac equation, Yang-Mill
equation and Einstein equation}

Here we would like to present the relationships for the three
fundamental equations. For convenience, we list all the related
relations here:

\begin{enumerate}
\item Dirac equation:
\begin{eqnarray}
(i\hat {\gamma }^{\alpha }\hat {D}_{\alpha }-\hat {m})|\Psi \rangle
=0. \nonumber
\end{eqnarray}
\item Yang-Mills equation:
\begin{eqnarray}
\hat {D}_{\alpha }\hat {F}_a^{\alpha \beta }=\hat {J}_a^{\beta }.
\nonumber
\end{eqnarray}
\item Einstein equation:
\begin{eqnarray}
\hat {R}^{\alpha }_{\beta }-\frac {1}{2}\delta _{\beta }^{\alpha
}\hat{R}=-8\pi G\hat {T}^{\alpha }_{\beta }. \nonumber
\end{eqnarray}
\item Total gauge current density:
\begin{eqnarray}
\hat {J}_a^{\beta }=\hat {j}_a^{\beta }+M_a^b\hat {A}_b^{\beta }.
\nonumber
\end{eqnarray}
\item Gauge current density of matter fields:
\begin{eqnarray}
j^{\beta }_a(x)=\langle \Psi |\hat {\varepsilon }(x) \hat {\gamma
}^{\beta }\hat {T}_a|\Psi \rangle . \nonumber
\end{eqnarray}
\item Total energy-momentum tensor:
\begin{eqnarray}
\hat {T}^{\alpha }_{\beta }=\hat {\tau }^{\alpha }_{\beta }+\hat
{t}^{\alpha }_{\beta }. \nonumber
\end{eqnarray}
\item Energy-momentum tensor of matter fields:
\begin{eqnarray}
t^{\alpha }_{\beta }(x)&=&\frac {i}{4}\{ \langle \Psi |\hat
{\varepsilon }(x)[(\hat {\gamma }^{\alpha }\hat {D}_{\beta }+\hat
{\gamma }_{\beta }\hat {D}^{\alpha })|\Psi \rangle ]+[\langle \Psi
|(\hat {D}_{\beta }\hat {\gamma }^{\alpha }+\hat {D}^{\alpha }\hat
{\gamma }_{\beta })]\hat {\varepsilon }(x)|\Psi \rangle \} \nonumber
\end{eqnarray}
\item Energy-momentum tensor of gauge fields:
\begin{eqnarray}
\hat {\tau }^{\alpha }_{\beta }=\hat {F}_a^{\alpha \rho }\hat
{F}^a_{\beta \rho }-\frac {1}{4}\delta ^{\alpha }_{\beta }\hat
{F}_a^{\rho \sigma }\hat {F}^a_{\rho \sigma }+M^a_b\hat {A}^{\alpha
}_a\hat {A}^b_{\beta }-\frac {1}{2}\delta ^{\alpha }_{\beta
}M^a_b\hat {A}^{\rho }_a\hat {A}^b_{\rho }. \nonumber
\end{eqnarray}
\end{enumerate}
The general covariance derivative operator has the elements of the
gravity field, gauge fields. It corresponds to force. The Dirac
equation means that the gravity force and the gauge forces can act
on the matter fields. For Yang-Mills equation, we find that the
gravity force has effect on gauge fields, and gauge fields have
effects on themselves. For Einstein equation, gravity can acts on
itself. From the representation of gravity field, we find that gauge
charge of graviton is zero. So gauge force does not have effect on
gravity field. We can thus find that all three fundamental equations
are involved together. One key point that those equations are
compatible is that from Einstein equation, we can prove the
energy-momentum conservation law. This result can also be obtained
from Dirac equation and Yang-Mills equation. The total
energy-momentum of matter fields and gauges fields in Einstein
equation can be understood as the source of gravity.

\section{Energy-momentum conservation law and physical quantities}
Based on the representations and the fundamental equations. We next
consider some physical quantities of the unified theory. By
tremendous calculations, we find one important result of our theory:
the energy-momentum conservation law. With this result, we confirm
that our theory is a compatible and combined form for three
fundamental equations.

\subsection{Particle current density and spin current density}
By using matter field $|\Psi \rangle $, its adjoint $\langle \Psi |$
and the antisymmetric matrix $\hat {\gamma }^{\alpha _1\cdots \alpha
_p}$ (p=0,1,2,3,), one can construct two antisymmetric tensors
listed as
\begin{eqnarray}
\rho ^{\alpha }(x)=\langle \Psi |\hat {\varepsilon }(x)\hat {\gamma
}^{\alpha }|\Psi \rangle ,
\end{eqnarray}
\begin{eqnarray}
\rho ^{\alpha \beta \gamma }(x)=\langle \Psi |\hat {\varepsilon
}(x)\hat {\gamma }^{\alpha \beta \gamma }|\Psi \rangle ,
\end{eqnarray}
The first one $\rho ^{\alpha }(x)$ is the particle current density,
the second one $\rho ^{\alpha \beta \gamma }(x)$ is related with the
spin current density. The antisymmetric tensor fields can be
represented in an unified form,
\begin{eqnarray}
\rho ^{\alpha _1\cdots \alpha _p }(x)=\langle \Psi |\hat
{\varepsilon }(x)\hat {\gamma }^{\alpha _1\cdots \alpha _p }|\Psi
\rangle .
\end{eqnarray}
This antisymmetric tensor field and its adjoint are the same due to
the properties of $\hat {\varepsilon}$ and $\hat {\gamma }^{\alpha
_1\cdots \alpha _p }$,
\begin{eqnarray}
\overline{\rho ^{\alpha _1\cdots \alpha _p }(x)}=\rho ^{\alpha
_1\cdots \alpha _p }(x).
\end{eqnarray}
It is also antisymmetric.

The results of covariance differential are presented as,
\begin{eqnarray}
\hat {D}_{\alpha }\rho ^{\alpha }(x)=0,
\end{eqnarray}
\begin{eqnarray}
\hat {D}_{\alpha }\rho ^{\alpha \beta \gamma }(x)=0,
\end{eqnarray}
The proof of the first one is like the following,
\begin{eqnarray}
\theta ^{\mu }_{\alpha }(x)\partial _{\mu }\rho ^{\alpha }&=&\theta
^{\mu }_{\alpha }(x)\partial _{\mu }\langle \Psi |\hat
{\varepsilon }(x)\hat {\gamma }^{\alpha }|\Psi \rangle \nonumber \\
&=&\langle \Psi |\hat {\varepsilon }(x)\left( \hat {D}_{\alpha }\hat
{\gamma }^{\alpha }|\Psi \rangle \right) -\left(\langle \Psi |\hat
{D}_{\alpha }\right)\hat {\varepsilon }(x)\hat {\gamma
}^{\alpha }|\Psi \rangle \nonumber \\
&=&\langle \Psi |\left( [\hat {D}_{\alpha }, \hat {\gamma }^{\alpha
}]\hat {\varepsilon }(x)|\Psi \rangle \right) -i\langle \Psi |\hat
{\varepsilon }(x)\left( i\hat {\gamma }^{\alpha }\hat {D}_{\alpha
}|\Psi \rangle \right) +i\left(\langle \Psi |i\hat {D}_{\alpha }\hat
{\gamma }^{\alpha }\right)\hat {\varepsilon }(x)|\Psi \rangle \nonumber \\
&=&-\hat {\Gamma }^{\alpha }_{\alpha ,\beta }(x)\langle \Psi |\hat
{\varepsilon }(x)\hat {\gamma }^{\beta }|\Psi \rangle -i\langle \Psi
|\hat {\varepsilon }(x)\left( \hat {m}|\Psi \rangle \right)+i\left(
\langle \Psi |\hat {m}\right) \hat {\varepsilon }(x)|\Psi \rangle
\nonumber \\
&=&-\hat {\Gamma }^{\alpha }_{\alpha ,\beta }(x)\rho ^{\beta }(x).
\end{eqnarray}
We thus have
\begin{eqnarray}
\hat {D}_{\alpha }\rho ^{\alpha }(x)=\theta ^{\mu }_{\alpha
}(x)\partial _{\mu }\rho ^{\alpha }(x)+\hat {\Gamma }^{\alpha
}_{\alpha ,\beta }(x)\rho ^{\beta }(x)=0.
\end{eqnarray}

The proof of the second one is like the proof of the first one and
is given as,
\begin{eqnarray}
\theta _{\alpha }^{\mu }(x)\partial _{\mu }\rho ^{\alpha \beta
\gamma }(x)&=&\theta ^{\mu }_{\alpha }(x)\partial _{\mu }\langle
\Psi |\hat {\varepsilon }(x)\hat {\gamma }^{\alpha \beta \gamma
}|\Psi \rangle \nonumber \\
&=&\langle \Psi |\hat {\varepsilon }(x)\left( \hat {D}_{\alpha }\hat
{\gamma }^{\alpha \beta \gamma }|\Psi \rangle \right) -\left(\langle
\Psi |\hat {D}_{\alpha }\right)\hat {\varepsilon }(x)\hat {\gamma
}^{\alpha \beta \gamma }|\Psi \rangle \nonumber \\
&=&\langle \Psi |\left( [\hat {D}_{\alpha }, \hat {\gamma }^{\alpha
\beta \gamma }]\hat {\varepsilon }(x)|\Psi \rangle \right) -i\langle
\Psi |\hat {\varepsilon }(x)\hat {\gamma }^{\beta \gamma }\left(
i\hat {\gamma }^{\alpha }\hat {D}_{\alpha }|\Psi \rangle \right)
+i\left(\langle \Psi |i\hat {D}_{\alpha }\hat {\gamma }^{\alpha
}\right)\hat {\gamma }^{\beta \gamma }
\hat {\varepsilon }(x)|\Psi \rangle \nonumber \\
&=&-\hat {\Gamma }^{\alpha }_{\alpha ,\rho }(x)\langle \Psi |\hat
{\varepsilon }(x)\hat {\gamma }^{\rho \beta \gamma }|\Psi \rangle
-\hat {\Gamma }^{\beta }_{\alpha ,\rho }(x)\langle \Psi |\hat
{\varepsilon }(x)\hat {\gamma }^{\alpha \rho \gamma }|\Psi \rangle
-\hat {\Gamma }^{\gamma }_{\alpha ,\rho }(x)\langle \Psi |\hat
{\varepsilon }(x)\hat {\gamma }^{\alpha \beta \rho }|\Psi \rangle
\nonumber \\
&& -i\langle \Psi |\hat {\varepsilon }(x)\hat {\gamma }^{\beta
\gamma }\left( \hat {m}|\Psi \rangle \right)+i\left( \langle \Psi
|\hat {m}\right) \hat {\gamma }^{\beta \gamma }\hat {\varepsilon
}(x)|\Psi \rangle
\nonumber \\
&=&-\hat {\Gamma }^{\alpha }_{\alpha ,\rho }(x)\rho ^{\rho \beta
\gamma }(x)-\hat {\Gamma }^{\beta }_{\alpha ,\rho }(x)\rho ^{\alpha
\rho \gamma }(x)-\hat {\Gamma }^{\gamma }_{\alpha ,\rho }(x)\rho
^{\alpha \beta \rho }(x).
\end{eqnarray}
We thus have
\begin{eqnarray}
&&\hat {D}_{\alpha }\rho ^{\alpha \beta \gamma }(x) \nonumber \\
&=&\theta ^{\mu }_{\alpha }(x)\partial _{\mu }\rho ^{\alpha \beta
\gamma }(x)+\hat {\Gamma }^{\alpha }_{\alpha ,\rho }(x)\rho ^{\rho
\beta \gamma }(x)+\hat {\Gamma }^{\beta }_{\alpha ,\rho }(x)\rho
^{\alpha \rho \gamma }(x)+\hat {\Gamma }^{\gamma }_{\alpha ,\rho
}(x)\rho ^{\alpha \beta \rho }(x) \nonumber \\
&=&0.
\end{eqnarray}

The particle current density in 4D is defined by $\rho ^{\alpha
}(x)$, where $\rho ^0(x)$ is the particle number density, $\rho
^i(x), (i=1,2,3)$ are the particle current density in 3D. So
equation
\begin{eqnarray}
\hat {D}_{\alpha }\rho ^{\alpha }(x)=0
\end{eqnarray}
means that the number of particles is conserved. The 4D current
density of particles with gauge indices $t$ can be defined as,
\begin{eqnarray}
\rho ^{\alpha }_{(t)}=\langle \Psi _t|\hat {\varepsilon }(x)\hat
{\gamma }^{\alpha }|\Psi ^t\rangle ,
\end{eqnarray}
where $|\Psi ^t\rangle $ is the $t$ gauge element of matter fields,
note that no summation is assumed in the above equation.

The spin of the matter particles is related with $S^{\alpha \beta
\gamma }(x)\equiv \frac {1}{2}\rho ^{\alpha \beta \gamma }(x)$, so
$S^{\alpha \beta \gamma }(x)$ can be considered as the 4D spin
current density. By definition, the spin current density is
\begin{eqnarray}
S^{\alpha \beta \gamma }&=&\frac {1}{2}\langle \Psi |\hat
{\varepsilon }(x)(\hat {\gamma }^{\alpha }\hat {s}^{\beta \gamma
}+\hat {s}^{\beta \gamma }\hat {\gamma }^{\alpha })|\Psi \rangle
\nonumber \\
&=&\frac {1}{2}\langle \Psi |\hat {\varepsilon }(x)\hat {\gamma
}^{\alpha \beta \gamma }|\Psi \rangle \nonumber \\
\end{eqnarray}
We can find that this spin current density is antisymmetric,
\begin{eqnarray}
S_{\alpha \beta \gamma }=S_{\beta \gamma \alpha }=S_{\gamma \alpha
\beta }=-S_{\gamma \beta \alpha }=-S_{\beta \alpha \gamma
}=-S_{\alpha \gamma \beta }
\end{eqnarray}
The divergence equations of the spin current is zero as we have
proved,
\begin{eqnarray}
\hat {D}_{\alpha }S^{\alpha \beta \gamma }&=&0.
\end{eqnarray}

\subsection{The gauge current density of matter fields}

The gauge current density and gauge charge production rate density
of matter fields are defined respectively as,
\begin{eqnarray}
j^{\beta }_a(x)=\langle \Psi |\hat {\varepsilon }(x) \hat {\gamma
}^{\beta }\hat {T}_a|\Psi \rangle , \nonumber
\end{eqnarray}
\begin{eqnarray}
u_a(x)&=&-i\langle \Psi |\hat {m}_a\hat {\varepsilon }(x)|\Psi
\rangle \nonumber \\
&=&-i\langle \Psi |\hat {\varepsilon }(x)[\hat {T}_a,\hat {m}]|\Psi
\rangle .
\end{eqnarray}
Recall the definition of the gauge current density of matter fields
(\ref{ym-gauge-current}), we have an equation,
\begin{eqnarray}
\hat {D}_{\alpha }j^{\alpha }_a=u_a.
\end{eqnarray}
It means that for matter fields, the changing of the gauge current
density equals to the gauge charge production rate density.

The proof is presented below.
\begin{eqnarray}
\hat {\theta }_{\alpha }^{\mu }(x)\partial _{\mu }j^{\alpha }_a(x)
&=&\hat {\theta }_{\alpha }^{\mu }(x)\partial _{\mu }\langle \Psi
|\hat {\varepsilon }(x)\hat {\gamma }^{\alpha }\hat {T}_a|\Psi
\rangle \nonumber
\\
&=&\langle \Psi |\hat {\varepsilon}(x)\left( \hat {D}_{\alpha }\hat
{\gamma }^{\alpha }\hat {T}_a|\Psi \rangle \right)-\left( \langle
\Psi |\hat {D}_{\alpha }\right) \hat {\varepsilon }(x)\hat {\gamma
}^{\alpha }\hat {T}_a|\Psi \rangle \nonumber
\\
&=&\langle \Psi |\hat {\varepsilon }(x) \left( [\hat {D}_{\alpha
},\hat {\gamma }^{\alpha }]\hat {T}_a|\Psi \rangle \right) + \langle
\Psi |\hat {\varepsilon }(x) \left( \hat {\gamma }^{\alpha
}[\hat {D}_{\alpha },\hat {T}_a]|\Psi \rangle \right) \nonumber \\
&&-i\langle \Psi |\hat {\varepsilon }(x)\hat {T}_a\left( i\hat
{\gamma }^{\alpha }\hat {D}_{\alpha } |\Psi \rangle \right) +i\left(
\langle \Psi |\hat {D}_{\alpha }\hat {\gamma }^{\alpha }i\right)
\hat {\varepsilon}(x)\hat {T}_a|\Psi \rangle \nonumber
\\
&=&-\hat {\Gamma }^{\alpha }_{\alpha ,\beta }(x)\langle \Psi |\hat
{\varepsilon }(x)\hat {\gamma }^{\beta }\hat {T}_a|\Psi \rangle
+iC^c_{ab}\hat {A}^b_{\alpha }(x)\langle \Psi |\hat {\varepsilon
}(x)\hat {\gamma }^{\alpha }\hat {T}_c|\Psi \rangle \nonumber \\
&&-i\langle \Psi |\hat {\varepsilon }(x)\hat {T}_a\hat {m}|\Psi
\rangle +i\langle \Psi |\hat {m}\hat {T}_a\hat {\varepsilon
}(x)|\Psi \rangle \nonumber
\\
&=&-\hat {\Gamma }^{\alpha }_{\alpha ,\beta }(x)j^{\beta
}_a(x)+iC_{ab}^c\hat {A}_{\alpha }^b(x)j_c^{\alpha }(x)+u_a(x)
\end{eqnarray}
Thus we can find,
\begin{eqnarray}
\hat {D}_{\alpha }j^{\alpha }_a=\theta ^{\mu }_{\alpha }\partial
_{\mu }j^{\alpha }_a+\hat {\Gamma }^{\alpha }_{\alpha ,\beta
}j^{\alpha }_a-iC^c_{ab}\hat {A}^b_{\alpha }j^{\alpha }_c=u_a.
\end{eqnarray}

\subsection{Energy-momentum conservation law of matter fields and gauge fields}

Recall the definition of the energy-momentum tensor of gauge fields,
\begin{eqnarray}
\hat {\tau }^{\alpha }_{\beta }=\hat {F}_a^{\alpha \rho }\hat
{F}^a_{\beta \rho }-\frac {1}{4}\delta ^{\alpha }_{\beta }\hat
{F}_a^{\rho \sigma }\hat {F}^a_{\rho \sigma }+M^a_b\hat {A}^{\alpha
}_a\hat {A}^b_{\beta }-\frac {1}{2}\delta ^{\alpha }_{\beta
}M^a_b\hat {A}^{\rho }_a\hat {A}^b_{\rho }.
\end{eqnarray}
We can find the following important property of the energy-momentum
tensor of gauge fields with the help of Yang-Mills equation and Eq.
(\ref{ym-mass-comm}),
\begin{eqnarray}
\hat {D}_{\alpha }\hat {\tau }^{\alpha }_{\beta }=-\hat {j}^{\alpha
}_a\hat {F}^a_{\alpha \beta }-\hat {A}^a_{\beta }\hat {u}_a.
\label{ym-en-mo}
\end{eqnarray}
The proof is presented below,
\begin{eqnarray}
&&\hat {D}_{\alpha }(\hat {F}_a^{\alpha \rho }\hat {F}^a_{\beta \rho
}-\frac {1}{4}\delta ^{\alpha }_{\beta }\hat {F}_a^{\rho \sigma
}\hat {F}^a_{\rho \sigma }) \nonumber \\
&=&(\hat {D}_{\alpha }\hat {F}_a^{\alpha \rho })\hat {F}^a_{\beta
\rho }+\hat {F}_a^{\alpha \rho }(\hat {D}_{\alpha }\hat {F}^a_{\beta
\rho })-\frac {1}{2}\hat {F}_a^{\rho \sigma }(\hat {D}_{\beta
}\hat {F}^a_{\rho \sigma }) \nonumber \\
&=&(\hat {D}_{\alpha }\hat {F}_a^{\alpha \rho })\hat {F}^a_{\beta
\rho }+\frac {1}{2}\hat {F}_a^{\alpha \rho }(\hat {D}_{\alpha }\hat
{F}^a_{\beta \rho }+\hat {D}_{\rho }\hat {F}^a_{\alpha \beta }+\hat
{D}_{\beta }\hat {F}^a_{\rho \alpha }) \nonumber \\
&=&-\hat {J}_a^{\alpha }\hat {F}^a_{\alpha \beta }.
\end{eqnarray}
From the definition of gauge curvature and the relation between
masses of gauge fields and matter fields, also $M_{ab}=M_{ba},
M_{ab}=G_{bb'}M_a^{b'}$, we have
\begin{eqnarray}
&&\hat {D}_{\alpha }(M_a^b\hat {A}^{\alpha }_b\hat {A}^a_{\beta
}-\frac {1}{2}\delta ^{\alpha }_{\beta }M_a^b\hat {A}^{\gamma
}_b\hat {A}^a_{\gamma }) \nonumber \\
&=&\hat {D}_{\alpha }(M_a^b\hat {A}^{\alpha }_b)\hat {A}^a_{\beta
}+M_a^b\hat {A}^{\alpha }_b\hat {D}_{\alpha }\hat {A}^a_{\beta
}-M_a^b\hat {A}^{\alpha }_b\hat {D}_{\beta }\hat {A}^a_{\alpha
}+iC_{cd}^aM_a^b
\hat {A}_b^{\alpha }\hat {A}^c_{\alpha }\hat {A}^d_{\beta } \nonumber \\
&=&-\hat {A}^a_{\beta }\hat {u}_a+M_a^b\hat {A}^{\alpha }_b\hat
{F}^a_{\alpha \beta }.
\end{eqnarray}
Thus we end the proof.

As we presented, the energy-momentum tensor of matter fields takes
the form
\begin{eqnarray}
t^{\alpha }_{\beta }(x)&=&\frac {i}{4}\{ \langle \Psi |\hat
{\varepsilon }(x)[(\hat {\gamma }^{\alpha }\hat {D}_{\beta }+\hat
{\gamma }_{\beta }\hat {D}^{\alpha })|\Psi \rangle ]+[\langle \Psi
|(\hat {D}_{\beta }\hat {\gamma }^{\alpha }+\hat {D}^{\alpha }\hat
{\gamma }_{\beta })]\hat {\varepsilon }(x)|\Psi \rangle \} \nonumber
\\
&=&\frac {1}{2}(\theta ^{\mu }_{\beta }t^{\alpha }_{\mu }+\theta
_{\mu }^{\alpha }t_{\beta }^{\mu }+\hat {\Gamma }^{\rho \sigma
}_{\beta }S^{\alpha }_{\rho \sigma }+\hat {\Gamma }^{\alpha }_{\rho
\sigma }S_{\beta }^{\rho \sigma }+\hat {A}^{\alpha }_{\beta
}j^{\alpha }_a+\hat {A}^{\alpha }_aj^{\alpha }_{\beta }).
\end{eqnarray}
We can find that the energy-momentum tensor is symmetric,
\begin{eqnarray}
t_{\alpha ,\beta }=t_{\beta ,\alpha }.
\end{eqnarray}
By Dirac equation, the divergence equation for the energy-momentum
tensor can be calculated as,
\begin{eqnarray}
\hat {D}_{\alpha }t^{\alpha }_{\beta }=F^a_{\alpha \beta }j^{\alpha
}_a+A^a_{\beta }u_a. \label{dirac-en-mo}
\end{eqnarray}
The proof  is presented in the following. We have used the following
notations,
\begin{eqnarray}
t'^{\alpha }_{\beta }(x)&=&\frac {i}{2}\{ \langle \Psi |\hat
{\varepsilon }(x)(\hat {\gamma }^{\alpha }\hat {D}_{\beta }|\Psi
\rangle ) +(\langle \Psi |\hat {D}_{\beta }\hat {\gamma }^{\alpha
})\hat {\varepsilon }(x)|\Psi
\rangle \} \nonumber \\
t''^{\alpha }_{\beta }(x)&=&\frac {i}{2}\{ \langle \Psi |\hat
{\varepsilon }(x)(\hat {\gamma }_{\beta }\hat {D}^{\alpha }|\Psi
\rangle ) +(\langle \Psi |\hat {D}^{\alpha }\hat {\gamma }_{\beta
})\hat {\varepsilon }(x)|\Psi \rangle \} .
\end{eqnarray}
So we have the expression,
\begin{eqnarray}
t^{\alpha }_{\beta }=\frac {1}{2}\left( t'^{\alpha }_{\beta
}+t''^{\alpha }_{\beta }\right) .
\end{eqnarray}
We next consider terms $t'^{\alpha }_{\beta }$ and $t''^{\alpha
}_{\beta }$, respectively.

\begin{eqnarray}
&&\theta ^{\mu }_{\alpha }\partial _{\mu }t'^{\alpha }_{\beta }(x)
\nonumber \\
&=&\frac {i}{2} \left\{ \theta _{\alpha }^{\mu }(x)\partial _{\mu
}\langle \Psi |\hat {\varepsilon }(x)(\hat {\gamma }^{\alpha }\hat
{D}_{\beta }|\Psi \rangle )+\theta _{\alpha }^{\mu }(x)\partial
_{\mu } (\langle \Psi |\hat {D}_{\beta }\hat {\gamma }^{\alpha
})\hat {\varepsilon }(x)|\Psi \rangle )\right\}
\nonumber \\
&=& \frac {i}{2} \left\{ \langle \Psi |\hat {\varepsilon }(x)\left(
\hat {D}_{\alpha }\hat {\gamma }^{\alpha }\hat {D}_{\beta }|\Psi
\rangle \right) - \left( \langle \Psi |\hat {D}_{\alpha }\right)
\hat {\varepsilon }(x)\left( \hat {\gamma }^{\alpha }\hat {D}_{\beta
}|\Psi \rangle \right) \right.
\nonumber \\
&& +\left. \left( \langle \Psi |\hat {D}_{\beta }\hat {\gamma
}^{\alpha }\right) \hat {\varepsilon }(x)\left( \hat {D}_{\alpha
}|\Psi \rangle \right) - \left( \langle \Psi |\hat {D}_{\beta }\hat
{\gamma }^{\alpha }\hat {D}_{\alpha }\right) \hat {\varepsilon
}(x)|\Psi \rangle \right\}
\nonumber \\
&=& \frac {i}{2} \left\{ \langle \Psi |\hat {\varepsilon }(x)\left[
\left( [\hat {D}_{\alpha },\hat {\gamma }^{\alpha }]\hat {D}_{\beta
}+\hat {\gamma }^{\alpha }[\hat {D}_{\alpha },\hat {D}_{\beta }]-
[\hat {D}_{\beta },\hat {\gamma }^{\alpha }]\hat {D}_{\alpha }+\hat
{D}_{\beta }\hat {\gamma }^{\alpha }\hat {D}_{\alpha }\right) |\Psi
\rangle \right] \right.
\nonumber \\
&&-\left( \langle \Psi |\hat {D}_{\alpha }\hat {\gamma }^{\alpha
}\right) \hat {\varepsilon }(x)\left( \hat {D}_{\beta }|\Psi \rangle
\right)+\left( \langle \Psi |\hat {D}_{\beta }\right) \hat
{\varepsilon }(x)\left( \hat {\gamma }^{\alpha }\hat {D}_{\alpha
}|\Psi \rangle \right)
\nonumber \\
&& \left. -\left[ \langle \Psi |\left( \hat {D}_{\alpha }\hat
{\gamma }^{\alpha }\hat {D}_{\beta } -\hat {D}_{\beta }[\hat
{D}_{\alpha },\hat {\gamma }^{\alpha }]-[\hat {D}_{\alpha },\hat
{D}_{\beta }]\hat {\gamma }^{\alpha }+\hat {D}_{\alpha }[\hat
{D}_{\beta },\hat {\gamma }^{\alpha }]\right) \right]\hat
{\varepsilon }(x)|\Psi \rangle \right \} \nonumber \\
&=& \frac {i}{2} \left\{ \langle \Psi |\hat {\varepsilon }(x)\left[
\left( -\hat {\Gamma }^{\alpha }_{\alpha ,\gamma }\hat {\gamma
}^{\gamma }\hat {D}_{\beta }+\hat {\gamma }^{\alpha }(-i\hat {\Omega
}_{\alpha \beta }+\hat {f}^{\gamma }_{\alpha \beta }\hat {D}_{\gamma
}+\hat {\Gamma }^{\gamma }_{\beta ,\alpha }\hat {\gamma }^{\alpha
}\hat {D}_{\gamma }\right) |\Psi \rangle \right] \right.
\nonumber \\
&&-i\langle \Psi |\hat {\varepsilon }(x)\left( \hat {D}_{\beta }\hat
{m}|\Psi \rangle \right) +i\left( \langle \Psi |\hat {m}\right) \hat
{\varepsilon }(x)\left( \hat {D}_{\beta }|\Psi \rangle \right)
\nonumber \\
&&-i\left( \langle \Psi |\hat {D}_{\beta }\right) \hat {\varepsilon
}(x)\left( \hat {m}|\Psi \rangle \right) +i\left( \langle \Psi |\hat
{m}\hat {D}_{\beta }\right) \hat {\varepsilon }(x)|\Psi \rangle
\nonumber \\
&&\left. +\left[\langle \Psi |\left( -\hat {D}_{\beta }\hat {\gamma
}^{\gamma }\hat {\Gamma }^{\alpha }_{\alpha ,\gamma }+(-i\hat
{\Omega }_{\alpha \beta }+\hat {D}_{\gamma }\hat {f}^{\gamma
}_{\alpha \beta })\hat {\gamma }^{\alpha }+\hat {D}_{\gamma }\hat
{\gamma }^{\alpha }\hat {\Gamma }^{\gamma }_{\beta ,\alpha }\right)
\right] \hat {\varepsilon }(x)|\Psi \rangle \right\} \nonumber \\
&=&-\frac {i}{2}\hat {\Gamma }^{\alpha }_{\alpha ,\gamma }(x)\left\{
\langle \Psi |\hat {\varepsilon }(x)\left( \hat {\gamma }^{\gamma
}\hat {D}_{\beta }|\Psi \rangle \right) +\left( \langle \Psi | \hat
{D}_{\beta }\hat {\gamma }^{\gamma }\right) \hat {\varepsilon
}(x)|\Psi \rangle \right\}
\nonumber \\
&&+\frac {i}{2}\hat {\Gamma }^{\gamma }_{\alpha ,\beta }(x)\left\{
\langle \Psi |\hat {\varepsilon }(x)\left( \hat {\gamma }^{\alpha
}\hat {D}_{\gamma }|\Psi \rangle \right) +\left( \langle \Psi | \hat
{D}_{\gamma }\hat {\gamma }^{\alpha }\right) \hat {\varepsilon
}(x)|\Psi \rangle \right\}
\nonumber \\
&&+\frac {1}{4}\hat {R}^{\rho \sigma }_{\alpha \beta }(x)\langle
\Psi |\hat {\varepsilon }(x)(\hat {\gamma }^{\alpha }\hat {s}_{\rho
\sigma }+\hat {s}_{\rho \sigma }\hat {\gamma }^{\alpha })|\Psi
\rangle +F^a_{\alpha \beta }(x)\langle \Psi |\hat {\varepsilon
}(x)\hat {\gamma }^{\alpha }\hat {T}_a |\Psi \rangle
\nonumber \\
&&-iA^a_{\alpha }(x)\langle \Psi |\hat {\varepsilon }(x)[\hat
{T}_a,\hat {m}]|\Psi \rangle
\nonumber \\
&=&-\hat {\Gamma }^{\alpha }_{\alpha ,\gamma }(x)t'^{\gamma }_{\beta
}(x)+\hat {\Gamma }^{\gamma }_{\alpha ,\beta }(x)t'^{\beta }_{\gamma
}(x)+\frac {1}{2}R^{\rho \sigma }_{\alpha \beta }S^{\alpha }_{\rho
\sigma }(x)+F^a_{\alpha \beta }(x)j^{\alpha }_a(x)+A^a_{\alpha
}(x)u_a(x)
\end{eqnarray}
Thus we have
\begin{eqnarray}
\hat {D}_{\alpha }t'^{\alpha }_{\beta }(x)&=&\theta _{\alpha }^{\mu
}(x)\partial _{\mu }t'^{\alpha }_{\beta }(x)+ \hat {\Gamma }^{\alpha
}_{\alpha ,\gamma }(x)t'^{\gamma }_{\beta }(x)-\hat {\Gamma
}^{\gamma }_{\alpha ,\beta }(x)t'^{\beta }_{\gamma }(x) \nonumber \\
&&=\frac {1}{2}R^{\rho \sigma }_{\alpha \beta }(x)S^{\alpha }_{\rho
\sigma }(x)+F^a_{\alpha \beta }(x)j^{\alpha }_a(x)+A^a_{\alpha
}(x)u_a(x).
\end{eqnarray}

Similarly we have,

\begin{eqnarray}
&&\theta ^{\mu }_{\alpha }\partial _{\mu }t''^{\alpha }_{\beta }(x)
\nonumber \\
&=&\frac {i}{2} \left\{ \theta _{\alpha }^{\mu }(x)\partial _{\mu
}\langle \Psi |\hat {\varepsilon }(x)(\hat {\gamma }_{\beta }\hat
{D}^{\alpha }|\Psi \rangle )+\theta _{\alpha }^{\mu }(x)\partial
_{\mu} (\langle \Psi |\hat {D}^{\alpha }\hat {\gamma }_{\beta })\hat
{\varepsilon }(x)|\Psi \rangle \right\}
\nonumber \\
&=& \frac {i}{2} \left\{ \langle \Psi |\hat {\varepsilon }(x)\left(
\hat {D}_{\alpha }\hat {\gamma }_{\beta }\hat {D}^{\alpha }|\Psi
\rangle \right) - \left( \langle \Psi |\hat {D}_{\alpha }\right)
\hat {\varepsilon }(x)\left( \hat {\gamma }_{\beta }\hat {D}^{\alpha
}|\Psi \rangle \right) \right.
\nonumber \\
&& +\left. \left( \langle \Psi |\hat {D}^{\alpha }\hat {\gamma
}_{\beta }\right) \hat {\varepsilon }(x)\left( \hat {D}_{\alpha
}|\Psi \rangle \right) - \left( \langle \Psi |\hat {D}^{\alpha }\hat
{\gamma }_{\beta }\hat {D}_{\alpha }\right) \hat {\varepsilon
}(x)|\Psi \rangle \right\}
\nonumber \\
&=& \frac {i}{2} \left\{ \langle \Psi |\hat {\varepsilon }(x)\left[
\left( [\hat {D}_{\alpha },\hat {\gamma }_{\beta }]\hat {D}^{\alpha
}|\Psi \rangle \right) \right] \right\} +\langle \Psi |\hat
{\varepsilon }(x)(\hat {\gamma }_{\beta }\hat {D}_{\alpha }\hat
{D}^{\alpha }|\Psi \rangle )
\nonumber \\
&&\left. +(\langle \Psi |\hat {D}^{\alpha }[\hat {D}_{\alpha },\hat
{\gamma }_{\beta }])\hat {\varepsilon }(x)|\Psi \rangle \right
\}-(\langle \Psi |\hat {D}^{\alpha }\hat {D}_{\alpha }\hat {\gamma
}_{\beta  })\hat {\varepsilon }(x)|\Psi \rangle
\nonumber \\
&=& \frac {i}{2} \left\{ \hat {\Gamma }^{\gamma }_{\alpha ,\beta }
\langle \Psi |\hat {\varepsilon }(x)(\hat {\gamma }_{\gamma }\hat
{D}^{\alpha }|\Psi \rangle )+\langle \Psi |\hat {\varepsilon }(x)[
\hat {\gamma }_{\beta }(\hat {R}+\hat {F} -\hat {\Gamma }_{\alpha ,
\gamma }^{\alpha }\hat {D}^{\gamma }-\hat {A}^a_{\alpha }\hat
{\gamma }^{\alpha }\hat {m}_a-\hat {m}^2)|\Psi \rangle ] \right.
\nonumber
\\
&&+\left. \hat {\Gamma }^{\gamma }_{\alpha ,\beta }(\langle \Psi
|\hat {D}^{\alpha }\hat {\gamma }_{\gamma }\hat {\varepsilon
}(x)|\Psi \rangle -[\langle \Psi |( \hat {R}+\hat {F} +\hat
{D}^{\gamma }\hat {\Gamma }_{\alpha ,\gamma }^{\alpha }+\hat
{A}^a_{\alpha }\hat {\gamma }^{\alpha }\hat {m}_a-\hat {m}^2)\hat
{\gamma }_{\beta }]\hat {\varepsilon }(x)|\Psi \rangle \right\}
\nonumber
\\
&=&\frac {i}{2}\hat {\Gamma }^{\gamma }_{\alpha ,\beta }\left\{
\langle \Psi |\hat {\varepsilon }(x)(\hat {\gamma }_{\gamma }\hat
{D}^{\alpha }|\Psi \rangle ) +(\langle \Psi | \hat {D}^{\alpha }\hat
{\gamma }_{\gamma })\hat {\varepsilon }(x)|\Psi \rangle \right\}
\nonumber \\
&&-\frac {i}{2}\left\{ \hat {\Gamma }^{\alpha }_{\alpha ,\gamma
}\langle \Psi |\hat {\varepsilon }(x)(\hat {\gamma }_{\beta }\hat
{D}^{\gamma }|\Psi \rangle ) +(\langle \Psi | \hat {D}^{\gamma }\hat
{\gamma }_{\beta })\hat {\varepsilon }(x)|\Psi \rangle \right\}
\nonumber \\
&&+\frac {1}{4}\hat {R}^{\rho \sigma }_{\alpha \beta }\langle \Psi
|\hat {\varepsilon }(x)(\hat {\gamma }^{\alpha }\hat {s}_{\rho
\sigma }+\hat {s}_{\rho \sigma }\hat {\gamma }^{\alpha })|\Psi
\rangle +F^a_{\alpha \beta }\langle \Psi |\hat {\varepsilon }(x)\hat
{\gamma }^{\alpha }\hat {T}_a |\Psi \rangle -iA^a_{\alpha
}(x)\langle \Psi |\hat {\varepsilon }(x)\hat {m}_a|\Psi \rangle
\nonumber \\
&=&-\hat {\Gamma }^{\alpha }_{\alpha ,\gamma }(x)t''^{\gamma
}_{\beta }(x)+\hat {\Gamma }^{\gamma }_{\alpha ,\beta }(x)t''^{\beta
}_{\gamma }(x)+\frac {1}{2}R^{\rho \sigma }_{\alpha \beta }(x)\hat
{S}^{\alpha }_{\rho \sigma }(x)+F^a_{\alpha \beta }(x)j^{\alpha
}_a(x)+A^a_{\alpha }(x)u_a(x)
\end{eqnarray}
And we have
\begin{eqnarray}
\hat {D}_{\alpha }t''^{\alpha }_{\beta }(x)&=&\theta _{\alpha }^{\mu
}(x)\partial _{\mu }t''^{\alpha }_{\beta }(x)+ \hat {\Gamma
}^{\alpha }_{\alpha ,\gamma }(x)t''^{\gamma }_{\beta }(x)-\hat
{\Gamma
}^{\gamma }_{\alpha ,\beta }(x)t''^{\beta }_{\gamma }(x) \nonumber \\
&&=\frac {1}{2}R^{\rho \sigma }_{\alpha \beta }(x)S^{\alpha }_{\rho
\sigma }(x)+F^a_{\alpha \beta }(x)j^{\alpha }_a(x)+A^a_{\alpha
}(x)u_a(x).
\end{eqnarray}

Spin current density $S_{\rho \sigma \alpha }$ is completely
anti-symmetric tensor, the summation of gravity curvature $R^{\rho
\sigma ,\alpha }_{\beta }$ for cyclic indices is zero, so we have
\begin{eqnarray}
R^{\rho \sigma }_{\alpha \beta }S^{\alpha }_{\rho \sigma }&=&R^{\rho
\sigma ,\alpha }_{\beta }S_{\rho \sigma \alpha } \nonumber \\
&=&\frac {1}{3}(R^{\rho \sigma ,\alpha }_{\beta }+R^{\sigma \alpha
,\rho }_{\beta }+R^{\alpha \rho ,\sigma }_{\beta })\hat {S}_{\rho
\sigma \alpha } \nonumber \\
&=&0
\end{eqnarray}

Summarize the above three equations, we have
\begin{eqnarray}
D_{\alpha }t^{\alpha }_{\beta }&=& \theta ^{\mu }_{\alpha }\partial
_{\mu }t^{\alpha }_{\beta }+\Gamma ^{\alpha }_{\alpha ,\gamma
}t^{\gamma }_{\beta }-\Gamma ^{\gamma }_{\alpha ,\beta }t^{\beta
}_{\gamma } \nonumber \\
&=&F^a_{\alpha \beta }j^{\alpha }_a+A^a_{\alpha }u_a.
\end{eqnarray}

Comparing the Eq.(\ref{ym-en-mo}) of gauge fields and
Eq.(\ref{dirac-en-mo}) of matter fields, we immediately find that
for the total energy-momentum tensor $\hat {T}^{\alpha }_{\beta
}=\hat {\tau }^{\alpha }_{\beta }+\hat {t}^{\alpha }_{\beta }$,
there is the energy-momentum conservation law:
\begin{eqnarray}
\hat {D}_{\alpha }\hat {T}^{\alpha }_{\beta }=0.
\end{eqnarray}

\subsection{Leptons and quarks and related projectors}

We can define two projectors $\hat {P}_{(l)}$ and $\hat {P}_{(q)}$
which can project a quantum state onto special states for leptons or
quarks, respectively.
\begin{eqnarray}
\hat {P}_{(l)}&=&1-\frac {3}{4}\hat {\lambda }^2, \\
\hat {P}_{(q)}&=&\frac {3}{4}\hat {\lambda }^2,
\end{eqnarray}
And the matter fields are decomposed as the superposition of two
states for leptons and quarks.

The lepton field and quark field are defined as
\begin{eqnarray}
|\Psi _{(l)}\rangle &=&\hat {P}_{(l)}|\Psi \rangle , \\
|\Psi _{(q)}\rangle &=&\hat {P}_{(q)}|\Psi \rangle .
\end{eqnarray}
There is no overlaps for lepton field and quark field, so we have
the form of superposition,
\begin{eqnarray}
|\Psi \rangle =|\Psi _{(l)}\rangle +|\Psi _{(q)}\rangle .
\end{eqnarray}
Here we note that
\begin{eqnarray}
\hat {\lambda }^2|\Psi _{(l)}\rangle =0|\Psi _{(l)}\rangle ,
\\
\hat {\lambda }^2|\Psi _{(q)}\rangle =\frac {4}{3}|\Psi
_{(q)}\rangle .
\end{eqnarray}

Some properties of the lepton and quark projectors can be found as
\begin{eqnarray}
&&[\hat {P}_{(l)}, \hat {D}_{\alpha }]=0, \nonumber \\
&&[\hat {P}_{(q)}, \hat {D}_{\alpha }]=0, \nonumber \\
&&[\hat {P}_{(l)}, \hat {\gamma }^{\alpha }]=0, \nonumber \\
&&[\hat {P}_{(q)}, \hat {\gamma }^{\alpha }]=0, \nonumber \\
&&[\hat {P}_{(l)}, \hat {m}]=0, \nonumber \\
&&[\hat {P}_{(q)}, \hat {m}]=0.
\end{eqnarray}

From the Dirac equation for matter fields, with the help of the
properties of the projectors $\hat {P}_{(l)}$ and $\hat {P}_{(l)}$,
we can find that the lepton field and the quark field satisfy the
Dirac equation, respectively.
\begin{eqnarray}
&&(i\hat {\gamma }^{\alpha }\hat {D}_{\alpha }-\hat {m})|\Psi
_{(l)}\rangle =0, \nonumber \\
&&(i\hat {\gamma }^{\alpha }\hat {D}_{\alpha }-\hat {m})|\Psi
_{(q)}\rangle =0, \nonumber \\
&&(\hat {D}_{\alpha }\hat {D}^{\alpha }-\hat {R}-\hat {F}+\hat
{\Gamma }^{\alpha }_{\alpha ,\beta }\hat {D}^{\beta }+\hat
{A}^a_{\alpha }\hat {\gamma }^{\alpha }\hat {m}_a+\hat {m}^2)|\Psi
_{(l)}\rangle =0, \nonumber \\
&&(\hat {D}_{\alpha }\hat {D}^{\alpha }-\hat {R}-\hat {F}+\hat
{\Gamma }^{\alpha }_{\alpha ,\beta }\hat {D}^{\beta }+\hat
{A}^a_{\alpha }\hat {\gamma }^{\alpha }\hat {m}_a+\hat {m}^2)|\Psi
_{(q)}\rangle =0.
\end{eqnarray}

Similarly for the adjoint states of lepton and quark, we also have
the following results,
\begin{eqnarray}
&&\langle \Psi _{(l)}|(i\hat {D}_{\alpha }\hat {\gamma }^{\alpha }-\hat {m})=0, \nonumber \\
&&\langle \Psi _{(q)}|(i\hat {D}_{\alpha
}\hat {\gamma }^{\alpha }-\hat {m})=0, \nonumber \\
&&\langle \Psi _{(l)}|(\hat {D}_{\alpha }\hat {D}^{\alpha }-\hat
{R}-\hat {F}-\hat {D}^{\beta }\hat {\Gamma }^{\alpha }_{\alpha
,\beta }-\hat {A}^a_{\alpha }\hat {\gamma }^{\alpha }\hat
{m}_a+\hat {m}^2)=0, \nonumber \\
&&\langle \Psi _{(q)}|(\hat {D}_{\alpha }\hat {D}^{\alpha }-\hat
{R}-\hat {F}-\hat {\Gamma }^{\alpha }_{\alpha ,\beta }\hat
{D}^{\beta }-\hat {A}^a_{\alpha }\hat {\gamma }^{\alpha }\hat
{m}_a+\hat {m}^2)=0.
\end{eqnarray}

We then can discuss the current densities for leptons and quarks,
respectively. The lepton current density is defined as
\begin{eqnarray}
\rho _{(l)}^{\alpha }(x)=\langle \Psi _{(l)}|\hat {\varepsilon
}(x)\hat {\gamma }^{\alpha }|\Psi _{(l)}\rangle .
\end{eqnarray}
We can find the the number of leptons is conserved which is
expressed as the following,
\begin{eqnarray}
\hat {D}_{\alpha }\rho _{(l)}^{\alpha }=0.
\end{eqnarray}
Similarly for quarks, the current density takes the form
\begin{eqnarray}
\rho _{(q)}^{\alpha }(x)=\langle \Psi _{(q)}|\hat {\varepsilon
}(x)\hat {\gamma }^{\alpha }|\Psi _{(q)}\rangle .
\end{eqnarray}
Also the number of quarks is conserved,
\begin{eqnarray}
\hat {D}_{\alpha }\rho _{(q)}^{\alpha }=0.
\end{eqnarray}
The total current density of matter fields can be expressed as the
summation of lepton current density and the quark current density,
\begin{eqnarray}
\rho ^{\alpha }=\rho _{(l)}^{\alpha }+\rho _{(q)}^{\alpha }
\end{eqnarray}

\subsection{Gauge current densities for leptons and quarks}
The current densities with different gauges for leptons and quarks
are also important properties. The definition can be simply realized
by gauge operators $\hat {T}_a$. We now define the gauge current
density of lepton field as,
\begin{eqnarray}
{j_{(l)}}^{\alpha }_a(x)=\langle \Psi _{(l)}|\hat {\varepsilon
}(x)\hat {\gamma }^{\alpha }\hat {T}_a|\Psi _{(l)}\rangle .
\end{eqnarray}
The gauge charge production rate density for lepton field takes the
form,
\begin{eqnarray}
{u_{(l)}}_a(x)=-i\langle \Psi _{(l)}|\hat {m}_a\hat {\varepsilon
}(x)|\Psi _{(l)}\rangle .
\end{eqnarray}
One can find that for lepton field, the divergence of the gauge
current density is equal to the gauge charge production rate
density,
\begin{eqnarray}
\hat {D}_{\alpha }{j_{(l)}}^{\alpha }_a={u_{(l)}}_a.
\end{eqnarray}

Similar properties are also satisfied for quark field.
\begin{eqnarray}
{j_{(q)}}^{\alpha }_a(x)&=&\langle \Psi _{(q)}|\hat {\varepsilon
}(x)\hat {\gamma }^{\alpha }\hat {T}_a|\Psi _{(q)}\rangle ,
\nonumber \\
{u_{(q)}}_a(x)&=&-i\langle \Psi _{(q)}|\hat {m}_a\hat {\varepsilon
}(x)|\Psi _{(q)}\rangle , \nonumber \\
\hat {D}_{\alpha }{j_{(q)}}^{\alpha }_a&=&{u_{(q)}}_a.
\end{eqnarray}

The total gauge current density is the summation of lepton gauge
current density and the quark gauge current density. The total gauge
charge production rate density is the summation of lepton and quark
gauge charges production rate densities,
\begin{eqnarray}
j^{\alpha }_a(x)&=&{j_{(l)}}^{\alpha }_a(x)+{j_{(q)}}^{\alpha
}_a(x), \nonumber \\
u_a&=&{u_{(l)}}_a+{u_{(l)}}_a
\end{eqnarray}

The energy-momentum tensor of lepton field takes the form
\begin{eqnarray}
{t_{(l)}}^{\alpha }_{\beta }(x) &=&\frac {i}{4}\{ \langle \Psi
_{(l)}|\hat {\varepsilon }(x)[(\hat {\gamma }^{\alpha }\hat
{D}_{\beta }+\hat {\gamma }_{\beta }\hat {D}^{\alpha })|\Psi
_{(l)}\rangle ]+[\langle \Psi _{(l)}|(\hat {D}_{\beta }\hat {\gamma
}^{\alpha }+\hat {D}^{\alpha }\hat {\gamma }_{\beta })]\hat
{\varepsilon }(x)|\Psi _{(l)}\rangle \} .
\end{eqnarray}
The lepton energy-momentum tensor satisfies the equation,
\begin{eqnarray}
\hat {D}_{\alpha }{t_{(l)}}^{\alpha }_{\beta }=-F^a_{\alpha \beta
}{j_{(l)}}^{\alpha }_a-A^a_{\beta }{u_{(l)}}_a.
\end{eqnarray}
Similarly the energy-momentum tensor of quarks is
\begin{eqnarray}
{t_{(q)}}^{\alpha }_{\beta }(x) &=&\frac {i}{4}\{ \langle \Psi
_{(q)}|\hat {\varepsilon }(x)[(\hat {\gamma }^{\alpha }\hat
{D}_{\beta }+\hat {\gamma }_{\beta }\hat {D}^{\alpha })|\Psi
_{(q)}\rangle ]+[\langle \Psi _{(q)}|(\hat {D}_{\beta }\hat {\gamma
}^{\alpha }+\hat {D}^{\alpha }\hat {\gamma }_{\beta })]\hat
{\varepsilon }(x)|\Psi _{(q)}\rangle \} .
\end{eqnarray}
The lepton energy-momentum tensor satisfies the equation,
\begin{eqnarray}
\hat {D}_{\alpha }{t_{(q)}}^{\alpha }_{\beta }=-F^a_{\alpha \beta
}{j_{(q)}}^{\alpha }_a-A^a_{\beta }{u_{(q)}}_a.
\end{eqnarray}
The total energy-momentum tensor is the summation of lepton
energy-momentum and the quark energy-momentum,
\begin{eqnarray}
t^{\alpha }_{\beta }={t_{(l)}}^{\alpha }_{\beta }+{t_{(q)}}^{\alpha
}_{\beta }.
\end{eqnarray}

\section{Quantum system with discrete space-time,
observable quantities and particles scattering}
We have presented
above the unified description of the three fundamental equations. In
this section, we hope to discuss conceptually the quantum system,
and how to define the observable physical quantities in 3D space
from the present 4D theory. We will also present the particle
scattering results from the representations in this work.

\subsection{Quantum system}
We next present some features of our unified quantum theory. The
quantum field we discussed is generally in all 4D coordinate space
or in all 4D momentum space. This means that the coordinate and
momentum take values in all 4D spaces, $x\in R^4$ and $p\in R^4$. In
our formulas, the integral area should be all 4D space which
sometimes is omitted in the presentation. Here we would like to
point out that the integral area should be in the following form,

\begin{eqnarray}
&&|\Psi \rangle =\int _{R^4}\Psi ^{st}(x)|x_{st}\rangle d^4x=\int
_{R^4}\widetilde {\Psi }^{st}(p)|p_{st}\rangle d^4p \nonumber \\
&&\hat {e}=\int _{R^4}\hat {\theta }^{\mu }_{\alpha }(x)\hat
{\varepsilon }_{\mu }^{\alpha }(x)d^4x=\int _{R^4}\widetilde {\hat
{\theta }^{\mu }_{\alpha }}(p)\hat {\varepsilon }_{\mu
}^{\alpha }(p)d^4p \nonumber \\
&&\hat {A}=\int _{R^4}A^a_{\alpha }(x)\hat {\varepsilon }_a^{\alpha
}(x)d^4x=\int _{R^4}A^a_{\alpha }(p)\hat {\varepsilon }_a^{\alpha
}(p)d^4p .
\end{eqnarray}

In application of our theory, what we consider are 4D coordinate
space $M_x$ with a boundary or 4D momentum space $M_p$ with a
boundary. The quantum systems in 4D coordinate space with boundaries
and the 4D momentum space with boundaries all are systems we study.

For boundary 4D coordinate or momentum spaces, the matter fields
$|\Psi \rangle $, gravity orthogonal vierbein $\hat {e}$ and the
gauge connection $\hat {A}$ can be expressed as,
\begin{eqnarray}
&&|\Psi \rangle =\int _{M_x}\Psi ^{st}(x)|x_{st}\rangle d^4x=\int
_{M_p}\widetilde {\Psi }^{st}(p)|p_{st}\rangle d^4p \nonumber \\
&&\hat {e}=\int _{M_x}\hat {\theta }^{\mu }_{\alpha }(x)\hat
{\varepsilon }_{\mu }^{\alpha }(x)d^4x=\int _{M_p}\widetilde {\theta
}^{\mu }_{\alpha }(p)\hat {\varepsilon }_{\mu
}^{\alpha }(p)d^4p \nonumber \\
&&\hat {A}=\int _{M_x}A^a_{\alpha }(x)\hat {\varepsilon }_a^{\alpha
}(x)d^4x=\int _{M_p}A^a_{\alpha }(p)\hat {\varepsilon }_a^{\alpha
}(p)d^4p .
\end{eqnarray}
The 4D quantum theory, in which the space-time is in the sense of
general relativity, can be considered as a generalization of 3D
space plus time quantum theory. All physical quantities are
considered in the framework of 4D theory.

Here we have some remarks about the 4D unified quantum theory. (1).
A 4D quantum system can be described completely by the matter wave
function $\Psi _{st}(x)$, orthogonal vierbein function $e^{\mu
}_{\alpha }(x)$ and the gauge connection function $A^a_{\alpha }$.
The Dirac equation for matter fields, Einstein equation of gravity
and the Yang-Mill gauge equation constitute a complete description
of this quantum system. If we know the boundary condition $\partial
M_x$, the solution of these equations are fixed then. (2). The 4D
unified quantum theory is a local theory in the sense that the
interactions are local which just depend on their neighbors. (3).
For a 4D coordinate space $M_x$ with a boundary, the matter fields
$|\Psi \rangle $ and the interactions $\hat {F}$ can be expanded
discretely in 4D momentum space,
\begin{eqnarray}
|\Psi \rangle &=&\int _{M_x}\Psi (x)|x\rangle d^4x=\sum
_{n=1}^{\infty }\widetilde
{\Psi }(p_n)|p_n\rangle . \nonumber \\
\hat {F} &=&\int _{M_x}F(x)\hat {\varepsilon }(x)d^4x=\sum
_{n=1}^{\infty }\widetilde {F}(p_n)\hat {\varepsilon }(p_n).
\end{eqnarray}
And vise versa, the matter fields $|\Psi \rangle $ and the
interaction fields $\hat {F}$ in 4D momentum space with a boundary,
the coordinate expansion is discrete,
\begin{eqnarray}
|\Psi \rangle &=&\sum _{m=1}^{\infty }\widetilde
\Psi (x_m)|x_m\rangle =\int _{M_p}\widetilde {\Psi }(p)|p\rangle d^4p \nonumber \\
\hat {F} &=&\sum _{m=1}^{\infty }F(x_m)\hat {\varepsilon }(x_m)=\int
_{M_p}\widetilde {F}(p)\hat {\varepsilon }(p)d^4p.
\end{eqnarray}
For a matter field $|\Psi \rangle $ and interaction field $\hat {F}$
existing simultaneously in 4D coordinate space $M_x$ with a boundary
and in 4D momentum space $M_p$ with a boundary, the expansions are
discrete at the same time in coordinate space and the momentum
space, and the terms in the expansion are finite,
\begin{eqnarray}
|\Psi \rangle &=&\sum _{m=1}^{M}\Psi (x_m)|x_m\rangle =\sum _{n=1}^N
\widetilde {\Psi }(p_n)|p_n\rangle ,\nonumber \\
\hat {F} &=&\sum _{m=1}^{M}F(x_m)\hat {\varepsilon }(x_m)=\sum
_{n=1}^N\widetilde {F}(p_n)\hat {\varepsilon }(p_n), \nonumber \\
M&=&N=\int _{M_p}d^4x\int _{M_x}d^4p.
\end{eqnarray}
There are only finite number of quantum events for finite coordinate
space $M_x$ and finite momentum space $M_p$. When the term {\it
quantum} was first proposed, it means that the energy has an
smallest unit and is discrete. In our 4D unified theory, we can find
that the energy, momentum, space-time are all discrete.

\subsection{Observable quantities}
We next consider the real world observable quantities from this
theory.

In our unified theory, we can define several current densities,in
which, three important quantities are the densities of particle
current $\hat {\rho }$, total gauge current $\hat {J}$ defined in
Eq. (\ref{ym-total-current}) and the energy-momentum tensor $\hat
{T}$. The particle current density is related with the number of
particles, the gauge current density can be considered as the source
of gauge fields, the energy-momentum tensor can be considered as the
source of gravity. For 4D coordinate space $M_x$ with a boundary and
4D momentum space $M_p$ with a boundary, the density of particle
current, the gauge current density, the energy-momentum tensor are
defined respectively as,
\begin{eqnarray}
\hat {\rho }&=& \rho ^{\alpha }(\hat {x})\hat {\gamma }_{\alpha
}\nonumber \\
&=& \int _{M_x}\rho ^{\alpha }(x)\hat {\gamma }_{\alpha }\hat
{\varepsilon }(x)d^4x \nonumber \\
&=&\int _{M_p}\widetilde {\rho }^{\alpha }(p)\hat {\gamma }_{\alpha
}\hat {\varepsilon }(p)d^4p .
\end{eqnarray}
\begin{eqnarray}
\hat {J}&=&J^{\alpha }_a(\hat {x})\hat {\gamma }_{\alpha } \hat {T}^a\nonumber \\
&=&\int _{M_x}J_a^{\alpha }(x)\hat {\gamma }_{\alpha }\hat {T}^a\hat
{\varepsilon }(x)d^4x \nonumber \\
&=&\int _{M_p}\widetilde {J}^{\alpha }_a(p)\hat {\gamma }_{\alpha
}\hat {T}^a\hat {\varepsilon }(p)d^4p .
\end{eqnarray}
\begin{eqnarray}
\hat {T}&=& T^{\alpha }_{\beta }\hat {\gamma }_{\alpha }\otimes \hat
{\gamma }^{\beta }\nonumber \\
&=&\int _{M_x}T_{\beta }^{\alpha }(x) \hat {\gamma }_{\alpha
}\otimes \hat {\gamma }^{\beta }\hat {\varepsilon }(x)d^4x \nonumber
\\
&=&\int _{M_p}\widetilde {T}^{\alpha }_{\beta }(p)\hat {\gamma
}_{\alpha }\otimes \hat {\gamma }^{\beta }\hat {\varepsilon }(p)d^4p
.
\end{eqnarray}

We then define three production rate densities related with particle
numbers, gauge charges and energy-momentum

\begin{eqnarray}
\hat {\theta }^{\mu }_{\alpha }(\hat {x})\partial _{\mu }\hat {\rho
}^{\alpha }(\hat {x})+\hat {\Gamma }^{\beta }_{\beta ,\alpha }\hat
{\rho }^{\alpha }(x)=\hat {n}(x)
\end{eqnarray}
\begin{eqnarray}
\hat {\theta }^{\mu }_{\alpha }(\hat {x})\partial _{\mu }\hat
{J}^{\alpha }_a(\hat {x})+\hat {\Gamma }^{\beta }_{\beta ,\alpha
}\hat {J}^{\alpha }_a(\hat {x})=\hat {q}_a(x)
\end{eqnarray}
\begin{eqnarray}
\hat {\theta }^{\mu }_{\alpha }(\hat {x})\partial _{\mu }\hat
{T}^{\alpha }_{\beta }(\hat {x})+\hat {\Gamma }^{\sigma }_{\sigma
,\alpha }\hat {T}^{\alpha }_{\beta }=\hat {p}_{\beta }(x)
\end{eqnarray}

We define an operator on manifold corresponding to coordinate 4D
space $M_x$ and momentum 4D space $M_p$,
\begin{eqnarray}
\hat {\omega }&=&{\rm det}[\hat {e}^{\alpha }_{\mu }(x)] \nonumber
\\
&=&\int _{M_x}\omega (x)\hat {\varepsilon }(x)d^4x \nonumber
\\
&=& \int _{M_p}\widetilde {\omega }(p)\hat {\varepsilon }(p)d^4p.
\end{eqnarray}
We may notice vierbein $\hat {e}^{\alpha }_{\mu }$ appears here,
this is an indication of this theory is in general relativity curved
space-time. The trace of this operator is the 4D volume of $M_x$,
\begin{eqnarray}
\omega ={\rm tr}\hat {\omega }.
\end{eqnarray}
In 4D space $M_x$, the number of particles created is denoted as
$N$, it takes the form
\begin{eqnarray}
N&=&\langle \hat {n},\hat {\omega }\rangle ={\rm tr}(\hat {n}\hat
{w}) \nonumber \\
&=&\int _{M_x}n(x)\omega (x)d^4x \nonumber \\
&=&(2\pi )^2\widetilde {n}(0)*\widetilde {\omega }(0)
\end{eqnarray}
Similarly, the number of gauge charges created, the energy-momentum
created in 4D space $M_x$ can be written as,
\begin{eqnarray}
Q_a&=&\langle \hat {q}_a,\hat {\omega }\rangle ,\nonumber \\
P_{\beta }&=&\langle \hat {p}_{\beta },\hat {\omega }\rangle .
\end{eqnarray}

Generally if we would like to describe a physical quantities in the
real world 3D space with time, for example, in order to find the
total particle numbers in 3D area $V$, we need to make the integral
of the 3D area $V$ at a fixed time point $t_0$,
\begin{eqnarray}
N_{3D,t_0}=\int _{V}\rho ^0(x)d^3\vec {x}.
\end{eqnarray}
However, the problems of this equation are: first, one element of
particle current density is not general covariance; second, the 3D
manifold $V$ is not covariance and a fixed $t_0$ is not allowed in
general relativity. So the calculation for quantity $N_{3D,t_0}$ is
not allowed in the 4D unified theory.

The proper method to find the observable quantities of this theory
should be in covariance condition. We can consider to construct a
covariance 4D space $M_x$ by using the real world 3D space $V$ and a
real world short time period $[t_0,t_0+\Delta t]$,
\begin{eqnarray}
M_x=V\otimes [t_0,t_0+\Delta t].
\end{eqnarray}
The observable quantity like particle number takes the form,
\begin{eqnarray}
\langle N\rangle =\frac {\int _{M_x}\rho ^0(x)\omega (x)d^4x
}{\Delta t}.
\end{eqnarray}
Thus from 4D unified theory, all observable quantities in the real
world can be understood as the average quantities by a short time
period $\Delta t$. Similarly, the gauge charge and energy-momentum
in 3D space $V$ can be written as,
\begin{eqnarray}
\langle Q_a\rangle &=&\frac {\int _{M_x}J^0_a(x)\omega (x)d^4x
}{\Delta t}, \nonumber \\
\langle P_{\beta }\rangle &=&\frac {\int _{M_x}T^0_{\beta }(x)\omega
(x)d^4x }{\Delta t}.
\end{eqnarray}

\subsection{Particles scattering}
We next will first briefly present a summary of the representations
of matter particles, gauge particles and graviton. Based on those
representations, we will then consider the results of scattering of
those particles. As we all know the particle scattering process can
be described well by Feynman diagram method, our representations are
based on those results and on the other hand those representation
can deduce easily the particles scattering results. Those results
agree with the results by Feynman diagram method.

The following is a summary of the representations of particles.

\begin{enumerate}

\item
The state of matter particles takes the form,
\begin{eqnarray}
|e_{st}(p)\rangle =|p\rangle \otimes |s\rangle \otimes |t \rangle ,
\label{matter-particle}
\end{eqnarray}
where $p\in R^4$ is the momentum, $s=1,2,3,4$ is the spin, the spin
representation space is $V_S(\frac {1}{2},0)\oplus V_S(0,\frac
{1}{2})$, so the matter particles are spin-$\frac {1}{2}$,
$t=1,2,\cdots ,48$ corresponds to 48 classes of matter particles.

\item
The gauge particles are denoted as,
\begin{eqnarray}
\hat {\varepsilon }^{\alpha }_a(p)=\hat {\varepsilon }(p)\otimes
\hat {\gamma }^{\alpha }\otimes \hat {T}_a, \label{gauge-particle}
\end{eqnarray}
where $p\in R^4$ is the momentum, $\alpha =0,1,2,3$, the spin
representation space is $V_S(\frac {1}{2},\frac {1}{2})$ and so they
are spin-1, $a=1,2,\cdots ,12$, this means that there are 12 classes
of gauge particles.

\item The graviton is denoted as,
\begin{eqnarray}
\hat {\varepsilon }^{\alpha }_{\rho \sigma }(p)=\hat {\varepsilon
}(p)\otimes \hat {\gamma }^{\alpha }\otimes \hat {s}_{\rho \sigma },
\end{eqnarray}
where $p\in R^4$, $\alpha ,\rho ,\sigma =0,1,2,3$,  the graviton is
spin-1. Here we have a particle corresponding to vierbein, its
representation is written as,
\begin{eqnarray}
\hat {\varepsilon }^{\alpha }_{\mu }(p)=\hat {\varepsilon
}(p)\otimes \hat {\gamma }^{\alpha }\otimes \hat {p}_{\mu },
\end{eqnarray}
where $\alpha =0,1,2,3$, the spin representation space is $V_S(\frac
{1}{2},\frac {1}{2})$, the gauge charge of this vierbein particle
are 0, it is spin-1.
\end{enumerate}

With those representations, we next present the scattering matrices.
The 3-vertex scattering matrices for matter particles and action
particles, which include gauge particles, graviton and vierbein
particles, are written as,
\begin{eqnarray}
\langle e^{s_2t_2}(p_2)|\hat {\varepsilon }_a^{\alpha
}(p_3)|e_{s_1t_1}(p_1)\rangle =(\hat {\gamma }^{\alpha
})_{s_1}^{s_2} (\hat {T}_{a})^{t_2}_{t_1}\delta ^4(p_1+p_3-p_2).
\end{eqnarray}
This is the case of scattering between matter particles which are
fermions with the gauge bosons. Explicitly, we can see the initial
state is $|e_{s_1t_1}(p_1)\rangle $ defined in
(\ref{matter-particle}) which has $4\times 48$ choices for gauge and
spin altogether, the gauge boson (\ref{gauge-particle}) has
$12\times 4$ choices, the final state is still a matter particle.
The right hand side of the equation is the scattering result.

Also we have,
\begin{eqnarray}
\langle e^{s_2t_2}(p_2)|\hat {\varepsilon }_{\rho \sigma }^{\alpha
}(p_3)|e_{s_1t_1}(p_1)\rangle =(\hat {\gamma }^{\alpha }\hat
{s}_{\rho \sigma })_{s_1}^{s_2} \delta ^4(p_1+p_3-p_2),
\end{eqnarray}
This is the case of scattering between matter particles with the
graviton.
\begin{eqnarray}
\langle e^{s_2t_2}(p'')|\hat {\varepsilon }_{\mu }^{\alpha
}(p''')|e_{s_1t_1}(p')\rangle =(\hat {\gamma }^{\alpha
})_{s_1}^{s_2}p'_{\mu }\delta ^4(p'+p'''-p'').
\end{eqnarray}
The case describes the scattering between matter particles with the
vierbein particle related with the propagation of matter particles
in curved space-time.

The 4-vertex scattering matrices represent the scattering between
action particles and action particles,
\begin{eqnarray}
\langle \hat {\varepsilon }_{a_4}^{\alpha _4}(p_4),\hat {\varepsilon
}_{a_3}^{\alpha _3}(p_3),\hat {\varepsilon }_{a_2}^{\alpha
_2}(p_2),\hat {\varepsilon }_{a_1}^{\alpha _1}(p_1)\rangle = \eta
^{\alpha ^4\alpha _1}\eta ^{\alpha ^3\alpha
_2}G_{a_4c}C_{a_3b}^cC_{a_2a_1}^b \delta ^4(p_1+p_2+p_2-p_4).
\end{eqnarray}
The middle two particles $\hat {\varepsilon }_{a_3}^{\alpha
_3}(p_3),\hat {\varepsilon }_{a_2}^{\alpha _2}(p_2)$ are two gauge
particles, each of them can also be graviton and vierbein particle,
so altogether there are 9 cases. For example, we can change the
second particle as vierbein particle, the 4-vertex scattering matrix
is,
\begin{eqnarray}
\langle \hat {\varepsilon }_{a_4}^{\alpha _4}(p_4),\hat {\varepsilon
}_{a_3}^{\alpha _3}(p_3),\hat {\varepsilon }_{\mu }^{\alpha
_2}(p_2),\hat {\varepsilon }_{a_1}^{\alpha _1}(p_1)\rangle = \eta
^{\alpha ^4\alpha _1}\eta ^{\alpha ^3\alpha
_2}G_{a_4c}C_{a_3b}^cp_{1\mu } \delta ^4(p_1+p_2+p_2-p_4).
\end{eqnarray}

Those are the basic scattering matrices, other cases can be deduced
from those results.

\section{Symmetries and symmetries broken and dark energy}
The energy-momentum conservation law provides us a concrete
foundation to confirm our unified theory. Based on our theory, some
fundamental problems may be studied. We next consider several
questions.

Symmetries and symmetries broken play a key role in modern physics
\cite{Nobel}. They are also important in studying the physical
implications of our theory. In this work, the key parameters are
those involved in the three fundamental equations
(\ref{dirac-eq},\ref{ym-eq},\ref{einstein-eq}). Explicitly they are:
$|\Psi \rangle ,\hat {D}, \hat {M}, \hat {m}$. For definitions, see
$|\Psi \rangle $ in (\ref{state}), derivative operator $\hat {D}$ in
(\ref{deri-operator}), and mass related operators $\hat {m}$ for matter particles
in (\ref{mmass}), $\hat {M}$ for gauge fields in (\ref{Mmass}).

\subsection{Unitary transformation}
According to the representations, we can define the general
transformations as
\begin{eqnarray}
\hat {U}=\exp [-i(a_{\mu }\hat {x}^{\mu }+b^{\mu }\hat {p}_{\mu
}+\alpha +\frac {1}{2}\omega ^{\alpha \beta }\hat {s}_{\alpha \beta
}+\xi ^a\hat {T}_a)].
\end{eqnarray}
Under this transformation, $|\Psi \rangle ,\hat {D}, \hat {M}, \hat
{m}$ will be changed as,
\begin{eqnarray}
|\Psi \rangle &\rightarrow &\hat {U}|\Psi \rangle , \nonumber \\
\hat {D}&\rightarrow &\hat {U}\hat {D}\hat {U}^{-1}, \nonumber \\
\hat {M}&\rightarrow &\hat {U}\hat {M}\hat {U}^{-1}, \nonumber \\
\hat {m}&\rightarrow &\hat {U}\hat {m}\hat {U}^{-1}.
\end{eqnarray}
Here we define the unitary condition as
\begin{eqnarray}
\hat {U}^{-1}=\overline{\hat {U}}.
\end{eqnarray}
With this unitary transformation, the three fundamental equations
remains the same.



\subsection{Mass matrices of elementary particles,
color confinement}

For gauge theory, physical quantities should be gauge invariance.
For the unified theory of this work, all fields represented by
vectors and operators are gauge covariance. Still by proper
representations, physical process and physical quantities are gauge
invariance.

A key feature of our theory is that the general mass matrices are
defined in gauge space as,
\begin{eqnarray}
\hat {m}&=&m^t_{t'}|e_t\rangle \otimes \langle e^{t'}, \nonumber \\
\hat {M}&=&M^a_b\hat {T}_a\otimes \hat {T}^b.
\end{eqnarray}
We do not need each element of $\hat {m}$ and $\hat {M}$ be fixed.
However, the observable physical quantities including mass are still
gauge invariance. By comparison in conventional gauge theory, each
element $m^t_{t'}$ and $M^a_b$ of the mass matrices are gauge
invariance and mass in gauge theory is introduced by Higgs mechanism
by gauge symmetries breaking.

Mass matrices are representations in gauge space and the only
restrictions are three fundamental equations. Thus in principle, any
kind of mass matrices are allowed if no experimental facts are
violated. In this sense, no gauge symmetries breaking is necessary
to create mass. So we mean Higgs mechanism is not necessary in this
theory. We remark that the mass matrix itself is gauge covariant
while the observable mass which corresponding to eigenvalues of mass
matrix are of course should be gauge invariance. From the proof of
the energy-momentum conservation law, what we use is only the
conditions $M_{ab}=M_{ba}, \hat {m}^{\dagger }=\hat {m}$, no Higgs
particles are necessary to play a role in the energy-momentum
tensor.

From the form of mass matrices, we may notice that there exist a
symmetry in color charge space. Explicitly, the mass matrix is
$SU(3)$ invariant, for color charges $\hat {T}_a, a=5,6,\cdots ,12$
or $\hat {\lambda }_p, p=1,2,\cdots ,8$, as we already seen that,
\begin{eqnarray}
[\hat {T}_a, \hat {m}]=0.
\end{eqnarray}
Thus the solution of the three fundamental equations has this
$SU(3)$ symmetry for free boundary condition. Suppose we have a
solution presented as,
\begin{eqnarray}
\hat {A}&=&A^a_{\alpha }(\hat {x})\hat {T}_a\otimes \hat {\gamma
}^{\alpha }, \nonumber \\
|\Psi \rangle &=&\int _{R^4}\Psi ^{st}(x)|e_{st}\rangle d^4x .
\end{eqnarray}
We know that the unitary transformations $\hat {U}$ by the color
charges on the solution is still a solution,
\begin{eqnarray}
\hat {A}'&=&\hat {U} \hat {A}\hat {U}^{-1}, \nonumber \\
|\Psi '\rangle &=&\hat {U}|\Psi \rangle .
\end{eqnarray}
Since the existing of this $SU(3)$ symmetry in color charge space,
we will never be able to distinguish a single color state by only
mass since this will break this $SU(3)$ symmetry. This also means, a
single color quantum state which corresponds to a single color quark
will never been separated and observed. What we observed is always a
coherence of all three color states. This is the result of color
confinement. The color confinement is due to the symmetry of mass
matrices. While for weak interactions, no such symmetry exists so
there is no confinement. The reason of confinement is based on the
symmetry of mass matrices whose definitions are based on
experimental results in gauge space as in Eq.(\ref{mmass}) and in
Eq.(\ref{Mmass}), and also no Higgs mechanism is necessary, it is
thus easy to understand why there is confinement in color space but
no confinement in weak space. Gluon is connected only with quarks,
the color confinement also implies the confinement of gluon.
Explicitly, in the above equations, $\hat {A}'=\hat {U} \hat {A}\hat
{U}^{-1}$ implies the confinement of gluons, and $|\Psi '\rangle
=\hat {U}|\Psi \rangle$ implies the confinement of quarks.

Mass is always a basic question in physics. Einstein's mass-energy
equation $E=mc^2$ is well known. In quantum field theory, we have
$\hat {p}_{\mu }\hat {p}^{\mu }=m^2$. By this equation, we may
understand that mass is defined in momentum space. However, this
form is in general not true in the present theory. What we have is
due to the square differential of Dirac equation as in
Eq.(\ref{square-dirac}), it reduces to form $\hat {p}_{\mu }\hat
{p}^{\mu }=m^2$ only in special case for a free field.

\subsection{Parity violation for weak interactions}
The left chiral
projector and the right chiral projector are defined respectively
as,
\begin{eqnarray}
\hat {P}_L&=&\frac {1}{2}(1+\hat {\gamma }^5),\nonumber \\
\hat {P}_R&=&\frac {1}{2}(1-\hat {\gamma }^5).
\end{eqnarray}
Those two projectors satisfy the relations,
\begin{eqnarray}
&&\overline{\hat {P}_{(L)}}=\hat {P}_{(R)}, \nonumber \\
&&\overline{\hat {P}_{(R)}}=\hat {P}_{(L)}, \nonumber \\
&&\hat {P}_{(L)}\hat {\gamma }^{\alpha }=\hat
{\gamma }^{\alpha }\hat {P}_{(R)},\nonumber \\
&&\hat {P}_{(R)}\hat {\gamma }^{\alpha }=\hat
{\gamma }^{\alpha }\hat {P}_{(L)},\nonumber \\
&&[\hat {P}_{(L)}, \hat {D}_{\alpha }]=[\hat {P}_{(R)},\hat
{D}_{\alpha }]=0,\nonumber \\
&&[\hat {P}_{(L)},\hat {m}]=[\hat {P}_{(R)},\hat {m}]=0.
\end{eqnarray}
So the left chiral field and the right chiral field can then be
defined as,
\begin{eqnarray}
|\Psi _{(L)}\rangle &=&\hat {P}_{(L)}|\Psi \rangle ,\nonumber \\
|\Psi _{(R)}\rangle &=&\hat {P}_{(R)}|\Psi \rangle .
\end{eqnarray}
The properties can be found as,
\begin{eqnarray}
\hat {\gamma }^5|\Psi _{(L)}\rangle =(+1)|\Psi _{(L)}\rangle ,
\nonumber \\
\hat {\gamma }^5|\Psi _{(R)}\rangle =(-1)|\Psi _{(L)}\rangle .
\end{eqnarray}
The adjoint states have following relations, note that $L$ and $R$
are exchanged,
\begin{eqnarray}
\langle \Psi _{(L)}|&=&\langle \Psi |\hat {P}_{(R)},\nonumber \\
\langle \Psi _{(R)}\rangle &=&\langle \Psi |\hat {P}_{(L)},
\nonumber \\
\hat \langle \Psi _{(L)}|{\gamma }^5&=&\langle \Psi _{(L)}|(-1),
\nonumber \\
\langle \Psi _{(R)}|\hat {\gamma }^5&=&\langle \Psi _{(R)}|(+1).
\end{eqnarray}
When applying those projectors onto the Dirac equation, we have the
equations,
\begin{eqnarray}
i\hat {\gamma }^{\alpha }\hat {D}_{\alpha }|\Psi _{(R)}\rangle =\hat
{m}|\Psi _{(L)}\rangle ,\\
i\hat {\gamma }^{\alpha }\hat {D}_{\alpha }|\Psi _{(L)}\rangle =\hat
{m}|\Psi _{(R)}\rangle .
\end{eqnarray}
For adjoint states, we have
\begin{eqnarray}
\langle \Psi _{(R)}|\hat {D}_{\alpha }\hat {\gamma }^{\alpha
}i=\langle \Psi _{(L)}|\hat {m},\\
\langle \Psi _{(L)}|\hat {D}_{\alpha }\hat {\gamma }^{\alpha
}i=\langle \Psi _{(R)}|\hat {m}.
\end{eqnarray}

The isospin doublet projector and the isospin singlet projector are
defined as,
\begin{eqnarray}
\hat {P}_D&=&\frac {4}{3}\hat {I}^2,\nonumber \\
\hat {P}_S&=&1-\frac {4}{3}\hat {I}^2,
\end{eqnarray}
here, $\hat {I}^2$ is the summation of squares of three elements of
isospin. We can prove that,
\begin{eqnarray}
&&\overline{\hat {P}_{(D)}}=\hat {P}_{(D)}, \nonumber \\
&&\overline{\hat {P}_{(S)}}=\hat {P}_{(S)}, \nonumber \\
&&[\hat {P}_{(D)},\hat {\gamma }^{\alpha }]=[\hat {P}_{(S)},\hat
{\gamma }^{\alpha }]=0,\nonumber \\
&&[\hat {P}_{(D)},\hat {D}_{\alpha }]=[\hat {P}_{(S)},\hat
{D}_{\alpha }]=0,\nonumber \\
&&\hat {P}_{(D)}\hat {m}=\hat {m}\hat {P}_{(S)},\nonumber \\
&&\hat {P}_{(S)}\hat {m}=\hat {m}\hat {P}_{(D)}.
\end{eqnarray}

The isospin doublet state and the isospin singlet state are
\begin{eqnarray}
|\Psi _D\rangle &=&\hat {P}_D|\Psi \rangle ,\nonumber \\
|\Psi _S\rangle &=&\hat {P}_S|\Psi \rangle .
\end{eqnarray}

Applying the projection onto the Dirac equation, we can find two
equations,
\begin{eqnarray}
i\hat {\gamma }^{\alpha }\hat {D}_{\alpha }|\Psi _{(D)}\rangle =\hat
{m}|\Psi _{(S)}\rangle ,
\\
i\hat {\gamma }^{\alpha }\hat {D}_{\alpha }|\Psi _{(S)}\rangle =\hat
{m}|\Psi _{(D)}\rangle .
\end{eqnarray}

Next we define two new projectors corresponding to normal particles
and anomaly particles, respectively,
\begin{eqnarray}
\hat {P}_{(a)}&=&\hat {P}_{(L)}\hat {P}_{(D)}+\hat {P}_{(R)}\hat
{P}_{(S)}=\frac {1}{2}(1+\hat {\gamma }^5\hat {H}),
\\
\hat {P}_{(n)}&=&\hat {P}_{(R)}\hat {P}_{(D)}+\hat {P}_{(L)}\hat
{P}_{(S)}=\frac {1}{2}(1-\hat {\gamma }^5\hat {H}).
\end{eqnarray}

We can find the following properties for those two projectors,
\begin{eqnarray}
&&\overline{\hat {P}_{(a)}}=\hat {P}_{(n)}, \nonumber \\
&&\overline{\hat {P}_{(n)}}=\hat {P}_{(a)}, \nonumber \\
&&\hat {P}_{(a)}\hat {\gamma }^{\alpha }=\hat
{\gamma }^{\alpha }\hat {P}_{(n)},\nonumber \\
&&\hat {P}_{(n)}\hat {\gamma }^{\alpha }=\hat
{\gamma }^{\alpha }\hat {P}_{(a)},\nonumber \\
&&[\hat {P}_{(a)},\hat {D}_{\alpha }]=[\hat {P}_{(n)},\hat
{D}_{\alpha }]=0,\nonumber \\
&&\hat {P}_{(a)}\hat {m}=\hat {m}\hat {P}_{(n)},\nonumber \\
&&\hat {P}_{(n)}\hat {m}=\hat {m}\hat {P}_{(a)}.
\end{eqnarray}

The quantum states of normal particles and anomaly particles are,
\begin{eqnarray}
&&|\Psi _{(a)}\rangle =\hat {P}_{(a)}|\Psi \rangle  , ~~~\langle
\Psi
_{(a)}|=\langle \Psi |\hat {P}_{(a)}  , \\
&&|\Psi _{(n)}\rangle =\hat {P}_{(n)}|\Psi \rangle  , ~~~\langle
\Psi _{(n)}|=\langle \Psi |\hat {P}_{(n)}  .
\end{eqnarray}

We can find that the normal particle state and the anomaly particle
state satisfy the Dirac equation, respectively,
\begin{eqnarray}
i\hat {\gamma }^{\alpha }\hat {D}_{\alpha }|\Psi _{(a)}\rangle =\hat
{m}|\Psi _{(a)}\rangle ,\\
i\hat {\gamma }^{\alpha }\hat {D}_{\alpha }|\Psi _{(n)}\rangle =\hat
{m}|\Psi _{(n)}\rangle .
\end{eqnarray}
Since normal particles and the anomaly particles independently
satisfy their Dirac equations, the type of normal particles and the
type of anomaly particles do not evolve into each other. The
particles in our world are all normal particles, there does not
exist anomaly particles. We know that chiral left state only
corresponds to isospin doublet state, while chiral right state only
corresponds to isospin singlet state. Since the chiral left state
involves into the weak interactions, chiral right state does not
involves into the weak interactions, the consequence is that for
weak interactions, there is only chiral left state. This is the
conclusion that there is no parity conservation law for weak
interactions \cite{LeeYang}. Here we can see that the parity
violation can be explained naturally in our theory.

\subsection{Gravity and CPT violation}
CPT is a combined transformation of charge conjugation, space
reversal and time reversal. It is known that the Standard Model are
symmetric under CPT transformation. Here we would like to discuss
the relationships between gravity and CPT violation.

Since the theory of this work is in the framework of general
relativity, there is no independent space reversal and time
reversal. What we have is a combined PT reversal, the momentum
operator transforms as, $\hat {p}_{\mu }\rightarrow -\hat {p}_{\mu
}, \mu =0,1,2,3$. We define the charge conjugation transformation
as,
\begin{eqnarray}
\hat {T}_a\rightarrow -\hat {T}_a.
\end{eqnarray}
Under CPT transformation, we have $\hat {\gamma }^{\alpha
}\rightarrow -\hat {\gamma }^{\alpha }$, while spin operators remain
the same $\hat {s}_{\alpha \beta }\rightarrow \hat {s}_{\alpha \beta
}$. Recall the definition of the general covariance derivative
operator in (\ref{deri-operator}), $\hat {D}_{\alpha }=-i\hat
{\theta }_{\alpha }^{\mu }\otimes \hat {p}_{\mu }+\frac {i}{2}\hat
{\Gamma }_{\alpha }^{\rho \sigma }\otimes \hat {s}_{\rho \sigma
}-i\hat {A}_{\alpha }^a\otimes \hat {T}_a$, and also consider that
always the form $\hat {\gamma }^{\alpha }\hat {D}_{\alpha }$
appeares in fundamental equations, it transforms as
\begin{eqnarray}
\hat {\gamma }^{\alpha }\hat {D}_{\alpha }\rightarrow -i\hat {\theta
}_{\alpha }^{\mu }\hat {\gamma }^{\alpha }\otimes \hat {p}_{\mu
}-\frac {i}{2}\hat {\Gamma }_{\alpha }^{\rho \sigma }\hat {\gamma
}^{\alpha }\otimes \hat {s}_{\rho \sigma }-i\hat {A}_{\alpha }^a\hat
{\gamma }^{\alpha }\otimes \hat {T}_a.
\end{eqnarray}
One may notice that the first and the third terms are invariant
under CPT transformation, while the second term which is related
with gravity field changes its symbol from positive to negative.
Also we know that $\hat {m}, \hat {M}$ are invariant. Consider that
$\hat {\gamma }^{\alpha }\hat {D}_{\alpha }$ involves into Dirac
equation and Yang-Mills equation, we can conclude that the gravity
field will cause CPT violation. In case $\hat {\Gamma }_{\alpha
}^{\rho \sigma }=0$, which means there is no gravity field, the CPT
symmetry remains.

In general it is considered that the CPT symmetry represents the
symmetry between particle and antiparticle. The result that gravity
will cause CPT violation means that gravity is not symmetric for
particles and antiparticles. This is understandable since gravity
force is always attractive for particle and antiparticle.

\subsection{A dark energy solution}

The observed evidences \cite{accelerate1,accelerate2} show that the
expansion of the universe is accelerating. Also, by the data from
the Wilkinson Microwave Anisotropy Probe (WMAP) \cite{WMAP} and
other teams, the universe is flat, homogeneous, and isotropic and
the distribution of baryonic matter and radiation, dark matter and
dark energy is approximately: $4.5\% $, $23\% $ and $72\% $. The
dark energy is homogeneous, nearly independent of time and the
density is very small which is around $\rho _{\Lambda
}=10^{-29}g/cm^{3}$. The dark energy is not known to interact
through any of the fundamental forces except gravity. The
gravitational effect of dark energy approximates that of Einstein's
cosmological constant, it has a strong negative pressure which can
explain the observed accelerating universe. Still there are some
other models about it. Presently it seems that there is no general
accepted quantum theory of dark energy, actually the Planck energy
density which is the expected candidate is about 120 orders of
magnitude larger than the dark energy.

Based on the unified theory of the present work, here we propose a
dark energy solution. We assume: (a) There is no matter or matter particles,
$|\Psi \rangle =0, j^a_{\alpha }=0$; (b) There is no gravity $\Gamma
^{\rho \sigma }_{\alpha }=0$; (c) The gauge potential is constant, $\partial _{\mu }
A^a_{\alpha }=0$. Recall the definition in Eq.
(\ref{gauge-curvature}), we find the gauge curvature now takes the
form
\begin{eqnarray}
F^a_{\alpha \beta }=-iC^a_{bc}A^b_{\alpha }A^c_{\beta }.
\end{eqnarray}
Substitute the gauge curvature into Yang-Mills equation in
(\ref{ym-eq}), we have,
\begin{eqnarray}
D^{\alpha }F^a_{\alpha \beta }=M^a_bA^b_{\beta },
\end{eqnarray}
so,
\begin{eqnarray}
-iC^a_{bc}A^{b,\alpha }F^c_{\alpha \beta }=M^a_bA^b_{\beta }.
\end{eqnarray}
Thus the equation is,
\begin{eqnarray}
-C^a_{bc}C^c_{de}A^{b,\alpha }A^d_{\alpha }A^e_{\beta }=M^a_eA^e_{
\beta }.
\end{eqnarray}
One solution of this equation is $A^e_{\beta }=0$, it is trivial and
we do not discuss it. Next, we shall consider the solution,
\begin{eqnarray}
-C^a_{bc}C^c_{de}A^{b,\alpha }A^d_{\alpha }=M^a_e.
\label{dark-solution}
\end{eqnarray}
Consider that,
\begin{eqnarray}
F^{\alpha \rho }_aF^a_{\beta \rho }&=&-C_{abc}C^a_{de}A^{b,\alpha
}A^{c,\rho } A^d_{\beta }A^e_{\rho } \nonumber \\
&=&\left( C^b_{ca}C^a_{ed}A^{c,\rho }A^e_{\rho }\right)A^{\alpha
}_bA^d_{\beta } \nonumber \\
&=&-M^b_dA^{\alpha }_bA^d_{\beta },
\end{eqnarray}
where the solution (\ref{dark-solution}) is used in the last
equation. Also we have,
\begin{eqnarray}
F^{\rho \sigma }_{a}F^a_{\rho \sigma }=-M^b_dA^{\rho }_bA^d_{\rho }.
\end{eqnarray}
Substituting those results to the energy-momentum tensor of gauge
fields (\ref{en-mo-gauge}), we can find that,
\begin{eqnarray}
\tau ^{\alpha }_{\beta }&=&-\frac {1}{4}\delta ^{\alpha }_{\beta
}M^a_bA^{\rho }_aA^b_{\rho }. \label{dark-all-choice}
\end{eqnarray}
The explicit form of this energy-momentum tensor depends on the
solution of equation (\ref{dark-solution}) if the exact mass matrix
$M^a_e$ is given. Please note here this energy-momentum tensor will
provide a cosmological constant.

Next, we assume: (d) The candidate of dark energy is related with
the elementary particle gluon. For this case, the mass matrix of
gluon takes the form,
\begin{eqnarray}
M^a_e=m^2\delta ^a_e,
\end{eqnarray}
where $m$ is the mass of gluon, with the help of the solution of
Yang-Mills equation (\ref{dark-solution}), we have,
\begin{eqnarray}
m^2&=&-C^a_{bc}C^c_{de}A^{b,\alpha }A^d_{\alpha } \nonumber \\
&=&-g_3^2G_{bd}A^{b,\alpha }A^d_{\alpha } \nonumber \\
&=&-g_3^2A^{\alpha }_dA^d_{\alpha },
\end{eqnarray}
where $g_3$ is the coupling constant as presented in
(\ref{coup-constant}). Consider also the form of mass matrix of
gluon, the energy-momentum tensor of gauge field is,
\begin{eqnarray}
\tau ^{\alpha }_{\beta }=\delta ^{\alpha }_{\beta }\frac
{m^4}{4g_3^2}. \label{dark-final-solution}
\end{eqnarray}
Recall our assumption (a) which means the energy-momentum tensor of
matter fields is zero, the total energy-momentum tensor is
\begin{eqnarray}
T_{\alpha ,\beta }=\eta _{\alpha \beta }\frac {m^4}{4g_3^2},
\label{dark-en-mo}
\end{eqnarray}
where $\eta _{\alpha \beta }$ is the Minkowski metric. Please note
that no boundary condition is used to obtain this solution, and it
can be assumed to be valid for universe.  Recall the Einstein
equation in (\ref{einstein-eq}), it is now clear that the
energy-momentum tensor here (\ref{dark-en-mo}) corresponds to
Einstein's cosmological constant. So we conclude that the density of
the dark energy $\rho _{\Lambda }$ is,
\begin{eqnarray}
\rho _{\Lambda }=\frac {m^4}{4g_3^2}.
\end{eqnarray}
In Standard Model and also in our theory, $g_3\approx 1.22$,
consider that the density of
the dark energy $\rho _{\Lambda }\approx 10^{-29}g/cm^{3}$, we can
estimate that the mass of gluon is around $10^{-3}eV\sim 10^{-2}eV$.
It is theoretically accepted and experimentally confirmed that the
mass of gluon is zero, here we can see that the estimated mass of
gluon is very small thus it should not have detectable effects on
present experiments.

Let's list some properties of the present theory of dark energy and
show that this result is reasonable.
\begin{enumerate}

\item
The elementary particle gluon does not interact through electrical
force, does not have weak interactions, its mass is very small. And
further there is the confinement of gluon due to color confinement,
A free gluon is thus impossible or hard to be detected directly in
experiments. The solution of gluon is a 0-mode solution, the
momentum is zero since of the constant gauge potential (c) and
correspondingly it does not change in 4D coordinate space. That
means it is invariant in 4D space-time and the gluon will not cause
any energy excitation, it is like the vacuum state. Except
gravitational effects, our theory shows that there will be no
interactions available now and in the future. All of those agree
with the properties of dark energy. Here we would like to emphasize
again that the unified theory of the present work itself does not
assume that the gluon is the only candidate for dark energy since
Eq.(\ref{dark-all-choice}) always provides a cosmological constant
like energy-momentum tensor. However, the dark energy is generally
not assumed to be any matter particles including neutrino, it seems
not photon by observation. We may consider the weak interaction
bosons $W_{\pm }, Z_0$, but they seem too heavy to be related with
the dark energy. We may roughly estimate that $\rho _{\Lambda
}\approx m_Z^4/4g_2^2\approx 10^{43} eV^4$, it is around
$10^{25}g/cm^3$ which could be a candidate for black hole or the
early state of universe but not the dark energy, where $g_2\approx
0.65$, Because of the above reasons, we consider that gluon should
be the most possible candidate.

\item
We can find that the energy-momentum tensor with our assumptions
plays the role of Einstein's cosmological constant. The present
observations for dark energy, for example in WMAP \cite{WMAP},
prefer to the cosmos model of cosmological constant with equation of
state parameter $w=-1$ within $14\% $ level. Our theory of
dark-energy explains well the origin of the cosmological constant.

\item
The mass of gluon in Standard Model is generally assumed to be zero,
here we find that the estimated mass of gluon is very small which is
reasonable for present particle experiments. The Standard Model which
is very successful theoretically and experimentally still remains
correct. Of course as we mentioned, a massive gluon is allowed in
our 4D unified theory.

\item
We would like to remark that our solution of Yang-Mills equation
does not use any boundary condition thus it can be applied to case
of universe. It is a fixed constant solution of an equation. On the
other hand, for this existing solution of Yang-Mills equation, the
cosmological constant is a reasonable explanation.

\item
To find a solution from Yang-Mills equation, we make several
assumptions based on the observable facts of universe.  The fact
that there is no matter available corresponds to assumption (a). The
universe is flat, homogeneous and isotropic, this is corresponding
to assumptions (b) and (c). Those assumptions also lead to the fact
that cosmological constant has an origin from quantum effects and is
independent of gravity.

\item
We finally comment that the present theory of dark energy shows that
our 4D unified theory which is a unification of quantum mechanics
and general relativity can simply explain the dark energy well. This
theory should be a complete theory. Let us emphasize again the
structure of our theory: we first have representations, then we use
three fundamental equations to find the results.
\end{enumerate}

\section{Conclusions and discussions}
Our theory based on the cornerstones of physics, such as the Dirac
equation of quantum mechanics, Einstein general relativity of
gravity and Yang-Mills gauge theory. The results deduced from our
theory agree well with many basic facts of physics: such as those
already presented in abstract of this work: (1) Graviton is
massless; (2) Mass problem of gauge theory is explained well; (3)
Color confinement; (4) Parity violation; (5) CPT violation; (6) Dark
energy can be explained well. Also please note that our gauge
representations agree with the basic results of the Feynman diagram
quantum field theory.

Our theory is a unified theory which combines the general relativity
and quantum mechanics. This is not only a long dream in physics, but
also it provides a theory for things like black hole or early stage
of universe where both general relativity and quantum mechanics are
important. One may already see, as we have presented, the dark
energy is a large scale phenomenon of universe where the general
relativity is important, we provide a reasonable explanation from
quantum scale of Yang-Mills equation. The origin of cosmological
constant is from the elementary particle.

Two foundations of our theory are; first we formulate three
fundamental equations in the framework of general relativity; second
we provide a proper definition of mass; the problems solved in this
work generally depends on a correct concept of mass. We may also
notice that gauge condition in gauge theory is in general not
arbitrary except for massless particles such as photon for
electromagnetic field.

Here let us discuss further the problem of mass. Our opinion is that
the Higgs mechanism is not necessary. The reasons are the following:
(1). From our theory, we may find that mass is defined naturally in
gauge space instead of defining by momentum. In particular, no Higgs
particle is introduced, actually from the proof of energy-momentum
conservation law, as we mentioned, there is no role of Higgs
particles. (2). Our theory does not allow a spin-0 elementary
particle since it does not interact through gravity, while the Higgs
particle is spin-0. (3). We have three kind of gauge conditions
respectively for massless photons, massive weak interaction bosons,
and gluons. The gauge is arbitrary only for massless photons case of
electromagnetic field, while the gauge conditions are equations
which should be satisfied for weak interaction bosons and gluons,
the necessity of Higgs mechanism does not exist. We thus prefer the
idea that Higgs particle does not exist. On the other hand, one may
argue that the mass matrices $\hat {m}, \hat {M}$ defined in our
theory might also be explained as from the Higgs mechanism. Our idea
is that since mass matrices are defined in gauge space depending on
experiments, if still an explanation of those matrices is necessary,
or further to discuss the origin of mass, Higgs mechanism is now not
a must. Let us note that the experiments for searching of Higgs
particles are on going at Fermilab and will soon become operational
in Large Hadron Collider (LHC) at CERN. As we know the masses of
Higgs bosons below $114.4GeV$ were previously excluded. The result
of Fermilab excludes Higgs bosons between $160GeV\sim 170GeV$
\cite{fermilab}. The LHC is designed to make proton-proton
collisions at an energy of $7TeV$ per beam which can scan
$10^{11}eV$ of Higgs particles in electroweak theory thus it should
provide a definite answer whether Higgs particles exist or not. If
no Higgs particle is found in LHC while we still believe in it, one
might go to $10^{25}eV$ or higher to test grand-unification-theory
or super-symmetry theory with gravity.

Our theory provides a new framework of quantum theory. We expect
there may be two directions for the future of our theory. On the one
hand, we need to check whether our theory agrees well with all
well-established physical facts, experimentally and theoretically.
On the other hand, while our theory is proved to be powerful, we
still need to find some new predictions from our theory. Our comment
is that with a new foundation of quantum physics, we expect that
there will be a lot of new results.

We understand that quantum field theory has already been well
established. However, its problem is also well-accepted, the main
results are in general relies on the approximation methods. Our
theory is exact: we have the representations of fields and
interactions, their properties are governed by three equations,
Dirac equation, Einstein equation and Yang-Mills equation, then we
are led to observable quantities. The equations themselves are
exact. In this sense, our theory is quite simple. This theory turns
us to the similar route as the Newton classical mechanics: all are
governed by equations.

A brief introduction of the quantization of gravity is presented in
Ref.\cite{ZF1}. The present work is a detailed presentation of the
whole theory. A brief presentation about mass of gauge field will be
published elsewhere \cite{ZF2}.

Finally let us recall that the proposed theory provides an unified
framework for three fundamental equations. It could be falsified,
but it could not be fudged. To end this paper, we comment that we
intend to use the little to get the big, in Chinese, it is to throw
out a brick to attract a jade.

\acknowledgements

We thank Mr. Shi-Ping Ding for consistent supporting and
discussions.  This work is supported by grants of National Natural
Science Foundation of China (NSFC) Nos.(10674162,10974247), ``973''
program (2010CB922904) of Ministry of Science and Technology (MOST),
China, and Hundred-Talent Project of Chinese Academy of Sciences
(CAS), China.

\newpage

\appendix

\section{Appendix}

For self-consistent and convenient, we present the notations,
foundations and some detailed calculations of this work in this appendix.

\subsection{Tensor representation}
The 1st-order tensor space $T(1)$ is a $n$-dimensional space spanned
by basis $\{ e_i\}, (i=1,2,\cdots, n)$. A 1st-order tensor $K$ can
be expanded by this basis as
\begin{eqnarray}
K=K^ie_i,
\end{eqnarray}
where summation over  repeated lower-upper indices is assumed.

The $r$-th order tensor space $T(r)$ is the tensor product of $r$
1st-order tensor spaces,
\begin{eqnarray}
T(r)=\prod _r\otimes T(1)
\end{eqnarray}
The space  $T(r)$ is spanned by basis $\{ e_{i_1\cdots i_r}  \}$
expressed as,
\begin{eqnarray}
e_{i_1\cdots i_r}=e_{i_1}\otimes \cdots \otimes e_{i_r},
\end{eqnarray}
where $i_1,\cdots ,i_r=1,2,\cdots ,n$. The $r$-th order tensor can
be represented by the this basis as,
\begin{eqnarray}
K=K^{i_1\cdots i_r}e_{i_1\cdots i_r}.
\end{eqnarray}
The tensor product of a $r$-th tensor $K$ and a $s$-th tensor $M$ is
a $r+s$-th tensor which takes the form
\begin{eqnarray}
K\otimes M=K^{i_1\cdots i_r}e_{i_1\cdots i_r}\otimes M^{j_1\cdots
j_s}e_{j_1\cdots j_s}=K^{i_1\cdots i_r}M^{j_1\cdots j_s}e_{i_1\cdots
i_rj_1\cdots j_s}.
\end{eqnarray}
The tensor $K$ is symmetric if the tensor remains invariant by
permuting the indices of its elements $K^{i_1\cdots i_r}e_{i_1\cdots
i_r}$. It is antisymmetric if there is an negative symbol by odd
permuting of the indices and remains invariant by even permutation
of the indices. The basis for anti-symmetric tensor can be expressed
as
\begin{eqnarray}
\theta _{i_1\cdots i_r}=\delta _{i_1\cdots i_r}^{k_1\cdots
k_r}e_{k_1\cdots k_r},
\end{eqnarray}
where please note the definition of the generalized Kronecker symbol
$\delta _{i_1\cdots i_r}^{k_1\cdots k_r}$ takes the form
\begin{eqnarray}
\delta _{i_1\cdots i_r}^{k_1\cdots k_r}=\left| \begin{array}{ccc}
\delta _{i_1}^{k_1}&\cdots &\delta _{i_r}^{k_1}\\
\vdots &\vdots &\vdots \\
\delta _{i_1}^{k_r}&\cdots &\delta _{i_r}^{k_r} \end{array} \right|
.
\end{eqnarray}
When the upper indices is an even permutation of the lower indices,
$\delta _{i_1\cdots i_r}^{k_1\cdots k_r}$ is 1, it is $-1$ for odd
permutations and 0 for other cases. Any antisymmetric tensor can be
expressed as,
\begin{eqnarray}
a_r=\frac {1}{r!}a^{i_1\cdots i_r}\theta _{i_1\cdots i_r},
\end{eqnarray}
where parameter $a^{i_1\cdots i_r}$ is antisymmetric. For
$n$-dimensional spaces tensor product together, the highest
antisymmetric tensor is rank-$n$ antisymmetric tensor. The direct
summation of all rank antisymmetric spaces constitute a
antisymmetric space. The summation of the ranks is $2^n$,
\begin{eqnarray}
\Lambda &=&\Lambda ^0\oplus \cdots \oplus \Lambda ^n, \nonumber \\
{\rm dim}\Lambda &=&{\rm dim}\Lambda ^0+\cdots +{\rm dim}\Lambda ^n
\nonumber \\
&=&C_n^0+C_n^1+\cdots +C_n^n \nonumber \\
&=&2^n.
\end{eqnarray}
Besides computing multiplication and summation, we can define {\it
wedge} in this antisymmetric space $\Lambda $,
\begin{eqnarray}
\alpha _p\wedge \beta _q=\frac {(p+q)!}{p!q!}A_{p+q}(\alpha
_p\otimes \beta _q),
\end{eqnarray}
where
\begin{eqnarray}
A_{p+q}=\frac {1}{(p+q)!}\sum _{\sigma \in P(p+q)}{\rm sgn}(\sigma
)\sigma
\end{eqnarray}
is the anti-symmetrized operator, for odd permutation $\sigma $,
${\rm sgn}(\sigma )=-1$; for even permutation $\sigma $, ${\rm
sgn}(\sigma )=1$. Wedge has the property,
\begin{eqnarray}
\alpha _p\wedge \beta _q=(-1)^{pq}\beta _q\wedge \alpha _p
\end{eqnarray}

\subsection{Metric, covariance tensor and contra-variance tensor}
Suppose $K,M$ are the rank-1 tensors in space $T(1)$, the scalar
product of those two tensors is a rank-0 tensor, i.e., a number, and
satisfy the equation,
\begin{eqnarray}
K\cdot M=M\cdot K.
\end{eqnarray}
The scalar products between $n$ rank-1 tensor basis in $T(1)$ can
constitute a $n\times n$ matrix which is named metric:
\begin{eqnarray}
g_{ij}=e_i\cdot e_j.
\end{eqnarray}
It is obvious that $g_{ij}$ is symmetric,
\begin{eqnarray}
g_{ij}=g_{ji}.
\end{eqnarray}
Generally, we demand that the rank of the metric matrix is full,
that is its determinant is non-zero,
\begin{eqnarray}
{\rm det }\{ g_{ij}\} \not= 0.
\end{eqnarray}
With the help of the metric, the scalar product of two rank-1
tensors can be represented as,
\begin{eqnarray}
K\cdot M=g_{ij}K^iM^j.
\end{eqnarray}
For a tensor, there is two different indices, the upper indices and
the lower indices. The lower indices is called covariance indices,
and the upper indices is called contra-variance indices. If a tensor
only has covariance indices, it is called covariance tensor. A
tensor which has only contra-variance indices is called
contra-variance indices. If a tensor has both covariance indices and
the contra-variance indices, it is called mixed tensor. The metric
$g_{ij}$ is called covariance metric, $g^{ij}$ is called
contra-variance metric. They satisfy the equations,
\begin{eqnarray}
g_{ik}g^{jk}=g^{jk}g_{ki}=\delta ^j_i.
\end{eqnarray}
Metric can be represented as a rank-2 symmetric tensor as,
\begin{eqnarray}
G=g^{ij}e_{ij}.
\end{eqnarray}
By using contra-variance metric $g^{ij}$ and the covariance metric
$g_{ij}$, we can realize the rising or lowering the indices of a
tensor. For example, for a rank-1 tensor, we have,
\begin{eqnarray}
&&e^i=g^{ij}e_j,\\
&&K_i=g_{ij}K^j,\\
&&K=K^ie_i=K_ie^i.
\end{eqnarray}
And for a rank-2 tensor, we have,
\begin{eqnarray}
&&e^i_{i'}=g^{ij}e_{ji'}, ~~~K_i^{i'}=g_{ij}K^{ji'}, \\
&&e^{ii'}=g^{ij}g^{i'j'}e_{jj'}, ~~~K_{ii'}=g_{ij}g_{i'j'}K^{jj'},\\
&&K=K^{ii'}e_{ii'}=K_i^{i'}e_{i'}^i=K_{ii'}e^{ii'}.
\end{eqnarray}
If the rank of a tensor is larger than 2, the contraction
calculation can be made to this tensor and the rank will decrease 2.
For example, the contraction of a rank-2 tensor may be performed as,
\begin{eqnarray}
{\rm con}K=g_{ij}K^{ij}=K^i_i=g^{ij}K_{ij}.
\end{eqnarray}
The contraction for a rank-3 tensor may be performed as,
\begin{eqnarray}
{\rm con}K=g_{ij}K^{ijk}e_k=K^{ik}_ie_k=g^{ij}K^k_{ij}e_k.
\end{eqnarray}
In general, the contraction for a rank-$r$ $(r\ge 3)$ can be
calculated as,
\begin{eqnarray}
{\rm con}K={\rm con}(K^{i_1i_2i_3\cdots i_r}e_{i_1i_2i_3\cdots
i_r})=g_{i_1i_2}K^{i_1i_2i_3\cdots i_r}e_{i_1i_2i_3\cdots i_r}.
\end{eqnarray}

The tensor transformation is generally defined by the basis
transformation. Suppose there are two sets of tensor basis $\{
e_i\}$ and $\{ e_{i'}\}$, $(i,i'=1,2,\cdots ,n)$. The transformation
between these two sets basis is defined by the transformation matrix
$\{ L_i^{i'}\} $,
\begin{eqnarray}
e_i=L_i^{i'}e_{i'}'.
\end{eqnarray}
The inverse transformation can be found to be
\begin{eqnarray}
e_{i'}'={L^{-1}}^i_{i'}e_{i},
\end{eqnarray}
where ${L^{-1}}^i_{i'}$ is the inverse transformation matrix and
satisfy the equation,
\begin{eqnarray}
{L^{-1}}^i_{j}L^j_k=\delta ^i_k.
\end{eqnarray}
A general tensor can be expanded in two sets of basis,
\begin{eqnarray}
K=K^{k\cdots }_{i\cdots }e_{k\cdots }^{i\cdots }={K'}^{k'\cdots
}_{i'\cdots }e_{k'\cdots }^{i'\cdots }.
\end{eqnarray}
The transformation between two basis takes the form
\begin{eqnarray}
e_{k'\cdots }^{i'\cdots }={L^{-1}}^k_{k'}L^{i'}_i\cdots e^{i\cdots
}_{k\cdots }.
\end{eqnarray}
So the transformation between the elements of the tensor is,
\begin{eqnarray}
{K'}^{k'\cdots }_{i'\cdots }=L^{k'}_k{L^{-1}}^i_{i'}\cdots
K^{k\cdots }_{i\cdots }.
\end{eqnarray}

\subsection{Matrix}
The matrix calculation obey the standard law of mathematics. Suppose
$\hat {A}, \hat {B}$ are two $n\times n$ matrices, the inner product
of the two matrices is defined as,
\begin{eqnarray}
\langle \hat {A}, \hat {B}\rangle ={\rm tr}\left( \hat {A}^{\dagger
}\hat {B}\right)=a^*_{ij}b_{ij}
\end{eqnarray}
where the upper index $\dagger $ means to take a matrix
transposition plus complex conjugation. The inner product of
matrices has the properties:
\begin{eqnarray}
&&\langle \hat {A}, \hat {A}\rangle \ge 0, \nonumber \\
&&\langle \hat {A}, \hat {B}\rangle =\langle \hat {B}, \hat
{A}\rangle ^*.
\end{eqnarray}
If the inner product of two matrices is zero, $\langle \hat {A},
\hat {B}\rangle =0$, we say these two matrices are orthogonal.

In $n\times n$ matrix representation space, we can introduce $n^2$
ortho-normal basis,
\begin{eqnarray}
&&\hat {e}_l,~~~l=1,2,\cdots ,n^2, \nonumber \\
&& \langle \hat {e}_l, \hat {e}_{l'}\rangle ={\rm tr}(\hat
{e}_l^{\dagger }\hat {e}_{l'})=\delta _{ll'}.
\end{eqnarray}
An arbitrary $n\times n$ matrices can be represented in terms of
this basis as,
\begin{eqnarray}
\hat {A}=A_l\hat {e}_l,
\end{eqnarray}
where the coefficients in the expansion can be calculated as,
\begin{eqnarray}
A_l=\langle \hat {e}_l,\hat {A}\rangle ={\rm tr}(\hat {e}_l^{\dagger
}\hat {A}).
\end{eqnarray}

Suppose $F(x)$ is an analytical function, that is $F(x)$ can be
expanded by Taylor expansion,
\begin{eqnarray}
F(x)=\sum _{l=0}^{\infty }f(l)x^l.
\end{eqnarray}
the matrix function $F(\hat {A})$ is defined as,
\begin{eqnarray}
F(\hat {A})=\sum _{l=0}^{\infty }f(l)\hat {A}^l.
\end{eqnarray}

The matrix similarity transformation is defined as,
\begin{eqnarray}
\hat {A}'=\hat {L}\hat {A}\hat {L}^{-1}.
\end{eqnarray}
The matrix unitary transformation is defined as,
\begin{eqnarray}
\widetilde {\hat {A}}=\hat {U}\hat {A}\hat {U}^{\dagger },
\end{eqnarray}
where $\hat {U}$ is an unitary matrix $UU^{\dagger }=I$, $I$ is the
identity.

\subsection{Pauli matrices, Gell-Mann matrices}
The Paluli matrices are defined as,
\begin{eqnarray}
&&\hat {\sigma }_0=\left( \begin{array}{cc} 1&0\\
0&1\end{array}\right) ,~~~
\hat {\sigma }_1=\left( \begin{array}{cc} 0&1\\
1&0\end{array}\right) ,\nonumber \\
&&\hat {\sigma }_2=\left( \begin{array}{cc} 0&-i\\
i&0\end{array}\right) ,~~~
\hat {\sigma }_3=\left( \begin{array}{cc} 1&0\\
0&-1\end{array}\right) .
\end{eqnarray}
Some properties of Pauli matrices can be listed, for example, as,
\begin{eqnarray}
&&{\rm tr}\left( \sigma _{\mu }\sigma _{\nu }\right) =2\delta _{\mu
\nu }, ~~~~{\rm tr}\sigma _i=0, \nonumber \\
&& [\hat {\sigma }_i,\hat {\sigma }_i]=2i\varepsilon _{ijk}\hat
{\sigma }_k, ~~~\{ \hat {\sigma }_i,\hat {\sigma }_j\} =2i\delta
_{ij}\hat {\sigma }_0,\nonumber \\
&&\hat {\sigma }_i\hat {\sigma }_j=\delta _{ij}\hat {\sigma
}_0+i\varepsilon _{ijk}\hat {\sigma }_k,~~~ \hat {\sigma }_1\hat
{\sigma }_2\hat {\sigma }_3=i\sigma _0.
\end{eqnarray}

The $3\times 3$ Gell-Mann matrices take the forms,

\begin{eqnarray}
\hat {\lambda }_0=\frac {1}{\sqrt {6}}\left( \begin{array}{ccc} 1&0&0\\
0&1&0 \\
0&0&1\end{array}\right) , ~~~
\hat {\lambda }_1=\frac {1}{2}\left( \begin{array}{ccc} 0&1&0\\
1&0&0 \\
0&0&0\end{array}\right) , ~~~
\hat {\lambda }_2=\frac {1}{2}\left( \begin{array}{ccc} 0&-i&0\\
i&0&0 \\
0&0&0\end{array}\right) , \nonumber
\end{eqnarray}

\begin{eqnarray}
\hat {\lambda }_3=\frac {1}{2}\left( \begin{array}{ccc} 1&0&0\\
0&-1&0 \\
0&0&0\end{array}\right) , ~~~
\hat {\lambda }_4=\frac {1}{2}\left( \begin{array}{ccc} 0&0&1\\
0&0&0 \\
1&0&0\end{array}\right) , ~~~
\hat {\lambda }_5=\frac {1}{2}\left( \begin{array}{ccc} 0&0&-i\\
0&0&0 \\
i&0&0\end{array}\right) ,\nonumber
\end{eqnarray}

\begin{eqnarray}
\hat {\lambda }_6=\frac {1}{2}\left( \begin{array}{ccc} 0&0&0\\
0&0&1 \\
0&1&0\end{array}\right) , ~~~
\hat {\lambda }_7=\frac {1}{2}\left( \begin{array}{ccc} 0&0&0\\
0&0&-i \\
0&i&0\end{array}\right), ~~~
\hat {\lambda }_8=\frac {1}{2\sqrt {3}}\left( \begin{array}{ccc} 1&0&0\\
0&1&0 \\
0&0&-2\end{array}\right) .
\end{eqnarray}
Note that for convenience, the Gell-Mann matrices here is slightly
different from the standard Gell-Mann matrices by a whole factor
$\frac {1}{2}$.

The properties of the Gell-Mann matrices can be listed as,
\begin{eqnarray}
&&{\rm tr}\left( \hat {\lambda }_p\hat {\lambda }_q\right) =\frac
{1}{2}\delta _{pq}, \\
&&{\rm tr}\hat {\lambda }_p=0, ~(p\not= 0),
\\
&&[\hat {\lambda }_p, \hat {\lambda }_q]=if_{pqr}\hat {\lambda }_r,
\\
&&\{ \hat {\lambda }_p, \hat {\lambda }_q\} =d_{pqr}\hat {\lambda
}_r+\frac {\sqrt {6}}{3}\delta _{pq}\hat {\lambda }_0,
\end{eqnarray}
where $f_{pqr}$ and $d_{pqr}$ are completely anti-symmetric, the
non-zero elements are,
\begin{eqnarray}
&&f_{123}=1, \nonumber \\
&&f_{458}=f_{678}=\frac {\sqrt {3}}{2}, \nonumber \\
&&f_{147}=-f_{156}=f_{246}=f_{257}=f_{345}=-f_{367}=\frac {1}{2},
\end{eqnarray}
and
\begin{eqnarray}
&&d_{118}=d_{228}=d_{338}=-d_{888}=\frac {1}{\sqrt {3}}, \nonumber \\
&&d_{146}=d_{157}=-d_{247}=d_{256}=d_{344}=d_{355}=-d_{366}=-d_{377}=\frac
{1}{2}, \nonumber \\
&&d_{448}=d_{558}=d_{668}=-d_{778}=-\frac {1}{2\sqrt {3}}.
\end{eqnarray}

\subsection{Vector and its adjoint}
A vector $|a\rangle $ has an adjoint $\langle a|$ represented as,
\begin{eqnarray}
\langle a|=\overline{|a\rangle }.
\end{eqnarray}
It has the property,
\begin{eqnarray}
\overline{\alpha _1|a_1\rangle +\alpha _2|a_2\rangle }=\alpha ^*
_1\langle a_1|+\alpha _2^*\langle a_2|.
\end{eqnarray}
Each vector of basis $\{ |e_i\rangle \} $ has its adjoint and they
all constitute a basis $\{ \langle e_i| \} $ for the adjoint space
$\overline{V}$, so the adjoint of $|a\rangle =a^i|e_i\rangle $ takes
the form
\begin{eqnarray}
\langle a|=a^{*i}\langle e_i|.
\end{eqnarray}
The metric of a linear space $V$ is defined by a basis
$\{|e_i\rangle \}$ and the adjoint basis $\{ \langle e_i|\}$ by
their inner products,
\begin{eqnarray}
g_{ij}=\langle e_i|e_j\rangle .
\end{eqnarray}
We know that $g_{ij}=g^*_{ji}$, and also we demand that the metric
matrix is non-degenerate,
\begin{eqnarray}
{\rm det}\{ g_{ij}\} \not= 0.
\end{eqnarray}
If all eigenvalues of metric $g_{ij}$ are positive, it is the
positive metric, and all eigenvalues are negative, it is negative
metric. If the eigenvalues have both positive and negative symbols,
it is the indefinite metric. If the metric can be expressed as
$g_{ij}=\pm \delta _{ij}$, we say the metric is normalized.

The inner product of vectors $|a\rangle =a^i|e_i\rangle ,|b\rangle
=b^i|e_i\rangle $ can be expressed by the metric matrix,
\begin{eqnarray}
\langle a|b\rangle =a^{*i}\langle e_i|e_j\rangle b^j
=a^{*i}g_{ij}b^j.
\end{eqnarray}
As usual, the metric with lower indices is called the covariance
metric, the metric with upper indices is called contra-variance
metric which is defined below,
\begin{eqnarray}
\{ g^{ij}\} =\{ g_{ij}\} ^{-1}=\hat {g}^{-1}.
\end{eqnarray}
So we have
\begin{eqnarray}
g_{ik}g^{kj}=g^{jk}g_{ki}=\delta ^j_i.
\end{eqnarray}
One may find that
\begin{eqnarray}
\langle a|\hat {g}^{-1}|a\rangle \ge 0,
\end{eqnarray}
this is because,
\begin{eqnarray}
\langle a|\hat {g}^{-1}|a\rangle =g_{ik}g^{kj}a^{*i}a^j=\delta
^j_ia^{*i}a^j\ge 0
\end{eqnarray}
The metric $\hat {g}$ and its inverse can be used to realize the
rising and lower the lower and upper indices. For simplicity, we
denote the vector is expressed in terms of covariance basis $\{
|e_i\rangle \}$, the adjoint vector is expressed in terms of the
contra-variance basis $\{ \langle e^j| \}$, so we have
\begin{eqnarray}
&&\langle e^i|=g^{ij}\langle e_j|,
\nonumber \\
&&\langle e^i|e_j\rangle =\delta ^i_j, \nonumber \\
&&\langle a|=a^{*i}\langle e_i|=a_i^*\langle e^i|, \nonumber \\
&&a_i^*=a^{*j}g_{ji},\nonumber \\
&&\langle a|b\rangle =a_i^*b^i.
\end{eqnarray}

\subsection{Operator space}
A $m$-dimensional linear space $V$ tensor product with its adjoint
space $\overline{V}$ constitute an operator space. The basis of the
operator space can be constructed by the basis $\{ |e_j\rangle \} $
and the adjoint basis $\{ \langle e^j|\}$ as the following,
\begin{eqnarray}
|e_i\rangle \otimes \langle e^j|, ~~~(i,j=1,2,\cdots ,m),
\end{eqnarray}
where $\langle e^i|=g^{ij}\langle e_j|=g^{ij}\overline{ |e_j\rangle
}$. An operator $A$ can be expanded in terms of this basis,
\begin{eqnarray}
\hat {A}=A^i_j|e_i\rangle \otimes \langle e^j|.
\end{eqnarray}
The behaviors of the operator is the same as the matrix. The follows
are some of the properties and some notations,
\begin{eqnarray}
&&|a\rangle \otimes \langle b|=a^ib_j^*|e_i\rangle \otimes \langle
e^j|,
\nonumber \\
&&\hat {A}\hat {B}=(A^i_kB_j^k)|e_i\rangle \otimes \langle e^j|, \nonumber \\
&&[\hat {A}, \hat {B}]=\hat {A}\hat {B}-\hat {B}\hat {A}, \nonumber
\\
&&\{ \hat {A}, \hat {B}\}=\hat {A}\hat {B}+\hat {B}\hat {A}.
\end{eqnarray}
The adjoint of the basis of operator space takes the form,
\begin{eqnarray}
\overline{|e_i\rangle \otimes \langle e^j|}=|e^j\rangle \otimes
\langle e_i|=g^{lj}g_{ik}|e_l\rangle \otimes \langle e^k|.
\end{eqnarray}
The adjoint of an operator can be expressed as,
\begin{eqnarray}
\overline{\hat {A}}=g^{ik}{A^*}^l_kg_{lj}|e_i\rangle \otimes \langle
e^j|=\overline{A}^i_j|e_i\rangle \otimes \langle e^j|,
\end{eqnarray}
where we define,
\begin{eqnarray}
\overline{A}^i_j=g^{ik}{A^*}^l_kg_{lj}.
\end{eqnarray}
The adjoint of operators satisfy the properties,
\begin{eqnarray}
&&\overline{\overline{\hat {A}}}=\hat {A},\nonumber \\
&&\overline{\hat {A}|a\rangle }=\langle a|\overline{\hat {A}},
\nonumber \\
&&\overline{\alpha \hat {A}}=a^*\overline{\hat {A}}, \nonumber \\
&&\overline{\hat {A}\hat {B}}=\overline{\hat {B}\hat {A}}.
\end{eqnarray}
We can find the adjoint of an operator is related with the hermitian
conjugation of this operator by the metric matrix as,
\begin{eqnarray}
\overline{\hat {A}}=\hat {g}^{-1}\hat {A}^{\dagger }\hat {g}.
\end{eqnarray}
The inner product of the operators is defined as,
\begin{eqnarray}
\langle \hat {A},\hat {B}\rangle &=&{\rm tr}\left( \overline{\hat
{A}} \hat {B}\right)=\overline{A}^i_jB_i^j \nonumber \\
&=&g^{ik}g_{lj}{A^*}^l_kB_i^j.
\end{eqnarray}
The properties of the operator inner product calculation are listed
as,
\begin{eqnarray}
&&\langle \hat {A},\hat {B}\rangle =\langle \hat {B},\hat {A}\rangle
^*, \nonumber \\
&&\langle \hat {A},\beta \hat {B}+\gamma \hat {C}\rangle =\beta
\langle \hat {A},\hat {B}\rangle +\gamma
\langle \hat {A},\hat {C}\rangle , \nonumber \\
&&\langle \alpha \hat {A}+\beta \hat {B},\hat {C}\rangle =\alpha ^*
\langle \hat {A},\hat {C}\rangle +\beta ^* \langle \hat {B},\hat
{C}\rangle .
\end{eqnarray}
Operators $\hat {A},\hat {B}$ are orthogonal if there inner product
is zero, $\langle \hat {A},\hat {B}\rangle =0$.

The identity operator is denoted as,
\begin{eqnarray}
\hat {I}&=&|e_i\rangle \otimes \langle e^i|=|e^i\rangle \otimes
\langle e_i|=g^{ij}|e_i\rangle \otimes \langle e_j| \nonumber \\
&=&g_{ij}|e^i\rangle \otimes \langle e^j|=\hat {g}.
\end{eqnarray}
The inverse of an operator is defined as,
\begin{eqnarray}
\hat {A}^{-1}\hat {A}=\hat {I}.
\end{eqnarray}
And as usual the power zero of an operator is the identity,
\begin{eqnarray}
\hat {A}^0=\hat {I}.
\end{eqnarray}
Suppose function $F(x)$ is analytic, the function of an operator is
defined through Taylor expansion as,
\begin{eqnarray}
F(\hat {A})=\sum _{l=0}^{\infty }f(l)\hat {A}^l
\end{eqnarray}
By applying the identity operator $I=|e_i\rangle \otimes \langle
e^i|$ on a vector $|a\rangle $and on its adjoint, or on an operator,
they will be expanded by basis $|e_i\rangle $,
\begin{eqnarray}
&&|a\rangle =(|e_i\rangle \otimes \langle e^i|)|a\rangle
=a^i|e_i\rangle , ~~~a^i=\langle e^i|a\rangle , \nonumber \\
&& \langle a|=\langle a|(|e_i\rangle \otimes \langle
e^i|)=a_i^*\langle e^i|, ~~~a_i^*=\langle a|e_i\rangle
=a^{*j}g_{ji}, \nonumber \\
&&\hat {A}=(|e_i\rangle \otimes \langle e^i|)\hat {A}(|e_j\rangle
\otimes \langle e^j|)=A^i_j|e_i\rangle \otimes \langle e^j|,
~~~A^i_j=\langle e^i|\hat {A}|e_j\rangle .
\end{eqnarray}
For the identity operator in another basis $I=|e_i'\rangle \otimes
\langle e'^{i}|$, we have similar results. The transformation
between those two basis can be obtained as,
\begin{eqnarray}
|e_i\rangle =(|e_j'\rangle \otimes \langle e'^j|)|e_i\rangle
=L^j_i|e_j'\rangle ,
\end{eqnarray}
where $L_i^j=\langle e^{'j}|e_i\rangle $ gives the definition of the
transformation elements, and also we have
\begin{eqnarray}
\hat {L}=\hat {I}'\hat {I}=L^i_j|e_i'\rangle \otimes \langle e^j|.
\end{eqnarray}
The inverse of the basis transformation takes the form,
\begin{eqnarray}
|e_i'\rangle =(|e_j\rangle \otimes \langle e^j|)|e_i'\rangle
={L^{-1}}_i^j|e_j\rangle ,
\end{eqnarray}
where ${L^{-1}}^j_i=\langle e^j|e_i'\rangle $ and can be represented
as,
\begin{eqnarray}
\hat {L}^{-1}=\hat {I}\hat {I}'={L^{-1}}^i_j|e_i\rangle \otimes
\langle e'^j|.
\end{eqnarray}
The transformation matrix and its inverse have the relations,
\begin{eqnarray}
{L^{-1}}^i_jL^j_k=\langle e^i|e_j'\rangle \otimes \langle
e'^j|e_k\rangle =\langle e^i|e_k\rangle =\delta ^i_k,
\end{eqnarray}
that is $\hat {L}^{-1}\hat {L}=\hat {I}$. We also have,
\begin{eqnarray}
{L^{-1}}^i_j&=&\langle e^i|e_j'\rangle =\langle e_j'|e^i\rangle ^*
\nonumber \\
&=&g'_{lj}g^{ik}\langle e'^l|e_k\rangle
=g^{ik}L^{*l}_kg_{lj}'=\overline{L}^i_j,
\end{eqnarray}
\begin{eqnarray}
\overline{\hat {L}}&=&\hat {L}^{-1}.
\end{eqnarray}

\subsection{Metric transformation}
With basis $\{ |e_i'\rangle \}$, the metric can be defined as,
\begin{eqnarray}
g'^{ij}=\langle e'^i|e'^j\rangle .
\end{eqnarray}
By inserting an identity operator $\hat {I}=g^{kl}|e_k\rangle
\otimes \langle e_l|$ into this equation, we can have,
\begin{eqnarray}
g'^{ij}=g^{kl}\langle e'|e_k\rangle \otimes \langle e_l|e'^j\rangle
=g^{kl}\langle e'^i|e_k\rangle \otimes \langle e'^j|e_l\rangle ^*.
\end{eqnarray}
For different basis, the metric transformation takes the form,
\begin{eqnarray}
g'^{ij}=L^i_kg^{kl}L^{*j}_l.
\end{eqnarray}
In a concise form, it can be rewritten as,
\begin{eqnarray}
\hat {g}'=\hat {L}\hat {g}\hat {L}^{\dagger }.
\end{eqnarray}
In a different basis, a vector can be expressed as the following,
\begin{eqnarray}
|a\rangle =a^i|e_i\rangle =a'^i|e_i'\rangle =|a'\rangle ,
\end{eqnarray}
where the coefficients $a^i$ and $a'^i$
\begin{eqnarray}
a'^{i}=\langle e'^i|a\rangle =\langle e'^i|e_j\rangle \otimes
\langle e^j|a\rangle =L^i_ja^j,
\\
a^{i}=\langle e^i|a\rangle =\langle e^i|e_j'\rangle \otimes \langle
e'^j|a\rangle ={L^{-1}}^i_ja'^j.
\end{eqnarray}
Thus the vector transformation has the form,
\begin{eqnarray}
|a'\rangle =\hat {L}|a\rangle .
\end{eqnarray}
Please note that vectors $|a\rangle $ and $|a'\langle $ are actually
one vector in different basis. Similarly for adjoint vector $\langle
a|$, we have
\begin{eqnarray}
\langle a|=a_i^*\langle e^i|=a'^*_i\langle e'^i|=\langle a'|,
\end{eqnarray}
where $a_i'^*=a_j^*{L^{-1}}_i^j$, $a_i^*=a'^*_jL_i^j$, and
\begin{eqnarray}
\langle a'|=\langle a|\hat {L}^{-1}.
\end{eqnarray}
The operator transformation has the following results,
\begin{eqnarray}
\hat {A}=A^i_j|e_i\rangle \otimes \langle e^j|=A'^i_j|e_i'\rangle
\otimes \langle e'^j|=\hat {A}'.
\end{eqnarray}
We can find that
\begin{eqnarray}
A'^i_j=L^i_kA^k_l{L^{-1}}^l_j, \nonumber \\
A^i_j={L^{-1}}^i_kA'^k_lL^l_j, \nonumber \\
\hat {A}'=\hat {L}\hat {A}\hat {L}^{-1}.
\end{eqnarray}

Next we will present some results concerning about the
transformation for some calculations,
\begin{eqnarray}
&&(\alpha |a\rangle +\beta |b\rangle )'=\alpha |a'\rangle +\beta
|b'\rangle ; \nonumber \\
&&(\alpha \langle a|+\beta \langle b|)'=\alpha \langle a'|+\beta \langle b'|;
\nonumber \\
&&\langle a'|b'\rangle = \langle a|b\rangle , \nonumber \\
&&(\alpha \hat {A}+\beta \hat {B})'=\alpha \hat {A}'+\beta \hat
{B}', \nonumber \\
&&(\hat {A}|a\rangle )'=\hat {A}'|a\rangle , \nonumber \\
&&(\langle a|\hat {A})'=\langle a'|\hat {A}', \nonumber \\
&&(\hat {A}\hat {B})'=\hat {A}'\hat {B}', \nonumber \\
&&[\hat {A},\hat {B}]'=[\hat {A}',\hat {B}'], \nonumber \\
&&\{ \hat {A},\hat {B}\} '=\{ \hat {A}',\hat {B}'\}, \nonumber \\
&&{\rm det}\hat {A}'={\rm det}\hat {A}, \nonumber \\
&&{\rm tr}\hat {A}'={\rm tr}\hat {A}, \nonumber \\
&&(\hat {A}^{-1})'=\hat {A}'^{-1}, \nonumber \\
&&[F(\hat {A})]'=F(\hat {A}'), \nonumber \\
&&\left( \overline{\hat {A}}\right) '= \overline{\hat {A}'},
\nonumber \\
&& \langle \hat {A}', \hat {B}'\rangle =\langle \hat {A}, \hat
{B}\rangle .
\end{eqnarray}
Those results can be easily proved, for example,
\begin{eqnarray}
&&\langle a'|b'\rangle =\langle a|\hat {L}^{-1}\hat {L}|b\rangle
=\langle a|b\rangle ,\nonumber \\
&&\overline{\hat {A}'}=\overline{\hat {L}^{-1}\hat {A}\hat {L}}=\hat
{L}\overline{\hat {A}}\hat {L}^{-1}=\left( \overline{\hat
{A}}\right) ',\nonumber \\
&&\langle \hat {A}',\hat {B}'\rangle ={\rm tr}(\overline{\hat
{A}'}\hat {B}')={\rm tr}[\hat {L}\overline{\hat {A}}\hat
{L}^{-1}\hat {L}\hat {B}\hat {L}^{-1}]={\rm tr}(\overline{\hat
{A}}\hat {B})=\langle \hat {A},\hat {B}\rangle .
\end{eqnarray}

\subsection{Direct sum and direct product}
The direct sum of the $m$-dimensional space $V_1$ and the
$n$-dimensional space $V_2$ is a $(m+n)$-dimensional space,
\begin{eqnarray}
V=V_1\oplus V_2.
\end{eqnarray}
Similarly the adjoint can have the same result,
\begin{eqnarray}
\overline{V}=\overline{V_1}\oplus \overline{V_2}.
\end{eqnarray}
For operator space $O_1=V_1\otimes \overline{V_1}$ and
$O_2=V_2\otimes \overline{V_2}$, the operator space $O=V\otimes
\overline{V}$ has the decomposition,
\begin{eqnarray}
O=O_1\oplus O_2.
\end{eqnarray}
The vector space $V_1$ has basis $\{ |e_i\rangle \}$, the metric is
denoted as $g_{ii'}$, $(i,i'=1,2,\cdots ,m)$, additionally vector
space $V_2$ has basis $\{ |E_j\rangle \}$, the metric is $G_{jj'}$,
$(j,j'=1,2,\cdots ,n)$. The basis for the direct sum space
$V=V_1\oplus V_2$ should be,
\begin{eqnarray}
\{ |\varepsilon _k\rangle \}=\{ |e_i\rangle \}\cup \{ |E_j\rangle
\}=\{ |\varepsilon _k\rangle ;|\varepsilon _i\rangle =|e_i\rangle ,
|\varepsilon _{m+j}\rangle \}=\{ |E_j\rangle \},
\end{eqnarray}
where $k=1,2,\cdots ,m, m+1, \cdots ,m+n$. The metric $\{ \eta
_{kk'}\} $ of direct sum space $V$ can be found to be,
\begin{eqnarray}
\eta _{kk'}=\{ g_{ii'}\} \oplus \{ G_{jj'}\} .
\end{eqnarray}
The basis of the adjoint space $\overline{V}$ can be written as,
\begin{eqnarray}
\langle \varepsilon ^{k'}|=\eta ^{k'k}\overline{|\varepsilon
_k\rangle }.
\end{eqnarray}
The operator basis of the operator space $O$ takes the form
$|\varepsilon ^{k'}\rangle \otimes \langle \varepsilon _k|$.

The direct product calculations for spaces and operators obey the
standard method, like the following,
\begin{eqnarray}
&&O=O_1\otimes O_2, \nonumber \\
&&\{ |\varepsilon _k\rangle \}=\{ |\varepsilon _k\rangle
;|\varepsilon _{n(i-1)+j}\rangle =|e_i\rangle \otimes |E_j\rangle
\}, \nonumber
\\
&&\eta _{kk'}=\{ g_{ii'}\} \otimes \{ G_{jj'}\} .
\end{eqnarray}

\subsection{Lie algebra}
We next present some basics of Lie algebra and some particular
concepts and calculations which have been used in this paper.

Suppose $A$ is a finite-dimensional linear space over $F$, and we
also defined the commutation calculations on $A$, then $A$ is a Lie
algebra. The dimension of the Lie algebra is the dimension of the
linear space, denoted as ${\rm dim }A$.

We can define three kind of calculations, plus, number
multiplication and commutation. This defined plus and the number
multiplication constitute the linear space. The commutator of two
elements of $A$ is denoted as $[\hat {X}, \hat {Y}]$. For arbitrary
$\hat {X},\hat {Y}, \hat {Z}\in A$ and arbitrary numbers $a,b\in F$,
the commutator calculations have the properties, respectively listed
are closed condition, linear and anti-commuting,
\begin{eqnarray}
&&[\hat {X},\hat {Y}]\in A, \nonumber \\
&&[a\hat {X}+b\hat {Y},\hat {Z}]=a[\hat {X},\hat {Z}]+b[\hat
{Y},\hat
{Z}], \nonumber \\
&&[\hat {X},\hat {Y}]=-[\hat {Y},\hat {X}].
\end{eqnarray}
The Jacobi equation takes the form,
\begin{eqnarray}
[\hat {X},[\hat {Y},\hat {Z}]]+[\hat {Y},[\hat {Z},\hat {X}]]+[\hat
{Z},[\hat {X},\hat {Y}]]=0.
\end{eqnarray}
If all elements of the Lie algebra are commuting, then $A$ is Abel
algebra.

$A_1$ is a subspace of Lie algebra $A$, and the commutation
calculation of elements in $A_1$ is closed, then $A_1$ is a
subalgebra of $A$.

If $A_1$ is a subalgebra of $A$, and for arbitrary $\hat {X}_1\in
A_1$ and $\hat {X}\in A$, we have $[\hat {X}_1,\hat {X}]\in A_1$,
then $A_1$ is called the ideal of $A$. Lie algebra always has two
ideals, $\{ 0\}$ and itself $A$.

If $A$ has no fixed ideal, it is called simple Lie algebra.

If there is no non-Abel ideal which is not $\{ 0\}$, this Lie
algebra is called semi-simple Lie algebra. The semi-simple Lie
algebra can be decomposed as the direct sum of simple Lie algebras.

Given two Lie algebras $A_1$ and $A_2$, suppose $A_1\cap {A}_2=\{
0\}$ and for arbitrary $\hat {X}_1\in A_1, \hat {X}_2\in A_2$, we
have $[\hat {X}_1,\hat {X}_2]$, the direct sum can thus take the
form,
\begin{eqnarray}
A=A_1\oplus A_2=\{ \hat {X}, \hat {X}=\hat {X}_1+\hat {X}_2, \hat
{X}_1\in A_1, \hat {X}_2\in A_2\} .
\end{eqnarray}
We can find that $A$ is a Lie algebra, the dimension of $A$ is the
sum of dimensions $A_1$ and $A_2$, ${\rm dim}A={\rm dim}A_1+{\rm
dim}A_2$, both $A_1$ and $A_2$ are ideals of Lie algebra $A$.

Suppose $A$ and $A'$ are two Lie algebras, $f$ maps $A$ to $A'$, and
for arbitrary $\hat {X}, \hat {Y}\in A$, there exist $f(\hat {X}),
f(\hat {Y})\in A'$ and satisfy
\begin{eqnarray}
&&f(a\hat {X}+b\hat {Y})=af(\hat {X})+bf(\hat {Y}), \nonumber \\
&&f([\hat {X},\hat {Y}])=[f(\hat {X}),f(\hat {Y})],
\end{eqnarray}
we mean $A$ and $A'$ are homomorphism, $f$ is a homomorphism map. If
there exists an one to one homomorphism map, $A$ and $A'$ are
isomorphism.

For a given Lie algebra $A$, we can introduce a complete set basis,
\begin{eqnarray}
\{ \hat {t}_a\}, a=1,2,\cdots, {\rm }A.
\end{eqnarray}
An arbitrary element $X$ can be expanded by this basis,
\begin{eqnarray}
\hat {X}=x^a\hat {t}_a,
\end{eqnarray}
where $x^a\in F$ are the coefficients in the expansion. The
commutation relation of the basis can be represented as,
\begin{eqnarray}
[\hat {t}_a, \hat {t}_b]=C_{ab}^c\hat {t}_c,
\end{eqnarray}
where $C_{ab}^c$ are the structure constants. There are altogether
$({\rm dim}A)^3$ structure constants. One can check that the
structure constants satisfy the relations,
\begin{eqnarray}
&&C_{ab}^c=C_{ba}^c, \nonumber \\
&&C_{ab}^dC_{cd}^e+C_{bc}^dC_{ad}^e+C_{ca}^dC_{bd}^e=0.
\end{eqnarray}
For a given basis, the structure constants are fixed. A given Lie
algebra $A$ can be defined by a set of basis $\{ \hat {t}_a\} $ and
the structure constants.

With the help of the structure constants, the Cartan metric of a Lie
algebra can be defined as,
\begin{eqnarray}
g_{ab}=g_{ba}=C_{ac}^dC_{bd}^c=C_{bd}^cC_{ac}^d.
\end{eqnarray}
A Lie algebra is semi-simple, the necessary and sufficient condition
is that the Cartan metric is non-degenerate, ${\rm det}\{ g_{ab}\}
\not =0$. For an Abel algebra, Cartan metric is always zero,
$g_{ab}=0$.

In our 4D unified quantum theory, we can define a generalized metric
$G_{ab}$ for Lie algebra.

\begin{enumerate}
\item For a simple algebra, the generalized metric $G_{ab}$ is
defined as,
\begin{eqnarray}
G_{ab}=G_{ba}=gg_{ab}=gC_{ac}^dC_{bd}^c,
\end{eqnarray}
where $g\in R,\not= 0$ is a non-zero real constant, $g_{ab}$ is the
Cartan metric of simple algebra.

\item For a semi-simple algebra, if it can be constitute as the
direct sum of $n$ semi-simple algebras, and the Cartan metrics for
each simple algebra are $g_{a_1b_1}, \cdots ,g_{a_nb_n}$, the
generalized metric is defined as,
\begin{eqnarray}
G_{a_1b_1}=g_1g_{a_1b_1},\nonumber \\
\cdots ~~~\cdots , \nonumber \\
G_{a_nb_n}=g_ng_{a_nb_n},
\end{eqnarray}
where $g_1, \cdots ,g_n\in R, \not= 0.$

\item For an Abel algebra, the generalized metric can be defined as,
\begin{eqnarray}
&&G_{ab}=G_{ba}, \nonumber \\
&&{\rm det}\{ G_{ab}\} \not= 0,
\end{eqnarray}
where $G_{ab}$ can be arbitrary real number.
\end{enumerate}

Thus, for arbitrary Lie algebra including semi-simple Lie algebra
and the Abel Lie algebra, the generalized metric is non-degenerate,
${\rm det}\{ G_{ab}\} \not= 0$. Thus we can introduce an inverse for
this generalized metric,
\begin{eqnarray}
&&G^{ab}G_{bc}=\delta ^a_c, \nonumber \\
&&G^{ab}=\frac {A^{ab}}{{\rm det}\{ G_{ab}\} },
\end{eqnarray}
where $A^{ab}$ is the algebraic complement of element $G_{ab}$.

We define the covariance structure constant as,
\begin{eqnarray}
C_{abc}=G_{ad}C_{bc}^d.
\end{eqnarray}
We can find that the covariance structure constants are completely
antisymmetric,
\begin{eqnarray}
C_{abc}=C_{bca}=C_{cab}=-C_{cba}=-C_{bac}=-C_{acb}.
\end{eqnarray}
The contra-variance algebraic basis is defined as,
\begin{eqnarray}
\hat {t}^a=G^{ab}\hat {t}_b.
\end{eqnarray}
The rank-$n$, $(n\ge 2)$, Casimir operators take the form,
\begin{eqnarray}
\hat {C}_n=C_{a_1b_1}^{b_2}C_{a_2b_2}^{b_3}\cdots
C_{a_nb_n}^{b_1}\hat {t}^{a_1}\hat {t}^{a_2}\cdots \hat {t}^{a_n}.
\end{eqnarray}
In particular,for $n=2$, the Casimir operator is,
\begin{eqnarray}
\hat {C}_2=C_{a_1b_1}^{b_2}C_{a_2b_2}^{b_1}\hat {t}^{a_1}\hat
{t}^{a_2}.
\end{eqnarray}
One can prove that,
\begin{eqnarray}
[\hat {C}_n, \hat {t}_a]=0.
\end{eqnarray}
The number of independent Casimir operators is equal to the rank of
this Lie algebra.

\subsection{Vectors and operators associated with Lie algebra}
Lie algebra $A$ is defined as a linear space, it can be considered
as 1-form Lie algebra tensor space,
\begin{eqnarray}
T(1)=A.
\end{eqnarray}
The tensor basis is $\{ \hat {t}_a\}$ in this apace.

We then can define the direct product space of $r$ Lie algebra
spaces $A$ as the $r$-form Lie algebra tensor space,
\begin{eqnarray}
T(r)=\prod _r\otimes A.
\end{eqnarray}
The tensor basis takes the form,
\begin{eqnarray}
\hat {t}_{a_1\cdots a_r}=\hat {t}_{a_1}\otimes \cdots \otimes \hat
{t}_{a_r}
\end{eqnarray}
The direct sum of all $r$-form$(r=0,1,\cdots )$ spaces constitute
the Lie algebra tensor space,
\begin{eqnarray}
T=\sum _{r=0}^{\infty }\oplus T(r).
\end{eqnarray}
We next present several examples of the Lie algebra tensor:

A constant $\alpha $ can be considered as the 0-form Lie algebra
tensor.

An element $\hat {X}$ of Lie algebra can be considered as the form-1
Lie algebra tensor, $\hat {X}=x^a\hat {t}_a$.

The metric of a Lie algebra can be considered as the form-2 Lie
algebra tensor,
\begin{eqnarray}
\hat {G}=G^{ab}\hat {t}_a\otimes \hat {t}_b=G_{ab}\hat {t}^a\otimes
\hat {t}^b.
\end{eqnarray}
As we already known, the rising and lowering of the indices can be
realized by the metric matrix.

Structure constant can be considered as the type-3 Lie algebra
tensor,
\begin{eqnarray}
\hat {C}=C^c_{ab}\hat {t}^a\otimes \hat {t}^b\otimes \hat {t}_c.
\end{eqnarray}

The Lie algebra tensor is similar as the convention tensor, and we
can define calculations like plus, number multiplication, tensor
product, contraction. Additionally, we can define the commuting
calculation for the Lie algebra tensor the convention tensor. The
commutation relation for $\hat {t}_a$ and $\hat {t}_{a_1\cdots a_r}$
is,
\begin{eqnarray}
[\hat {t}_a,\hat {t}_{a_1\cdots a_r}]=C^c_{aa_1}\hat {t}_{c\cdots
a_r}+\cdots +C^c_{aa_r}\hat {t}_{a_1\cdots c}.
\end{eqnarray}

Two different basis of the Lie algebra $A$ can be transformed to
each other as,
\begin{eqnarray}
&&\hat {t}_a=\Lambda _a^b\hat {t}_b', \nonumber \\
&&\hat {t}_a'={\Lambda ^{-1}}^b_a\hat {t}_b.
\end{eqnarray}
By the transformation between two different basis, the Lie algebra
expansion coefficients, structure constants and metric changed as
the conventional tensor transformations,
\begin{eqnarray}
\hat {X}&=&x^a\hat {t}_a=x'^a\hat {t}'_a, \nonumber \\
x'^a&=&\Lambda ^a_bx^b, \nonumber \\
C'^c_{ab}&=&{\Lambda ^{-1}}^d_a{\Lambda ^{-1}}^e_b\Lambda
^c_fC^f_{de}, \nonumber \\
G'_{ab}&=&{\Lambda ^{-1}}^c_a{\Lambda ^{-1}}^c_bG_{cd}, \nonumber \\
G'^{ab}&=&\Lambda ^a_c\Lambda ^b_dG^{cd}.
\end{eqnarray}
Under this transformation, the calculations of plus, number
multiplication and commuting are invariant. The definition forms of
metric and the Casimir operator are invariant. That means that the
calculations of Lie algebra and the properties are independent of
the choice of basis.

\subsection{The representation theory of Lie algebra}
Suppose $M$ is a set of $n\times n$ matrices on $F$, with the
definition of the matrix plus and number multiplication, $M$
constitute a $n^2$ dimension linear space. For $\hat {X}, \hat
{Y}\in M$, the matrices commuting calculation can be defined as,
\begin{eqnarray}
[\hat {X}, \hat {Y}]=\hat {X}\hat {Y}-\hat {Y}\hat {X}.
\end{eqnarray}
The commutation calculation defined above satisfy the conditions
like closed, linear, antisymmetric and Jacobi relation, and thus $M$
is a $n^2$-dimension Lie algebra. It can be named as the matrix
algebra.

Given a Lie algebra $A$, if we can find a homomorphism map $f$ from
$A$ to $n\times n$ matrix algebra $M$, this map $f$ can be
considered as a linear representation of $A$ if additionally the
following conditions are satisfied: consider for arbitrary $\hat
{X}, \hat {Y}\in A$, there exist $f(\hat {X}),f(\hat {Y})\in M$, and
\begin{eqnarray}
f(a\hat {X}+b\hat {Y})&=&af(\hat {X})+bf(\hat {Y}), \nonumber \\
f([\hat {X},\hat {Y}])&=&[f(\hat {X}),f(\hat {Y})].
\end{eqnarray}
$f(\hat {X})$ is the matrix representation of $\hat {X}$. The linear
space $V$ of the matrix algebra $M$ is the representation space of
Lie algebra $A$. The dimension $n$ of space $V$ is the
representation dimension of Lie algebra $A$. The basis of $V$ is the
representation basis. Lie algebra $A$ can be denoted as $V(f)$ or
simply $V$.

In order to give Lie algebra $A$ a representation $V(f)$, we need
three conditions:

(1). There is a representation space $V$.

(2). The representation matrix $f(\hat {X})$ can be given.

(3). The representation can be found to be a homomorphism.

Suppose $V(f)$ is a representation of Lie algebra $A$, for arbitrary
different two elements of $A$, $M$ has two different corresponding
matrices, this representation is called faithful. Otherwise it is
not a faithful representation.

Suppose $V_1(f_1)$ and $V_1(f_1)$ are two representations of Lie
algebra $A$, the direct sum of those two representations can be
defined as: First, the representation space $V(f)$ of the direct sum
is the direct sum of the two representation spaces of $V_1(f_1)$ and
$V_2(f_2)$,
\begin{eqnarray}
V=V_1\oplus V_2.
\end{eqnarray}

Correspondingly, the dimension of the direct sum representation
space is the sum of the two dimensions $V_1(f_1)$ and $V_2(f_2)$,
\begin{eqnarray}
{\rm dim}V(f)={\rm dim }V_1(f_1)+{\rm dim }V_2(f_2).
\end{eqnarray}
The representation matrix is the direct sum of two representation
matrices,
\begin{eqnarray}
f(\hat {X})=f_1(\hat {X})\oplus f_2(\hat {X}).
\end{eqnarray}
The homomorphism of the representation $V(f)$ can be proved in
following,
\begin{eqnarray}
[f(\hat {X}),f(\hat {Y})]&=&f(\hat {X})f(\hat {Y})-f(\hat {Y})f(\hat
{X})
\nonumber \\
&=&\left( f_1(\hat {X})\oplus f_2(\hat {X})\right)\left( f_1(\hat
{Y})\oplus f_2(\hat {Y})\right) -\left( f_1(\hat {Y})\oplus f_2(\hat
{Y})\right) \left( f_1(\hat {X})\oplus f_2(\hat {X})\right)
\nonumber \\
&=& \left( f_1(\hat {X})f_1(\hat {Y})\right)\oplus \left( f_2(\hat
{X}) f_2(\hat {Y})\right) -\left( f_1(\hat {Y})f_1(\hat
{X})\right)\oplus \left( f_2(\hat {Y}) f_2(\hat {X})\right)
\nonumber \\
&=&[f_1(\hat {X}),f_1(\hat {Y})]\oplus [f_2(\hat {X}), f_2(\hat
{Y})]
\nonumber \\
&=&f_1[\hat {X},\hat {Y}]\oplus f_2[\hat {X},\hat {Y}]
\nonumber \\
&=&f[\hat {X},\hat {Y}].
\end{eqnarray}
Here we have used the following relations,
\begin{eqnarray}
&&[f_1(\hat {X}),f_1(\hat {Y})]=f_1[\hat {X},\hat {Y}],
\nonumber \\
&&[f_2(\hat {X}),f_2(\hat {Y})]=f_2[\hat {X},\hat {Y}],
\nonumber \\
&&V(f)=V_1(f_1)\oplus V_2(f_2).
\end{eqnarray}
If a representation can not be expressed as the direct sum of two
representations, this representation is a irreducible
representation, otherwise it is a reducible representation. If all
irreducible representations are given for a Lie algebra, it is
equivalent that all representation of this Lie algebra are given.

Suppose $V_1(f_1)$ and $V_2(f_2)$ are two representations of Lie
algebra $A$, the direct product $V(f)$ can be defined as follows:

The direct product representation space is the direct product of two
representation spaces of $V_1(f_1)$ and $V_2(f_2)$,
\begin{eqnarray}
V=V_1\otimes V_2.
\end{eqnarray}
The dimension of the multiplication of two dimensions,
\begin{eqnarray}
{\rm dim}V(f)={\rm dim}V_1(f_1)\times {\rm dim}V_2(f_2).
\end{eqnarray}
The matrix representation is defined as,
\begin{eqnarray}
f(\hat {X})=f_1(\hat {X})\otimes \hat {I}_2+\hat {I}_1\otimes
f_2(\hat {X}),
\end{eqnarray}
where $\hat {I}_1$ and $\hat {I}_2$ are identity operators on space
$V_1$ and $V_2$, respectively.

The homomorphism of the direct product representation can be proved
as follows,
\begin{eqnarray}
[f(\hat {X}),f(\hat {Y})]&=&f(\hat {X})f(\hat {Y})-f(\hat {Y})f(\hat
{X})
\nonumber \\
&=&\left( f_1(\hat {X})\otimes \hat {I}_2+\hat {I}_1\otimes f_2(\hat
{X})\right) \left( f_1(\hat {Y})\otimes \hat {I}_2+\hat {I}_1\otimes
f_2(\hat {Y})\right)
\nonumber \\
&&- \left( f_1(\hat {Y})\otimes \hat {I}_2+\hat {I}_1\otimes
f_2(\hat {Y})\right) \left( f_1(\hat {X})\otimes \hat {I}_2+\hat
{I}_1\otimes f_2(\hat {X})\right)
\nonumber \\
&=&\left( [f_1(\hat {X})f_1(\hat {Y})]\otimes \hat {I}_2+f_1(\hat
{X})\otimes f_2(\hat {Y}) +f_1(\hat {Y})\otimes f_2(\hat {X})+\hat
{I}_1\otimes [f_2(\hat {X})f_2(\hat {Y})] \right)
\nonumber \\
&&-\left( [f_1(\hat {Y})f_1(\hat {X})]\otimes \hat {I}_2-f_1(\hat
{Y})\otimes f_2(\hat {X}) -f_1(\hat {X})\otimes f_2(\hat {Y})-\hat
{I}_1\otimes [f_2(\hat {Y})f_2(\hat {X})] \right)
\nonumber \\
&=&[f_1(\hat {X}),f_1(\hat {Y})]\otimes \hat {I}_2+\hat {I}_1\otimes
[f_2(\hat {X}),f_2(\hat {Y})]
\nonumber \\
&=&f_1[\hat {X},\hat {Y}]\otimes \hat {I}_2+\hat {I}_1\otimes
f_2[\hat {X},\hat {Y}]
\nonumber \\
&=&f[\hat {X},\hat {Y}],
\end{eqnarray}
where we have used,
\begin{eqnarray}
&&[f_1(\hat {X}),f_1(\hat {Y})]=f_1[\hat {X},\hat {Y}], \nonumber \\
&&[f_2(\hat {X}),f_2(\hat {Y})]=f_2[\hat {X},\hat {Y}], \nonumber \\
&&V(f)=V_1(f_1)\otimes V_2(f_2).
\end{eqnarray}

Suppose a Lie algebra is a direct sum of two Lie algebras
$A=A_1\oplus A_2$, $V_1(f_1)$ $V_2(f_2)$ are representations of Lie
algebra $A_1$ and $A_2$, respectively,
\begin{eqnarray}
V(f)=V_1(f_1)\otimes V_2(f_2).
\end{eqnarray}
$V(f)$ is the representation of Lie algebra $A$, the dimension is
\begin{eqnarray}
{\rm dim}V(f)={\rm dim}V_1(f_1)\times {\rm dim}V_2(f_2).
\end{eqnarray}
Suppose $\hat {X}_1\in A_1$ and has the matrix representation
$f_1(\hat {X}_1)$, similarly $\hat {X}_2\in A_2$, the matrix
representation is $f_2(\hat {X}_2)$, we can find that $\hat {X}=\hat
{X}_1+\hat {X}_2\in A$, the matrix representation takes the form,
\begin{eqnarray}
f(\hat {X})=f_1(\hat {X}_1)\otimes \hat {I}_2+\hat {I}\otimes
f_2(\hat {X}_2),
\end{eqnarray}
where $\hat {I}_1$ and $\hat {I}_2$ are identity operators of spaces
$V_1$ and $V_2$, respectively. The homomorphism of this
representation can be proved in the following,
\begin{eqnarray}
[f(\hat {X}),f(\hat {Y})]&=&f(\hat {X})f(\hat {Y})-f(\hat {Y})f(\hat
{X})
\nonumber \\
&=&\left( f_1(\hat {X}_1)\otimes \hat {I}_2+\hat {I}_1\otimes
f_2(\hat {X}_2)\right) \left( f_1(\hat {Y_1})\otimes \hat {I}_2+\hat
{I}_1\otimes f_2(\hat {Y_2})\right)
\nonumber \\
&&- \left( f_1(\hat {Y}_1)\otimes \hat {I}_2+\hat {I}_1\otimes
f_2(\hat {Y}_2)\right) \left( f_1(\hat {X}_1)\otimes \hat {I}_2+\hat
{I}_1\otimes f_2(\hat {X}_2)\right)
\nonumber \\
&=&[f_1(\hat {X}_1)f_1(\hat {Y}_1)]\otimes \hat {I}_2+f_1(\hat
{Y}_1)\otimes f_2(\hat {X}_2) +f_1(\hat {X}_1)\otimes f_2(\hat
{Y}_2)+\hat {I}_1\otimes [f_2(\hat {X}_2)f_2(\hat {Y}_2)]
\nonumber \\
&&-[f_1(\hat {Y}_1)f_1(\hat {X}_1)]\otimes \hat {I}_2-f_1(\hat
{Y}_1)\otimes f_2(\hat {X}_2) -f_1(\hat {X}_1)\otimes f_2(\hat
{Y}_2)-\hat {I}_1\otimes [f_2(\hat {Y}_2)f_2(\hat {X}_2)]
\nonumber \\
&=&[f_1(\hat {X}_1),f_1(\hat {Y}_1)]\otimes \hat {I}_2+\hat
{I}_1\otimes [f_2(\hat {X}_2),f_2(\hat {Y}_2)]
\nonumber \\
&=&f_1[\hat {X}_1,\hat {Y}_1]\otimes \hat {I}_2+\hat {I}_1\otimes
f_2[\hat {X}_2,\hat {Y}_2]
\nonumber \\
&=&f[\hat {X},\hat {Y}],
\end{eqnarray}
where
\begin{eqnarray}
&&[f_1(\hat {X}_1),f_1(\hat {Y}_1)]=f_1[\hat {X}_1,\hat {Y}_1], \nonumber \\
&&[f_2(\hat {X}_2),f_2(\hat {Y}_2)]=f_2[\hat {X}_2,\hat {Y}_2].
\end{eqnarray}

\subsection{Vector representation and operator representation}
Suppose $A$ is a Lie algebra, the algebraic basis is $\{ \hat
{t}_a\}$, the structure constant is $C_{ab}^c$, consider that $V(f)$
is a representation of $A$, $V$ can be considered as a vector space,
and thus $V(f)$ can always be considered as a vector representation.
We can introduce a set of vectors as a basis of space $V$,
\begin{eqnarray}
\{|e_i\rangle \}, i=1,2,\cdots ,{\rm dim}V.
\end{eqnarray}
With this vector basis, the algebraic basis have the corresponding
matrix representations,
\begin{eqnarray}
f(\hat {t}_a)&=&(\hat {t}_a)^i_j|e_i\rangle \otimes \langle e^j|,
\nonumber \\
(\hat {t}_a)^i_j&=&\langle e^i|\hat {t}_a|e_j\rangle .
\end{eqnarray}
The commutation relations can be expressed as,
\begin{eqnarray}
&&[f(\hat {t}_a),f(\hat {t}_b)]=C_{ab}^cf(\hat {t}_c),
\nonumber \\
&&(\hat {t}_a)^i_j(\hat {t}_b)^j_k-(\hat {t}_b)^i_j(\hat
{t}_a)^j_k=C_{ab}^c(\hat {t}_c)^i_k,
\end{eqnarray}
The relation between the algebraic basis and the vector basis takes
the form,
\begin{eqnarray}
\hat {t}_a|e_i\rangle =(\hat {t}_a)^j_i|e_j\rangle .
\end{eqnarray}

We can have the adjoint vector representation $\overline
{V}(\overline {f})$. The adjoint basis in space $\overline {V}$
takes the form,
\begin{eqnarray}
\{ \langle e^i|\} , i=1,2,\cdots ,{\rm dim}V .
\end{eqnarray}
In the adjoint basis, the algebraic basis has a matrix
representation as,
\begin{eqnarray}
\overline {f}(\hat {t}_a)&=&(\hat {t}_a)^i_j|e_i\rangle \otimes
\langle e^j|,
\nonumber \\
(\hat {t}_a)^i_j&=&\langle e^i|\hat {t}_a|e_j\rangle .
\end{eqnarray}
The commutation relation takes the form,
\begin{eqnarray}
&&[\overline {f}(\hat {t}_a), \overline {f}(\hat
{t}_b)]=C_{ab}^c\overline {f}(\hat {t}_c),
\nonumber \\
&&(\hat {t}_a)^i_j(\hat {t}_b)^j_k-(\hat {t}_b)^i_j(\hat
{t}_a)^j_k=C_{ab}^c(\hat {t}_c)^i_k.
\end{eqnarray}
The algebraic basis and the vector basis has a relation,
\begin{eqnarray}
\langle e^i|\hat {t}_a=\langle e^j|(\hat {t}_a)^i_j.
\end{eqnarray}
We can find that actually the vector representation matrix $f(\hat
{t}_a)$ and the adjoint vector representation matrix $\overline
{f}(\hat {t}_a)$ are actually the same. For simplicity, we may
directly denote he matrix representations and the operator by the
same notations,
\begin{eqnarray}
f(\hat {t}_a)=\overline {f}(\hat {t}_a)=\hat {t}_a=(\hat
{t}_a)^i_j|e_i\rangle \otimes \langle e^j|.
\end{eqnarray}

The operator representation $\hat {O}(\hat {f})$ is in the
representation space $\hat {O}=V\otimes \overline {V}$ constituted
by the vector representation space $V$ and the adjoint vector
representation space $\overline {V}$. The basis of the operator
space is denoted as,
\begin{eqnarray}
\{ |e_i\rangle \otimes \langle e^j|\} , ~~~i,j=1,2,\cdots ,{\rm dim
}V.
\end{eqnarray}
For operator representation $\hat {O}(\hat {f})$, the action of
algebraic basis representation $\hat {f}(\hat {t}_a)$ is defined by
the following commutation relation,
\begin{eqnarray}
\hat {f}(\hat {t}_a)|e_i\rangle \otimes \langle e^j|&=&[\hat {t}_a,
|e_i\rangle \otimes \langle e^j|]
\nonumber \\
&=&\hat {t}_a|e_i\rangle \otimes \langle e^j|-|e_i\rangle \otimes
\langle e^j|\hat {t}_a
\nonumber \\
&=&(\hat {t}_a)^k_i|e_k\rangle \otimes \langle e^j|-|e_i\rangle
\otimes \langle e^l|(\hat {t}_a)^j_l
\nonumber \\
&=&[(\hat {t}_a)^k_i\delta ^j_l-\delta ^k_i(\hat
{t}_a)^j_l]|e_k\rangle \otimes \langle e^l|.
\end{eqnarray}
So we can find the algebraic basis matrix representation takes the
form,
\begin{eqnarray}
\hat {f}(\hat {t}_a)&=&[(\hat {t}_a)^k_i\delta ^j_l-\delta ^k_i(\hat
{t}_a)^j_l]\left[ |e_k\rangle \otimes \langle e^l|\right] \otimes
\left[ |e_j\rangle \otimes \langle e^i|\right]
\nonumber \\
&=&\hat {t}_a\otimes \hat {I}-\hat {I}\otimes \hat {t}_a^T,
\end{eqnarray}
where the upper indices $^T$ means the matrix transposition. Note
that for matrix representation, the operator representation and the
basis representation is not the same.

With help of the Jacobi relation deduced from the commutation
relation, we can find that,
\begin{eqnarray}
[\hat {t}_a,[\hat {t}_b, |e_i\rangle \otimes \langle e^j|]]-[\hat
{t}_b,[\hat {t}_a, |e_i\rangle \otimes \langle e^j|]]&=&[[\hat
{t}_a,\hat {t}_b], |e_i\rangle \otimes \langle e^j|]
\nonumber \\
&=&C_{ab}^c[\hat {t}_c,|e_k\rangle \otimes \langle e^l|].
\end{eqnarray}
Thus we can find that $\hat {f}(\hat {t}_a)$ satisfy the
homomorphism condition,
\begin{eqnarray}
\hat {f}(\hat {t}_a)\hat {f}(\hat {t}_b)-\hat {f}(\hat {t}_b)\hat
{f}(\hat {t}_a)&=&
[\hat {f}(\hat {t}_a),\hat {f}(\hat {t}_b)] \nonumber \\
&=&\hat {f}[\hat {t}_a,\hat {t}_b] \nonumber \\
&=&C_{ab}^c\hat {f}(\hat {t}_a).
\end{eqnarray}
The homomorphism can be proved directly by the matrix
representation,
\begin{eqnarray}
[\hat {f}(\hat {t}_a),\hat {f}(\hat {t}_b)]&=&\hat {f}(\hat
{t}_a)\hat {f}(\hat {t}_b)-\hat {f}(\hat {t}_b)\hat {f}(\hat {t}_a)
\nonumber \\
&=&(\hat {t}_a\otimes I-I\otimes \hat {t}_a^T)(\hat {t}_b\otimes
I-I\otimes \hat {t}_b^T)- (\hat {t}_b\otimes I-I\otimes \hat
{t}_b^T)(\hat {t}_a\otimes I-I\otimes \hat {t}_a^T)
\nonumber \\
&=&(\hat {t}_a\hat {t}_b)\otimes I-\hat {t}_b\otimes \hat {t}_a^T-
\hat {t}_a\otimes \hat {t}_b^T+I\otimes (\hat {t}_a^T\hat {t}_b^T)
\nonumber \\
&&-(\hat {t}_b\hat {t}_a)\otimes I+\hat {t}_a\otimes \hat {t}_b^T+
\hat {t}_b\otimes \hat {t}_a^T-I\otimes (\hat {t}_b^T\hat {t}_a^T)
\nonumber \\
&=&[\hat {t}_a, \hat {t}_b]\otimes \hat {I}+\hat {I}\otimes [\hat
{t}_a^T, \hat {t}_b^T]
\nonumber \\
&=&[\hat {t}_a, \hat {t}_b]\otimes \hat {I}-\hat {I}\otimes [\hat
{t}_a, \hat {t}_b]^T
\nonumber \\
&=&\hat {f}[\hat {t}_a, \hat {t}_b]
\nonumber \\
&=&C_{ab}^c\hat {f}(\hat {t}_c).
\end{eqnarray}

The adjoint representation space of Lie algebra is the Lie algebra
space itself,
\begin{eqnarray}
V=A.
\end{eqnarray}
The basis of the adjoint representation is the algebraic basis $\hat
{t}_b$,
\begin{eqnarray}
[\hat {t}_a,\hat {t}_b]=C_{ab}^c\hat {t}_c.
\end{eqnarray}
The matrix representation of the adjoint representation is the
structure constant,
\begin{eqnarray}
(\hat {t}_a)^c_b&=&C_{ab}^c, \nonumber \\
\hat {t}_a&=&\{ (\hat {t}_a)_b^c\} .
\end{eqnarray}
Since the structure constants satisfy the Jacobi relation, we can
find the adjoint representation satisfy the homomorphism condition,
\begin{eqnarray}
(\hat {t}_a)^d_e(\hat {t}_b)^e_c-(\hat {t}_b)^d_e(\hat
{t}_a)^e_c&=&C_{ae}^dC_{bc}^e-C_{be}^dC_{ac}^e
\nonumber \\
&=&C_{ab}^eC_{ec}^d
\nonumber \\
&=&C_{ab}^e(\hat {t}_e)^d_c.
\end{eqnarray}

The direct product of $r$ Lie algebra representation space of $A$
constitute $r$-formtensor space,
\begin{eqnarray}
T^{(r)}=\prod _r\otimes A.
\end{eqnarray}
The tensor basis of $T^{(r)}$ takes the form,
\begin{eqnarray}
\hat {t}_{a_1\cdots a_r}=\hat {t}_{a_1}\otimes \cdots \otimes \hat
{t}_{a_r}
\end{eqnarray}
The action of $\hat {t}_a$ on $\hat {t}_{a_1\cdots a_r}$ is defined
b the commutation relation,
\begin{eqnarray}
[\hat {t}_a, \hat {t}_{a_1\cdots a_r}]=C^c_{aa_1}\hat {t}_{c\cdots
a_r}+\cdots +C_{aa_r}^c \hat {t}_{a_1\cdots c} .
\end{eqnarray}
The representation matrix of $\hat {t}_a$ by tensor representation,
\begin{eqnarray}
\hat {t}_a=\{ (\hat {t}_a)_b^c\} \otimes \hat {I}_2\otimes \cdots
\otimes \hat {I}_r +\hat {I}_1\otimes \{ (\hat {t}_a)_b^c\} \otimes
\cdots \otimes \hat {I}_r+\cdots + \hat {I}_1\otimes \hat
{I}_2\otimes \cdots \otimes \{ (\hat {t}_a)_b^c\},
\end{eqnarray}
where $\hat {I}_1,\hat {I}_2, \cdots \hat {I}_r$ are ${\rm
dim}A\times {\rm dim}A$ identity matrices.

The relationships between Lie group and Lie algebra can be found in
textbooks of group theory, generally speaking, the behaviors of Lie
group $G$ near the unit identity $e$ is described by the
corresponding Lie algebra,
\begin{eqnarray}
A\rightarrow G:g=\exp (i\hat {X}), ~~~~\hat {X}\in A, g\in G.
\end{eqnarray}

\subsection{The Lie algebra used in 4D unified quantum theory}
In 4D unified quantum theory, we have used three Lie algebras, the
coordinate-momentum algebra $A_{xp}$, the spin algebra $A_S$ and the
gauge algebra $A_g$

The coordinate-momentum algebra $A_{xp}$ is a 9-dimension Lie
algebra, the algebraic basis is constituted by 4D coordinate $\hat
{x}^{\mu }$, 4D momentum $\hat {p}_{\nu }$ and the identity. The
coordinate-momentum algebra is a very special algebra, its not Abel,
cannot be represented as the direct sum of simple Lie algebras.

The spin algebra $A_S$ is a 6D Lie algebra. Its algebraic basis is
the 6 spin elements $\hat {s}_{\alpha \beta }$, it has the structure
of Lie algebra $D_2$. The spin algebra is a semi-simple algebra and
can be represented as a direct sum of two algebras $A_1$,
\begin{eqnarray}
A_S=A_1\oplus A_2.
\end{eqnarray}

The gauge algebra $A_g$ is a 12D Lie algebra. Its algebraic basis is
constituted by hypercharge $\hat {Y}$, isospin charge $\hat {I}_i$
and color charge $\hat {\lambda }_p$. The hypercharge $\hat {Y}$ is
the basis for a $u(1)$ algebra, three isospin charges $\hat {I}_i$
provide a basis for $su(2)$ algebra, eight color charges $\hat
{\lambda }_p$ constitute a basis for $su(3)$ algebra. The gauge
algebra is the direct sum of $u(1), su(2)$ and $su(3)$,
\begin{eqnarray}
A_g=u(1)\oplus su(2)\oplus su(3).
\end{eqnarray}

\subsection{4D $\delta $-function and convolution}
The 4D $\delta $-function is defined as,
\begin{eqnarray}
\delta ^4(x)=\delta (x^0)\delta (x^1)\delta (x^2)\delta (x^3).
\end{eqnarray}
Its properties have the following,
\begin{eqnarray}
\int \Psi (x')\delta ^4(x-x')d^4x'&=&\Psi (x)
\nonumber \\
\int \delta ^4(x)d^4x&=&1,
\nonumber \\
\delta ^4(-x)&=&\delta (x)
\end{eqnarray}

The definition of the partial differential of the $\delta $-function
and its properties can be found as,
\begin{eqnarray}
\delta ^{(1)}_{\mu }(x-x')&=&\frac {\partial }{\partial x'^{\mu
}}\delta (x-x'),
\nonumber \\
\int \Psi (x')\frac {\partial }{\partial x'^{\mu }}\delta
^4(x-x')d^4x'&=&-\frac {\partial }{\partial x^{\mu }}
\Psi (x), \nonumber \\
\int \frac {\partial }{\partial x^{\mu }}\delta ^4(x)d^4x&=&0, \nonumber \\
\frac {\partial }{\partial x^{\mu }}\delta ^4(-x)&=&-\frac {\partial
}{\partial x^{\mu }}\delta ^4(x)
\end{eqnarray}

We also have,
\begin{eqnarray}
\exp (ip\cdot x)&=&\exp (ip_{\mu }x^{\mu }),
\nonumber \\
\int \exp (ip\cdot x)d^4p&=&(2\pi )^4\delta ^4(x),
\nonumber \\
\int \exp (ip\cdot x)d^4x&=&(2\pi )^4\delta ^4(p).
\end{eqnarray}
The 4D Fourier transformations take the form
\begin{eqnarray}
\widetilde {\Phi }(p)&=&(2\pi )^{-2}\int \Phi (x)\exp (ip\cdot
x)d^4x,
\nonumber \\
\Phi (x)&=&(2\pi )^{-2}\int \widetilde {\Phi }(p)\exp (-ip\cdot
x)d^4p.
\end{eqnarray}
The derivative of the Fourier transformation and its multiplication
can be written as,
\begin{eqnarray}
\frac {\partial }{\partial p_{\mu }}\widetilde {\Phi }(p)&=&(2\pi
)^{-2}\int ix^{\mu }\Phi (x)
\exp (ip\cdot x)d^4x,\nonumber \\
\frac {\partial }{\partial x^{\mu }}\Phi (x)&=&(2\pi )^{-2}\int
-ip_{\mu }
\widetilde {\Phi }(p)\exp (-ip\cdot x)d^4p, \nonumber \\
p_{\mu }\widetilde {\Phi }(p)&=&(2\pi )^{-2}\int i\frac {\partial
}{\partial x^{\mu }}\Phi (x)
\exp (ip\cdot x)d^4x, \nonumber \\
x^{\mu }\Phi (x)&=&(2\pi )^{-2}\int -i\frac {\partial }{\partial
p_{\mu }}\widetilde {\Phi }(p) \exp (-ip\cdot x)d^4p.
\end{eqnarray}

The Fourier transformations of the $\delta $-function take the
following form,
\begin{eqnarray}
\exp (ip\cdot x')&=&\int \delta ^4(x-x')\exp (ip\cdot x)d^4x,
\nonumber \\
\delta ^4(x-x')&=&(2\pi )^{-4}\int \exp (ip\cdot x')\exp (-ip\cdot
x)d^4p,
\nonumber \\
\delta ^4(p-p')&=&(2\pi )^{-4}\int \exp (-ip'\cdot x)\exp (ip\cdot
x)d^4x,
\nonumber \\
\exp (-ip'\cdot x)&=&\int \delta ^4(p-p')\exp (-ip\cdot x)d^4p.
\end{eqnarray}

The 4D convolution is defined as,
\begin{eqnarray}
F(p)*G(p)=\int F(p')G(p-p')d^4p'.
\end{eqnarray}
The properties of the 4D convolution are listed as,
\begin{eqnarray}
F(p)*G(p)&=&G(p)*F(p)
\nonumber \\
F(p)*[G(p)*H(p)]&=&[G(p)*F(p)]*H(p)
\nonumber \\
F(p)*[G(p)+H(p)]&=&F(p)*G(p)+F(p)*H(p)
\end{eqnarray}
Suppose $F(p)*G(p)=K(p)$, we can find that,
\begin{eqnarray}
F(p-p')*G(p)=F(p)*G(p-p')=K(p-p').
\end{eqnarray}
The derivative of the convolution is,
\begin{eqnarray}
\frac {\partial }{\partial p_{\mu }}[F(p)*G(p)]= \left[ \frac
{\partial }{\partial p_{\mu }}F(p)\right] *G(p) =F(p)*\left[ \frac
{\partial }{\partial p_{\mu }}G(p)\right] .
\end{eqnarray}
The convolution of the 4D $\delta $-function take the form,
\begin{eqnarray}
F(p)*\delta (p)&=&F(p),
\\
F(p)*\frac {\partial }{\partial p_{\mu }}\delta (p)&=& \frac
{\partial }{\partial p_{\mu }}F(p).
\end{eqnarray}

The Fourier transformation of the convolution,
\begin{eqnarray}
F(x)G(x)&=&(2\pi )^{-2}\int _{R^4}\left[ \widetilde
{F}(p)*\widetilde {G}(p)\right] \exp (-ip\cdot x)d^4p,
\nonumber \\
\widetilde {F}(p)*\widetilde {G}(p)&=&(2\pi )^{-2}\int _{R^4}\left[
F(x)G(x)\right] \exp (ip\cdot x)d^4x,
\nonumber \\
F(x)*G(x)&=&(2\pi )^{-2}\int _{R^4}\left[ \widetilde
{F}(p)\widetilde {G}(p)\right] \exp (-ip\cdot x)d^4p,
\nonumber \\
\widetilde {F}(p)\widetilde {G}(p)&=&(2\pi )^{-2}\int _{R^4}\left[
F(x)*G(x)\right] \exp (ip\cdot x)d^4x.
\end{eqnarray}


\begin{thebibliography}{99}
\bibitem{einstein1}Einstein, A. ``Zur Elektrodynamik bewegter
K\"orper'', Annalen der Physik {\bf 17}, 891 (1905).

\bibitem{einstein2}Einstein, A.  ``Die Grundlage der allgemeinen Relativit\"atstheorie '',
Annalen der Physik {\bf 49}, 769 (1916).

\bibitem{dirac}Dirac, P. A. M. ``The Quantum Theory of the Electron'',
Proc. Roy. Soc. A {\bf 117}, 610 (1928).

\bibitem{schrodinger}Schr\"odinger, E. ``Quantisierung als
Eigenwertproblem'', Annalen der Physik {\bf 79}, 361 (1926).














\bibitem{QFT1}Feynman, R. P. ``Space-Time approach to Quantum
Electrodynamics'', Phys. Rev. {\bf 76}, 769 (1949).

\bibitem{QFT2}Schwinger, J. S. ``Quantum Electrodynamics. III: The Electromagnetic Properties
of the Electron: Radiative Corrections to Scattering'', Phys. Rev.
{\bf 76}, 790 (1949).

\bibitem{QFT3}Tomonaga, S. ``On a Relativistically Invariant Formulation of the Quantum Theory of Wave
Fields'', Prog. Theor. Phys. {\bf 1}, 27 (1946).

\bibitem{QFT4}Dyson, F. J. ``The Radiation Theories of Tomonaga, Schwinger and
Feynman'', Phys. Rev. {\bf 75}, 486 (1949).

\bibitem{QFT5}M. Gell-Mann, ``Isotopic spin and new unstable
particles,'' Phys. Rev. {\bf 92}, 833 (1953).

\bibitem{YM}Yang, C. N. and Mills, R. L. ``Conservation of Isotopic Spin and
Isotopic Gauge Invariance'',  Phys. Rev. {\bf 96}, 191 (1954).

\bibitem{standard1}Weinberg, S. ``A Model of Leptons'', Phys. Rev.
Lett. {\bf 19}, 1264 (1967).

\bibitem{standard2}Salam, A. ``Weak and Electromagnetic Interactions'',
Proc. Nobel Sym. 1968 at Lerum, Sweden, 267 (1968).

\bibitem{standard3}Glashow, S. L. ``Partial Symmetries of Weak
Interactions'', Nucl. Phys. {\bf 22}, 579 (1961).

\bibitem{standard4}Gross, D. J. and Wilczek, F. ``Ultraviolet Behavior of Non-Abelian Gauge
Theories'', Phys. Rev. Lett. {\bf 30}, 1343 (1973).

\bibitem{standard5}Politzer, H. D. ``Reliable Perturbative Results for Strong
Interactions?'', Phys. Rev. Lett. {\bf 30}, 1346 (1973).

\bibitem{nambu}Nambu, Y. ``A `superconductor' model of elementary
particles and its consequencies', talk given at a conference at
Purdue (1960); Nambu, Y. and Jona-Lasinio, G. ``A dynamical of
elementary particles based on an analogy with superconductivity I\&
II'', Phys. Rev. {\bf 122}, 345 (1961), {\it ibid} {\bf 124}, 246
(1961).

\bibitem{kmmatrix}Kobayashi, M. and Maskawa, K. ``CP violation in the renormalizable theory of
weak interactions'', Progr. Theor. Phys. {\bf 49}, 652 (1964).

\bibitem{thooft}'t Hooft, G. and Veltman, M. J. G. ``Regularization and renormalization
of gauge fields'', Nucl. Phys. B {\bf 44}, 189 (1972).

\bibitem{Nobel}Scientific Background of the Nobel Prize in Physics 2008, {\it Broken Symmtries},
available at http://nobelprize.org/

\bibitem{weinberg}Weinberg, S. {\it Dreams of a Final Theory: The
Search for the Fundamental Laws of Nature}, (Hutchinson Radius,
London, 1993).

\bibitem{LeeYang}Lee, T. D. and Yang, C. N. ``Question of parity conservation in weak
interaction'', Phys. Rev. {\bf 104}, 254 (1956).




\bibitem{accelerate1}Riess, A. G. {\it et al.} ``Observational
evidence from supernovae for an accelerating universe and a
cosmological constant'', Astronomical J. {\bf 116}, 1009 (1998).

\bibitem{accelerate2}Perlmutter, S. {\it et al.} ``Meassurements of
omega and lambda from 42 high-redshift supernovae'', Astronomical J.
{\bf 517}, 565 (1999).

\bibitem{WMAP}Komatsu, E. {\it et al.} ``Five-year Wilkinson
Microwave Anisotropy Probe observations: cosmological
interpretation'', available at http://map.gsfc.nasa.gov/

\bibitem{fermilab}Craig Group (CDF Collaboration), ``Higgs boson searches at CDF'',
eprint arXiv:0905.4267.

\bibitem{ZF1}Zhu, C. Y. and Fan, H. ``A new quantum theory of gravity in the
framework of general relativity'', eprint arXiv:0911.1401

\bibitem{ZF2}Zhu, C. Y. and Fan, H. ``Mass of gauge field'',
to appear.

\end{thebibliography}
\end{document}